\documentclass[traditabstract]{aa}
\pdfoutput=1
\usepackage{graphicx}
\usepackage{txfonts}
\usepackage{aas_macros}
\usepackage[sort]{natbib}
\usepackage{rotating}
\usepackage{subfig}
\usepackage{longtable}
\usepackage{fixltx2e}
\usepackage{amssymb}

\begin{document}

\title{Mosaiced wide-field VLBI observations of the Lockman Hole/XMM}

\author{
Enno Middelberg\inst{1}
\and Adam Deller\inst{2}
\and Ray Norris\inst{3}
\and Sotiria Fotopoulou\inst{4,5,6}
\and Mara Salvato\inst{5,6}
\and John Morgan\inst{7}
\and Walter Brisken\inst{8}
\and Dieter Lutz\inst{5}
\and Emmanouil Rovilos\inst{9}
}

\authorrunning{Middelberg et al.}

\institute{Astronomisches Institut, Ruhr-Universit\"at Bochum, Universit\"atsstr. 150, 44801 Bochum, Germany\\ \email{middelberg@astro.rub.de}\and
ASTRON, Dwingeloo, The Netherlands                                                                                                            \and
Australia Telescope National Facility, PO Box 76, Epping NSW 1710, Australia                                                                  \and
Max-Planck-Institut f\"ur Plasmaphysik, Boltzmannstrasse 2, 85478, Garching, Germany                                                          \and
Max-Planck-Institut f\"ur Extraterrestrische Physik, Giessenbachstrasse, Garching, Germany                                                    \and
Excellence Cluster ``Universe'', Boltzmannstr. 2, D-85748, Garching, Germany                                                                  \and
Curtin University of Technology, GPO BOX U1987, Perth, WA 6845, Australia                                                                     \and
National Radio Astronomy Observatory, PO Box 0, Socorro, NM, 87801, USA                                                                       \and
Department of Physics, Durham University, South Road, Durham, DH1 3LE, UK
}

\date{Received...}

\abstract{Active Galactic Nuclei (AGN) play a decisive role in galaxy
  evolution, particularly so when operating in a radiatively
  inefficient mode, where they launch powerful jets that reshape their
  surroundings. However, identifying them is difficult, since radio
  observations commonly have resolutions of between 1\,arcsec and
  10\,arcsec, which is equally sensitive to radio emission from
  star-forming activity and from AGN. Very Long Baseline
  Interferometry (VLBI) observations allow one to filter out all but
  the most compact non-thermal emission from radio survey data. The
  observational and computational demands to do this in large surveys
  have been, until recently, too high to make such undertakings
  feasible. Only the recent advent of wide-field observing techniques
  have facilitated such observations, and we here present the results
  from a survey of 217 radio sources in the Lockman Hole/XMM field. We
  describe in detail some new aspects of the calibration, including
  primary beam correction, multi-source self-calibration, and
  mosaicing. As a result, we detected 65 out of the 217 radio sources
  and were able to construct, for the first time, the source counts of
  VLBI-detected AGN. The source counts indicate that at least
  15\,\%--25\,\% of the sub-mJy radio sources are AGN-driven,
  consistent with recent findings using other AGN selection
  techniques. We have used optical, infrared and X-ray data to enhance
  our data set and to investigate the AGN hosts. We find that among
  the sources nearby enough to be resolved in the optical images,
  88\,\% (23/26) could be classified as early-type or bulge-dominated
  galaxies.  While 50\,\% of these sources are correctly represented
  by the SED of an early-type galaxy, for the rest the best fit was
  obtained with a heavily extinct starburst template. However, this is
  due to a degeneracy in the fit, as such extinction in the templates
  is mimicking early-type objects. Overall, the typical hosts of
  VLBI-detected sources are in good agreement with being early-type or
  bulge-dominated galaxies.}


\maketitle

\section{Introduction}

It has recently become clear that the formation and evolution of
galaxies is significantly influenced by the presence of active
galactic nuclei (AGN). The large amounts of radiation produced by AGN
can heat the interstellar gas in galaxies so that star formation is
slowed down (e.g., \citealt{DiMatteo2005}), but the ejecta from AGN
can also compress interstellar gas and trigger star formation (e.g.,
\citealt{Gaibler2012}). It is therefore important to determine if an
AGN is present or not, but this is a difficult
undertaking. Unambiguous identifications of AGN are difficult to make,
since the AGN can be shielded from our view at most wavelengths, and
so a non-detection does not imply that an AGN is not present.

An exception is the radio regime -- even the most dust-rich galaxies
are transparent at GHz frequencies, and so sensitive radio surveys of
large portions of the sky have become an indispensable ingredient in
the melting pot of contemporary extragalactic surveys. They provide
information about thermal and non-thermal emission and on kpc-scale
morphology, and so yield clues about the stellar and accretion
activity in galaxies. The angular resolution of surveys carried out
with compact interferometers, however, is insufficient to reliably
infer the emission processes at work if the objects are unresolved
(which most of them are with arcsec resolution), and so we are non the
wiser.

Fortunately, observations using the Very Long Baseline Interferometry
(VLBI) technique provide characteristics which help with the
identification of radio-emitting AGN. The long baselines provide
milli-arcsecond-scale resolution, which implies that the emission
required to make a detection comes from very small volumes, which in
turn implies that the brightness temperatures of these regions must be
high (of order $10^6$\,K). Such brightness temperatures generally can
only be reached by AGN or extremely bright supernovae
(\citealt{Kewley2000}), but if one observes objects at redshifts
larger than $z=0.1$ one finds that the luminosity required for a
detection exceeds that of stellar non-thermal sources, and one can be
confident that an AGN has been detected. In nearby objects,
non-thermal sources such as supernova remnants can be bright enough
for VLBI studies.

But VLBI observations traditionally target single, carefully selected
objects, because of the computational challenges inherent in imaging
larger fields (\citealt{Garrett2005,Lenc2008}) and because objects
which provide the high brightness temperatures required to make
detections are sparsely distributed on the sky. Recent technical
progress, however, have made it feasible to carry out the required
calculations to image larger fields (\citealt{Deller2007,Deller2011}),
and have facilitated bandwidth upgrades resulting in substantial
sensitivity improvements. Therefore imaging wide fields with VLBI
techniques has become feasible, and first exploratory projects have
proven to be successful (\citealt{Middelberg2011a,Morgan2011}). VLBI
observers finally have the chance to carry out surveys of large
portions of the sky.

However, the aforementioned experiments did fall short of matching
typical wide, deep radio surveys because they only used single
pointings and required sources with substantial flux density in their
fields to enable self-calibration of the data, required to reach the
full sensitivity of the observations. We therefore embarked on a
project to test the feasibility of surveying fields which are larger
than a single pointing and which do not contain suitable in-beam
calibrators. The Very Long Baseline Array (VLBA) was the instrument of
choice, since it has equal antennas, simplifying the calibration, and
the antennas are small (25\,m) resulting in comparatively large
primary fields of view.

This paper is structured as follows. Section~\ref{subsec:field}
describes the selection of the target field;
Sect.~\ref{sec:observations} describes the observations and
Sect.~\ref{sec:calibration} the calibration of the data, including
several new aspects required in wide-field VLBI
observations. Section~\ref{sec:imaging} presents the imaging and image
analysis and Sect.~\ref{sec:discussion} contains an analysis of the
results, detailing the fraction of detected sources, radio source
counts, and properties of the host
galaxies. Section~\ref{sec:conclusions} summarises these results, and
Appendix~A contains catalogues, contour plots and RGB images of the
detected sources.

\subsection{Field selection}
\label{subsec:field}

\begin{table*}
\caption{Candidate field parameters. Given are the number of sources
  found in arcsec resolution observations at 1.4\,GHz, $N_{\rm src}$,
  the field declination, quality of the multi-wavelength coverage, the
  separation to the nearest listed calibrator and the availability of
  a potential in-beam calibrator.}
\centering
\small
\begin{tabular}{lccccc}
\hline
Field name                                     & $N_{\rm src}$   & Declination &  coverage & ext. cal.    & in-beam cal.\\
\hline                                         
Lockman Hole/XMM     (\citealt{Ibar2009})      & 1450           & $+57^\circ$  & good     & 1.27$^\circ$  & possible\\ 
COSMOS               (\citealt{Schinnerer2007})& 3643           & $+2^\circ$   & superb   & 3.01$^\circ$  & no\\ 
ATLAS/CDFS           (\citealt{Norris2006a})   & 726            & $-28^\circ$  & superb   & 1.87$^\circ$  & yes\\ 
Subaru/XMM           (\citealt{Simpson2006})   & 512            & $-5^\circ$   & good     & 1.46$^\circ$  & possible\\ 
Lockman Hole/North   (\citealt{Owen2008})      & 2056           & $+59^\circ$  & poor     & 2.23$^\circ$  & no\\ 
ELAIS N2             (\citealt{Ciliegi1999})   & 305            & $+41^\circ$  & medium   & 1.49$^\circ$  & possible\\ 
ELAIS N1             (\citealt{Ciliegi1999})   & 361            & $+54^\circ$  & medium   & 2.24$^\circ$  & possible\\ 
\hline
\end{tabular}
\label{tab:fields}
\end{table*}

Even though this project aimed at further developing VLBI survey
techniques, care was taken to ensure an adequate interpretation of the
results. Hence, to maximise the output of this project a field was
selected based on (i) visibility, (ii) coverage at other wavelengths,
and (iii) availability of calibrators, inside and outside the field
(here a calibrator source is any source with an arcsec-scale flux
density of tens of mJy, but no information was available if that
source would be sufficiently compact for VLBI observations). The
candidates along with a few of their characteristics are shown in
Table~\ref{tab:fields}, ranked in descending order of suitability. The
COSMOS area is quite far away from a listed source suitable for
external calibration and there is no sufficiently strong source in the
field; ATLAS/CDFS is very far south; Subaru/XMM is far south and has
relatively shallow arcsec-resolution coverage; Lockman Hole/North does
not have good coverage at other wavelengths and was deliberately
chosen to avoid strong sources (i.e., potential calibrators); and the
ELAIS fields have too shallow arcsec-resolution coverage.

\begin{figure}[ht]
\centering
\includegraphics[width=\linewidth]{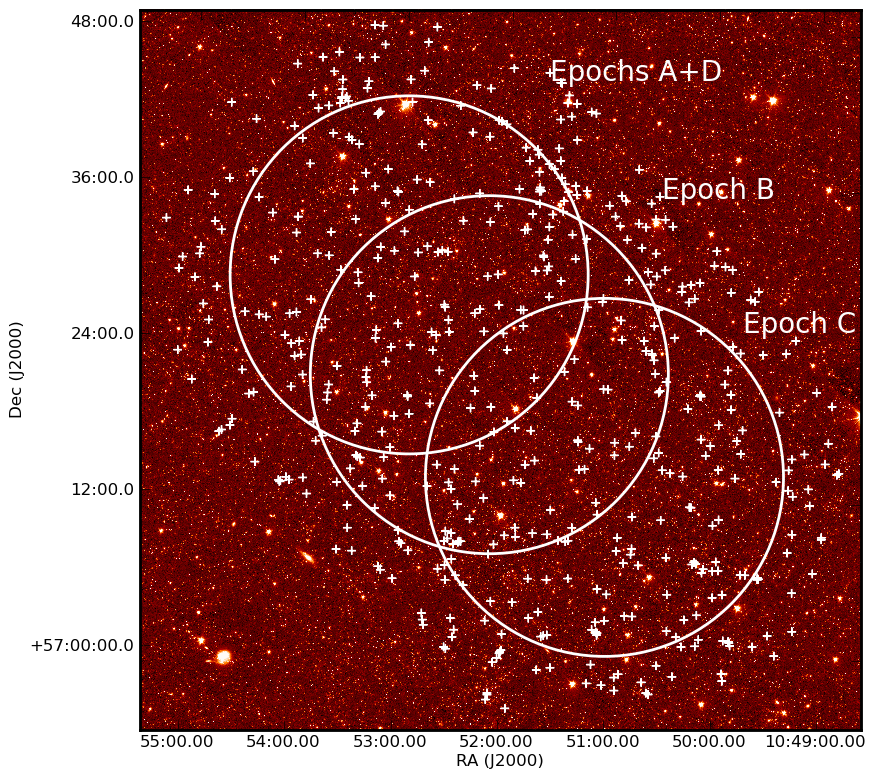}
\caption[Overview of the Lockman Hole/XMM and]{Overview of the Lockman
  Hole/XMM observations. Shown is the {\it Spitzer} 3.6\,$\mu$m image
  as colour scale in the background. The three circles denote the FWHM
  of the VLBA antennas' primary beams. The pointing coordinates have
  been taken from the VLA observations of this field reported in
  \cite{Ibar2009}. Also indicated are which pointings have been
  observed in epochs A to D. Crosses indicate radio sources from
  \cite{Ibar2009} with a flux density of 100\,$\mu$Jy or more,
    and located within 20\,arcmin of a pointing centre.}
\label{fig:bm332-overview}
\end{figure}

Therefore the Lockman Hole/XMM field was targeted, at declination
$+57^\circ$. An overview of the radio observations is shown in
Fig.~\ref{fig:bm332-overview}. Complementary data include GMRT
610\,MHz data, providing spectral indices (or limits) for all sources;
deep {\it Spitzer}/SWIRE data (\citealt{Lonsdale2003}); deep XMM data
(\citealt{Brunner2008}); and optical coverage with the Large Binocular
Telescope and the Subaru telescope
(\citealt{Barris2004,Rovilos2009,Fotopoulou2012}). Furthermore, very
deep 3.6\,$\mu$m and 4.5\,$\mu$m data from the {\it Spitzer}/SERVS
mission (\citealt{Mauduit2012}), and very deep 250\,$\mu$m-500\,$\mu$m
data from the HERMES will be available soon. These data are crucial
for selecting galaxies with matching properties and to disentangle the
contributions of AGN to the bolometric luminosity.

\begin{figure}[ht]
\centering
\includegraphics[width=\linewidth]{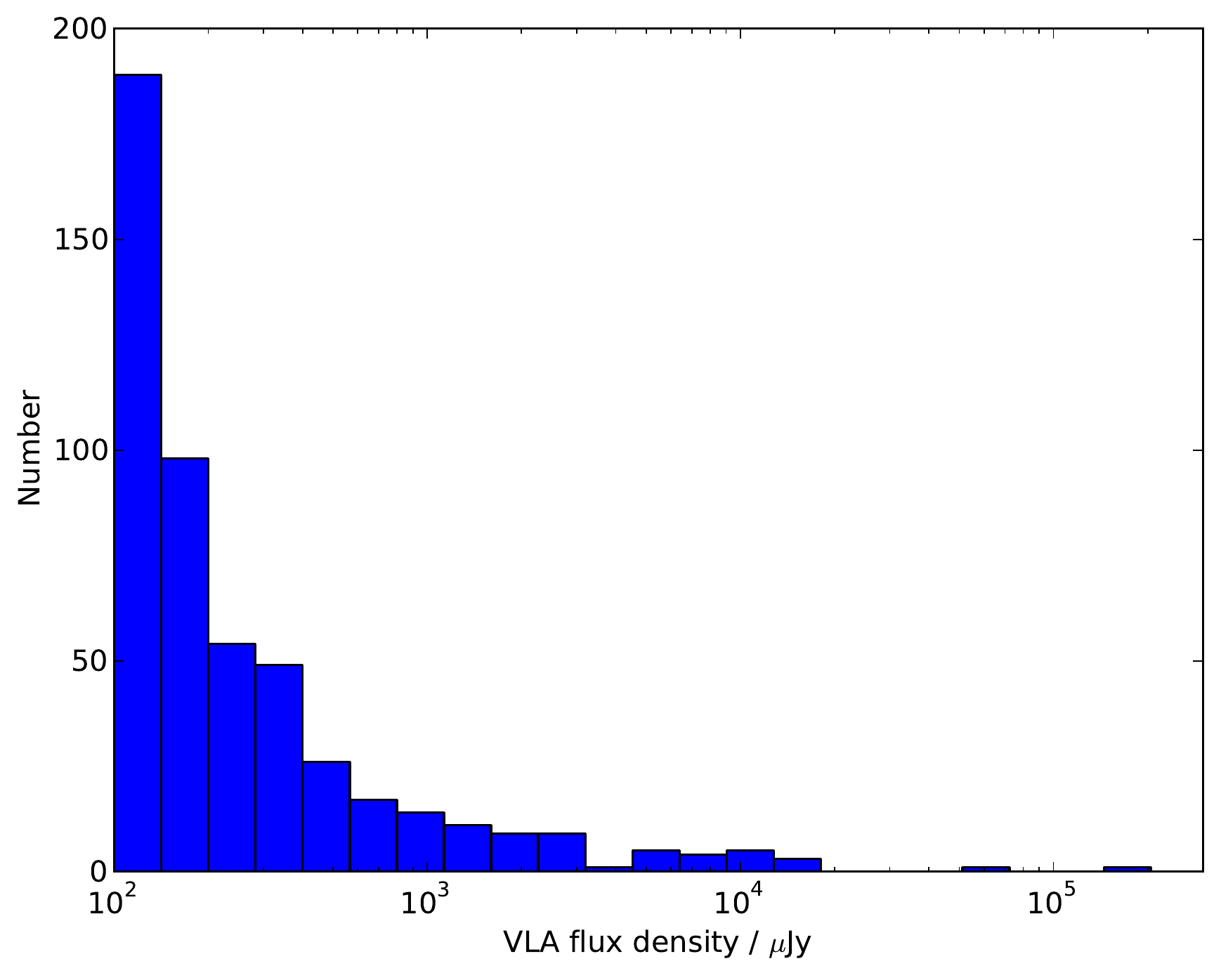}
\caption{Distribution of VLA source flux densities in bins with edges
  at 100\,$\mu$Jy$\times \sqrt{2}^N$, $N=0,1,2,...$}
\label{fig:fluxes}
\end{figure}

Only sources with integrated flux densities of more than 100\,$\mu$Jy
were targeted in our observations, because it was expected that the
noise in the combined VLBA images would reach around
20\,$\mu$Jy\,beam$^{-1}$. Of the input sample, 496 sources met this
criterion and were located within 20\,arcmin of any of the pointing
centres. However, after calibration only 217 were found to be in
principle detectable, because their VLA flux densities exceeded 6
times the local noise level (see Sect.~\ref{sec:discussion} for
details). An overview of the 3 VLBA pointings, along with the location
of target sources, is shown in Fig.~\ref{fig:bm332-overview}, and a
histogram of the source flux densities is shown in
Fig.~\ref{fig:fluxes}.

\section{Observations}
\label{sec:observations}

The observations for this project were carried out with the VLBA at
1.382\,GHz on 3 June (epoch A), 4 July (epoch B), 16 July (epoch C)
and 3 September (epoch D) of 2010. A recording bitrate of 512\,Mbps
was used, resulting in a bandwidth of 64\,MHz in two parallel-hand
polarisations, yielding a nominal sensitivity of around
24\,$\mu$Jy\,beam$^{-1}$ towards the pointing centre. The
fringe-finder 4C\,39.25 was observed every 2.5\,h for data consistency
checks, and the fields were observed for 4.5\,min, followed by a
1\,min observation of the phase-referencing source
NVSS\,J105837+562811. The elapsed time of the observations was 12\,h
per epoch, resulting in a total on-source time of around 326
baseline-hours per epoch. The VLBA observational status
summary\footnote{http://www.vlba.nrao.edu/astro/obstatus/current/obssum.html}
predicts a baseline sensitivity of 3.3\,mJy in a 2\,min observations
with a recording rate of 256\,Mbit, scaling to an image sensitivity of

\begin{equation}
\sigma=\frac{3.3\,{\rm mJy}}{\sqrt{\frac{512\,{\rm Mbps}}{256\,{\rm Mbps}}}\sqrt{\frac{60\,{\rm min}}{2\,{\rm min}}}\sqrt{326}}=23.6\,\mu{\rm Jy}
\end{equation}

A pre-production version of VLBA-DiFX was used to correlate the data,
using the new multi-field-centre mode. A memory leak in an associated
program (difx2fits, which converts the data from the internal DiFX
format to FITS-IDI) caused the loss of the first half of the data of
the epoch A observations, which was overwritten with the second
half. This was only discovered after the raw data had already been
deleted, and a recorrelation was not possible. A reobservation of
epoch A was requested and granted (epoch D), so that one of the three
pointings in the Lockman Hole/XMM region now has deeper coverage than
the other two. Correlation of the data resulted in approximately 320
data sets per epoch, each with a spectral resolution of 500\,kHz and
4\,s integrations. Since the field of view of each data set is centred
on the known position of a VLA-detected source the field of view to be
imaged is relatively small, and bandwidth and time smearing are not an
issue (see \citealt{Morgan2011} for a detailed description of these
effects). The data volume was around 1\,GB per source, per epoch.

\section{Calibration}
\label{sec:calibration}

\subsection{Standard steps}

The calibration followed standard procedures used in phase-referenced
VLBI observations, using the Astronomical Image Processing System,
AIPS\footnote{http://www.aips.nrao.edu}, with its Python interface,
ParselTongue\footnote{http://www.jive.nl/dokuwiki/doku.php?id=parseltongue:parseltongue}
(\citealt{Kettenis2006}). Amplitude calibration was carried out using
$T_{\rm sys}$ measurements and known gain curves. Fringe-fitting was
carried out on the phase calibrator directly, to compensate for
residual delays, without fringe-fitting the fringe finder (4C39.25)
first, as would normally be done (a bug in the VLBA electronics causes
unpredictable delay jumps in observations using a recording rate of
512\,Mbps and 8\,MHz IFs, and fringe finder observations are too
infrequent to keep track of these jumps). However, the phase
calibrator was bright enough to be detected separately in each IF
channel during the 1\,min scans, and so there was no need to use the
fringe finders.

\subsection{Multi-field self-calibration}

After fringe fitting the delay and phase corrections were copied to
the individual data files. In phase-referenced observations such as
this the SNR typically is limited by ionospheric and atmospheric
turbulence between calibrator scans, and therefore purely
phase-referenced images have reduced coherence. When the target is
sufficiently strong, self-calibration can be used to improve
coherence, but here the targets were too faint. However, the combined
flux density of the strongest few targets would be sufficient to carry
out phase self-calibration. This has already been pointed out by
\cite{Garrett2004}, but had never been demonstrated in practice. We
describe here a simple procedure which implements multi-field
self-calibration.

\begin{itemize}
\item Phase-referenced images were made of all targets, and the images
  were searched for emission. Typically around 30 sources were found
  with a SNR of more than 7, with the brightest reaching an SNR of
  almost 100. The data are later combined using weights which are
  proportional to the square of the SNR, hence a source with SNR=10
  contributes only 1/100 of the combined signal compared to a source
  with SNR=100. Therefore only the brightest 10 sources or so were
  used in the following steps.

\item The individual data sets were divided by the CLEAN model
  obtained during imaging. This results in data sets each showing a
  1\,Jy point source in the field centre. It is worth noting that in
  this process the data weights are modified by the inverse square of
  the amplitude adjustment, and so this procedure conveniently takes
  care of proper weighting when the data are combined.

\item The source coordinates in the data set headers were set to the
  same value. Subsequently, the data were concatenated into a single
  data set. For each baseline, time, and frequency the combined data
  now contains multiple measurements of a point source.

\item Self-calibration was used with a 1\,Jy point source model to
  improve the coherence of the combined data set. The phase
  corrections derived in this process were then copied to all original
  data sets, and improved images could be made.
\end{itemize}

\begin{figure*}
\center
\includegraphics[width=0.48\linewidth]{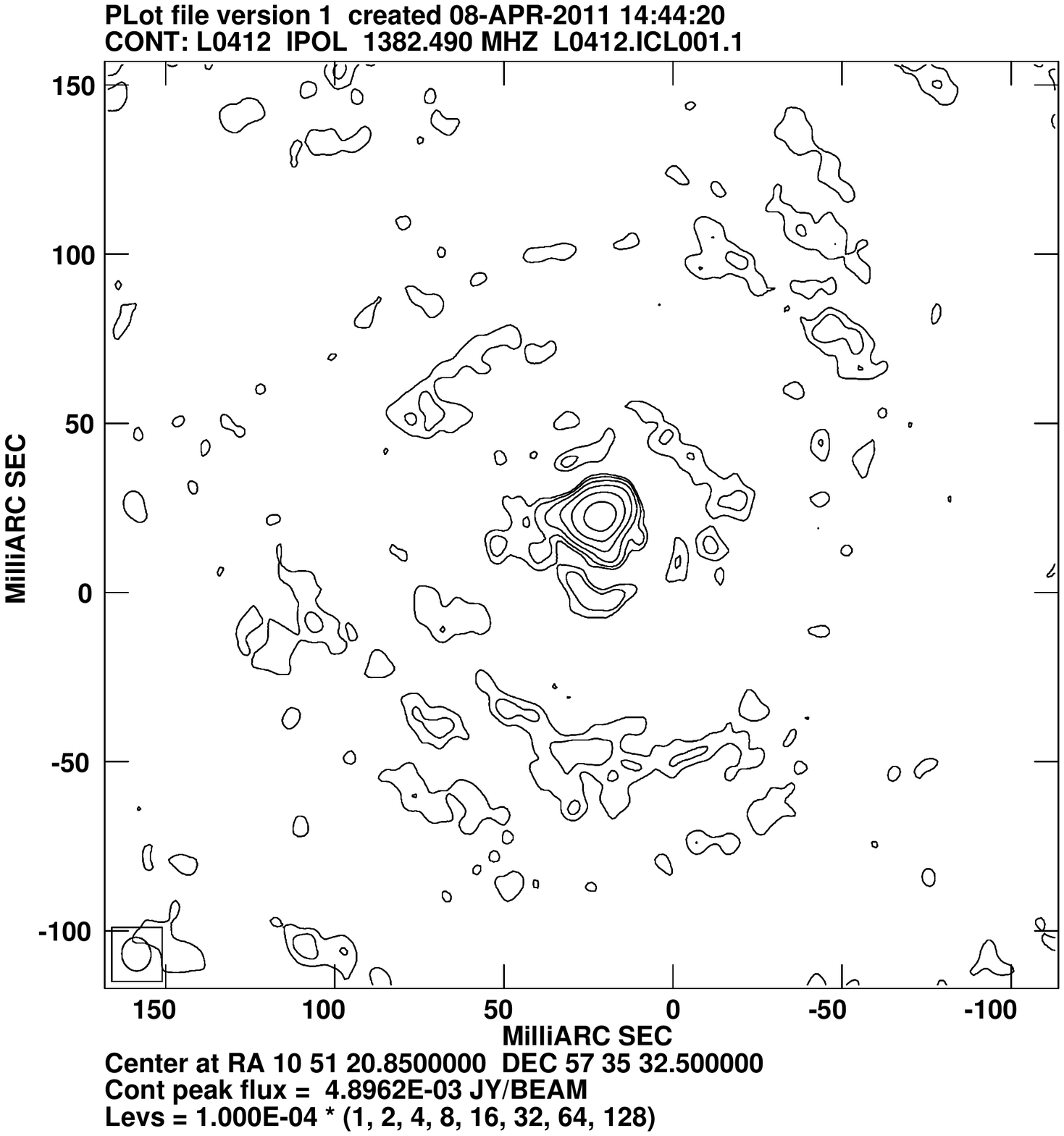}
\includegraphics[width=0.48\linewidth]{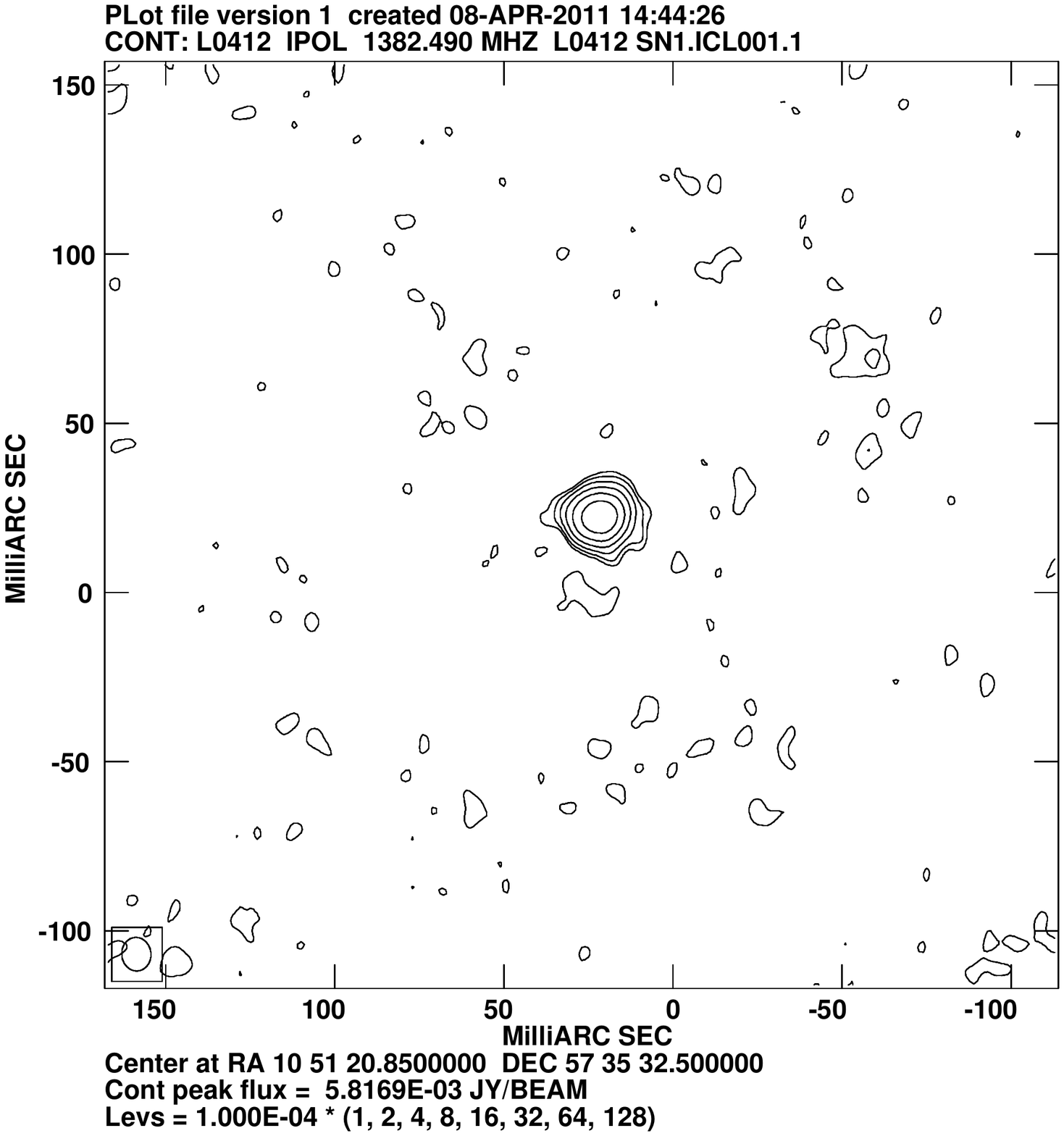}
\caption[Illustration of the effect of multi-source
  self-calibration]{Contour plot of the radio source L0412 before
  (left) and after (right) multi-source self-calibration has been
  applied. Contours are drawn at 0.1\,mJy$\times$(1, 2, 4, ...). The
  signal-to-noise ratio was improved from 72 to 115. The peak flux
  density has increased significantly, the noise has dropped, and
  image artifacts are much reduced. These effects will be more
  pronounced when the initial calibration is not as good as in this
  case, for example, when a field at low elevation is observed.}
\label{fig:multi-source-selfcal}
\end{figure*}

The improvement attainable with this technique is illustrated in
Fig.~\ref{fig:multi-source-selfcal}.

\subsection{Mosaicing}

After multi-field self-calibration the calibration of each individual
epoch was considered to be complete, and the data from the various
epochs needed to be combined to reach maximum sensitivity. Two effects
needed to be calibrated before combining the data: a potential
systematic astrometric offset, and the primary beam attenuation.

Astrometric offsets between the epochs arise from a number of
effects. First, unmodelled propagation delays (e.g., due to
tropospheric water vapour, or the ionosphere) will vary between epochs
and lead to a different residual phase error at the target field.
Secondly, any component of these unmodelled propagation effects which
is constant over long timescales will introduce further errors which
differ between epochs, since the separation between phase calibrator
and target field changes between epochs.  And third, the in-beam
calibrators will all have modelled positions and structure which differ
from their true properties due to the limited S/N during the initial
image reconstruction and differential phase calibration effects across
the target field.  The necessary usage of different sets of in-beam
calibrators for multi-source selfcal in the different pointings will
introduce different, small errors between the epochs.

To measure and calibrate a potential systematic offset a set of 18
sources was selected from epoch B which were also found in epochs A/D
and C (but not all sources were present in all data epochs). These
sources were imaged after multi-field self-calibration and the images
were searched for the brightest pixels. The position differences
relative to epoch B were calculated and the median used as the best
approximation of a systematic offset between epochs A/C/D and B. The
median offsets were found to be less than 1\,mas in all cases, while
the resolution of the images was around $10\,{\rm mas}\times8\,{\rm
  mas}$. Correction of the offsets before the data were combined
resulted in an increase of the peak flux densities of the calibrator
sources of around 1\,\%, compared to a trial run in which the data
were combined without offset correction. The mean offsets of epochs
A/C/D relative to B before and after correction are listed in
Tab.~\ref{tab:offsets}.

\begin{table}
\caption{Average position offsets relative to epoch B before and after
  position correction.}
\center
\begin{tabular}{lrrrr}
\hline
         & \multicolumn{2}{c}{before correction} & \multicolumn{2}{c}{after correction}\\

         & $\Delta$RA   & $\Delta$Dec & $\Delta$RA   & $\Delta$Dec\\
         & mas          & mas         & mas          & mas\\
\hline
B--A     & $-0.17$      & $ 1.40$     & $0.19$       & $-0.20$\\
B--C     & $-0.46$      & $-0.67$     & $0.00$       & $-0.08$\\
B--D     & $-0.17$      & $-1.00$     & $0.00$       & $ 0.00$\\
\hline
\end{tabular}
\label{tab:offsets}
\end{table}

\subsection{Primary beam corrections}

The primary beam correction scheme used here has been described in
detail in \cite{Middelberg2011a}, who noted that the accuracy of the
scheme was unknown, and they estimated its errors to be of the order
of 10\,\%. We present here the results of an experiment to improve the
quality of the corrections and to assess their accuracy.

On 21 September 2011, a pattern of pointing positions around 3C\,84
was observed with the VLBA at 1.382\,GHz to measure the primary beam
response of the antennas. The frequency setup was identical to the
VLBA observations of the CDFS by \cite{Middelberg2011a} and the
Lockman Hole/XMM.

\begin{figure}
\center
\includegraphics[width=\linewidth]{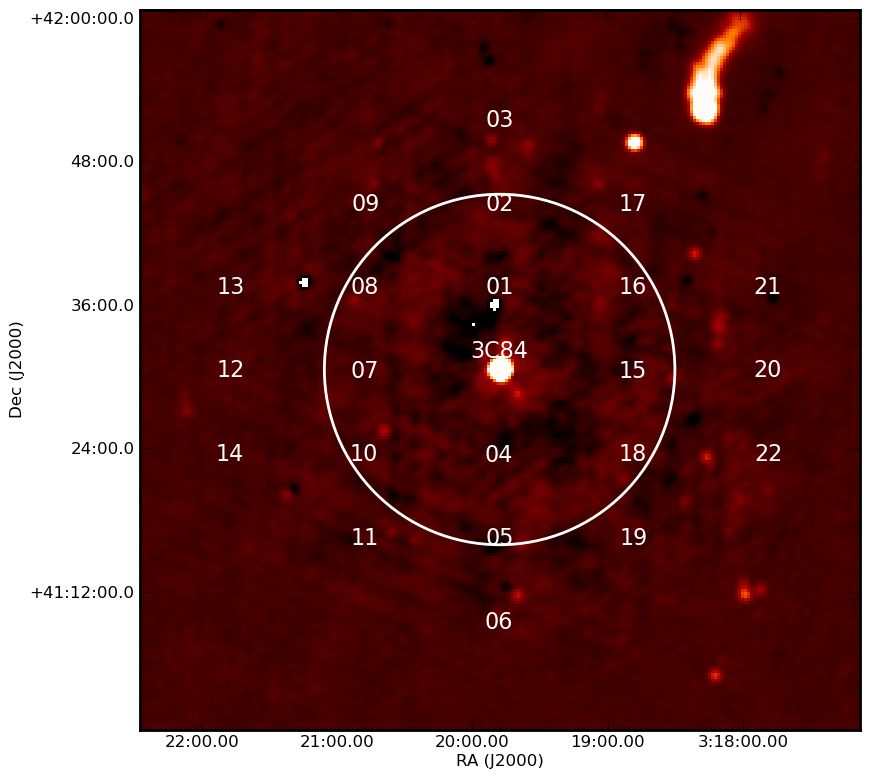}
\caption[Pointing pattern used for primary beam
  measurements]{Locations of the antenna pointing positions used to
  test the primary beam attenuation of the VLBA antennas, shown on top
  of an NVSS postage stamp. 3C\,84, in the image centre, has always
  been used as the field centre in the correlation, so that it was
  seen by the antennas through the various parts of the primary beam
  power pattern, the FWHM of which is indicated by the circle.}
\label{fig:to004-pointings}
\end{figure}

The pattern of pointings used for this experiment is shown in
Fig.~\ref{fig:to004-pointings}. The antennas were pointed towards each
position for 1\,min, except for the central pointing which was
observed five times for 1.5\,min for amplitude reference. Correlation
has been performed using the true position of 3C\,84 for all
pointings, hence the amplitudes were only affected by primary beam
effects. Initial amplitude calibration was carried out using $T_{\rm
  sys}$ measurements and known antenna gain curves. Fringe-fitting
using the data from all pointings was used to correct phase and delay
errors. The data from the central 1.5\,min scan of 3C\,84 was then
imaged to obtain an approximate Stokes I model for further
calibration. Amplitude self-calibration was subsequently carried out
using this model to correct for residual amplitude variations and in
particular to remove amplitude differences between RCP and LCP
(assuming 3C\,84 is circularly unpolarised, which, according to
\citealt{Homan2004}, is the case at frequencies below 15\,GHz). These
corrections were then applied to the data from all pointing positions,
so that amplitude variations should exclusively be caused by primary
beam attenuation. Images were then made of all pointings and the flux
densities of the brightest image pixels were extracted. Note that for
an image of the pointing centred on the true position of 3C\,84 only
the centre scan was used to eliminate effects arising from a better
$(u,v)$ coverage for this pointing.

\begin{figure*}
\center
\includegraphics[width=0.48\linewidth]{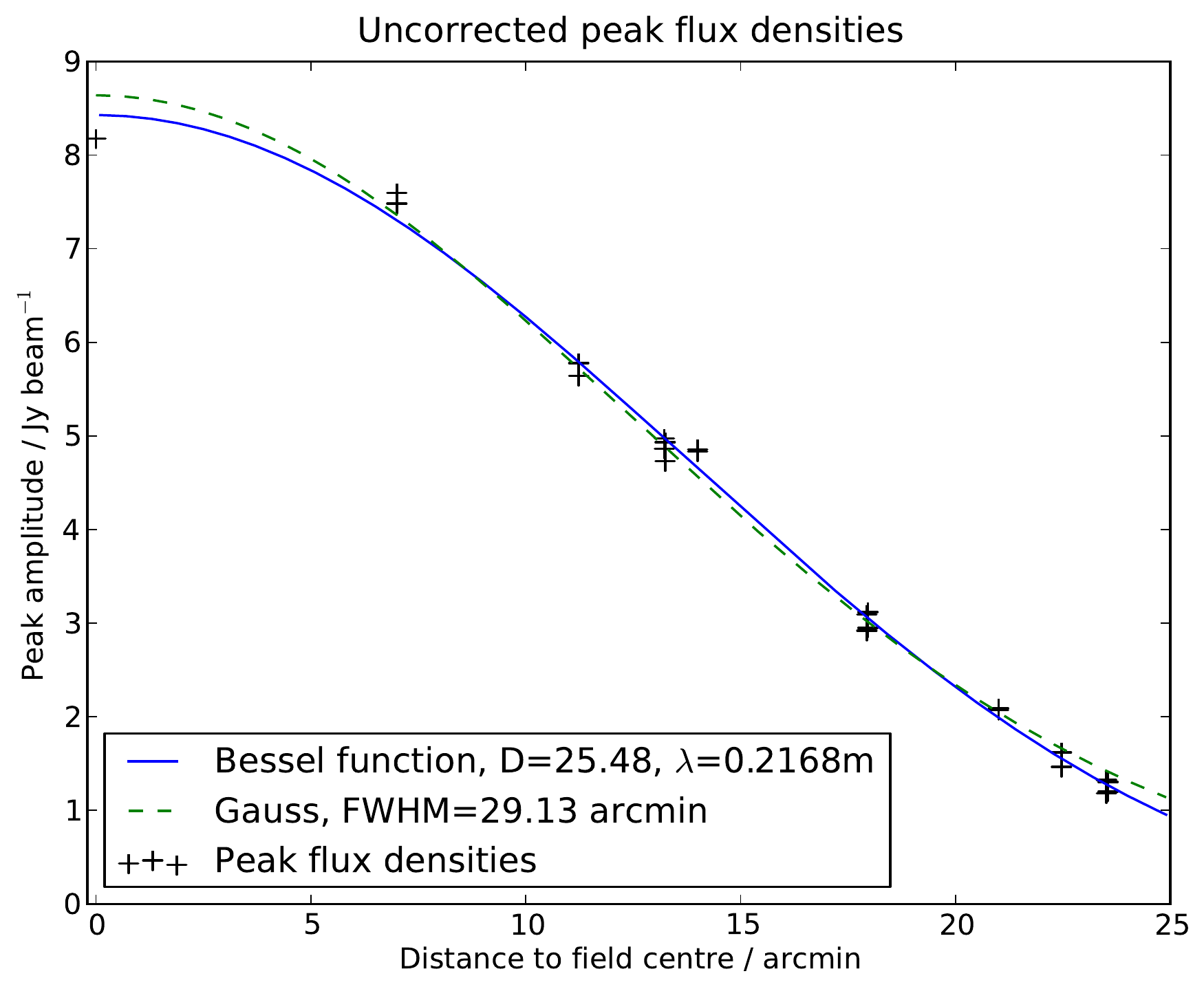}
\includegraphics[width=0.49\linewidth]{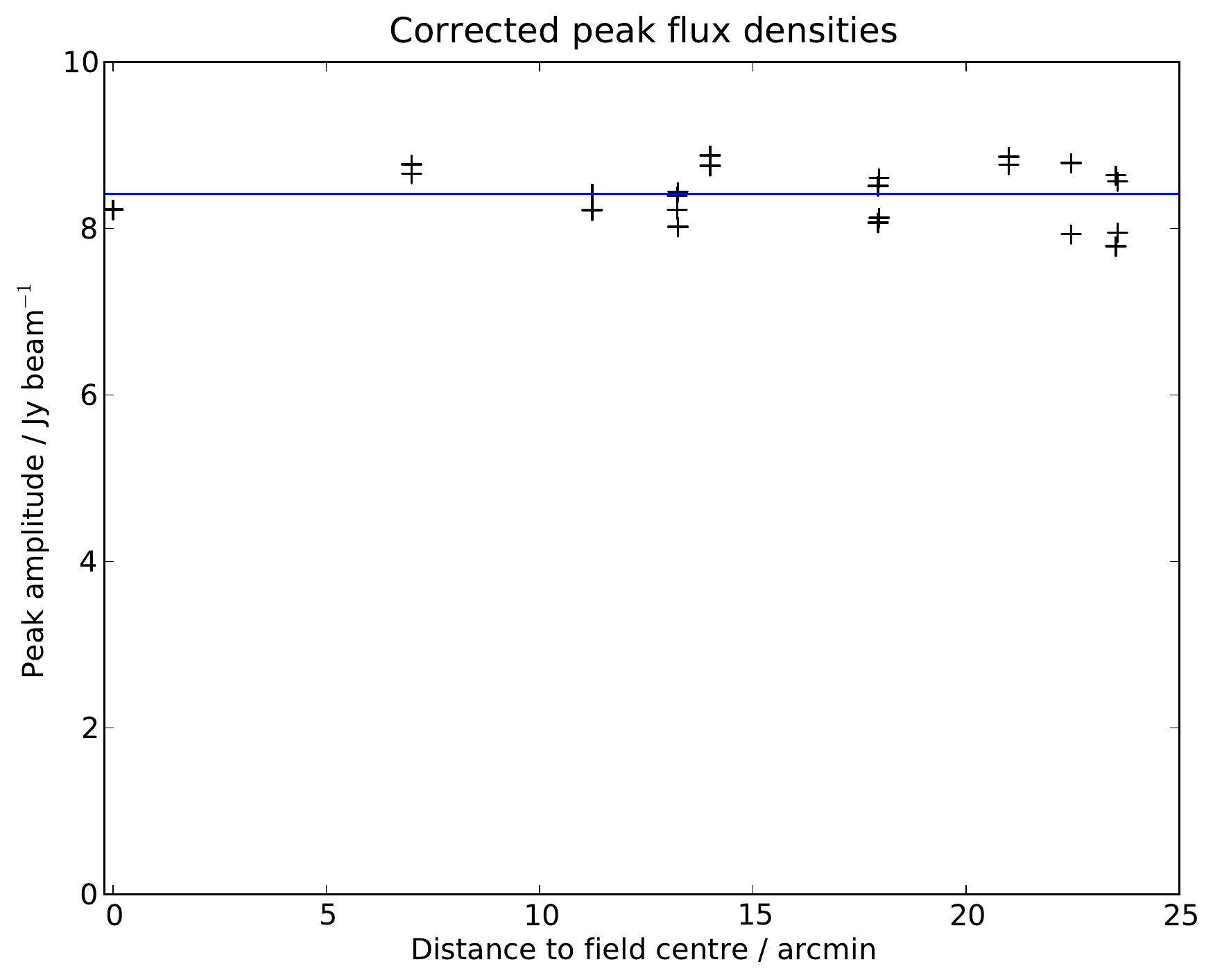}
\caption{Peak flux densities of 3C\,84 before and after primary
  beam correction. {\it Left panel:} Peak amplitudes measured using
  the pointings in Fig.~\ref{fig:to004-pointings} as a function of
  distance to the true source position. Also shown are two fits to the
  data, one using a Gaussian (dashed line) and one using an Airy disk
  (solid line). The Airy disk performs slightly better. {\it Right
    panel:} Peak amplitudes after applying the primary beam
  corrections using an Airy disk model fitted to the data in the left
  panel. The standard deviation of the measurements, normalised to
  their mean (shown as the solid line), is 0.039.}
\label{fig:peaks}
\end{figure*}

The uncorrected peak flux densities are shown in Fig.~\ref{fig:peaks}
as a function of distance to the 3C\,84 position. Two models have been
used to reproduce these measurements, a Gaussian and an Airy disk. A
Gaussian can be used effectively to model the inner portion of an
antenna's power pattern, since it is a good approximation and has a
convenient and familiar parameter to represent its properties, which
is the full width at half maximum (FWHM). However, it does a poor job
near and beyond the first null of the antenna power pattern, and
increasingly deviates from this simple model. A relatively obvious
starting point for an improved model is the Airy disk, which
essentially is a Bessel function. To first order an antenna can be
treated as a uniformly illuminated disk, the Fraunhofer diffraction
pattern of which is given by

\begin{equation}
I(\theta)=I_0 \times\left(\frac{2J_1(\frac{\pi}{\lambda} D \sin\theta)}{\frac{\pi}{\lambda} D \sin\theta}\right)^2
\end{equation}

where $J_1(x)$ is the Bessel function of order one, $\lambda$ is the
observing wavelength, $D$ is the diameter of the aperture, and
$\theta$ is the direction in which the intensity is to be
calculated. Since only deviations from the on-axis sensitivity are to
be calculated, $I_0$ can be set to one and ignored. Both the Gaussian
and the Airy disk model have been fitted to the data; in the case of
the Gaussian the FWHM has been allowed to vary and in the case of the
Airy disk model the antenna diameter was allowed to vary (and the
wavelength was kept constant). Since the antenna feed horns
potentially only illuminate a part of the dish or, conversely, can
illuminate a solid angle which extends beyond it, the effective dish
diameter can differ slightly from the geometric aperture. It therefore
is more appropriate to solve for the effective dish diameter rather
than for the observing wavelength. For the observations presented here
the centre observing wavelength was $\lambda=0.21685$\,m.

The best-fitting FWHM of the Gaussian model was found to be
29.13\,arcmin, and the best-fitting antenna diameter for the Airy disk
model was found to be $D=25.47$\,m. The slightly larger antenna
diameter is likely to be caused by the shadowing of the main dish by
the 4\,m secondary reflector. This situation leads to the outer radii
to contribute more than would be the case for uniform illumination. A
diameter of 25.47\,m was subsequently used to correct for the primary
beam attenuation, using the Airy disk model.

Since the receivers of the VLBA antennas are not aligned with the
optical axis of the system, but are mounted off-axis, the beam
patterns for the two orthogonal polarisations are offset on the sky by
approximately 5\,\% of the primary beam FWHM (see \citealt{Uson2008}
and references therein for a detailed assessment of the effect at the
VLA, which has similar antennas). This effect is called beam squint,
and it needs to be taken care of in particular when measuring circular
polarisation. In total intensity, however, the impact of beam squint
is rather small, since RCP and LCP tend to average out. Nevertheless
it was incorporated into the correction scheme here, using
measurements of the squint carried out by R. Craig Walker
(priv. comm., Tab.~\ref{tab:squint}). These measurements were carried
out at 1438\,MHz, a few percent above the observing frequency used
here. But beam squint is expected to be a fixed fraction of the
primary beam width, and therefore the squint measurements were scaled
linearly to the centre frequencies of the IFs used in our
observations.

\begin{table}
\caption{Beam squint parameters at 1.438\,GHz for the VLBA antennas
  (C. Walker, priv. comm.) The squint is angular separation of the RCP
  and LCP beams on the sky. The antennas are pointed towards the
  mid-point of the line connecting the two beams.}
\center
\begin{tabular}{lcc}
\hline
Antenna & Squint(R-L) Az & Squint(R-L) El\\
        & arcmin         & arcmin\\
\hline
St. Croix     & $-1.53$ &  $-0.56$\\
Hancock       & $-1.35$ &  $-0.67$\\
North Liberty & $-1.56$ &  $-0.62$\\
Fort Davis    & $-1.56$ &  $-0.58$\\
Los Alamos    & $-1.62$ &  $-0.67$\\
Pie Town      & $-1.59$ &  $-0.63$\\
Kitt Peak     & $-1.61$ &  $-0.59$\\
Owens Valley  & $-1.76$ &  $-0.95$\\
Brewster      & $-1.44$ &  $-0.50$\\
Mauna Kea     & $-1.56$ &  $-0.60$\\
\hline
\end{tabular}
\label{tab:squint}
\end{table}

\begin{figure*}
\center
\includegraphics[width=\linewidth]{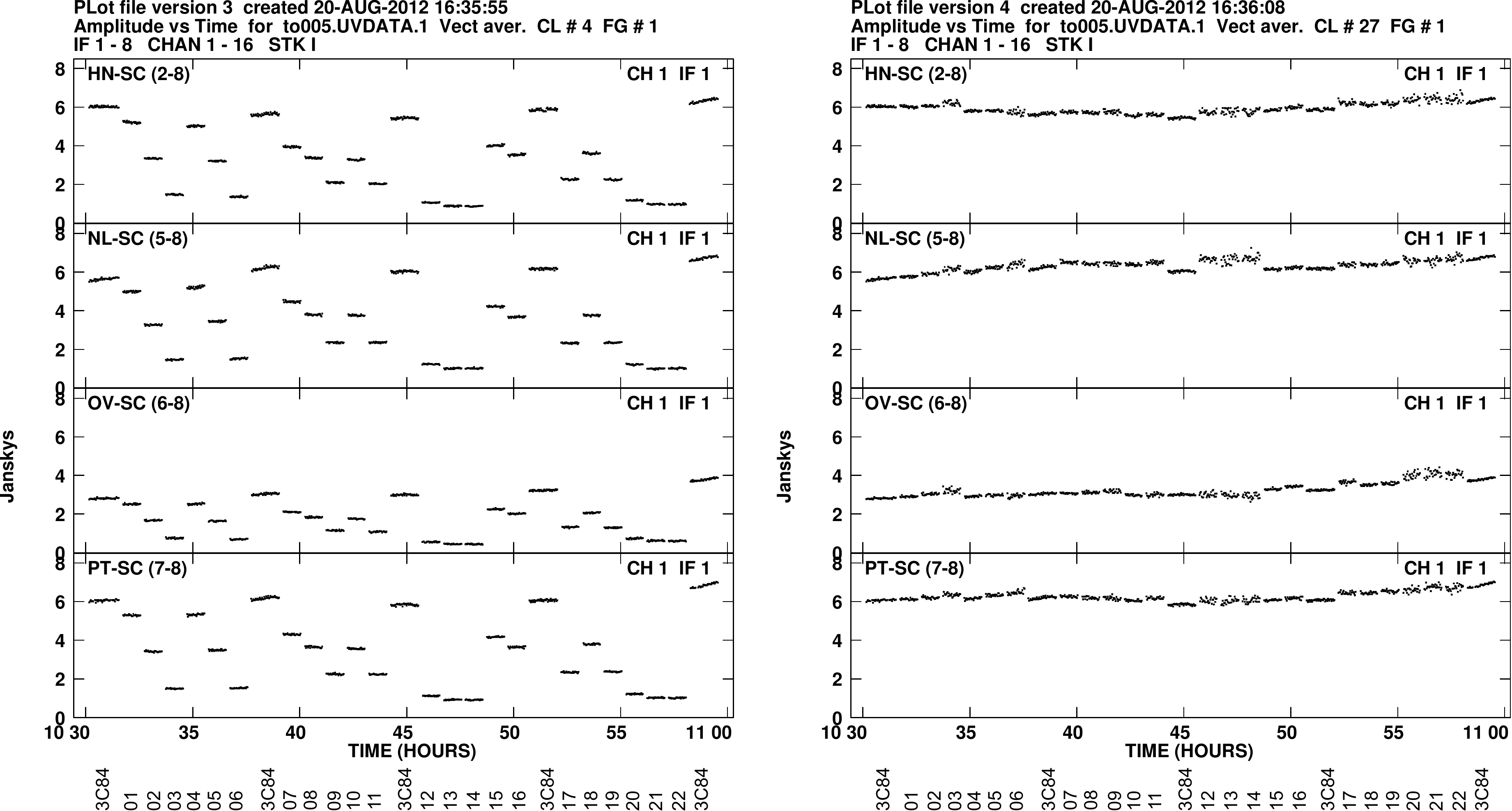}
\caption{Visibility amplitudes before and after primary beam
  correction. {\it Left panel:} Visibility amplitudes from all
  pointings on a few selected baselines before primary beam
  correction. The pointing centres are indicated at the bottom of the
  diagram. Shown is total intensity after averaging all frequencies
  and the two polarisations. {\it Right panel:} The same visibilities
  after primary beam correction. Whilst there are only small steps and
  discontinuities in the corrected visibilities (indicating that the
  correction scheme is working well), there are longer-term trends in
  the amplitudes, which may be attributed to the substantial structure
  of 3C\,84, which was not modelled due to the poor $(u,v)$ coverage
  of the data. The larger scatter during some pointings is a
  consequence of the reduced sensitivity towards directions well away
  from the pointing centre. }
\label{fig:visibilities}
\end{figure*}

\begin{figure*}
\center
\includegraphics[width=\linewidth]{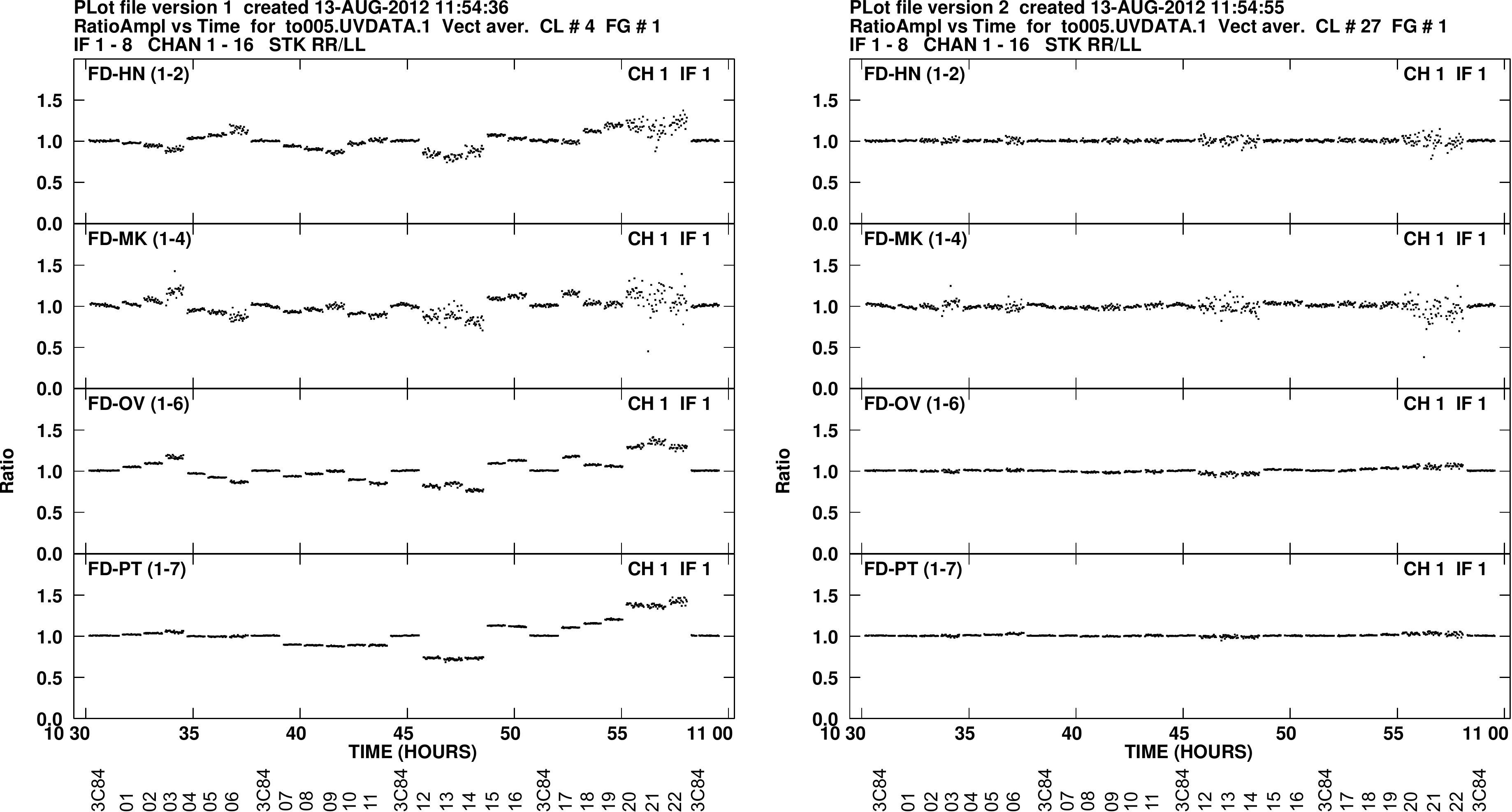}
\caption[Illustration of the effects of beam squint and its
  correction]{Illustration of the effects of beam squint and its
  correction. {\it Left panel:} The ratio of the RCP and LCP
  amplitudes on a few selected baselines before primary beam
  correction. The pointing centres are indicated at the bottom of the
  diagram. Deviations from unity are caused by the RCP and LCP beams
  pointed at different positions. {\it Right panel:} The same
  visibilities after primary beam correction. The differences have
  essentially be eliminated.}
\label{fig:ratios}
\end{figure*}

The results of the improved primary beam correction scheme can be
summarised as follows. After applying the corrections to the data, the
quality of the corrections was measured as the standard deviation of
the corrected peak flux densities, normalised to the average peak flux
density of the images from all pointings. This value was found to be
0.039, which is better than estimated by \cite{Middelberg2011a}, and
comparable to the typical amplitude calibration error assigned to VLBI
observations (Figs.~\ref{fig:peaks} and \ref{fig:visibilities}). The
squint correction can be illustrated by plotting the ratio of the RCP
and LCP amplitudes (Fig.~\ref{fig:ratios}). Overall the measurements
indicate that the primary beam correction scheme performs well and
does not introduce significant systematic errors.

\subsection{Amplitude consistency between epochs}

\begin{figure*}
\center
\includegraphics[width=0.49\linewidth]{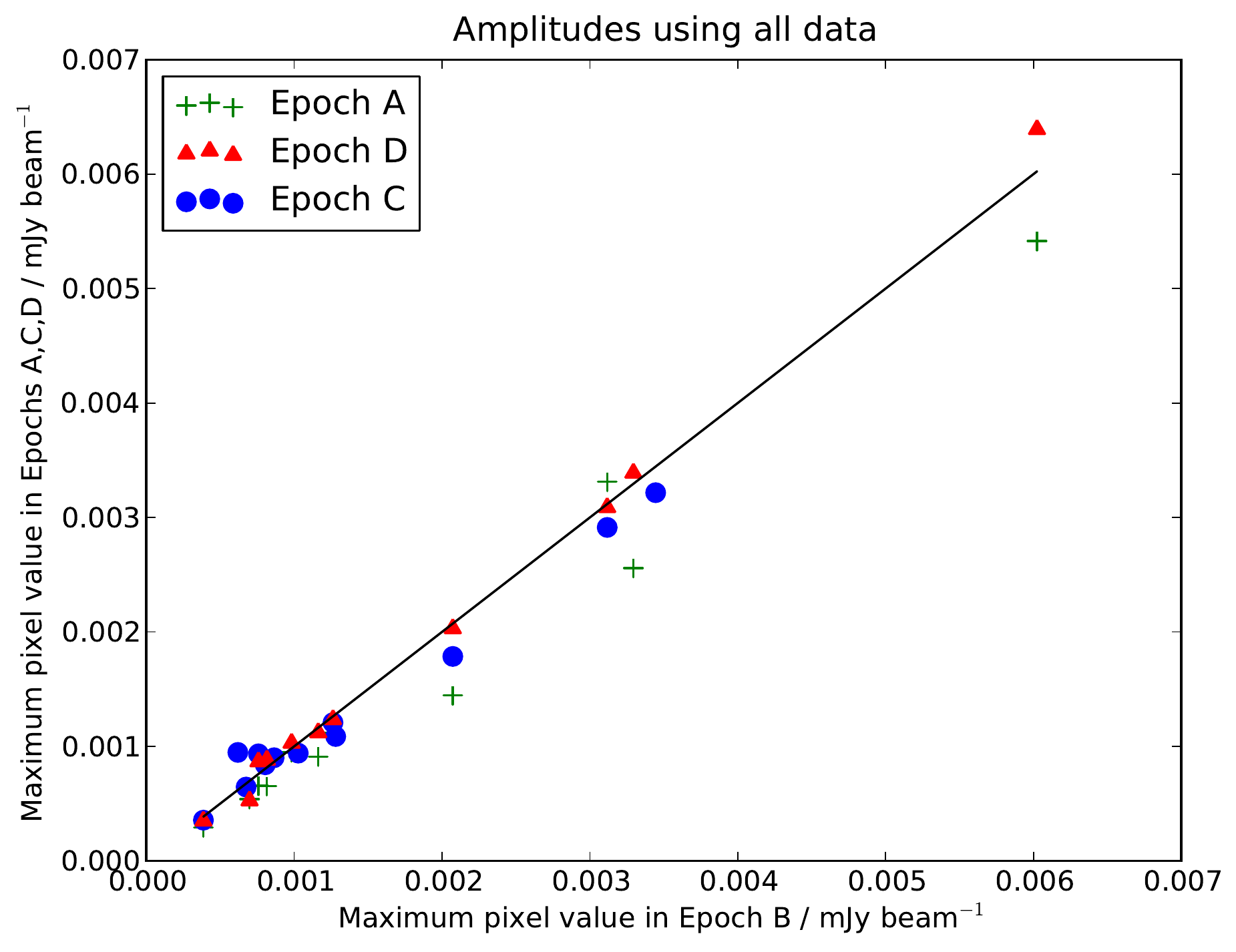}
\includegraphics[width=0.49\linewidth]{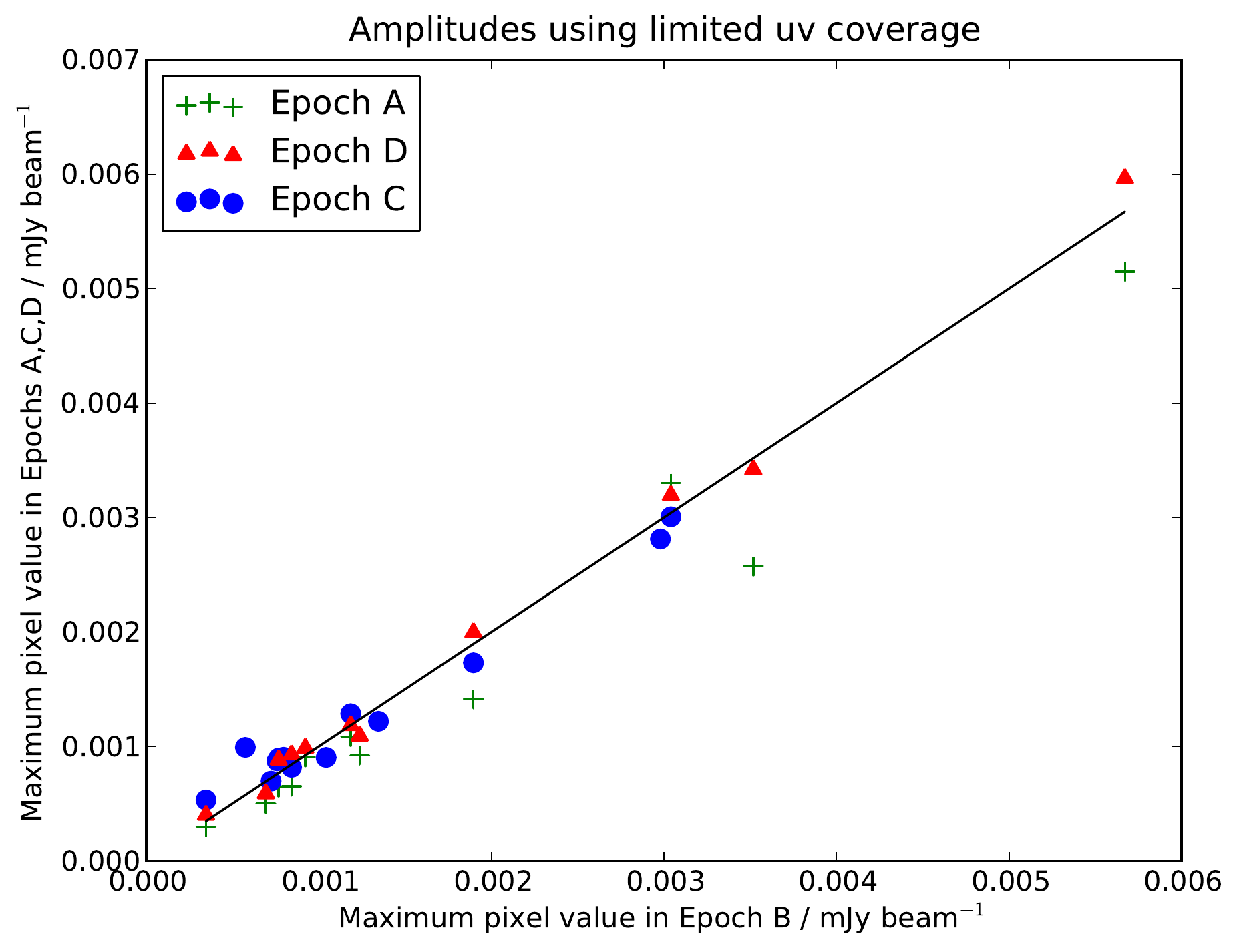}
\caption[Illustration of the amplitude consistency between
  epochs.]{Illustration of the amplitude consistency between
  epochs. {\it Left panel:} amplitudes of a sample of bright sources
  in epochs A/C/D as a function of the amplitude in epoch B. {\it
    Right panel:} the same diagram but amplitudes have been extracted
  from images with the same limited $(u,v)$ coverage.}
\label{fig:amps}
\end{figure*}

The amplitude errors can be estimated by imaging the data of a few
bright sources from each epoch separately, and comparing the
amplitudes between epochs. While this approach is susceptible to
variability, however, selecting a sufficient number of sources (we use
18) is likely to be a robust estimator for the average amplitude
error. Of this sample, 7 sources have been observed in two epochs only
(e.g., in the overlap region between A/B, or B/C), 5 have been
observed in three epochs (e.g., in A/D/B) and 6 in all four
epochs. All sources were observed in epoch B (the one between A/D and
C, see Fig.~\ref{fig:bm332-overview}) to maximise the overlap. The
brightest image pixel was extracted as a measure for the
amplitude. All amplitudes from epochs A/C/D were then plotted against
the amplitudes measured in epoch B (Fig.~\ref{fig:amps}). The
consistency of the amplitude calibration can be determined by
calculating the median of the ratios of the amplitudes between epochs,
and the standard deviation is a measure of the uncertainty when
measuring individual flux densities (including the effects of
variability). The median ratios of flux densities in the pairs A/B,
C/B, and D/B were found to be 0.799, 0.955, and 0.990, respectively,
and the standard deviations (i.e., errors for single measurements)
were 0.100, 0.129, and 0.084. Epochs B/C/D therefore appear to be
consistent within 1\,$\sigma$, whereas epoch A deviates by around
20\,\%. One potential cause of this is the limited $(u,v)$ coverage
during epoch A (where half the data were lost post-correlation due to
a software bug). The exercise was therefore repeated by limiting the
$(u,v)$ coverage in all epochs to the same extent, but the amplitude
ratios and standard deviations were found not to have changed by
much: the medians were determined for A/B, C/B, and D/B as 0.838,
0.989, 1.051, and the standard deviations 0.137, 0.143, and 0.078,
respectively. Hence epoch A appears to yield consistently lower
amplitudes by around 15\,\% to 20\,\%. However, epoch A contributed
only 6\,h of observing time out of a total of 42\,h, therefore its
contribution to the final sensitivity is only around 8\,\%, and the
final images will only marginally be affected by this systematic
error.

From this exercise one can conclude that the scatter of amplitudes
between epochs arising from amplitude calibration errors, combined
with potential source variability, is of order 10\,\%. This margin was
used later in source extraction to determine the errors of the
integrated flux densities.

\section{Imaging and image analysis}
\label{sec:imaging}

Three types of images were made: naturally-weighted, untapered images,
naturally-weighted images using a 10\,M$\lambda$ taper, and
uniformly-weighted, untapered images. Natural weighting was used for
source detection, whereas uniform weighting was used to measure
integrated flux densities.

\begin{figure}[ht]
\centering
\includegraphics[width=\linewidth]{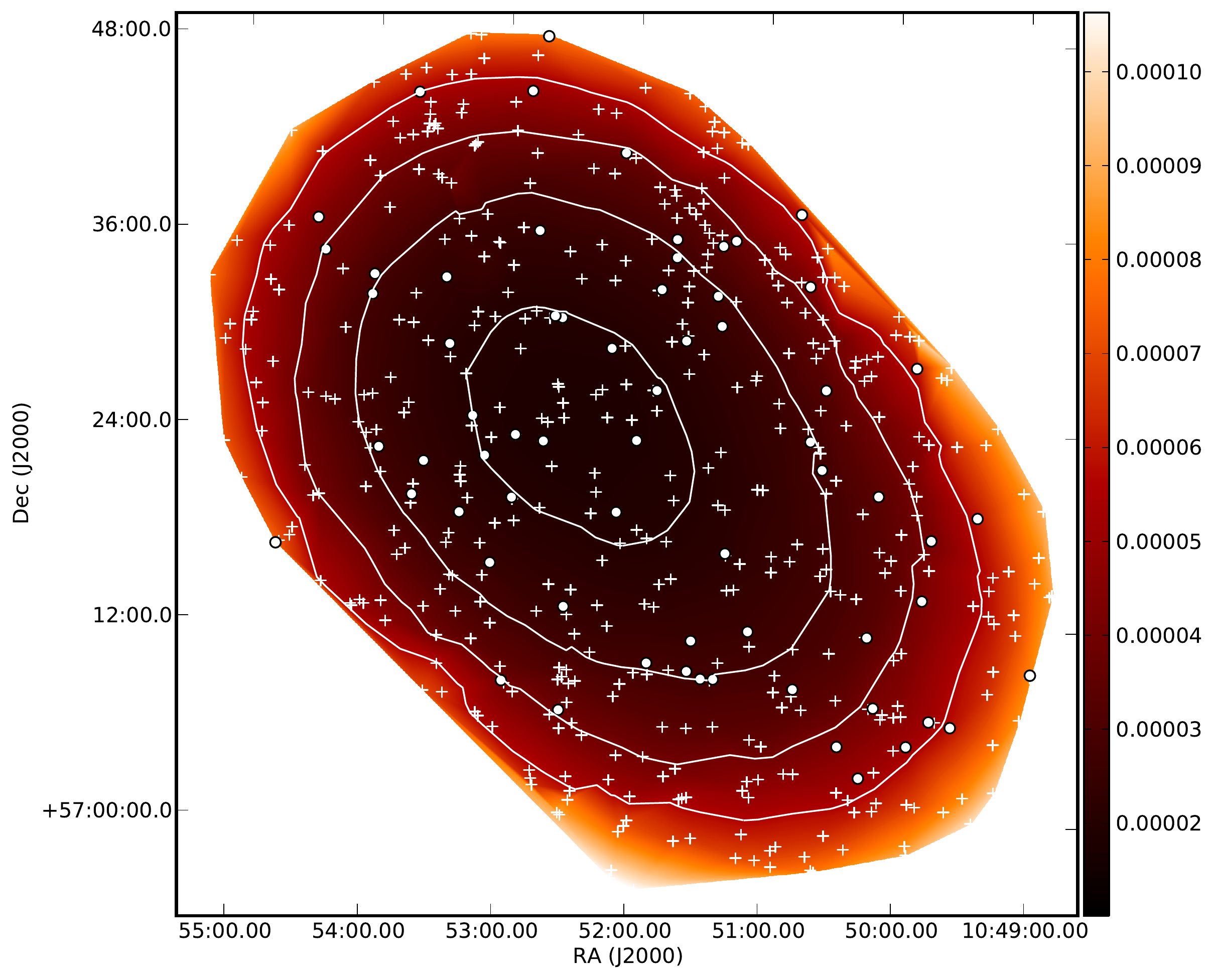}
\caption[Sensitivity map of the VLBA observations]{Sensitivity map of
  the VLBA observations. The rms noise of the individual images has
  been placed on a grid and then linearly interpolated to fill the
  map. The colour bar indicates Jy, and contours were drawn at
  $20\times\sqrt{2}^n\,{\rm \mu Jy/beam}$, $n=0...3$.}
\label{fig:sensitivity-map}
\end{figure}

An illustration of the sensitivity across the observed area is shown
in Fig.~\ref{fig:sensitivity-map}. The measured rms fluctuations in
the naturally-weighted images were placed on a regular grid and then
linearly interpolated to the empty regions. The most sensitive region
is slightly displaced from the centre of the field because the
pointing at the north-east corner was observed twice, in epoch A and
epoch D.

Natural weighting with no tapering was used to maximise the
sensitivity to search for emission. It was expected that in some cases
the VLA position was significantly offset from the VLBA core (because
the VLA emission is extended and therefore potentially offset from the
AGN), hence large images were made, covering 67.1\,arcsec$^2$. The
pixel size for these images was 1\,mas, and the median resolution was
11.7$\times$9.4\,mas$^2$. Deconvolution using the CLEAN algorithm was
stopped after the first negative model component.

Additionally, tapered images were made using a Gaussian taper falling
to 30\,\% at 10\,M$\lambda$. This allowed to increase the pixel size
to 2.5\,mas to cover a larger area, at the expense of a 25\,\%
increase of the image noise. Imaging a larger area was desirable for a
small number of targets where the catalogued VLA position was
significantly offset from the likely location of the nucleus. The
median resolution was 21.7$\times$19.9\,mas$^2$.

\begin{figure*}
\center
\includegraphics[width=\linewidth]{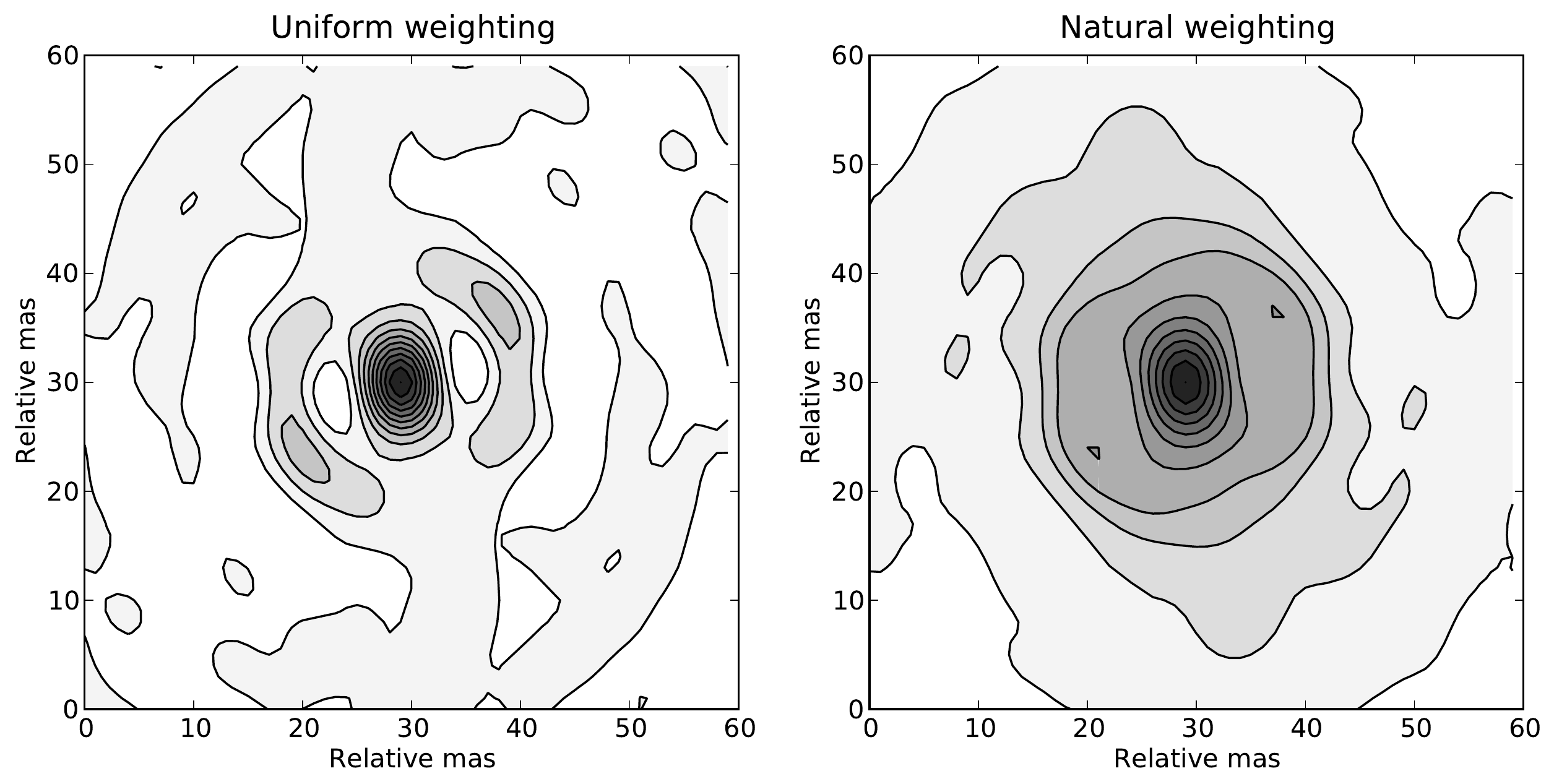}
\caption{Point spread functions using uniform (left panel) and natural
  (right panel) weighting. Contours are drawn at 10\,\%$\times N$
  ($N$=0, 1, 2, ..., 9) of the peak. The plateau in the
  naturally-weighted image at the 30\,\% level is obvious.}
\label{fig:beams}
\end{figure*}

Finally, the data of detected sources were imaged using uniform
weighting. In natural weighting, the distribution of the VLBA antennas
causes the point spread function (the ``dirty beam'') to exhibit a
significant plateau around the main peak, as illustrated in
Fig.~\ref{fig:beams}. This plateau was found to increase the
integrated flux density measurements, in particular at low SNR. Whilst
deconvolution using the CLEAN algorithm in principle should separate
the true emission from the dirty beam, we explored a variety of
CLEANed images, and remains of the plateau were found in all. It was
therefore decided that measurements of integrated flux densities were
to use uniformly-weighted images, which do not exhibit similar
artifacts. The pixel size for these images was 0.5\,mas, and the
median resolution was 7.4$\times$5.5\,mas$^2$.

\subsection{Identifying detections}
\label{subsec:detections}

\begin{figure*}
\center
\includegraphics[width=\linewidth]{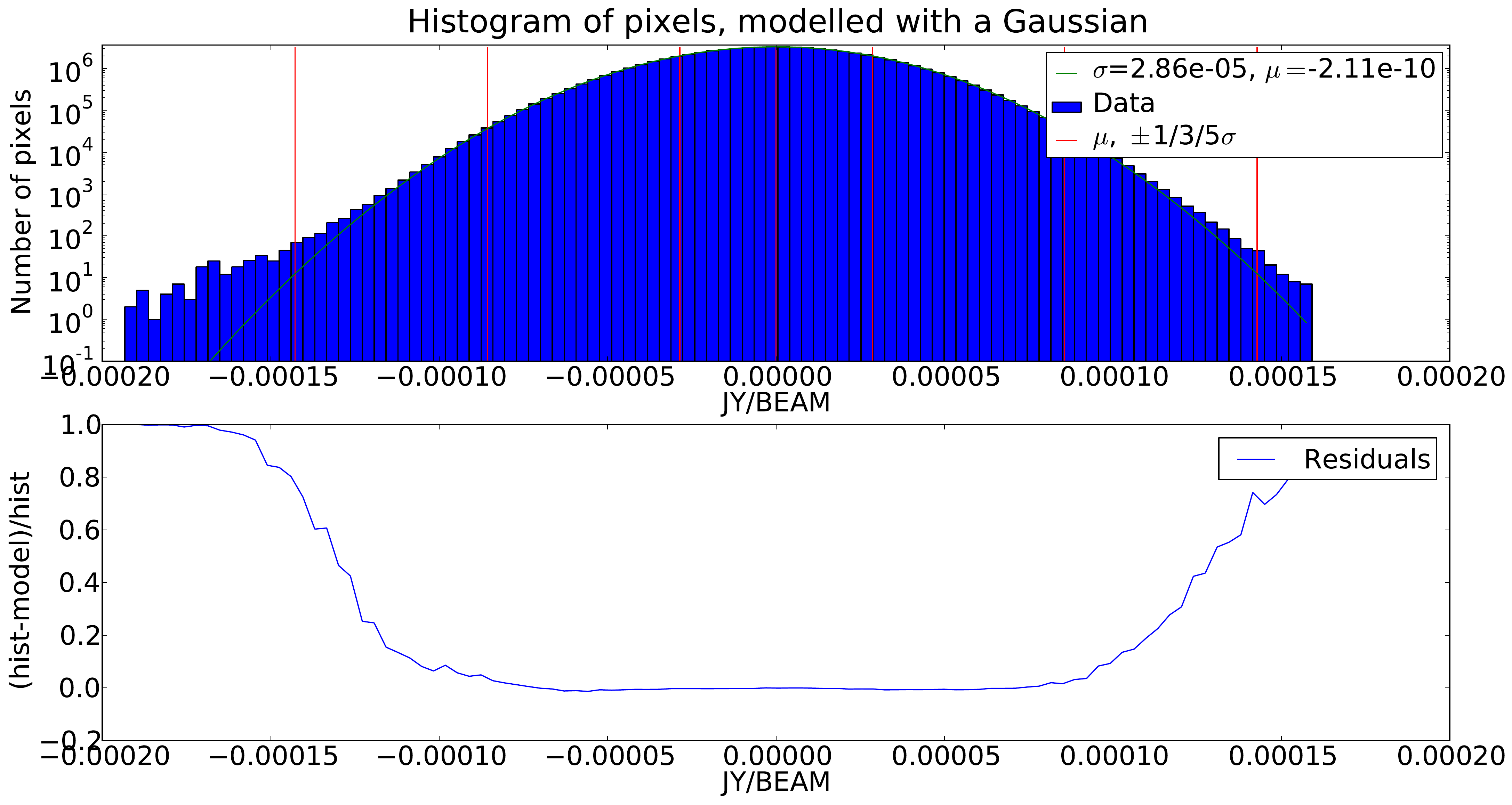}
\caption[Histogram of image pixels]{Histogram of the pixels in the
  image made from source L0211, which was not detected. The upper
  panel shows the histogram along with a model distribution calculated
  from the pixel's $\sigma$, and vertical lines indicating the mean
  and $\pm1\,\sigma$, $\pm3\,\sigma$, and $\pm5\,\sigma.$ The lower
  panel shows the residuals, normalised to the histogram. At the very
  low and very high end an excess of pixels can be clearly seen.}
\label{fig:pixhist}
\end{figure*}

A common practice in radio surveys is to treat pixels exceeding 5
times the local noise as a detection. Since the images are relatively
large, however, there is a substantial chance of having random noise
peaks exceeding this threshold. We estimate this chance as follows.

The number of independent resolution elements, $N$, can be estimated
following Eq.~3 in \cite{Hales2012}:

\begin{equation}
N=\frac{A \times \bar{d}}{\Omega_{\rm b}},
\end{equation}

where A is the covered area, $\bar{d}=\pi\sqrt{12}$ is a correction
factor for the packing of the beams, and $\Omega_{\rm b}$ is the beam
volume according to

\begin{equation}
\Omega_{\rm b} = \frac{\pi}{4{\rm ln}2}\Theta_{\rm maj}\Theta_{\rm min},
\end{equation}

with $\Theta_{\rm maj}$ and $\Theta_{\rm min}$ the beam's major and
minor axes. Therefore, in this case here, with $A=8.192^2\,{\rm
  arcsec}^2=67.1\,{\rm arcsec}^2$:

\begin{equation}
N=\frac{67.1\,{\rm arcsec}^2 \times \bar{d}}{\frac{\pi}{4{\rm ln}2}10\,{\rm mas}\times 8\,{\rm mas}}=671405.
\end{equation}

If the image noise is Gaussian, the number of beams exceeding
$5\,\sigma$, with $\sigma$ the standard deviation of the image pixels,
is

\begin{equation}
N_{+5}=N \times [1-{\rm erf}(5/\sqrt{2})]/2 = 0.193. 
\end{equation}

However, it was found that $N_{+5}$ was of the order of 1 to 3 and
sometimes was as high as 6. A typical pixel histogram clearly shows
that the distribution of pixel values is non-Gaussian at the edges
(Fig.~\ref{fig:pixhist}). Exploring a number of different weighting
schemes of the visibilities did not result in a reduction of the
effect. Furthermore, averaging all images, to investigate whether the
effect is a subtle structure present in all images, also did not
result in conclusive evidence. The cause of this effect therefore
remains unknown.

An unwelcome consequence of this is that the presence of a $5\,\sigma$
peak in an image is inadequate for deciding if a source has been
detected or not, and a $6\,\sigma$ threshold had to be used for
identifying true emission.

In five cases, $>6\,\sigma$ peaks were detected but deemed to be
noise spikes since there was no coincidence with the VLA
emission. Conversely, one has to conclude that a small number of the
$>6\,\sigma$ detections which coincide with VLA emission potentially
are chance coincidences. This fraction can be estimated as follows.

The median of the separation between the VLBA-detected emission and
the maximum of the VLA emission is 133\,mas, and in 88\,\% of the
cases the separation was smaller than 500\,mas (and in those cases
where the separation was larger it was quite obvious that the VLA
position was not a good indicator of the AGN position anyway). The
area covered by a circle of 500\,mas radius is only 1.2\,\% of that in
an image which is 8.2\,arcsec on a side. If one assumes that the
number of $>6\,\sigma$ peaks is of order 10, and if these noise spikes
are scattered randomly across the images, then the number of
6\,$\sigma$ noise spike within 500\,mas of the VLA position is,
following a binomial distribution:

\begin{equation}
B(k=1 | p=0.012; n=10) = {n \choose k} p^{k} (1-p)^{n-k} = 0.11,
\end{equation}

i.e., less than 1. Since the number of random $>6\,\sigma$ noise peaks
is likely to be smaller and visual inspection of images helps to weed
out chance coincidences, one can be reasonably certain that no
$>6\,\sigma$ peak coincident with a VLA and SWIRE source is due to
noise.

\subsection{Synthesis with VLA and SWIRE data}

For a visual inspection of each of the 496 targeted sources, plots
were generated to show a contour plot of the VLA emission superimposed
on the 3.6\,$\mu$m image from the SWIRE survey
(Fig.~\ref{fig:contour_plots}). The SWIRE data from data release 3
were retrieved from the NASA/IPAC Infrared Science
Archive\footnote{http://irsa.ipac.caltech.edu} and mosaiced using the
Montage package\footnote{http://montage.ipac.caltech.edu}, and then a
30''$\times$30'' cutout centred on each VLA position was
extracted. Postage stamps from the VLA observations were provided by
\cite{Ibar2009}. Contours were drawn from the $2\,\sigma$ level, to
accentuate faint emission, increasing by factors of 2. The two VLBA
images for each source -- untapered and tapered -- were analysed and
searched for emission.  Along with these data a contour plot was
produced of the untapered VLBA image, centred on the brightest image
pixel, starting at $2\,\sigma$ and increasing by factors of two. These
plots were used to group catalogued radio components to radio sources,
to identify the correct SWIRE counterparts to these sources and to
check the plausibility of a VLBA detection. The three sources shown in
Fig.~\ref{fig:contour_plots} illustrate this process.

As was noted in Sect.~\ref{subsec:field} and is further discussed in
Sect.~\ref{sec:discussion}, only sources with an integrated VLA flux
density exceeding 6 times the noise in the VLBA images were deemed to
be detectable, and after calibration only 217 sources were found to
satisfy this criterion. Nevertheless, we inspected all 496 targeted
sources, to ensure proper cross-identifications and to inspect the
data for errors (such as the false $>6\,\sigma$ peaks described in
Sect.~\ref{subsec:detections}) and unexpected results. However, none
of the remaining $(496-217)=279$ sources with insufficient sensitivity
in the VLBA data were found to yield a credible VLBA detection.

\begin{figure}[h!]
\centering
\subfloat[][]{
\includegraphics[width=\linewidth]{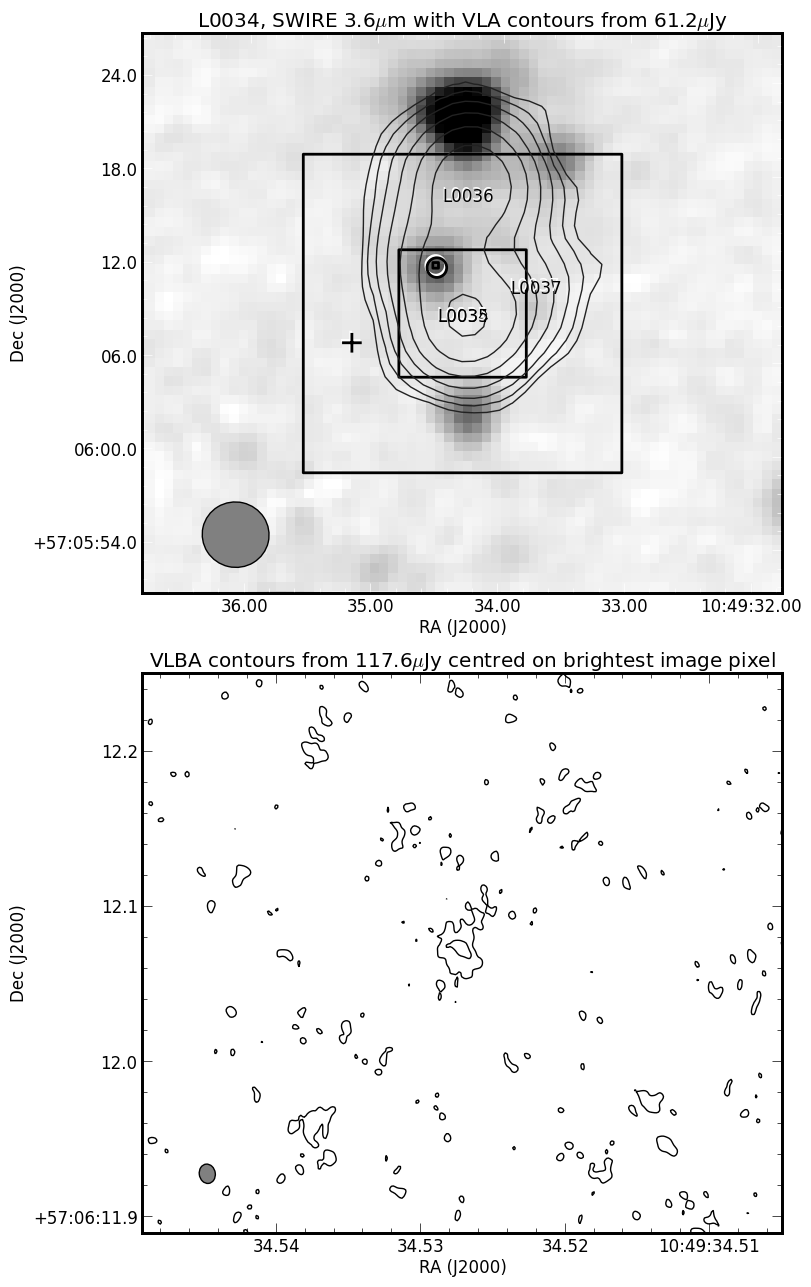}
}
\caption{Example contour plots of the VLBA, VLA, and SWIRE data. The
  upper panel shows the logarithm of the SWIRE 3.6\,$\mu$m emission as
  a pixel map, superimposed with contours of the VLA data of source
  L0034.  Contours start at 2 times the local rms (as indicated in
    the plots) and increase by factors of 2. This object has four
  entries in the VLA catalogue: one, L0034, representing it as a
  source and three for its constituents, L0035, L0036, and
  L0037. Their locations are marked as labels in the image. Three
  black squares indicate (from large to small) (i) the regions imaged
  with the VLBA using a 10\,M$\lambda$ taper, (ii) the region imaged
  using untapered data, and (iii) the region shown in the contour plot
  in the bottom panel. The maximum of the tapered VLBA data is marked
  with a cross, and the maximum of the untapered VLBA data is marked
  with an open circle. In this case, the maxima of the tapered and
  untapered VLBA images are not coincident, because the tapered image
  has reduced sensitivity. However, the maximum of the full-resolution
  (6.2\,$\sigma$) coincides with a faint SWIRE source located between
  L0035 and L0036, indicating that this is the location of the
  AGN. Furthermore, the value of the tapered image exceeds $5\,\sigma$
  at the location of the maximum of the untapered image, adding
  credibility to the detection. The bottom panel shows a contour map
  of a region 0.36\,arcsec across, centred on the maximum of the
  untapered VLBA image.  Contours start at 2 times the local rms (as
  indicated in the plots) and increase by factors of 2.}
\label{fig:contour_plots}
\end{figure}

\begin{figure}
\ContinuedFloat
\centering
\subfloat[][]{\includegraphics[width=\linewidth]{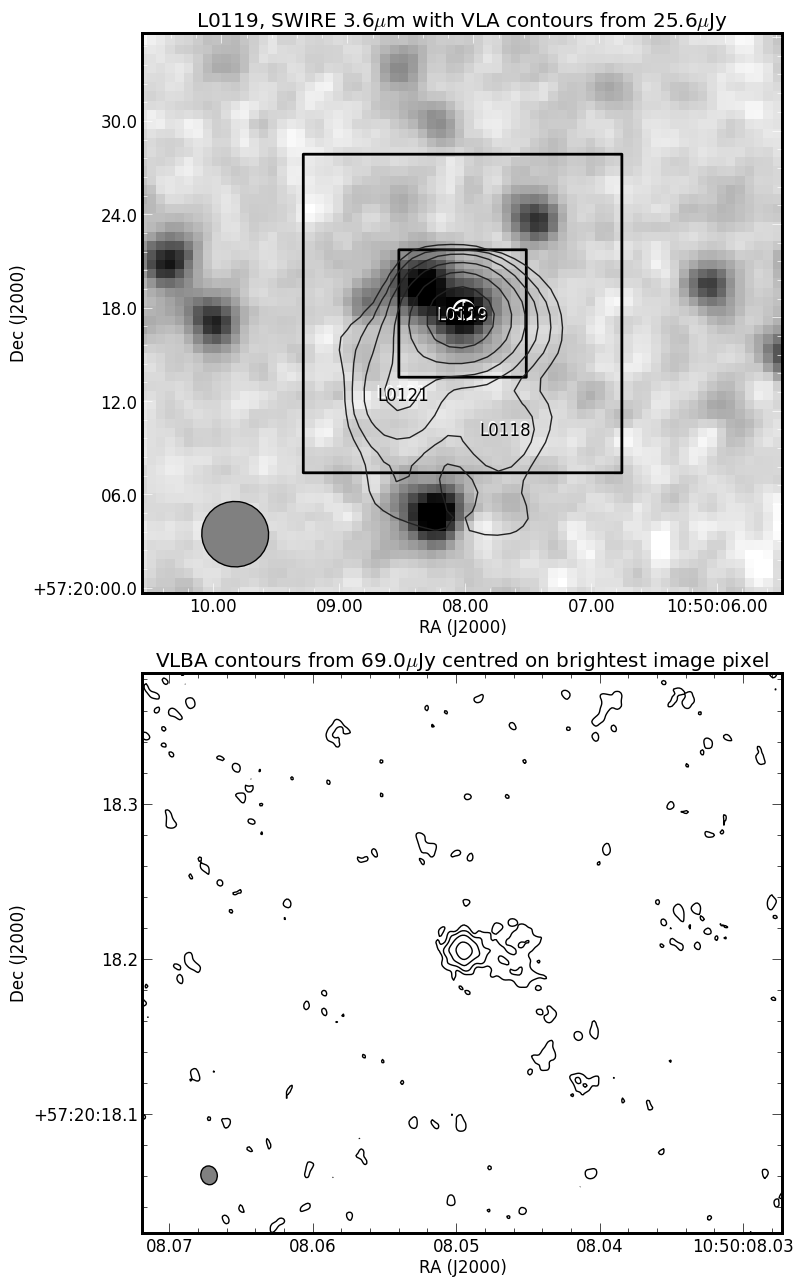}}
\caption{(continued) This plot shows data for source L0119. Three
  catalogued VLA components belong to this source. The detection at
  24\,$\sigma$ and cross-identification with a SWIRE source is
  unambiguous.}
\end{figure}

\begin{figure}
\ContinuedFloat
\centering
\subfloat[][]{\includegraphics[width=\linewidth]{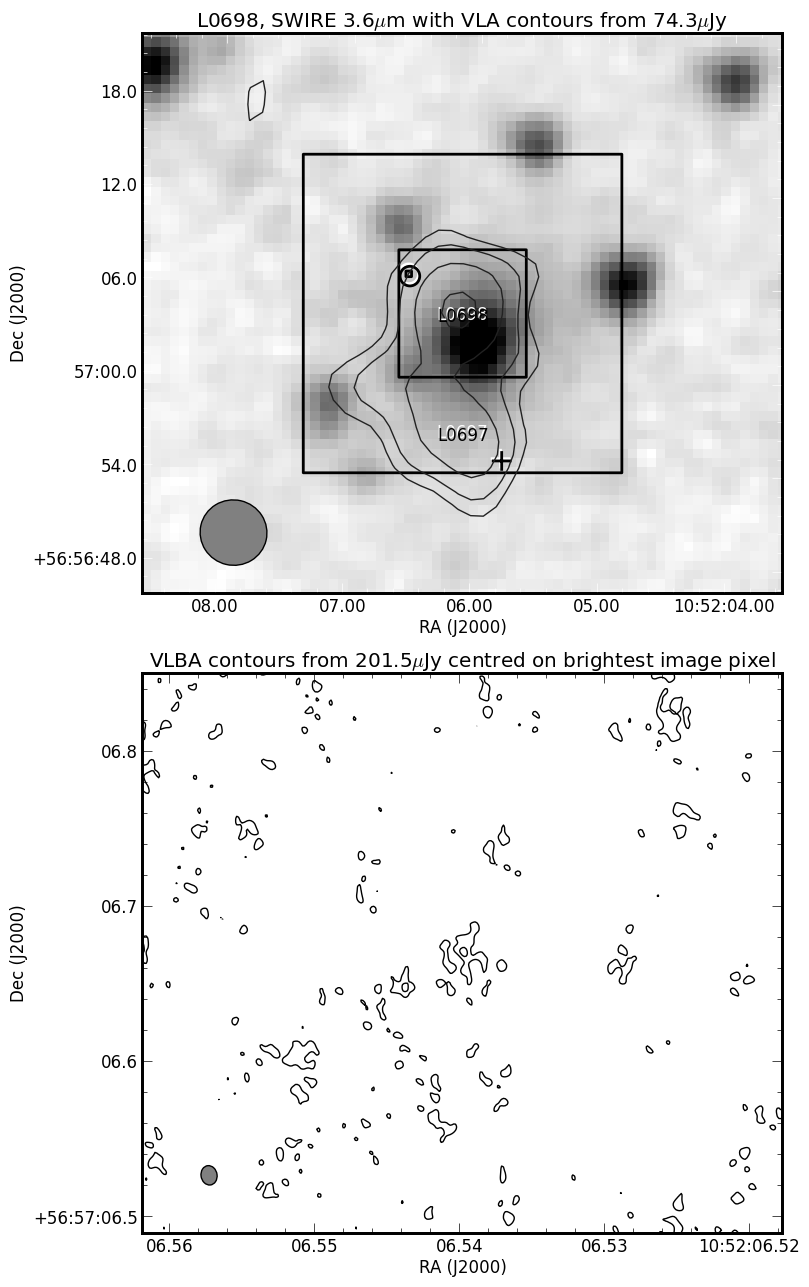}}
\caption{(continued) This plot shows data for source L0698, which has
  been grouped with source L0697. The maxima of the two VLBA images do
  neither coincide with one another, nor with an infrared source, and
  therefore this source with a maximum of 5.2\,$\sigma$ was deemed to
  be undetected.}
\end{figure}

\subsection{Flux density extraction}

In radio astronomy, source flux densities are commonly measured using
models of 2D Gaussians to the image pixels. Whilst this method is
suitable under well-constrained circumstances, it is known to yield
inaccurate results when left unconstrained. For example, if the SNR of
a source is below 10, then a 2D Gaussian is likely to overestimate the
flux density. Also if a source has a non-Gaussian shape then the model
is inappropriate and will yield incorrect results. In a recent
publication, \cite{Hales2012} presented a source extraction program,
\texttt{BLOBCAT}\footnote{http://blobcat.sourceforge.net}, which
uses the flood fill algorithm to measure the integrated flux density
of sources. It takes proper account of the various biases involved and
performs better than Gaussian fits when the source has low SNR or is
irregularly shaped. It was therefore used to measure the flux
densities of the VLBA-detected sources from uniformly-weighted images.

Two modifications to the default settings of \texttt{BLOBCAT} were
made. First, a pixellation error of only 1\,\% was assumed because of
the significant oversampling of the point spread function
(\texttt{--ppe=0.01}). Second, the error of the surface brightness in
the images was assumed to be 10\,\% (\texttt{--pasbe=0.1}), reflecting
the amplitude calibration errors.

%

\section{Discussion}
\label{sec:discussion}

The most basic result from the VLBA observations of the Lockman
Hole/XMM is whether or not a source is detected, and what its flux
density and morphology are on mas scales. Even though the initial
sample consisted of 496 radio sources with $S>100\,\mu$Jy it was
always clear that many would not be detectable with our observations,
because the VLA had a $1\,\sigma$ sensitivity of $6\,\mu{\rm
  Jy\,beam^{-1}}$, compared to a $1\,\sigma$ sensitivity of just under
$20\,\mu{\rm Jy\,beam^{-1}}$ of our VLBA data. For the subsequent
analysis a sub-sample was selected based on whether sources would have
been, in principle, detectable. The criterion we used was that the
noise in the full-resolution, naturally-weighted image was smaller
than 1/6 of the VLA peak flux density, because we used a 6\,$\sigma$
threshold for source detection. Out of the original sample, 217
sources satisfied this criterion and were used for analysis. Out of
these 217 detectable sources, 65 were deemed to be actually
detected. Note that whilst naturally-weighted images were used to
determine detections because of the higher sensitivity, flux densities
were measured from uniformly-weighted images to reduce the effect of
the plateau in the point spread function (see Sect.~\ref{sec:imaging}
and Fig.~\ref{fig:beams}).

\subsection{Fraction of detected sources}

\begin{figure*}
\center
\includegraphics[width=\linewidth]{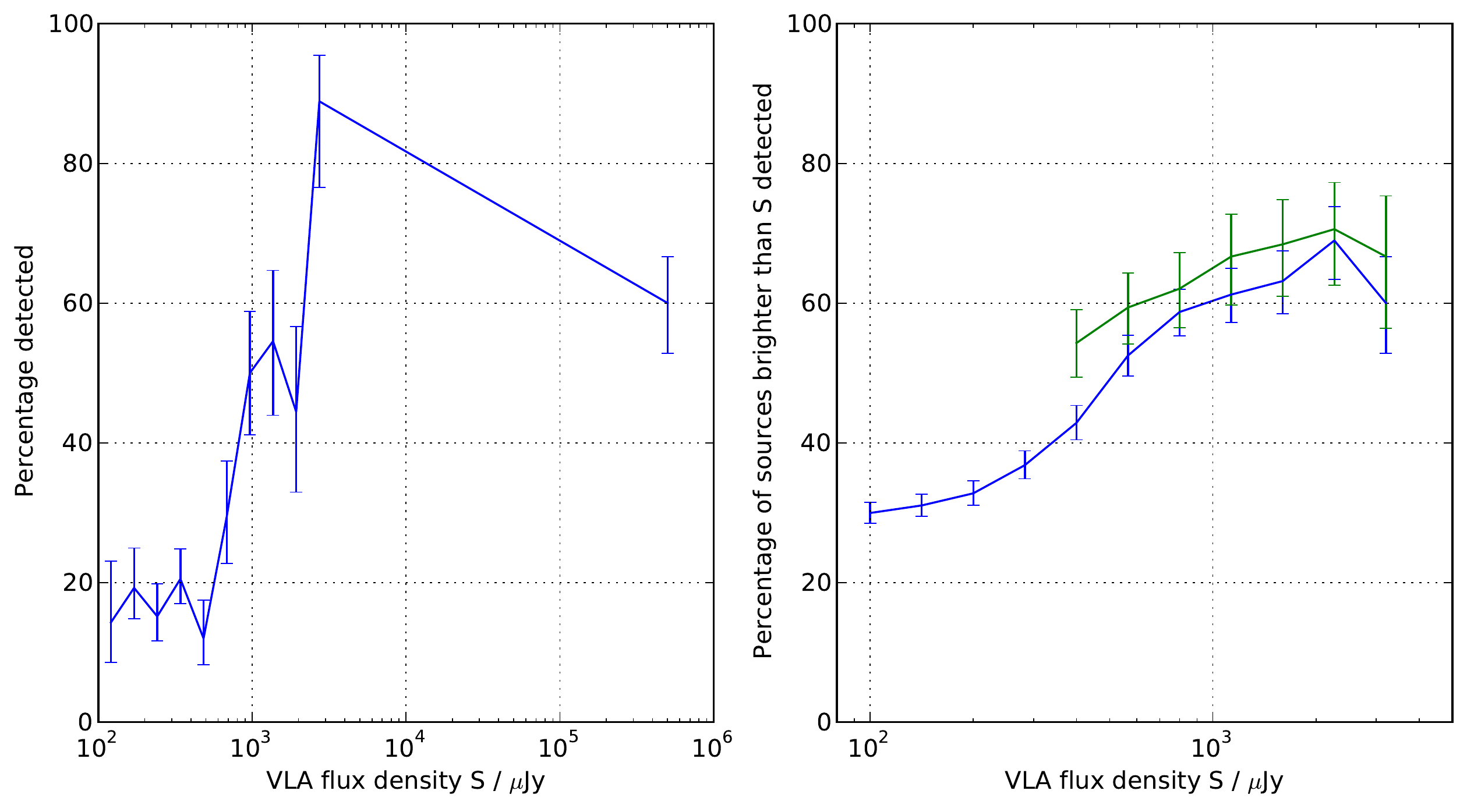}
\caption{Fraction of VLBA-detected sources. {\it Left panel:} The
  fraction of VLBA-detected sources as a function of flux
  density. Detectable sources have been grouped into bins with limits
  100\,$\mu$Jy$\times\sqrt{2}^N$, $N=0,1,2,...$. Error bars denote the
  68\,\% confidence interval for a small-sample binomial distribution,
  according to \cite{Clopper1934}. Whilst at the faintest flux density
  levels only around 15\,\% have been detected, this fraction rises
  sharply at a level of around 0.5\,mJy, and increases to more than
  80\,\% at the 2\,mJy level. {\it Right panel:} Using the same bins
  as for the left panel, the fraction of detected sources brighter
  than the bin limits has been calculated. Whilst of all sources
  brighter than 100\,$\mu$Jy 30\,\% have been detected, the fraction
  reaches 60\,\% when all sources brighter than 1\,mJy are
  considered. The green line indicates the data of the CDFS from
  \cite{Middelberg2011a}, which had a maximum sensitivity of
  $55\,\mu{\rm Jy/beam}$, compared to just under $20\,\mu{\rm
    Jy/beam}$ in our data. All curves are subject to small number
  statistics, which causes the Lockman Hole data to drop at the bright
  end and the discrepancy between the CDFS and Lockman Hole data.}
\label{fig:detected_fraction}
\end{figure*}

The fraction of VLBA-detected sources is an interesting quantity
because it essentially gives us a lower limit on the number of
radio-emitting AGN in the sample. Since our sample is large we can
subdivide it and determine the detection fraction as a function of
flux density. The results are shown in
Fig.~\ref{fig:detected_fraction} and Table~\ref{tab:detections}.

The fraction of VLBA-detected sources is obviously a function of
sensitivity. Towards faint flux density levels only sources which are
increasingly compact can be detected, yet at some flux density level
the detection fraction approaches completeness. One can estimate that
level as follows. Around 90\,\% of the targets have been observed with
a sensitivity of 60\,$\mu$Jy/beam$^{-1}$, allowing the detection of
sources brighter than 360\,$\mu$Jy, when a $6\,\sigma$ cutoff is
used. If one requires a minimum of 10\,\% of that flux density to come
from the core, providing emission for a VLBA detection, then sources
brighter than 3.6\,mJy can be detected across 90\,\% of the area. In
Fig.~\ref{fig:detected_fraction} (right panel), the nearest point is
at 3.2\,mJy, and the detection fraction is 60\,\% (12 out of 20
sources).

\begin{table}
\caption{Fraction of detected sources as a function of VLA flux
  density.}
\center
\begin{tabular}{rrrr}
\hline
Bin edges
&$N_{\rm VLA}$
&$N_{\rm detected}$
&\%\\
$\mu$Jy\\
\hline
 100.0-    141.4 & 14 &  2 & 14\\
 141.4-    200.0 & 26 &  5 & 19\\
 200.0-    282.8 & 33 &  5 & 15\\
 282.8-    400.0 & 39 &  8 & 21\\
 400.0-    565.6 & 25 &  3 & 12 \\
 565.6-    800.0 & 17 &  5 & 29\\
 800.0-   1131.3 & 14 &  7 & 50\\
1131.3-   1600.0 & 11 &  6 & 54\\
1600.0-   2262.7 &  9 &  4 & 44\\
2262.7-   3200.0 &  9 &  8 & 89\\
$>$3200.0        & 20 & 12 & 60.0\\
\hline
\end{tabular}
\label{tab:detections}
\end{table}

\subsection{Variability}
\label{sec:variability}

For an unresolved radio source, the interferometric visibility
amplitude is constant with baseline length; for resolved sources, it
must generally decrease with baseline length (although constructive
interference between separated, compact components can lead to local
increases). When comparing VLBA baselines with lengths of hundreds to
thousands of kilometres to VLA baselines with lengths of much less
than 100\,km, it can be safely assumed that the visibility amplitudes
on longer baselines must be less than, or at most equal to, the
visibility amplitudes on short baselines.

\begin{figure}
\center
\includegraphics[width=\linewidth]{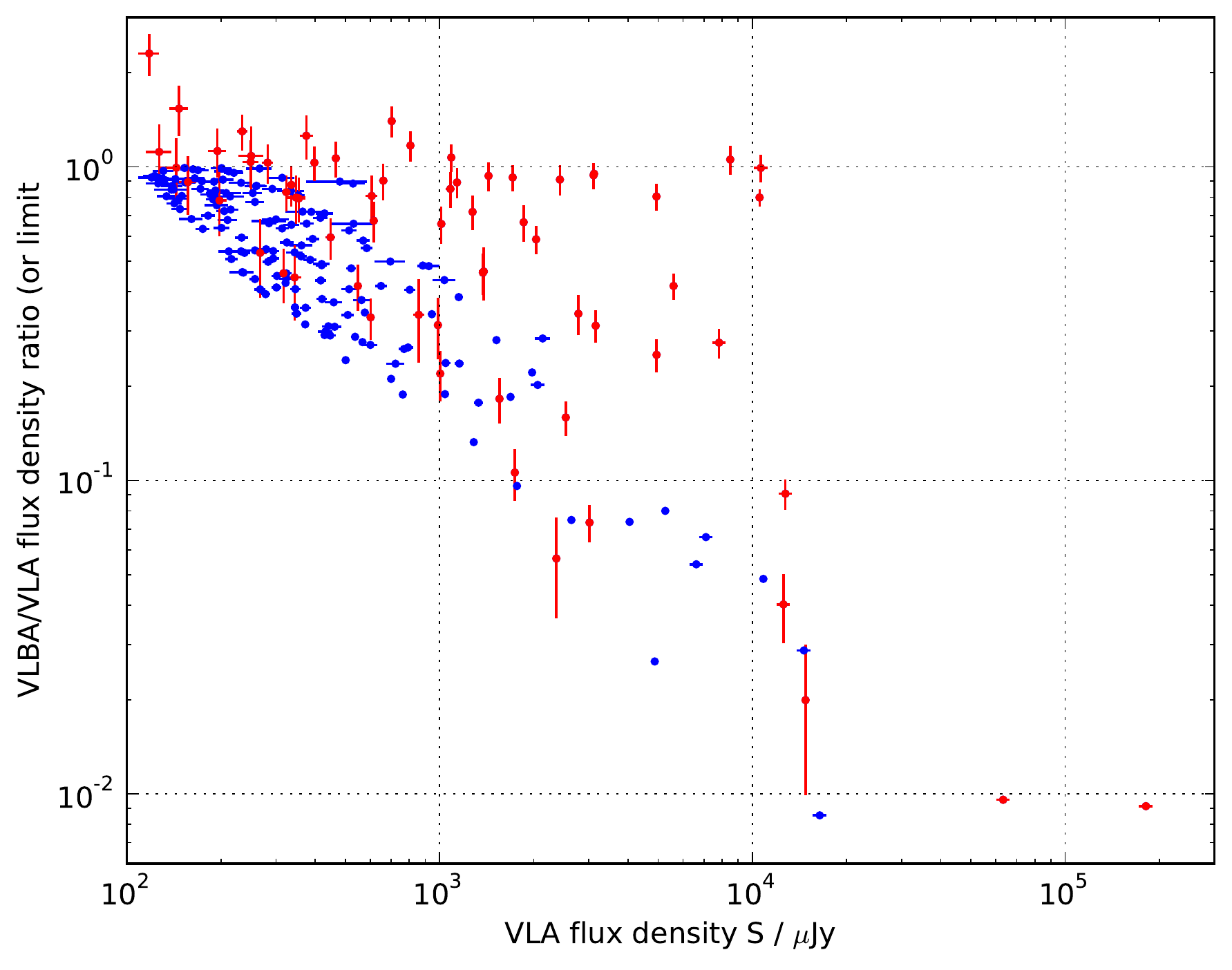}
\caption[Ratio of VLBA and VLA flux densities]{Ratio of the VLBA and
  VLA flux densities. Blue dots show the detectable sources, placed at
  $6\times{\rm rms}/S_{\rm VLA}$ (where rms is the local noise level
  of the VLBA observations), and red dots with error bars detected
  sources. Out of the 65 detected sources, 6 exceed a ratio of 1.0 by
  more than 1\,$\sigma$, but only one source exceeds $1+3\,\sigma$.}
\label{fig:variability}
\end{figure}

It is therefore a useful consistency check to compare the VLBA flux
densities to the VLA flux densities. In Fig.~\ref{fig:variability} the
ratios of the integrated VLBA peak and integrated flux densities to
the integrated VLA flux densities are plotted for all detectable
sources and the 65 detected ones. Out of the 65 detected targets, 6
have ratios which exceed 1.0 by more than the width of their error
bars ($1\,\sigma$), in particular at low flux densities: L0231, L0578,
L0593, L0648, L0997, and L1200. While it is clear that some sources
might have flux density ratios in excess of $1+\sigma$, one source
(L0593) exceeds $1+3\,\sigma$, which is significant given our sample
size. We interpret this result as source variability, for several
reasons:

\begin{itemize}
\item Although the variability of bright radio sources has long been
  studied, the variability of the mJy and sub-mJy radio source
  population is poorly constrained. \cite{Carilli2003} found that the
  density of variable sources in this flux density regime is less than
  $5\times10^{-3}\,{\rm arcmin}^{-2}$.

\item The elapsed time between the VLA and VLBA observations is
  significant. Whilst the VLBA observations where carried out in
  July-September 2010, the VLA data were obtained between 2002 and
  2005.

\item Faint sources are preferably detected if they are variable and
  in a state with high flux density, compared to their average flux
  density. Faint sources in a state with low flux density will be
  missed, and this is the reason why the fraction of sources with
  ratios greater than 1 increases towards fainter flux densities.

\item The VLA observations are sensitive to the combined emission from
  star formation and AGN, whereas the VLBA observations isolate the
  AGN emission. Hence the variability of the AGN is likely to be
  higher than determined by our observations, because only one of our
  two available flux density measurements (VLA and VLBA) is purely
  sensitive to the AGN emission.
\end{itemize}

The effect of interstellar scintillation on the source flux densities
is difficult to estimate with our data, but is likely to be small. The
galactic latitude of the Lockman Hole/XMM field is $53^\circ$, where,
according to \cite{Walker1998}, sources (or parts thereof) are
required to be smaller than around $6\,\mu$as (0.05\,pc at $z=1$) to
show significant scintillation. Most of the VLBA-measured flux
densities are lower than the VLA flux densities, implying that the
sources are partly resolved between the length scales probed by the
VLBA and VLA. We deem it safe to assume that most, if not all, of the
sources had been resolved further, had the resolution of our
observations been higher. Thus, whilst it is possible that small
fractions of the flux density recovered by our VLBA observations
originate in regions sufficiently small for scintillation, we do not
expect it to have a significant effect.

\subsection{Spectral index}

A common method to classify radio sources into AGN and non-AGN is to
use the spectral index, $\alpha$, with $S\propto\nu^\alpha$. Sources
with spectral indices of more than $-0.3$ (``flat'' or ``inverted'')
or less than around $-1.0$ (``steep'') are typically classified as
AGN, because flat and inverted spectral indices arise in compact,
optically thick synchrotron emitters typical of AGN, whereas steep
indices are attributed to optically thin emission from radio lobes,
with a potentially aged particle population. Sources with $\alpha$
around $-0.7$ can not be classified based on spectral index alone,
because both starburst galaxies and AGN can exhibit such values, which
arises from synchrotron emission with a continuous injection of fresh
particles (e.g., \citealt{Pacholczyk1970}).

Whilst the histogram of the spectral indices of detectable sources is
typical of the population found in sensitive, extragalactic radio
surveys, the histogram of detected sources exhibits a prominent
deficiency at low spectral indices. The lowest spectral index of any
detected source is $-0.88$, whereas the distribution of detectable
sources extends to well below $-1.0$. Almost all sources with
$\alpha>0.0$ are detected, but no source with $\alpha<-0.9$ is.

One might argue that this is a selection effect, which de-selects
steep spectral indices, because they tend to have lower 1.4\,GHz flux
densities than flat- or inverted-spectrum sources (the spectral
indices at our disposal were measured between 610\,MHz and
1.4\,GHz). However, we find only a mild correlation between the
spectral index and the flux density, with a product-moment correlation
coefficient of $-0.02$ (where $+1$ and $-1$ denote perfect linear
correlation and anti-correlation). We therefore conclude that, whilst
inverted spectra arise from compact synchrotron emission and therefore
are likely to be detected, steep spectral indices arise from larger
and therefore potentially older regions, which tend to be resolved
out. What follows from this exercise is that at ``normal'' spectral
indices of between $-1.0$ and $-0.3$ VLBI observations can cleanly
separate the AGN from star-forming activity, whereas in the regime of
$\alpha>-0.3$ spectral index alone is already a good indicator for
AGN.

\begin{figure}
\center
\includegraphics[width=\linewidth]{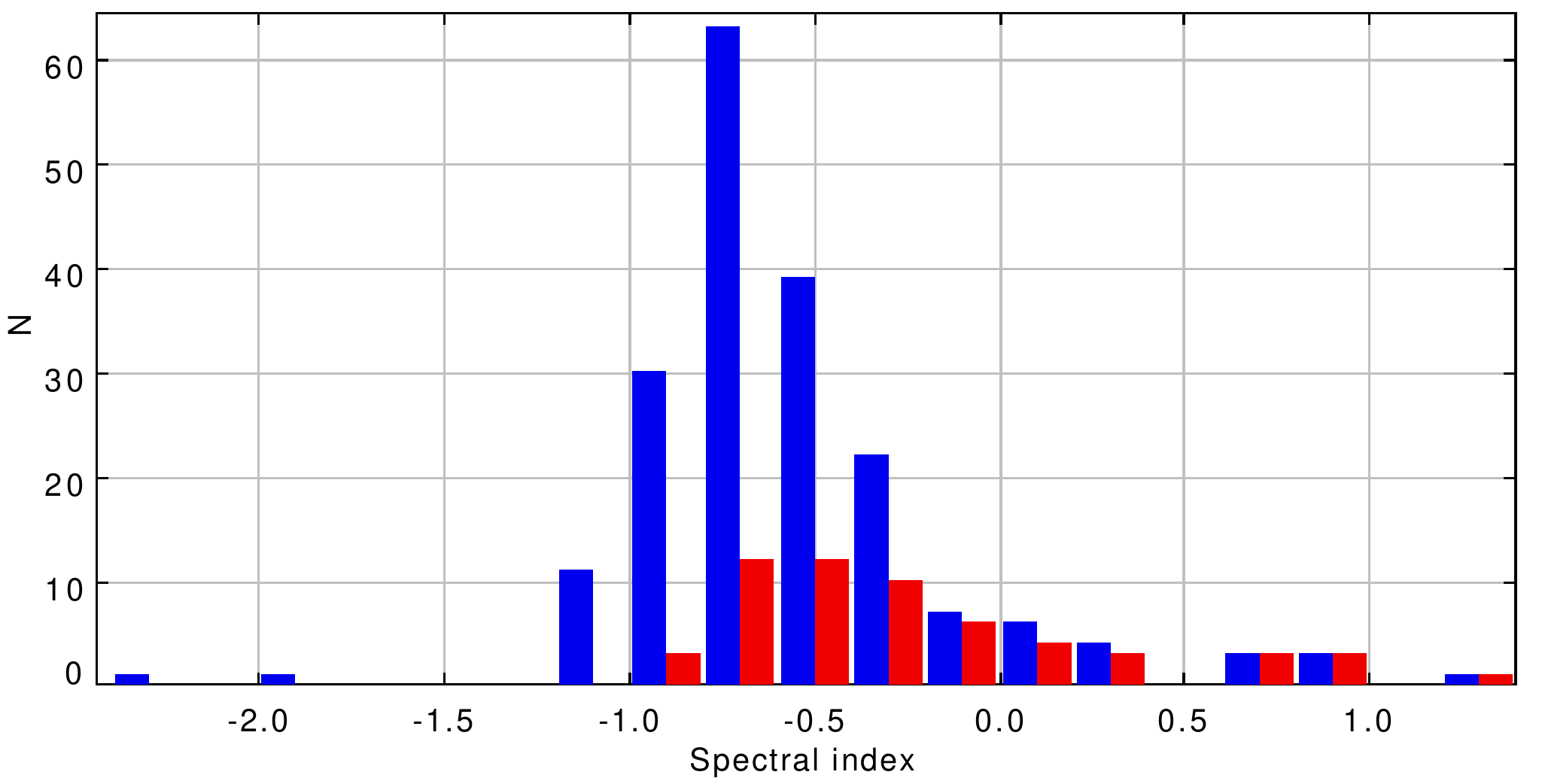}
\caption{A histogram of the spectral indices of detectable (blue bars)
  and detected (red bars).}
\label{fig:spec_index}
\end{figure}

\subsection{Source counts}
\label{sec:source_counts}

\begin{figure}[ht]
\centering
\includegraphics[width=\linewidth]{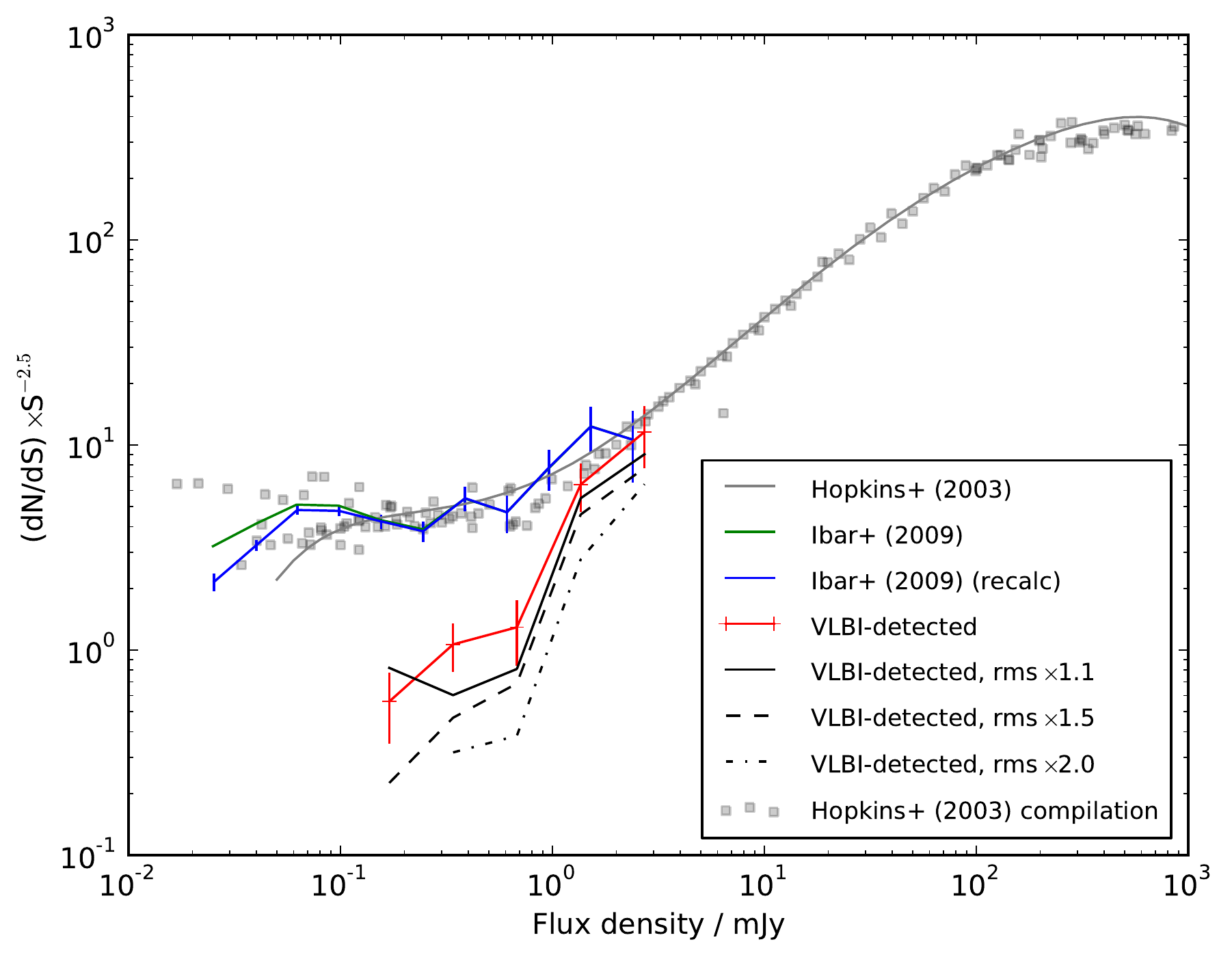}
\caption{Source counts in the Lockman Hole/XMM field. Shown are the
  euclidean-normalised source counts of the VLBA-detected sources,
  using their VLA flux densities (red line). Shown with black lines
  are source counts which would have been measured had the sensitivity
  of our observations been lower by factors of 1.1, 1.5, and 2.0. For
  comparison, we have plotted the source counts constructed from the
  \cite{Ibar2009} data using the same methods (blue line), the source
  counts as published by \cite{Ibar2009} (green line), and the data
  compilation from \cite{Hopkins2003} (light grey squares) along with
  their 6th order polynomial fit.}
\label{fig:source_counts}
\end{figure}

While radio source counts of VLBI-detected sources at flux density
levels of above 100\,mJy are in principle available from VLBI
calibrator searches, counts of the mJy and sub-mJy population are yet
unknown, because they would have required prohibitive amounts of
observing time using conventional methods. The work presented here
yielded only the detection of 65 sources in total (a number frequently
found in a single bin in other source count studies), yet this is, to
our knowledge, the first attempt to produce radio source counts of
VLBI-detected sources in the mJy and sub-mJy regime.

It is worth noting that in this paper we do not construct radio source
counts using VLBI flux densities, since such measurements would be
plagued by tremendous resolution effects which would render the source
counts useless. Instead, we use the VLBI observations to select a
sub-sample of radio sources with very compact cores from the parent
sample of known radio sources detected with the VLA, and then use the
VLA flux densities of this sample to construct the source counts.

When constructing source counts, two effects need to be taken into
account. The first is a weighting factor arising from the varying
sensitivity of the observations, i.e., each source flux density one
needs to determine over which area this source would have been
detected. Since observations typically have significantly smaller
areas with very high sensitivity than with lower sensitivity, this
correction can have a large effect on the source counts if not carried
out correctly. The second effect is resolution bias. Extended sources
can remain undetected when they are close to the detection limit of
the survey, and so the counts at low flux density levels will
underestimate the true source counts. This effect depends on the
resolution of the observations, and in our case is much smaller than
the correction for effective area, of order below 10\,\%
(\citealt{Ibar2009}). Hence the number of sources in each bin needs to
be corrected by $1+c$, where $c$ is of order 0 to 0.1.

As a consistency check, we initially constructed the source counts
from the VLA data published by \cite{Ibar2009}. From their catalogue
we have eliminated all entries marked as components (non-zero
``m\_Name'' column). Corrections for effective area and resolution
were extracted from their Figs.~6 and 7, from which we constructed
look-up tables using
Dexter\footnote{http://dexter.sourceforge.net}. Using the same bins as
\cite{Ibar2009} we determined

\begin{equation}
s=\sum_N S/A_{\rm eff}
\label{eq:counts}
\end{equation}

for the $N$ sources in each bin, where $A_{\rm eff}$ was determined
from a look-up table. This procedure ensures proper weighting of all
sources in a bin. Each bin was then corrected for resolution bias by
multiplying its value by $1+c$, where $c$ is a value extracted from
the other look-up table. The bin values were then divided by the bin
widths, and subsequently by the geometric mean of the bin edges raised
to the power of $-2.5$. Count errors of the area-weighted counts in
Eq.~\ref{eq:counts} were estimated as $\sqrt{N}/N\times s$ and then
propagated through the same steps as the counts themselves. This
procedure resulted in identical counts as published by
\cite{Ibar2009}, with the exception of the two lowest bins, where we
found agreement with previously published counts
(Fig.~\ref{fig:source_counts}). The reason for this discrepancy can in
principle arise from the way the effective area is dealt with. If one
uses, e.g., the lower edge of a bin to calculate the effective area
for the entire bin, then one underestimates the area over which a
source would be detected, resulting in an over-estimate of the source
counts.

We proceeded to calculate the source counts for the VLBA-detected
sources only, using the same methods as for the VLA source
counts. However, the determination of the effective area was
different. Two corrections are needed to properly account for the
effective area -- one to determine over which area each source would
in principle be detectable with the VLBA, another to determine the
area over which this source would in principle be detectable with the
VLA. The first correction was carried out as follows. For each source
the area over which it could have been detected with the VLBA was
determined using the sensitivity map shown in
Fig.~\ref{fig:sensitivity-map}. The number of pixels with values
smaller than 6 times the VLA source flux density was counted and
multiplied with the pixel area. The second correction would have
required knowledge about the rms distribution of the VLA observations,
which were not available (the catalogue of \citealt{Ibar2009} only
lists peak-to-noise ratio, but neither peak nor noise are
given). However, Fig.~6 in \cite{Ibar2009} indicates that for our
weakest source greater than the lowest bin edge at 120\,$\mu$Jy, at
$S=127\,\mu{\rm Jy}$, the area over which it was detectable is 94\,\%
of the entire observed area, and for stronger sources this percentage
comes even closer to 100\,\%. We therefore neglected this correction.

The source count bins were defined to begin at a total integrated flux
density of 120\,$\mu$Jy and to increase by a factor of two to
3.84\,mJy. This resulted in a reasonably even distribution of the
number of sources per bin.  The source counts of all sources, the
VLBA-detected sources, and the source counts by \cite{Hopkins2003} are
shown in Fig.~\ref{fig:source_counts}.

At the bright end of the distribution the source counts of
VLBA-detected sources appear to be an extrapolation of the general
radio source counts, indicating that this population is dominated by
AGN. Between 1\,mJy and 2\,mJy there is a sharp drop, which is also
visible in the percentage of detected sources in
Fig.~\ref{fig:detected_fraction} (left panel). However, the counts of
VLBA-detected sources exhibit a shoulder between 0.1\,mJy and 1\,mJy,
and this shoulder is a lower limit on the number of radio-emitting AGN
at sub-mJy flux density levels.

It is important to note here that this lower limit is subject to the
resolution of the VLBA observations. At high flux densities a
detection can be made even if only a small fraction of the radio flux
density originates from the AGN, whereas at flux density levels near
the detection limit only the most compact sources will be detected
(see Fig.~\ref{fig:variability}). Our counts near the VLBA detection
limit therefore underestimate the ``true'' detection rate by a larger
margin than at higher flux density levels. At higher sensitivity, the
counts of VLBA-detected sources would rise towards the general radio
source counts.

To illustrate this effect we have constructed source counts which
would have been measured with lower sensitivity. We multiplied the
sensitivity map in Fig.~\ref{fig:sensitivity-map} by factors of 1.1,
1.5, and 2.0 and selected only those sources which would have been
detected with this increased noise level. This was done by selecting
sources with a peak flux density in the VLBA observations of 6 times
the modified noise. The source counts from this exercise are shown in
Fig.~\ref{fig:source_counts} as black lines (solid, dashed, and
dash-dotted). One can see that the source counts drop with decreasing
sensitivity, as expected, because less sensitivity requires that only
the most compact sources can be detected. Conversely, if one
extrapolates this trend to {\it higher} sensitivity, one can conclude
that the source counts of VLBA-detected sources are likely to be
higher, above the red line, but below the general radio source
counts. 

We therefore conclude that at flux density levels of between 0.1\,mJy
and 1\,mJy, at least 15\,\% to 25\,\% of the radio sources contain
radio-emitting AGN. This finding adds new and independent support for
the recently established picture that a substantial fraction of the
sub-mJy radio populations is AGN-driven, rather than powered by
star-forming activity. For example, \cite{Smolcic2008} classified
radio sources from the COSMOS survey using optical colours and found
that 50\,\% to 60\,\% of the sub-mJy radio sources contain AGN. In
another study, \cite{Padovani2011} classified radio sources in the
{\it Chandra} Deep Field South using {\it Spitzer} data, and found
that 50\,\% contained AGN. A recent compilation of studies
investigating the AGN fraction of the sub-mJy radio source population
can be found in Fig.~4 of \cite{Norris2011c}.

\begin{table*}
\caption{Source counts of VLBA-detected sources.}
\center
\begin{tabular}{rrrrr}
\hline
Bin edges
&\multicolumn{1}{c}{$\langle S\rangle$}
&\multicolumn{1}{c}{$N$}  
&\multicolumn{1}{c}{$dN/dS$}          
&\multicolumn{1}{c}{$ (dN/dS)S^{-2.5}$}
\\
$\mu$Jy
&\multicolumn{1}{c}{$\mu$Jy}                                     
&
&\multicolumn{1}{c}{Jy$^{-1}$\,sterad$^{-1}$}
&\multicolumn{1}{c}{Jy$^{1.5}$\,sterad$^{-1}$}
\\
\hline
120.0-240.0   &   169.7  &    7 &   1.50e+09  &    0.561$\pm$0.212\\
240.0-480.0   &   339.4  &   14 &   5.02e+08  &    1.066$\pm$0.285\\
480.0-960.0   &   678.8  &    8 &   1.08e+08  &    1.292$\pm$0.457\\
960.0-1920.0  &  1357.6  &   14 &   9.45e+07  &    6.419$\pm$1.716\\
1920.0-3840.0 &  2715.3  &    9 &   3.02e+07  &   11.589$\pm$3.863\\
\hline
\end{tabular}
\label{tab:source_counts}
\end{table*}

\subsection{The extended emission}

As one can see in Fig.~\ref{fig:variability}, the flux density
recovered with the VLBA is in general a fraction of the VLA flux
density. Variability can only explain a small fraction of this trend,
but not the large differences found in many sources, nor the majority
of the VLBA non-detections. Moreover, Fig.~\ref{fig:spec_index} shows
that the spectral indices of the VLA sources (extended) tend to be
lower than those of the VLBA sources (compact), indicating that the
nature of the compact emission is different than that of the extended
emission. Here, we assume that the compact emission comes from the
AGN, whereas the extended emission comes from star-formation, either
circumnuclear, or from the host galaxy. In a recent study of
radio-quiet ($L\lesssim10^{24.5}\,{\rm W\,Hz^{-1}}$) VLA sources in
the {\it Chandra} Deep Field South (CDFS), \cite{Padovani2011} found
that the bulk of the their emission comes from star-formation
processes, by comparing the luminosity functions of radio-quiet AGN,
radio-loud AGN and star-forming galaxies. Also
Fig.~\ref{fig:source_counts} showed that the source counts of
VLBA-detected and VLBA-undetected sources follow a dichotomy, so it is
safe to assume that the extended part of the emission traces
star-formation, whereas the compact emission traces the AGN.

\begin{figure}[ht]
\centering
\includegraphics[width=\linewidth]{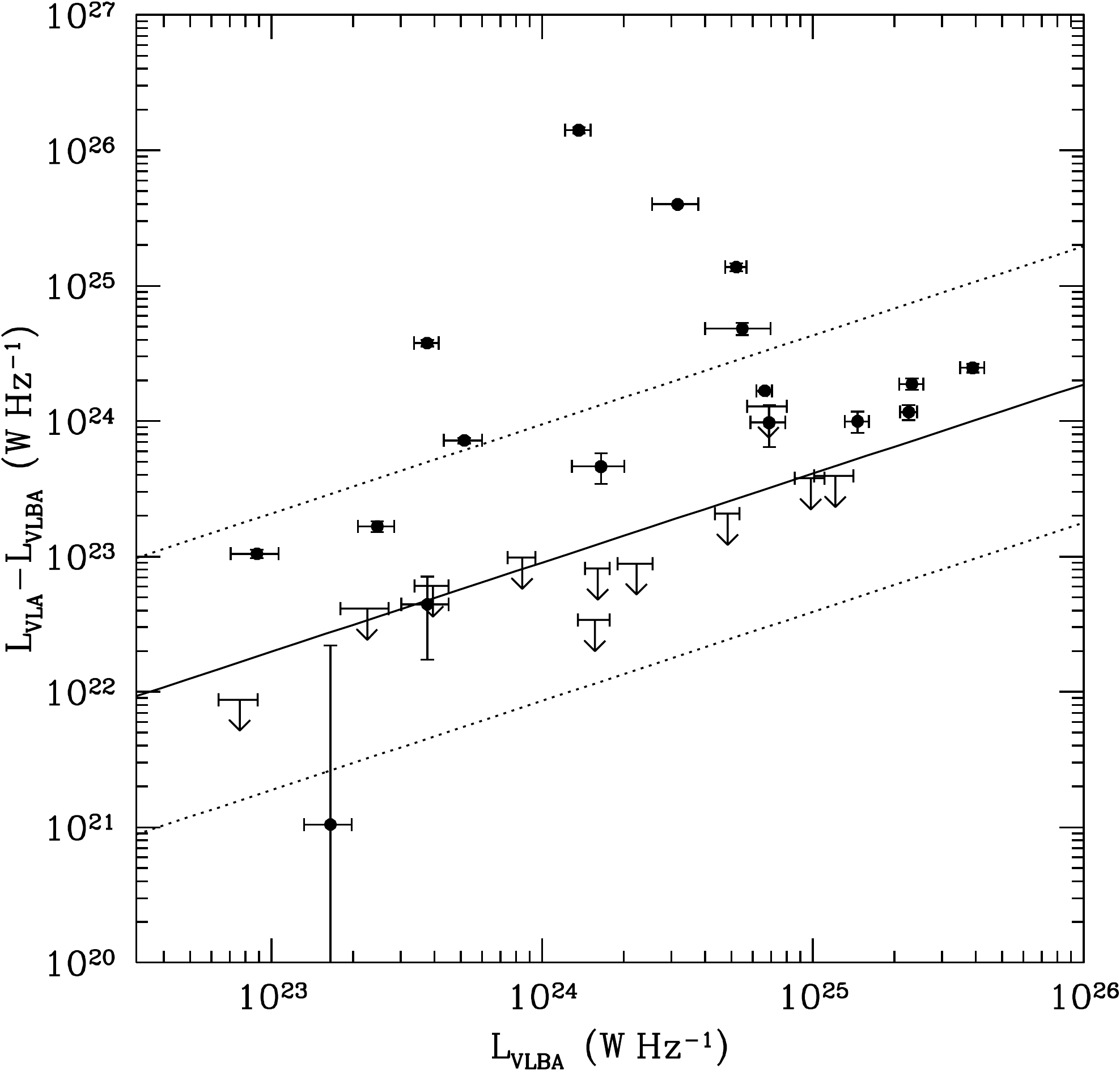}
\caption{The difference between the 1.4\,GHz radio luminosity measured
  with the VLA, $L_{\rm VLA}$ and measured with the VLBA, $L_{\rm
    VLBA}$, as a function of $L_{\rm VLBA}$. The solid line indicates
  the correlation as indicated in the text, and the dotted lines the
  error of the normalisation.}
\label{fig:VLA_VLBI_limits}
\end{figure}

A sample of AGN where we can measure the host galaxy properties
(star-formation rate) independently from the AGN emission provides us
with the opportunity to test the co-evolution of the AGN and their
hosts. There is observational evidence that some co-evolution between
the AGN and the host galaxy exists (e.g., the M-$\sigma$ relation,
\citealt{Ferrarese2000}), and recent observations link host SFRs to
their AGN for different luminosities and redshifts (see
\citealt{Shao2010,Mullaney2012,Rovilos2012,Rosario2012}, but also
\citealt{Page2012}). In Fig.\ref{fig:VLA_VLBI_limits} we plot the
difference between the VLA and the VLBA luminosities (see
Sect.~\ref{sec:redshifts} and Tab.~\ref{tab:redshifts}) as a function
of VLBA luminosity for our sample. There are a number of sources for
which the VLBA luminosity is higher than the VLA luminosity, and we
assume that this is because of variability (see
Sect.~\ref{sec:variability}). These cases have been treated as
follows: in cases where $L_{\rm VLA}+\Delta L_{\rm VLA}<L_{\rm
  VLBA}-\Delta L_{\rm VLBA}$, we assumed that the extended component
of the radio emission seen by the VLA but resolved out by the VLBA was
small. Since some extended emission is likely to be present, we
adopted an upper limit of the extended component of 5\,\% of the VLA
luminosity. In cases where $L_{\rm VLA}<L_{\rm VLBA}$ but $L_{\rm
  VLA}+\Delta L_{\rm VLA}>L_{\rm VLBA}-\Delta L_{\rm VLBA}$, we used
$L_{\rm VLA}+\Delta L_{\rm VLA}-(L_{\rm VLBA}-\Delta L_{\rm VLBA})$ as
an upper limit of the extended component.

One can see from Fig.~\ref{fig:VLA_VLBI_limits} that there is a
positive correlation between the two luminosities, and applying the
generalised Kendall's $\tau$ method using the ASURV package
(Rev.\,1.3, \citealt{Lavalley1992b}), we find that the significance of
the correlation is at the 96.2\% level. The solid and dotted lines
correspond to the equation:

\begin{eqnarray}\nonumber
\log\left(L_{\rm VLA}-L_{\rm VLBA}\right) &=& \\
(0.66&\pm&0.24)\log\left(L_{\rm VLBA}\right)+(7.2\pm1.0)
\end{eqnarray}

(only the standard deviation of the normalisation is shown in the
dotted lines). If we ignore the upper limits, the significance rises
to 97\%. However, one can also see that in a small number of sources
the extended luminosity is larger than $\rm 10^{24.5}\,W\,Hz^{-1}$, so
there might be a contribution from the AGN to the extended flux, from
structures which are resolved out by the VLBA observations. Ignoring
those cases, the correlation becomes even more statistically
significant, at the 99.81\,\% and 99.98\,\% levels, including and
excluding upper limits, respectively.

An issue that needs to be addressed is whether the correlation is
driven by a redshift effect. It is known that the mean star-formation
rate increases with redshift up to z$\approx$2 (see, e.g.,
\citealt{Daddi2007b,Elbaz2011}), and also the VLBA luminosity can
increase with redshift, because the survey is flux
density-limited. Such an effect can mimic an AGN-host correlation (see
\citealt{Mullaney2012}). As a test we determined the correlation in
the flux density domain and found that after excluding the upper
limits the correlation is at the 94.7\,\% level, but its significance
drops to 91.8\,\% if we include the upper limits. We therefore claim
that a tentative correlation exists. This tentative correlation shows
that the most luminous VLBA AGN tend to have higher levels of
star-forming activity and argues against quenching of the
star-formation by AGN feedback when the AGN luminosity reaches a
certain level (e.g., \citealt{Page2012}), at least for the case of
compact radio AGN. This shows that the mechanical power from the AGN,
which is detected at radio wavelengths, is not sufficient to disrupt
the host galaxy and stop the star-formation, a process which may be
attributed to radiation pressure (\citealt{King2005}). On the
contrary, radio AGN activity might even enhance the star-formation
rate, by creating disruptions of the hosts' density profiles
(\cite{Gaibler2012}).

\subsection{Multi-wavelength properties of the detected sources.}
\label{sec:redshifts}

In addition to the basic results from the VLBA survey, significantly
more information can be gathered from the literature, to obtain flux
densities at other wavebands. Therefore the VLA and VLBA observations
were cross-matched to the publicly available multi-band photometry and
photometric redshift catalogues of the Lockman Hole/XMM by
\cite{Fotopoulou2012}. In this catalogue the Large Binocular Telescope
(LBT) images presented in \cite{Rovilos2009} were combined to create a
new catalogue of optically and near-infrared detected sources in the
field. In addition to the $UBV$ bands this multi-wavelength photometry
catalogue includes the PSF-homogenised fixed-aperture photometry from
LBT $Y$ and $z$-band observations, the {\it Subaru} $R$, $I_c$, $z'$
bands from \cite{Barris2004}, the UKIDSS $J$ and $K$ bands, the
3.6\,$\mu$m to 8.0\,$\mu$m photometry from the {\it Spitzer} Wide-area
Infrared Extragalactic survey (SWIRE, \citealt{Lonsdale2003}) and
  GALEX photometry. Except for the near-infrared data, all the other
  bands provide very sensitive data (see Table~2 of
  \citealt{Fotopoulou2012} for the depth of the optical bands). The
  photometric redshift catalogue uses all the above-mentioned bands to
  produce accurate photometric redshifts for normal galaxies and for
  X-ray detected sources by using templates of normal galaxies and AGN
  hybrids, and luminosity and morphological priors to reduce the
  parameter space of possible redshift solutions (more details can be
  found in \citealt{Fotopoulou2012}).

\begin{figure}
\center
\includegraphics[width=\linewidth]{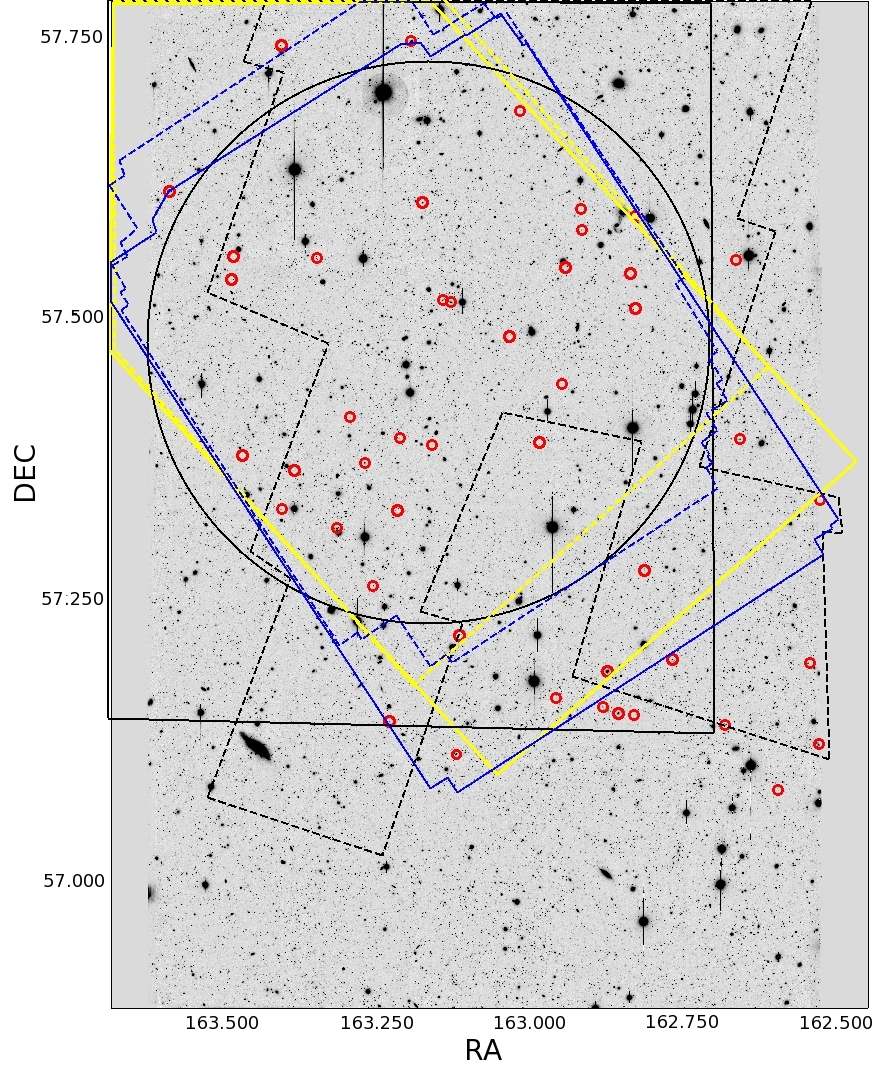}
\caption{VLBA-detected sources with optical counterpart (red
  circles). The multi-wavelength coverage of the field is also shown:
  Subaru $R_c$, $I_c$, $z'$ filters (image), LBT $U$ and $B$ filters
  (blue solid line), LBT $VYz'$ filters (blue dashed line), IRAC
  3.6\,$\mu$m and 5.8\,$\mu$m (yellow solid line), IRAC 4.5\,$\mu$m
  and 8\,$\mu$m (yellow dashed line), SDSS (black dashed line), $J$
  and $K$ UKIDSS filters (black solid lines). The black circle denotes
  the area observed by XMM-Netwon.}
\label{fig:fotopoulou}
\end{figure}

We cross-matched the catalogues\footnote{updated versions of the
  catalogues are available at
  http://www.rzg.mpg.de/~sotiriaf/surveys/LH} to our data by visual
inspection, and by overplotting the locations of the radio, {\it
  Spitzer}/IRAC, and optical positions to ensure that the emission
originated from the same object in each case. Because of the depth of
the available data and the highly accurate astrometry of the VLBA
observations we are confident that a simple match in coordinates
provides the correct counterpart.  It was found that the VLBA position
and the optical position were in very good agreement. The mean
separation between the two positions was 0.28\,arcsec, with a maximum
of 0.88\,arcsec. In a first attempt, optical counterparts were
searched using a 3\,arcsec radius. All sources with separations of
$>1$\,arcsec between the radio and optical positions were found to be
chance coincidences upon closer inspection of the images, and so a
1\,arcsec search radius, followed by visual inspection, was adopted
for cross-identification.  Of our 65 VLBA-detected sources, 10 were
located outside the region covered by the optical data, 47 had
reliable optical counterparts (shown in Fig.~\ref{fig:spectra+rgb}),
and 6 had either faint or uncatalogued counterparts, sometimes because
of blending, or were overlapping with image artifacts such as
diffraction spikes (L0199, L0251, L0386, L0506, L0708, L1148, see
Fig.~\ref{fig:rgb_borked}). The remaining 2 had no identifiable
counterparts at all (L0194, L1364, see Fig.~\ref{fig:rgb_invisible}).

In two objects, L0506 and L1148 (Fig.~\ref{fig:rgb_borked}), the
position of the VLBA source is at the very edge of a clearly visible
optical source, but a cross-identification with these objects was
deemed unlikely. In L0506, a small extension to the optical source can
be seen, indicating that two objects are blending and a reliable
cross-identification can not be made. In L1148 both the VLA and the
VLBA positions are offset from the foreground object by more than
1\,arcsec. The nature of the three sources without any optical
counterpart (Fig.~\ref{fig:rgb_invisible}) is puzzling. Radio sources
with extremely faint or invisble optical or infrared counterparts have
recently gained some attention (see, e.g.,
\citealt{Higdon2005,Middelberg2011b,Norris2011a}), and the speculation
is that they are similar to high-redshift radio galaxies. But our data
do not allow any further conclusions to be drawn.

\begin{figure}
\center
\includegraphics[width=\linewidth]{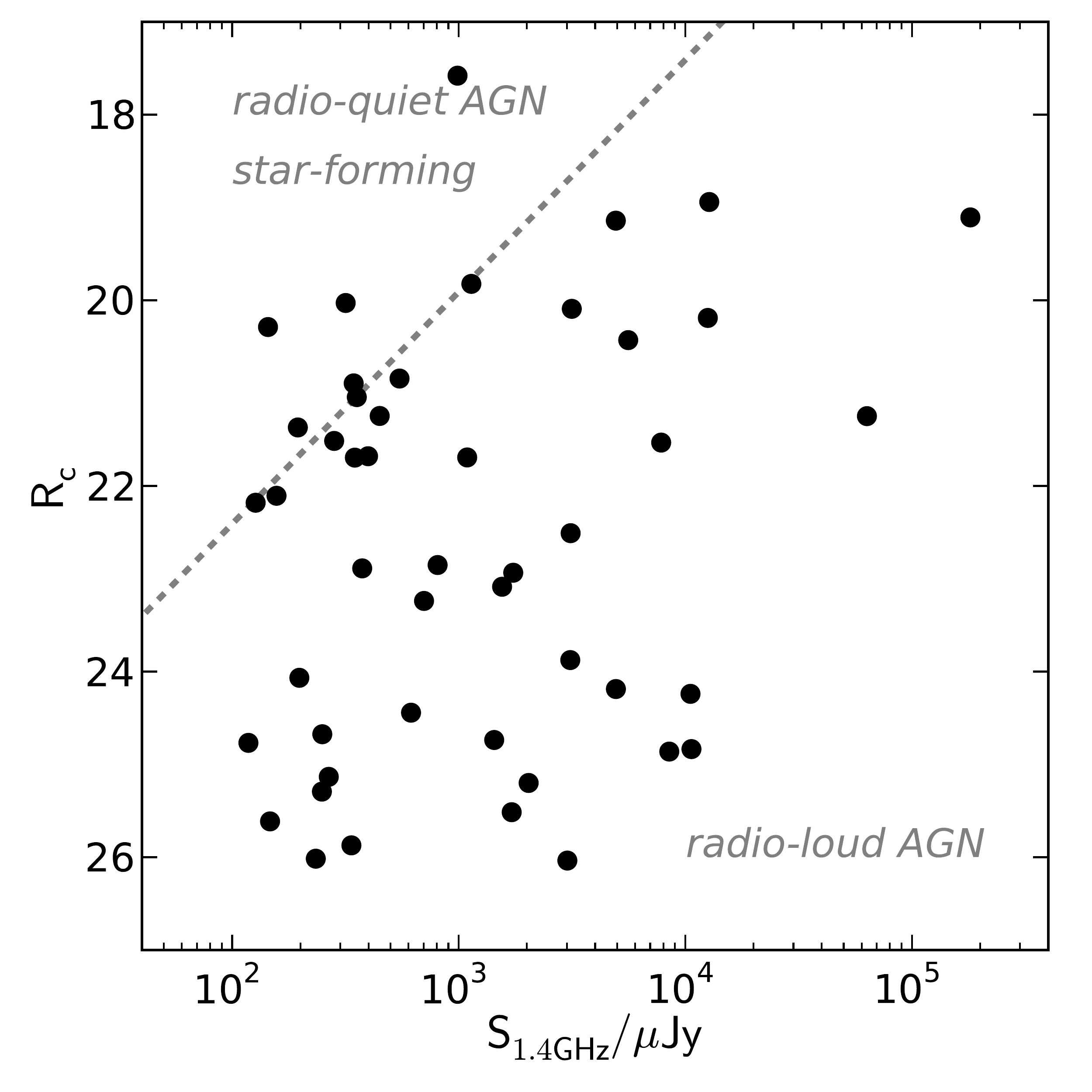}
\caption{$R_c$ magnitude vs. VLA flux density for the 47 sources with
  optical counterparts (for source L0953 the SDSS $r$ magnitude was
  used). The majority of the population is classified as
  radio-loud. The limit for the radio loud population is given by the
  grey dashed line indicating $\rm{\log(S_{\rm VLA}/R_c)=1.4}$.}
\label{fig:Rc_VLA}
\end{figure}

Figure~\ref{fig:Rc_VLA} presents the $R_c$ band distribution of the
optical counterparts as a function of the radio flux density. We also
plot the line of $\log(S_{1.4}/R_c)=1.4$ which, according to
\cite{Padovani2011b}, defines the separation between star-forming
galaxies and radio-quiet AGN in the top-left corner, and radio-loud
AGN in the bottom-right corner. We find that the majority of the
sources are classified as radio-loud (33/47), while 11 out of the 47
are located on the line. Only 3 objects are classified as
star-forming galaxies or radio-quiet AGN.

\begin{figure}
\center
\includegraphics[width=\linewidth]{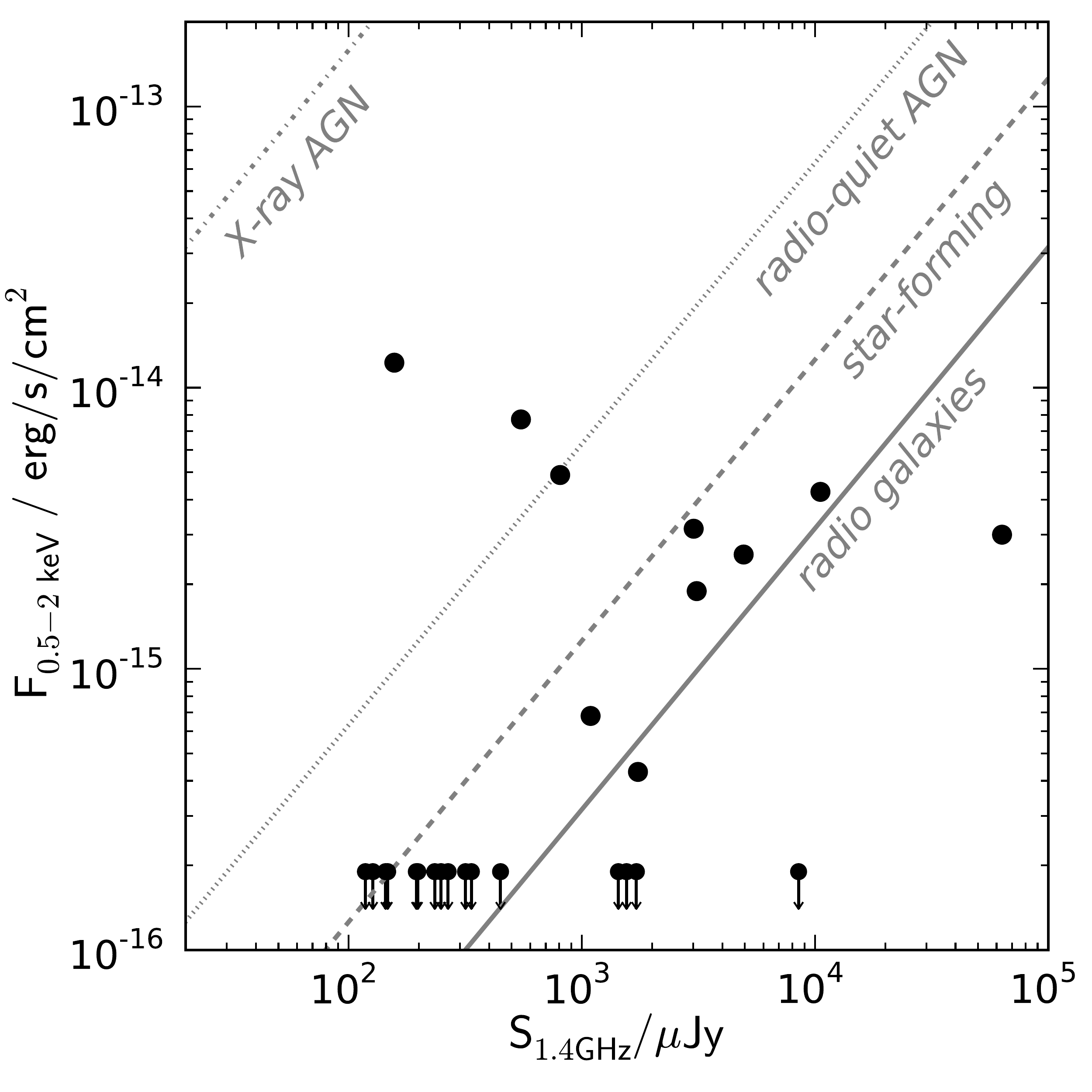}
\caption{X-ray flux versus VLA flux density for the 26 sources inside
  the area covered by XMM-Newton. The lines denote the area occupied
  by X-ray AGN (dash-dotted line), radio-quiet AGN (dotted line),
  star-forming objects (dashed line) and radio galaxies (solid line),
  as presented in \cite{Padovani2011b}. Upper limits are drawn for the
  16 sources below the X-ray flux limit.}
\label{fig:X_VLA}
\end{figure}

In Fig.~\ref{fig:X_VLA}, we plot the X-ray flux versus VLA flux
density for the sources located in the area observed by
XMM-Newton. The lines in the plot show the loci occupied by X-ray AGN
(dash-dotted line), radio-quiet AGN (dotted line), star-forming
objects (dashed line) and radio galaxies (solid line), as presented in
\cite{Padovani2011b}.  Out of the 47 sources with optical counterparts
26 are inside the area observed by XMM-Newton, and only 10 are
detected in the X-rays ($F_{\rm lim, 0.5-2\,keV}=1.9\times
10^{-16}\,{\rm erg/s/cm^2}$). The low number of X-ray dected sources
does not allow one to draw statistically meaningful conclusions about
the population. Roughly we can see that the X-ray sources are
preferentially occupying the locus between star-forming and radio
galaxies. We also include upper limits for the 16 sources inside the
XMM-Netwon area which fall below the X-ray flux limit. These sources
show similar behaviour to the dected sample, with the majority
occupying the area between star-forming and radio galaxies.

In order to classify properly the sources and rank the reliability of
each SED fit, colour images for all 55 sources with optical coverage
were produced. In Fig.~\ref{fig:spectra+rgb} we show all of them
togheter with the best SED fitting and photometric redshift
computation.  The following two subsections present the main result
from this exercise.

\subsubsection{Variety of radio source host galaxy morphologies}

The host galaxies of the detected sources display a great variety in
morphology.  We have attempted to coarsely group them by visual
appearance into 0 -- point-like (as L0227), 1 -- unresolved (as
L0119), 2 -- early type or bulge dominated (as L0449 and L0876,
respectively), 3 -- unclassified (as L0256) and 4 -- spiral (as L1374,
our only clear, though lopsided, spiral).
 
Sources in the ``unresolved'' or ``unclassified'' groups (coded 1 or
3, respectively) are just that -- unresolved, or barely larger than
the point spread function. However, they are mostly very faint, so one
cannot tell if the faint speck that is seen in an image is only the
brightest part of a much larger object. In an expanding universe, the
surface brightness of an object decreases with $(1+z)^4$, a situation
known as the ``Tolman effect'' (\citealt{Tolman1930}). Therefore
extended but faint parts of objects quickly become very difficult to
detect as their redshift increases. Some objects in this category
could also reasonably be assigned to the ``ellipticals'' subset
(L0423, L0997), but judging by the visual appearance this is a matter
of taste.

The ``bulge dominated' group (coded 2) mostly are circular,
extended objects with a more or less pronounced surface brightness
gradient towards the edges. Some have clearly extended halos as one
typically finds in cD galaxies (L0449, L0846), while others have less
pronounced halos (L0535).

In total 21 sources (45\,\%) were classified as
unresolved/unclassified, 23 sources (49\,\%) as bulge-dominated, 2
(4\,\%) as point-like and 1 (2\,\%) as a spiral. To determine if this
is correlated with the selection of VLBA-detected sources we have
selected a control sample of 66 VLBA-undetected radio sources which
were selected to have a similar distribution of VLA flux density as
the VLBA-detected sources. Since most bright sources were detected,
the control sample had a slight deficiency in bright sources and a
slight overabundance of fainter sources. Out of the 66 sources in the
control sample, only 36 had reliable counterparts, but these
counterparts displayed a similar variety in host galaxy types.  Since
22 out of the 25 morphologically classified objects are early-type
objects, we can conclude that the VLBA radio sources are found in the
expected hosts. This also agrees well with distribution of sources in
the $R_c$ vs. VLA flux density diagram in Fig.~\ref{fig:Rc_VLA}, where
the majority of the sources are in the radio-loud regime.

\subsubsection{The SED models of radio source host galaxies}

Using the photometric redshift estimation code
LePhare\footnote{http://www.cfht.hawaii.edu/$\sim$arnouts/LEPHARE/lephare.html},
normal galaxy and AGN templates were fitted to the photometric points
(see \citealt{Ilbert2009,Salvato2009,Salvato2011} for a detailed
description of the fitting procedure). The normal galaxy templates
include extinction values of E(B-V) = 0.00, 0.05, 0.10 - 0.50 (in
steps of 0.1, see \citealt{Fotopoulou2012} for details). Spiral and
starburst templates (Sb-SB3) include extinction according to the Small
Magellanic Cloud (SMC) law given in \cite{Prevot1984}. This law is
also used for the library of AGN and hybrids.  Bluer templates
(SB4-SB11) include extinction laws from \cite{Calzetti2000} and a
modified Calzetti law (\citealt{Ilbert2009}), while templates redder
than Sb (ellipticals and S0) include no further exctinction. In
addition to the best photometric redshift solution LePhare also
provides the full probability distribution function (PDFz), $F(z)=
\exp{(\chi^2_{\rm min}(z)/2)}$. This distribution is calculated for a
predefined redshift range ($0<z<7$ for this work) and is plotted as an
inset in Fig.~\ref{fig:spectra+rgb}. In general, a source with a broad
PDFz does not have a well-constrained redshift solution.

Clearly, the accuracy of the fit is depending on the number,
wavelength and depth of bands available (depending on the redshift of
the source a band is more or less crucial to identify specific
spectral features). The fit also depends on whether the source varied
in flux (typical AGN behaviour) within the time scale of the
observations,and on contamination of photometry from objects within
the aperture used to compute the flux (in this case 3 arcsec
diameter).  We could not identify quantifiable criteria, such as
number of available bands, or reduced $\chi^2$ of the fit, to judge
the quality of the models. Instead, we rated the fits according to the
following guidelines:

\begin{itemize}

\item 1: a bad fit due to a limited number of bands (less than 6) or
  because the source is at the border of the imaged field (see for
  example L0140).

\item 2: a doubtful fit because the photometry is effected by blending
  with nearby objects (see for example L0352).

\item 3: a potential AGN (doubtful fit): the lack of Xray information
  did not allow the use of proper template (see example L0227).

\item 4: the optical photometry is not contaminated and the number of
  bands is sufficient.

\end{itemize}

We do not discuss here sources with a SED fit rated 1 and 2 for
obvious reasons. Only two sources are rated 3 (L0227 and L1374),
classified as potential AGN due to their morphological appearance and
due to the typical power-law seen in the mid-infrared bands. These
objects lie outside the XMM area and thus were treated as normal
galaxies during the photometric redshift estimation. Following the
definition of SEDs as in \cite{Fotopoulou2012} and \cite{Salvato2011}
we classified the 30 sources with a good fit as following: 11 present
a typical SED of a early type galaxy, 6 are best modelled as spirals,
and 12 are classified as starburst (either pure starburst or X-ray
detected but still starburst dominated).  Three of the 12 objects
classified as starburst are clearly early-type objects by visual
appearance. The erroneous SED fit is a clear and well-known case of
degeneracy where a starburst template with heavy extinction (0.3, 0.35
and 0.4 of E(B-V), as visible in the plot of L0397 and L0911 and
L1165) can mimic an early-type template (see also the discussion in
\citealt{Onodera2012}).  In the case of L0973 the image clearly
indicates an early-type the best fit is obtained with the template of
M82. Unlike the other starburst templates that are theoretical, this
one is empirical and already includes dust. For this reason we group
this source with the 3 described above.  For the remaining 8 sources
fit by a starburst template the applied extinction is very high but we
can not prove that this are early type because they are
morphologically unresolved.

%

\section{Conclusions}
\label{sec:conclusions}

The main results from this work are summarised as follows:

\begin{itemize}

\item VLBI observations of 217 sources have been carried out with the
  VLBA. Primary beam corrections have been developed and tested using
  dedicated observations, and a novel multi-source self-calibration
  scheme has improved the coherence of the data significantly. Data
  from overlapping pointings were added to improve the sensitivity in
  the overlap regions (mosaicing). Images were made of all targets,
  and 65 sources were detected, a fraction of 30\,\%. Data from the
  literature were added for interpretation. VLBI observations with
  hundreds of pointings are now feasible and can be added to large
  surveys carried out at other wavelengths.

\item The VLBA-detected sources have larger spectral indices than the
  non-detected sources, which is consistent with their emission coming
  from more compact regions near the AGN. Above a spectral index of
  $\alpha=0.0$, almost all sources (14/17) are detected, confirming
  that spectral index is a strong selection criterion for radio
  AGN. Using the difference between the flux densities measured with
  the VLA and the VLBA as a measure for star forming activity, a
  positive correlation has been found between the star formation and
  AGN activity. This correlation indicates that interactions between
  the radio ejecta and the host galaxy tend to increase, and not
  disrupt, star formation.

\item Source counts of VLBA-detected, faint radio sources are
  presented for the first time. They form a lower limit on the
  fraction of radio AGN among the radio source population. At the
  bright end of the distribution the counts line up with literature
  values, indicating that the majority of mJy sources is
  AGN-powered. At the sub-mJy level, the counts drop below the sub-mJy
  plateau seen at arcsec-scale resolution but seem to exhibit a
  similar flattening. A significant fraction of the sub-mJy, 15\,\% to
  25\,\%, are AGN-driven. However, the high resolution of the VLBA
  observations implies that some AGN-driven radio sources are not
  detected, and so it is expected that the correct fraction of
  AGN-driven sources is higher. 

\item Using literature data we discuss the multiwave-length properties
  of the counterparts to 47 of the 65 VLBA-detected radio sources.
  For about 50\,\% of the sources the morphology suggests the host to
  be an early-type galaxy.  For the rest of the sources the image
  quality and the redshift do not allow to make any assessment of the
  morphology.  In many cases the best fit is obtained with the
  template of a heavily extinct startburst.  This is a well-known case
  of degeneracy in SED fitting where these kinds of templates plus
  extinction mimic an early-type object.  Globally, the typical hosts
  of VLBI sources are passive galaxies, in agreement with previous
  results.

\end{itemize}

With software correlators, accurate primary beam corrections and
multi-source self-calibration, wide-field VLBA observations have
become practical and versatile. The calibration strategies will
certainly improve, and larger bandwidths of future instruments will
make such observations easier because the instantaneous sensitivity
increases.

The significance of wide-field VLBI observations has been acknowledged
by NRAO during the last call for proposals, in accepting a 276\,h
proposal for a sensitive survey of the 2\,deg$^2$ field of the COSMOS
survey with the VLBA, to observe almost 3000 sub-mJy radio
sources. This survey will greatly simplify the data analysis since it
will result in a uniform sensitivity over a large region. Furthermore,
a 200\,h project has been accepted, aiming at carrying out snapshot
observations of more than 10\,000 FIRST sources over 100\,deg$^2$. The
data from these two projects taken together, and adding statistics
from all-sky surveys of VLBI calibrator sources, will allow a
construction of the source counts of VLBI-detected sources from the Jy
to the $\mu$Jy regime.

\begin{acknowledgements}
This work made ample use of Topcat (\citealt{Taylor2005}), written by
Mark Taylor and available at
http://www.star.bris.ac.uk/~mbt/topcat. We wish to thank its author
for support and added features which greatly simplified the data
analysis presented here. We also acknowledge the authors of APLpy, an
open-source plotting package for Python hosted at
http://aplpy.github.com, with which many of the figures shown here
were made. Finally we wish to thank the VLBA staff who greatly
supported the experimental observations in this project, and who
continue to do so. The National Radio Astronomy Observatory, which
operates the VLBA, is a facility of the National Science Foundation
operated under cooperative agreement by Associated Universities, Inc.
\end{acknowledgements}

\section{Appendix A}

\begin{table*}[h!]
\caption{Data of the 217 sources which were in principle detectable ($S_{\rm VLA}>6\times {\rm rms}$).
{\it Col 1:}  source ID used in this paper;
{\it Col 2:}  source designation used by \cite{Ibar2009};
{\it Cols 3,4:}  610\,MHz and 1.4\,GHz flux densities;
{\it Cols 5:}  spectral index calculated from cols. 3,4;
{\it Col 6:}  rms of the VLBA observations (uniform weighting);
{\it Col 7-10:}  peak and integrated flux densities of the VLBI observations and their associated errors (uniform weighting);
{\it Col 11,12:} flux-weighted position of the VLBI source (uniform weighting)
}
\center
\scriptsize
\begin{tabular}{|l|l|r|r|r|r|r|r|r|r|r|r|}
\hline
  \multicolumn{1}{|c|}{ID} &
  \multicolumn{1}{c|}{Name} &
  \multicolumn{1}{c|}{$S_{610}$} &
  \multicolumn{1}{c|}{$S_{\rm VLA}$} &
  \multicolumn{1}{c|}{$\alpha$} &
  \multicolumn{1}{c|}{rms} &
  \multicolumn{1}{c|}{$S_{\rm p}$} &
  \multicolumn{1}{c|}{$\Delta S_{\rm p}$} &
  \multicolumn{1}{c|}{$S_{\rm VLBA}$} &
  \multicolumn{1}{c|}{$\Delta S_{\rm VLBA}$} &
  \multicolumn{1}{c|}{RA} &
  \multicolumn{1}{c|}{Dec} \\

  \multicolumn{1}{|c|}{} &
  \multicolumn{1}{c|}{} &
  \multicolumn{1}{c|}{$\mu$Jy} &
  \multicolumn{1}{c|}{$\mu$Jy} &
  \multicolumn{1}{c|}{} &
  \multicolumn{1}{c|}{$\mu$Jy/beam} &
  \multicolumn{1}{c|}{$\mu$Jy/beam} &
  \multicolumn{1}{c|}{$\mu$Jy/beam} &
  \multicolumn{1}{c|}{$\mu$Jy} &
  \multicolumn{1}{c|}{$\mu$Jy} &
  \multicolumn{1}{c|}{deg} &
  \multicolumn{1}{c|}{deg} \\

 \multicolumn{1}{|c|}{(1)} &
  \multicolumn{1}{c|}{(2)} &
  \multicolumn{1}{c|}{(3)} &
  \multicolumn{1}{c|}{(4)} &
  \multicolumn{1}{c|}{(5)} &
  \multicolumn{1}{c|}{(6)} &
  \multicolumn{1}{c|}{(7)} &
  \multicolumn{1}{c|}{(8)} &
  \multicolumn{1}{c|}{(9)} &
  \multicolumn{1}{c|}{(10)} &
  \multicolumn{1}{c|}{(11)} &
  \multicolumn{1}{c|}{(12)} \\

\hline
  L0003 & LH1.4GHzJ104849.8+571415 & 19386 & 10840 & -0.70 &  &  &  &  &  &  & \\
  L0009 & LH1.4GHzJ104858.3+570925 & 1709 & 1384 & -0.25 & 102 & 462 & 112 & 642 & 121 & 162.242958361 & 57.157160902\\
  L0012 & LH1.4GHzJ104905.1+571151 & 26691 & 14612 & -0.73 &  &  &  &  &  &  & \\
  L0016 & LH1.4GHzJ104914.8+570508 & 1029 & 529 & -0.80 &  &  &  &  &  &  & \\
  L0025 & LH1.4GHzJ104922.9+571901 & 4994 & 2425 & -0.87 & 76 & 939 & 121 & 2209 & 234 & 162.345288947 & 57.317164956\\
  L0034 & LH1.4GHzJ104934.3+570608 &  & 14787 &  & 74 & 323 & 81 & 295 & 80 & 162.393862805 & 57.103353297\\
  L0049 & LH1.4GHzJ104943.8+571737 & 1008 & 1013 & 0.01 & 52 & 541 & 75 & 666 & 84 & 162.432542797 & 57.29366112\\
  L0051 & LH1.4GHzJ104944.1+570628 & 1823 & 1275 & -0.43 & 66 & 528 & 85 & 916 & 113 & 162.433765316 & 57.108035949\\
  L0055 & LH1.4GHzJ104947.8+571354 & 689 & 607 & -0.15 & 54 & 423 & 69 & 490 & 73 & 162.4489331 & 57.231806689\\
  L0061 & LH1.4GHzJ104950.4+570120c &  & 5266 &  &  &  &  &  &  &  & \\
  L0066 & LH1.4GHzJ104951.3+572812 & 967 & 1083 & 0.14 & 75 & 713 & 104 & 922 & 119 & 162.463891893 & 57.470082634\\
  L0077 & LH1.4GHzJ104954.2+570456 & 1363 & 1376 & 0.01 & 66 & 541 & 86 & 633 & 91 & 162.475858493 & 57.082238326\\
  L0081 & LH1.4GHzJ104956.7+570647 & 596 & 335 & -0.69 &  &  &  &  &  &  & \\
  L0082 & LH1.4GHzJ104956.8+571042 & 2053 & 1046 & -0.81 &  &  &  &  &  &  & \\
  L0087 & LH1.4GHzJ104958.0+570727 & 1082 & 650 & -0.61 &  &  &  &  &  &  & \\
  L0100 & LH1.4GHzJ105002.0+571622 & 213 & 232 & 0.10 &  &  &  &  &  &  & \\
  L0117 & LH1.4GHzJ105007.3+571652 & 452 & 257 & -0.68 &  &  &  &  &  &  & \\
  L0119 & LH1.4GHzJ105008.1+572018 & 3486 & 2035 & -0.65 & 43 & 647 & 78 & 1195 & 127 & 162.533539506 & 57.338390632\\
  L0130 & LH1.4GHzJ105010.4+570724b & 1876 & 5598 & 1.32 & 54 & 1291 & 141 & 2335 & 240 & 162.538743032 & 57.121401186\\
  L0140 & LH1.4GHzJ105012.6+571137 & 1280 & 616 & -0.88 & 42 & 235 & 48 & 415 & 59 & 162.552329219 & 57.193600719\\
  L0144 & LH1.4GHzJ105014.4+572844 & 689 & 416 & -0.61 &  &  &  &  &  &  & \\
  L0149 & LH1.4GHzJ105015.6+570258 & 5060 & 3155 & -0.57 & 64 & 687 & 94 & 983 & 117 & 162.564968684 & 57.048810568\\
  L0160 & LH1.4GHzJ105019.7+572812 & 662 & 266 & -1.10 &  &  &  &  &  &  & \\
  L0165 & LH1.4GHzJ105023.4+572439 & 485 & 253 & -0.78 &  &  &  &  &  &  & \\
  L0171 & LH1.4GHzJ105025.5+570453 & 266 & 250 & -0.07 & 55 & 275 & 62 & 271 & 61 & 162.606338811 & 57.081508924\\
  L0173 & LH1.4GHzJ105026.3+570744 & 1270 & 793 & -0.57 &  &  &  &  &  &  & \\
  L0181 & LH1.4GHzJ105029.2+571427 & 2450 & 1286 & -0.78 &  &  &  &  &  &  & \\
  L0183 & LH1.4GHzJ105029.6+571036 & 417 & 203 & -0.87 &  &  &  &  &  &  & \\
  L0189 & LH1.4GHzJ105032.0+572027 & 5184 & 723 & -2.37 &  &  &  &  &  &  & \\
  L0194 & LH1.4GHzJ105032.7+572646 & 371 & 466 & 0.27 & 41 & 364 & 55 & 497 & 64 & 162.636123187 & 57.446286836\\
  L0197 & LH1.4GHzJ105033.1+571703 & 388 & 169 & -1.00 &  &  &  &  &  &  & \\
  L0199 & LH1.4GHzJ105034.0+572152 & 1830 & 1005 & -0.72 & 36 & 214 & 42 & 220 & 42 & 162.641811126 & 57.364553024\\
  L0201 & LH1.4GHzJ105034.2+572922 & 712 & 336 & -0.90 &  &  &  &  &  &  & \\
  L0211 & LH1.4GHzJ105035.9+572318 & 4128 & 1768 & -1.02 &  &  &  &  &  &  & \\
  L0217 & LH1.4GHzJ105037.6+572844 & 701 & 420 & -0.62 &  &  &  &  &  &  & \\
  L0225 & LH1.4GHzJ105039.1+565806 & 2554 & 1520 & -0.62 &  &  &  &  &  &  & \\
  L0227 & LH1.4GHzJ105039.6+572336 & 4326 & 4936 & 0.16 & 36 & 2687 & 272 & 3972 & 399 & 162.664768172 & 57.393502619\\
  L0231 & LH1.4GHzJ105040.7+573308 & 959 & 703 & -0.37 & 54 & 740 & 92 & 984 & 112 & 162.669491607 & 57.552292637\\
  L0234 & LH1.4GHzJ105042.0+570706 & 1516 & 770 & -0.82 &  &  &  &  &  &  & \\
  L0251 & LH1.4GHzJ105045.2+573734 & 1286 & 858 & -0.49 & 76 & 348 & 84 & 290 & 81 & 162.688439094 & 57.626114051\\
  L0256 & LH1.4GHzJ105045.9+570822 & 430 & 344 & -0.27 & 39 & 164 & 42 & 153 & 42 & 162.691253621 & 57.139547122\\
  L0263 & LH1.4GHzJ105048.2+571614 & 336 & 165 & -0.86 &  &  &  &  &  &  & \\
  L0273 & LH1.4GHzJ105050.5+570715 & 856 & 386 & -0.96 &  &  &  &  &  &  & \\
  L0284 & LH1.4GHzJ105053.4+572426 & 568 & 279 & -0.86 &  &  &  &  &  &  & \\
  L0290 & LH1.4GHzJ105054.6+570810 & 841 & 344 & -1.08 &  &  &  &  &  &  & \\
  L0301 & LH1.4GHzJ105056.6+571532 & 1542 & 700 & -0.95 &  &  &  &  &  &  & \\
  L0302 & LH1.4GHzJ105056.6+571631 & 557 & 324 & -0.65 &  &  &  &  &  &  & \\
  L0320 & LH1.4GHzJ105100.9+572036 & 751 & 322 & -1.02 &  &  &  &  &  &  & \\
  L0328 & LH1.4GHzJ105101.9+573445 & 3371 & 1332 & -1.12 &  &  &  &  &  &  & \\
  L0339 & LH1.4GHzJ105104.9+572719 & 348 & 153 & -0.99 &  &  &  &  &  &  & \\
  L0342 & LH1.4GHzJ105104.7+570150b &  & 4050 &  &  &  &  &  &  &  & \\
  L0346 & LH1.4GHzJ105105.8+571057 & 699 & 421 & -0.61 &  &  &  &  &  &  & \\
  L0352 & LH1.4GHzJ105106.7+571152 & 255 & 355 & 0.40 & 32 & 163 & 36 & 282 & 43 & 162.777939135 & 57.197994694\\
  L0354 & LH1.4GHzJ105107.7+570205 &  & 314 &  &  &  &  &  &  &  & \\
  L0380 & LH1.4GHzJ105113.2+574156 & 2490 & 1152 & -0.93 &  &  &  &  &  &  & \\
  L0381 & LH1.4GHzJ105113.4+571426 & 1455 & 763 & -0.78 &  &  &  &  &  &  & \\
  L0382 & LH1.4GHzJ105113.5+572654 & 440 & 283 & -0.53 &  &  &  &  &  &  & \\
  L0386 & LH1.4GHzJ105115.0+573552 & 4841 & 2362 & -0.86 & 47 & 168 & 50 & 133 & 49 & 162.812383572 & 57.597906246\\
  L0397 & LH1.4GHzJ105117.6+571639 & 552 & 398 & -0.39 & 27 & 265 & 38 & 411 & 49 & 162.823463843 & 57.277592149\\
  L0399 & LH1.4GHzJ105118.0+571920 & 218 & 131 & -0.61 &  &  &  &  &  &  & \\
  L0408 & LH1.4GHzJ105120.0+570954 & 1209 & 601 & -0.84 &  &  &  &  &  &  & \\
  L0409 & LH1.4GHzJ105120.4+572253 & 399 & 238 & -0.62 &  &  &  &  &  &  & \\
  L0411 & LH1.4GHzJ105120.8+573037 & 439 & 336 & -0.32 & 32 & 205 & 38 & 294 & 44 & 162.83673954 & 57.510376172\\
  L0412 & LH1.4GHzJ105120.9+573532 & 9061 & 8496 & -0.08 & 45 & 4679 & 472 & 8966 & 898 & 162.836886524 & 57.592367389\\
  L0416 & LH1.4GHzJ105121.5+573613 &  & 563 &  &  &  &  &  &  &  & \\
  L0420 & LH1.4GHzJ105122.1+570854 & 18706 & 10644 & -0.68 & 36 & 2080 & 212 & 10562 & 921 & 162.841971832 & 57.148560655\\
  L0423 & LH1.4GHzJ105122.9+573228 & 1928 & 1434 & -0.36 & 34 & 642 & 73 & 1343 & 138 & 162.845349839 & 57.541242785\\
  L0427 & LH1.4GHzJ105123.6+570849 & 697 & 459 & -0.50 &  &  &  &  &  &  & \\
  L0437 & LH1.4GHzJ105125.7+573544 & 945 & 577 & -0.59 &  &  &  &  &  &  & \\
  L0447 & LH1.4GHzJ105127.5+573620 &  & 362 &  &  &  &  &  &  &  & \\
  L0449 & LH1.4GHzJ105127.8+570854 &  & 12570 &  & 36 & 337 & 49 & 506 & 62 & 162.86695653 & 57.150285403\\
  L0452 & LH1.4GHzJ105128.1+573502 & 585 & 361 & -0.58 &  &  &  &  &  &  & \\
  L0465 & LH1.4GHzJ105130.5+573808 & 444 & 259 & -0.65 &  &  &  &  &  &  & \\
  L0468 & LH1.4GHzJ105130.5+574407 &  & 924 &  &  &  &  &  &  &  & \\
  L0477 & LH1.4GHzJ105132.4+571114 & 22200 & 12756 & -0.67 & 32 & 823 & 89 & 1157 & 120 & 162.884787347 & 57.18764161\\
  L0484 & LH1.4GHzJ105134.1+570922 & 177 & 348 & 0.81 & 34 & 261 & 43 & 278 & 44 & 162.892178451 & 57.156140392\\
  L0492 & LH1.4GHzJ105135.5+572739 & 324 & 182 & -0.69 &  &  &  &  &  &  & \\
  L0498 & LH1.4GHzJ105136.1+572959 & 605 & 302 & -0.84 &  &  &  &  &  &  & \\
  L0500 & LH1.4GHzJ105136.2+573302 & 955 & 537 & -0.69 &  &  &  &  &  &  & \\
\hline\end{tabular}
\label{tab:results1}
\end{table*}

\begin{table*}[h!]
\ContinuedFloat
\caption{(Continued)}
\center
\scriptsize
\begin{tabular}{|l|l|r|r|r|r|r|r|r|r|r|r|}
\hline
  \multicolumn{1}{|c|}{ID} &
  \multicolumn{1}{c|}{Name} &
  \multicolumn{1}{c|}{$S_{610}$} &
  \multicolumn{1}{c|}{$S_{\rm VLA}$} &
  \multicolumn{1}{c|}{$\alpha$} &
  \multicolumn{1}{c|}{rms} &
  \multicolumn{1}{c|}{$S_{\rm p}$} &
  \multicolumn{1}{c|}{$\Delta S_{\rm p}$} &
  \multicolumn{1}{c|}{$S_{\rm VLBA}$} &
  \multicolumn{1}{c|}{$\Delta S_{\rm VLBA}$} &
  \multicolumn{1}{c|}{RA} &
  \multicolumn{1}{c|}{Dec} \\

  \multicolumn{1}{|c|}{} &
  \multicolumn{1}{c|}{} &
  \multicolumn{1}{c|}{$\mu$Jy} &
  \multicolumn{1}{c|}{$\mu$Jy} &
  \multicolumn{1}{c|}{} &
  \multicolumn{1}{c|}{$\mu$Jy/beam} &
  \multicolumn{1}{c|}{$\mu$Jy/beam} &
  \multicolumn{1}{c|}{$\mu$Jy/beam} &
  \multicolumn{1}{c|}{$\mu$Jy} &
  \multicolumn{1}{c|}{$\mu$Jy} &
  \multicolumn{1}{c|}{deg} &
  \multicolumn{1}{c|}{deg} \\

 \multicolumn{1}{|c|}{(1)} &
  \multicolumn{1}{c|}{(2)} &
  \multicolumn{1}{c|}{(3)} &
  \multicolumn{1}{c|}{(4)} &
  \multicolumn{1}{c|}{(5)} &
  \multicolumn{1}{c|}{(6)} &
  \multicolumn{1}{c|}{(7)} &
  \multicolumn{1}{c|}{(8)} &
  \multicolumn{1}{c|}{(9)} &
  \multicolumn{1}{c|}{(10)} &
  \multicolumn{1}{c|}{(11)} &
  \multicolumn{1}{c|}{(12)} \\

\hline
  L0502 & LH1.4GHzJ105136.4+570129 & 1623 & 803 & -0.85 &  &  &  &  &  &  & \\
  L0505 & LH1.4GHzJ105136.9+573751 & 1129 & 514 & -0.95 &  &  &  &  &  &  & \\
  L0506 & LH1.4GHzJ105137.0+572940 & 4179 & 2532 & -0.60 & 28 & 295 & 41 & 402 & 49 & 162.904118977 & 57.494500938\\
  L0507 & LH1.4GHzJ105137.7+574454 & 2746 & 1036 & -1.17 &  &  &  &  &  &  & \\
  L0513 & LH1.4GHzJ105139.1+573043 &  & 441 &  &  &  &  &  &  &  & \\
  L0529 & LH1.4GHzJ105141.4+571951 & 626 & 372 & -0.63 &  &  &  &  &  &  & \\
  L0534 & LH1.4GHzJ105142.0+571501 & 392 & 201 & -0.80 &  &  &  &  &  &  & \\
  L0535 & LH1.4GHzJ105142.0+573447 & 483 & 1090 & 0.98 & 35 & 885 & 96 & 1168 & 122 & 162.925189291 & 57.579945389\\
  L0536 & LH1.4GHzJ105142.1+573554 &  & 4940 &  & 38 & 982 & 106 & 1243 & 130 & 162.925392256 & 57.59848336\\
  L0548 & LH1.4GHzJ105143.7+572937 & 891 & 432 & -0.87 &  &  &  &  &  &  & \\
  L0553 & LH1.4GHzJ105144.5+570555 &  & 300 &  &  &  &  &  &  &  & \\
  L0555 & LH1.4GHzJ105144.5+571719 & 294 & 150 & -0.81 &  &  &  &  &  &  & \\
  L0578 & LH1.4GHzJ105148.7+573248 & 835 & 807 & -0.04 & 30 & 747 & 81 & 944 & 99 & 162.952962275 & 57.546820439\\
  L0593 & LH1.4GHzJ105150.1+572635 & 177 & 118 & -0.49 & 24 & 246 & 34 & 272 & 36 & 162.958786403 & 57.443275605\\
  L0595 & LH1.4GHzJ105150.3+573244 & 35615 & 16418 & -0.93 &  &  &  &  &  &  & \\
  L0603 & LH1.4GHzJ105151.7+572636 & 238 & 127 & -0.76 &  &  &  &  &  &  & \\
  L0604 & LH1.4GHzJ105152.0+570907 & 305 & 208 & -0.46 &  &  &  &  &  &  & \\
  L0607 & LH1.4GHzJ105152.4+570950 & 4951 & 3118 & -0.56 & 34 & 1005 & 107 & 2965 & 216 & 162.968238773 & 57.163967637\\
  L0644 & LH1.4GHzJ105158.3+570913 & 372 & 214 & -0.67 &  &  &  &  &  &  & \\
  L0648 & LH1.4GHzJ105158.9+572330 & 121 & 234 & 0.79 & 23 & 177 & 29 & 304 & 38 & 162.9954889 & 57.391705693\\
  L0653 & LH1.4GHzJ105159.4+570013b &  & 2059 &  &  &  &  &  &  &  & \\
  L0665 & LH1.4GHzJ105201.1+565947 & 2181 & 480 & -1.82 &  &  &  &  &  &  & \\
  L0667 & LH1.4GHzJ105201.3+572445 & 104 & 120 & 0.17 &  &  &  &  &  &  & \\
  L0669 & LH1.4GHzJ105201.9+574051 & 611 & 260 & -1.03 &  &  &  &  &  &  & \\
  L0679 & LH1.4GHzJ105204.2+572655 & 195 & 121 & -0.57 &  &  &  &  &  &  & \\
  L0687 & LH1.4GHzJ105205.0+572917 & 250 & 130 & -0.79 &  &  &  &  &  &  & \\
  L0691 & LH1.4GHzJ105205.5+571710 & 115 & 141 & 0.25 &  &  &  &  &  &  & \\
  L0698 & LH1.4GHzJ105206.1+565704 &  & 2136 &  &  &  &  &  &  &  & \\
  L0703 & LH1.4GHzJ105206.5+574109 & 13423 & 10542 & -0.29 & 50 & 2828 & 289 & 8420 & 555 & 163.026882769 & 57.686063698\\
  L0705 & LH1.4GHzJ105207.2+570745 & 560 & 287 & -0.80 &  &  &  &  &  &  & \\
  L0708 & LH1.4GHzJ105207.5+571903 &  & 324 &  & 24 & 275 & 37 & 269 & 36 & 163.031206078 & 57.31778852\\
  L0722 & LH1.4GHzJ105210.6+571203 & 289 & 172 & -0.62 &  &  &  &  &  &  & \\
  L0724 & LH1.4GHzJ105211.0+572908 & 2917 & 1713 & -0.64 & 24 & 1064 & 110 & 1585 & 160 & 163.045937557 & 57.485563316\\
  L0736 & LH1.4GHzJ105212.5+572453 & 466 & 278 & -0.62 &  &  &  &  &  &  & \\
  L0739 & LH1.4GHzJ105213.4+571605 & 574 & 301 & -0.78 &  &  &  &  &  &  & \\
  L0750 & LH1.4GHzJ105215.0+572635 & 299 & 161 & -0.75 &  &  &  &  &  &  & \\
  L0760 & LH1.4GHzJ105216.6+573530 & 296 & 194 & -0.51 &  &  &  &  &  &  & \\
  L0763 & LH1.4GHzJ105216.9+572017 & 215 & 126 & -0.64 &  &  &  &  &  &  & \\
  L0767 & LH1.4GHzJ105217.6+572127 & 445 & 216 & -0.87 &  &  &  &  &  &  & \\
  L0776 & LH1.4GHzJ105219.1+571857 & 219 & 123 & -0.69 &  &  &  &  &  &  & \\
  L0800 & LH1.4GHzJ105224.5+570838 & 3944 & 2639 & -0.48 &  &  &  &  &  &  & \\
  L0805 & LH1.4GHzJ105225.3+571131 & 275 & 174 & -0.55 &  &  &  &  &  &  & \\
  L0809 & LH1.4GHzJ105225.6+573322 & 8663 & 4873 & -0.69 &  &  &  &  &  &  & \\
  L0810 & LH1.4GHzJ105225.8+570152 & 975 & 696 & -0.41 &  &  &  &  &  &  & \\
  L0824 & LH1.4GHzJ105227.4+571415 & 268 & 154 & -0.67 &  &  &  &  &  &  & \\
  L0838 & LH1.4GHzJ105229.1+571242 & 236 & 189 & -0.27 &  &  &  &  &  &  & \\
  L0845 & LH1.4GHzJ105230.5+570853 & 551 & 314 & -0.68 &  &  &  &  &  &  & \\
  L0846 & LH1.4GHzJ105230.6+571312 & 373 & 282 & -0.34 & 30 & 198 & 36 & 291 & 42 & 163.12756064 & 57.220099911\\
  L0851 & LH1.4GHzJ105231.1+573501 & 278 & 233 & -0.21 &  &  &  &  &  &  & \\
  L0860 & LH1.4GHzJ105231.8+570650 & 539 & 1137 & 0.90 & 49 & 582 & 76 & 1015 & 113 & 163.132553769 & 57.114020856\\
  L0862 & LH1.4GHzJ105232.3+572447 & 293 & 234 & -0.27 &  &  &  &  &  &  & \\
  L0867 & LH1.4GHzJ105232.4+570842 & 565 & 212 & -1.18 &  &  &  &  &  &  & \\
  L0870 & LH1.4GHzJ105232.9+572542 & 320 & 148 & -0.93 &  &  &  &  &  &  & \\
  L0876 & LH1.4GHzJ105234.0+573057 &  & 144 &  & 25 & 113 & 28 & 143 & 29 & 163.141622298 & 57.515886497\\
  L0879 & LH1.4GHzJ105234.9+572641 & 255 & 134 & -0.77 &  &  &  &  &  &  & \\
  L0882 & LH1.4GHzJ105235.4+572652 & 484 & 236 & -0.86 &  &  &  &  &  &  & \\
  L0890 & LH1.4GHzJ105237.3+573103 &  & 63323 &  & 25 & 551 & 61 & 607 & 66 & 163.15566438 & 57.517692521\\
  L0891 & LH1.4GHzJ105237.4+572148 & 280 & 175 & -0.57 &  &  &  &  &  &  & \\
  L0902 & LH1.4GHzJ105239.6+572431 & 257 & 142 & -0.71 &  &  &  &  &  &  & \\
  L0903 & LH1.4GHzJ105239.7+573054 & 227 & 139 & -0.59 &  &  &  &  &  &  & \\
  L0911 & LH1.4GHzJ105241.4+572320 & 2435 & 1740 & -0.40 & 23 & 160 & 28 & 184 & 30 & 163.172665445 & 57.389069831\\
  L0917 & LH1.4GHzJ105242.4+572444 & 342 & 267 & -0.30 &  &  &  &  &  &  & \\
  L0918 & LH1.4GHzJ105242.4+571914 & 602 & 349 & -0.66 &  &  &  &  &  &  & \\
  L0927 & LH1.4GHzJ105243.3+574813 & 1013 & 1857 & 0.73 & 100 & 1313 & 166 & 1237 & 159 & 163.180471707 & 57.803888368\\
  L0933 & LH1.4GHzJ105245.3+573616 & 619 & 448 & -0.39 & 31 & 261 & 41 & 267 & 41 & 163.188933432 & 57.604521945\\
  L0953 & LH1.4GHzJ105250.0+574450 & 1682 & 988 & -0.64 & 65 & 334 & 73 & 309 & 72 & 163.208413838 & 57.747505888\\
  L0966 & LH1.4GHzJ105252.8+570753 & 553 & 292 & -0.77 &  &  &  &  &  &  & \\
  L0968 & LH1.4GHzJ105252.8+572859 & 538 & 212 & -1.12 &  &  &  &  &  &  & \\
  L0973 & LH1.4GHzJ105254.2+572341 & 170 & 157 & -0.10 & 24 & 143 & 28 & 140 & 28 & 163.226115327 & 57.39499839\\
  L0975 & LH1.4GHzJ105255.1+571944 &  & 140 &  &  &  &  &  &  &  & \\
  L0977 & LH1.4GHzJ105255.3+571950 & 4514 & 3107 & -0.45 & 26 & 1480 & 151 & 2921 & 293 & 163.23051676 & 57.330690586\\
  L0987 & LH1.4GHzJ105256.8+570825 & 910 & 522 & -0.67 &  &  &  &  &  &  & \\
  L0997 & LH1.4GHzJ105258.0+570834 & 529 & 375 & -0.41 & 52 & 284 & 59 & 471 & 70 & 163.241765558 & 57.14302443\\
  L1007 & LH1.4GHzJ105259.3+573226 & 298 & 232 & -0.30 &  &  &  &  &  &  & \\
  L1020 & LH1.4GHzJ105301.3+570543 & 1697 & 945 & -0.70 &  &  &  &  &  &  & \\
  L1022 & LH1.4GHzJ105301.7+572520 & 361 & 257 & -0.41 &  &  &  &  &  &  & \\
  L1025 & LH1.4GHzJ105302.6+571813 & 209 & 143 & -0.46 &  &  &  &  &  &  & \\
  L1028 & LH1.4GHzJ105303.4+573527 &  & 210 &  &  &  &  &  &  &  & \\
  L1029 & LH1.4GHzJ105303.7+571205 & 627 & 417 & -0.49 &  &  &  &  &  &  & \\
  L1031 & LH1.4GHzJ105303.9+573532 &  & 323 &  &  &  &  &  &  &  & \\
\hline\end{tabular}
\end{table*}

\begin{table*}[h!]
\ContinuedFloat
\caption{(Continued)}
\center
\scriptsize
\begin{tabular}{|l|l|r|r|r|r|r|r|r|r|r|r|}
\hline
  \multicolumn{1}{|c|}{ID} &
  \multicolumn{1}{c|}{Name} &
  \multicolumn{1}{c|}{$S_{610}$} &
  \multicolumn{1}{c|}{$S_{\rm VLA}$} &
  \multicolumn{1}{c|}{$\alpha$} &
  \multicolumn{1}{c|}{rms} &
  \multicolumn{1}{c|}{$S_{\rm p}$} &
  \multicolumn{1}{c|}{$\Delta S_{\rm p}$} &
  \multicolumn{1}{c|}{$S_{\rm VLBA}$} &
  \multicolumn{1}{c|}{$\Delta S_{\rm VLBA}$} &
  \multicolumn{1}{c|}{RA} &
  \multicolumn{1}{c|}{Dec} \\

  \multicolumn{1}{|c|}{} &
  \multicolumn{1}{c|}{} &
  \multicolumn{1}{c|}{$\mu$Jy} &
  \multicolumn{1}{c|}{$\mu$Jy} &
  \multicolumn{1}{c|}{} &
  \multicolumn{1}{c|}{$\mu$Jy/beam} &
  \multicolumn{1}{c|}{$\mu$Jy/beam} &
  \multicolumn{1}{c|}{$\mu$Jy/beam} &
  \multicolumn{1}{c|}{$\mu$Jy} &
  \multicolumn{1}{c|}{$\mu$Jy} &
  \multicolumn{1}{c|}{deg} &
  \multicolumn{1}{c|}{deg} \\

 \multicolumn{1}{|c|}{(1)} &
  \multicolumn{1}{c|}{(2)} &
  \multicolumn{1}{c|}{(3)} &
  \multicolumn{1}{c|}{(4)} &
  \multicolumn{1}{c|}{(5)} &
  \multicolumn{1}{c|}{(6)} &
  \multicolumn{1}{c|}{(7)} &
  \multicolumn{1}{c|}{(8)} &
  \multicolumn{1}{c|}{(9)} &
  \multicolumn{1}{c|}{(10)} &
  \multicolumn{1}{c|}{(11)} &
  \multicolumn{1}{c|}{(12)} \\

\hline
  L1034 & LH1.4GHzJ105304.5+571547 & 143 & 198 & 0.39 & 30 & 183 & 35 & 155 & 34 & 163.268708013 & 57.263232348\\
  L1036 & LH1.4GHzJ105304.8+573055 & 755 & 501 & -0.49 &  &  &  &  &  &  & \\
  L1040 & LH1.4GHzJ105305.3+572330 & 192 & 147 & -0.32 &  &  &  &  &  &  & \\
  L1052 & LH1.4GHzJ105308.1+572222 & 355 & 317 & -0.14 & 25 & 127 & 28 & 145 & 29 & 163.283666617 & 57.37303887\\
  L1059 & LH1.4GHzJ105309.3+571659 &  & 462 &  &  &  &  &  &  &  & \\
  L1061 & LH1.4GHzJ105309.5+573712 & 416 & 294 & -0.42 &  &  &  &  &  &  & \\
  L1062 & LH1.4GHzJ105309.5+570636 & 887 & 585 & -0.50 &  &  &  &  &  &  & \\
  L1067 & LH1.4GHzJ105310.6+573435 & 316 & 190 & -0.61 &  &  &  &  &  &  & \\
  L1077 & LH1.4GHzJ105312.2+571105 & 947 & 420 & -0.98 &  &  &  &  &  &  & \\
  L1080 & LH1.4GHzJ105312.7+573111 & 822 & 429 & -0.78 &  &  &  &  &  &  & \\
  L1092 & LH1.4GHzJ105314.0+572448 &  & 127 &  & 25 & 174 & 30 & 142 & 29 & 163.308246293 & 57.413486699\\
  L1094 & LH1.4GHzJ105314.2+573020 & 725 & 345 & -0.89 &  &  &  &  &  &  & \\
  L1106 & LH1.4GHzJ105316.8+573550 & 492 & 294 & -0.62 &  &  &  &  &  &  & \\
  L1108 & LH1.4GHzJ105317.4+572722 & 185 & 132 & -0.41 &  &  &  &  &  &  & \\
  L1112 & LH1.4GHzJ105318.8+574546 &  & 531 &  &  &  &  &  &  &  & \\
  L1113 & LH1.4GHzJ105318.9+572140 & 767 & 447 & -0.65 &  &  &  &  &  &  & \\
  L1115 & LH1.4GHzJ105319.0+571851 & 908 & 548 & -0.61 & 29 & 144 & 32 & 229 & 37 & 163.329236431 & 57.314387781\\
  L1132 & LH1.4GHzJ105322.4+573652 & 322 & 192 & -0.62 &  &  &  &  &  &  & \\
  L1136 & LH1.4GHzJ105322.8+571500 & 597 & 509 & -0.19 &  &  &  &  &  &  & \\
  L1139 & LH1.4GHzJ105323.5+571733 & 305 & 185 & -0.60 &  &  &  &  &  &  & \\
  L1144 & LH1.4GHzJ105324.6+571658 & 975 & 567 & -0.65 &  &  &  &  &  &  & \\
  L1148 & LH1.4GHzJ105325.3+572911 & 1031 & 602 & -0.65 & 26 & 122 & 29 & 199 & 33 & 163.355459276 & 57.486609371\\
  L1155 & LH1.4GHzJ105326.7+571405 & 527 & 325 & -0.58 &  &  &  &  &  &  & \\
  L1165 & LH1.4GHzJ105327.5+573316 &  & 195 &  & 29 & 260 & 39 & 219 & 36 & 163.364535455 & 57.554736564\\
  L1166 & LH1.4GHzJ105327.6+574543 & 12354 & 6618 & -0.75 &  &  &  &  &  &  & \\
  L1168 & LH1.4GHzJ105328.0+571115 & 685 & 393 & -0.67 &  &  &  &  &  &  & \\
  L1197 & LH1.4GHzJ105335.2+572921 & 578 & 373 & -0.53 &  &  &  &  &  &  & \\
  L1200 & LH1.4GHzJ105335.8+572157 & 88 & 147 & 0.62 & 30 & 194 & 36 & 226 & 38 & 163.399383286 & 57.365923937\\
  L1204 & LH1.4GHzJ105337.1+574315 &  & 354 &  &  &  &  &  &  &  & \\
  L1206 & LH1.4GHzJ105337.3+574240 & 1432 & 1157 & -0.26 &  &  &  &  &  &  & \\
  L1210 & LH1.4GHzJ105338.2+574211 & 702 & 365 & -0.79 &  &  &  &  &  &  & \\
  L1225 & LH1.4GHzJ105340.9+571952 & 2332 & 1554 & -0.49 & 34 & 262 & 43 & 283 & 44 & 163.420225248 & 57.331410611\\
  L1227 & LH1.4GHzJ105341.2+571920 & 262 & 163 & -0.57 &  &  &  &  &  &  & \\
  L1234 & LH1.4GHzJ105342.1+573026 & 765 & 346 & -0.96 &  &  &  &  &  &  & \\
  L1235 & LH1.4GHzJ105342.1+574436 & 12655 & 7825 & -0.58 & 72 & 755 & 105 & 2152 & 194 & 163.425629443 & 57.743542688\\
  L1239 & LH1.4GHzJ105343.2+571633 & 366 & 285 & -0.30 &  &  &  &  &  &  & \\
  L1240 & LH1.4GHzJ105343.3+572530 &  & 257 &  &  &  &  &  &  &  & \\
  L1254 & LH1.4GHzJ105346.9+571609 & 260 & 201 & -0.31 &  &  &  &  &  &  & \\
  L1266 & LH1.4GHzJ105348.8+573034 & 303 & 205 & -0.47 &  &  &  &  &  &  & \\
  L1280 & LH1.4GHzJ105354.1+574244 & 718 & 429 & -0.62 &  &  &  &  &  &  & \\
  L1286 & LH1.4GHzJ105356.5+572244 & 462 & 249 & -0.74 & 35 & 145 & 38 & 258 & 44 & 163.485255414 & 57.379118349\\
  L1299 & LH1.4GHzJ105359.9+572600 & 397 & 215 & -0.74 &  &  &  &  &  &  & \\
  L1302 & LH1.4GHzJ105400.5+573321 & 4646 & 3016 & -0.52 & 37 & 255 & 45 & 222 & 43 & 163.501801452 & 57.556002324\\
  L1306 & LH1.4GHzJ105401.2+573207 & 410 & 267 & -0.52 & 36 & 143 & 39 & 142 & 39 & 163.505027094 & 57.535413357\\
  L1325 & LH1.4GHzJ105406.1+572413 & 311 & 190 & -0.59 &  &  &  &  &  &  & \\
  L1328 & LH1.4GHzJ105406.8+571256 & 3558 & 1686 & -0.90 &  &  &  &  &  &  & \\
  L1344 & LH1.4GHzJ105415.2+573336 & 359 & 220 & -0.59 &  &  &  &  &  &  & \\
  L1361 & LH1.4GHzJ105421.2+572544 & 1725 & 1041 & -0.61 &  &  &  &  &  &  & \\
  L1364 & LH1.4GHzJ105423.3+573446 & 1109 & 661 & -0.62 & 50 & 415 & 65 & 598 & 78 & 163.597162837 & 57.579559074\\
  L1371 & LH1.4GHzJ105425.7+571937 & 695 & 376 & -0.74 &  &  &  &  &  &  & \\
  L1374 & LH1.4GHzJ105427.0+573644 & 322140 & 181160 & -0.69 & 58 & 1096 & 124 & 1658 & 176 & 163.609453312 & 57.613618513\\
  L1402 & LH1.4GHzJ105440.6+571647 & 2859 & 1976 & -0.44 &  &  &  &  &  &  & \\
  L1404 & LH1.4GHzJ105442.1+571639 & 3366 & 2778 & -0.23 & 98 & 769 & 125 & 945 & 136 & 163.675525388 & 57.277615166\\
  L1409 & LH1.4GHzJ105445.9+572747 & 999 & 389 & -1.14 &  &  &  &  &  &  & \\
  L1423 & LH1.4GHzJ105456.3+573017 & 662 & 514 & -0.30 &  &  &  &  &  &  & \\
  L1424 & LH1.4GHzJ105458.5+572035 & 1104 & 884 & -0.27 &  &  &  &  &  &  & \\
  L1425 & LH1.4GHzJ105458.6+572828 & 689 & 570 & -0.23 &  &  &  &  &  &  & \\
  L1436 & LH1.4GHzJ105516.0+573257 &  & 7104 &  &  &  &  &  &  &  & \\
\hline\end{tabular}
\end{table*}

\begin{table*}[h!]
\caption{Multi-wavelength data for the VLBI-detected sources.
{\it Column 1:} source ID used in this paper; 
{\it Column 2:} ID from \cite{Fotopoulou2012}, $-99$: outside imaged area, -1: uncatalogued/blended counterpart, 0: no visible counterpart; 
{\it Column 3:} $R_c$ AB magnitude; 
{\it Column 4:} $3.6\,\mu$m AB magnitude from the SWIRE survey; 
{\it Column 5:} spectroscopic redshift, where available; 
{\it Column 6:} best photometric redshift; 
{\it Column 7:} morphology flag, 0:  point-like, 1 -- unresolved, 2 -- early type or bulge-dominated, 3 -- unclassified, 4 -- spiral; 
{\it Column 8:} quality of the fit, 1: too few bands, 2: blended photometry, 3: doubful because of a lack of X-ray information, 4: a good fit; 
{\it Column 9:} model type used in the best fit for photometric redshift; 
{\it Column 10:} X-ray flux in the 0.5--2\,keV band; 
{\it Column 11:} X-ray luminosity in the 0.5--2\,keV band, $-99$: outside imaged area, 0: imaged, but undetected; 
{\it Column 12:} a ``$<$'' indicates that the X-ray luminosity is an upper limit;
{\it Column 13:} hardness ratio calculated between the 0.5--2\,keV and 2.0--4.5\,keV bands.}
\center
\scriptsize

\begin{tabular}{|l|r|r|r|r|r|r|r|r|r|r|r|r|}
\hline
  \multicolumn{1}{|c|}{ID} &
  \multicolumn{1}{c|}{opt. ID} &
  \multicolumn{1}{c|}{$R_c$} &
  \multicolumn{1}{c|}{3.6\,$\mu$m} &
  \multicolumn{1}{c|}{${\rm z_{spec}}$} &
  \multicolumn{1}{c|}{${\rm z_{phot}}$} &
  \multicolumn{1}{c|}{morph.} &
  \multicolumn{1}{c|}{fit} &
  \multicolumn{1}{c|}{model} &
  \multicolumn{1}{c|}{$F_X$(0.5--2\,keV)} &
  \multicolumn{1}{c|}{flag} &
  \multicolumn{1}{c|}{$\log{L_{\rm X}}$(0.5--2\,keV)} &
  \multicolumn{1}{c|}{HR1} \\

  \multicolumn{1}{|c|}{} &
  \multicolumn{1}{c|}{} &
  \multicolumn{1}{c|}{mag} &
  \multicolumn{1}{c|}{mag} &
  \multicolumn{1}{c|}{} &
  \multicolumn{1}{c|}{} &
  \multicolumn{1}{c|}{} &
  \multicolumn{1}{c|}{} &
  \multicolumn{1}{c|}{} &
  \multicolumn{1}{c|}{$10^{-16}\,{\rm erg/s/cm^2}$} &
  \multicolumn{1}{c|}{} &
  \multicolumn{1}{c|}{erg/s} &
  \multicolumn{1}{c|}{} \\

 \multicolumn{1}{|c|}{(1)} &
  \multicolumn{1}{c|}{(2)} &
  \multicolumn{1}{c|}{(3)} &
  \multicolumn{1}{c|}{(4)} &
  \multicolumn{1}{c|}{(5)} &
  \multicolumn{1}{c|}{(6)} &
  \multicolumn{1}{c|}{(7)} &
  \multicolumn{1}{c|}{(8)} &
  \multicolumn{1}{c|}{(9)} &
  \multicolumn{1}{c|}{(10)} &
  \multicolumn{1}{c|}{(11)} &
  \multicolumn{1}{c|}{(12)} &
  \multicolumn{1}{c|}{(13)} \\

\hline
  L0009 & -99 &  &  &  &  &  &  &  &  &  &  & \\
  L0025 & -99 &  &  &  &  &  &  &  &  &  &  & \\
  L0034 & -99 &  &  &  &  &  &  &  &  &  &  & \\
  L0049 & -99 &  &  &  &  &  &  &  &  &  &  & \\
  L0051 & -99 &  &  &  &  &  &  &  &  &  &  & \\
  L0055 & -99 &  &  &  &  &  &  &  &  &  &  & \\
  L0066 & -99 &  &  &  &  &  &  &  &  &  &  & \\
  L0077 & -99 &  &  &  &  &  &  &  &  &  &  & \\
  L0119 & 166492 & 25.20 &  &  & 1.70 & 2 & 1 & 110 & -99 &  &  & \\
  L0130 & 42435 & 20.43 &  & 0.4239 & 0.34 & 2 & 1 & 131 & -99 &  &  & \\
  L0140 & 55191 & 24.44 &  &  & 4.20 & 1 & 1 & 126 & -99 &  &  & \\
  L0149 & 28661 & 20.09 &  & 3.2687 & 0.29 & 2 & 1 & 131 & -99 &  &  & \\
  L0171 & 34378 & 24.67 &  &  & 0.97 & 1 & 1 & 123 & -99 &  &  & \\
  L0194 & 0 &  &  &  &  &  &  &  &  &  &  & \\
  L0199 & -1 &  &  &  &  &  &  &  &  &  &  & \\
  L0227 & 89171 & 19.14 & 18.69 & 1.439 & 0.27 & 0 & 3 & 129 & -99 &  &  & \\
  L0231 & 117241 & 23.24 &  &  & 0.87 & 2 & 1 & 108 & -99 &  &  & \\
  L0251 & -1 &  &  &  &  &  &  &  &  &  &  & \\
  L0256 & 45485 & 20.90 &  & 0.5396 & 0.60 & 3 & 1 & 101 & -99 &  &  & \\
  L0352 & 55842 & 21.04 &  & 0.593 & 0.91 & 2 & 2 & 129 & -99 &  &  & \\
  L0386 & -1 &  &  &  &  &  &  &  &  &  &  & \\
  L0397 & 69623 & 21.68 & 18.95 &  & 0.69 & 2 & 4 & 120 & -99 &  &  & \\
  L0411 & 187543 & 25.87 & 21.54 &  & 1.82 & 1 & 4 & 123 & 0 & $<$ & 42.95 & \\
  L0412 & 124257 & 24.86 & 21.39 &  & 2.12 & 1 & 4 & 122 & 0 & $<$ & 43.14 & \\
  L0420 & 46934 & 24.83 &  &  & 1.13 & 1 & 1 & 131 & -99 &  &  & \\
  L0423 & 114988 & 24.74 & 19.72 &  & 1.34 & 2 & 4 & 107 & 0 & $<$ & 42.57 & \\
  L0449 & 47368 & 20.19 &  &  & 0.55 & 2 & 1 & 109 & -99 &  &  & \\
  L0477 & 53986 & 18.94 & 17.83 & 0.3177 & 0.40 & 2 & 4 & 107 & -99 &  &  & \\
  L0484 & 48731 & 21.69 &  &  & 0.59 & 3 & 1 & 114 & -99 &  &  & \\
  L0506 & -1 &  &  &  &  &  &  &  &  &  &  & \\
  L0535 & 206628 & 21.69 & 18.89 &  & 0.58 & 2 & 4 & 2 & 6.8 &  & 42.11 & -0.69\\
  L0536 & 206644 & 24.19 & 20.47 &  & 1.73 & 1 & 2 & 114 & 25.5 &  & 44.02 & -0.5\\
  L0578 & 206598 & 22.85 & 18.99 & 0.99 & 1.00 & 1 & 4 & 5 & 48.9 &  & 43.61 & 0.03\\
  L0593 & 97588 & 24.77 & 19.55 &  & 1.04 & 3 & 4 & 113 & 0 & $<$ & 42.25 & \\
  L0607 & 49780 & 22.51 & 19.27 &  & 1.16 & 3 & 4 & 119 & -99 &  &  & \\
  L0648 & 88715 & 26.02 &  &  & 2.07 & 1 & 4 & 130 & 0 & $<$ & 43.12 & \\
  L0703 & 206690 & 24.24 & 20.18 & 0.462 & 1.53 & 1 & 4 & 114 & 42.6 &  & 42.65 & -0.62\\
  L0708 & -1 &  &  &  &  &  &  &  &  &  &  & \\
  L0724 & 167757 & 25.51 & 20.13 &  & 1.51 & 3 & 4 & 113 & 0 & $<$ & 42.72 & \\
  L0846 & 59617 & 21.51 & 19.14 &  & 0.58 & 2 & 4 & 108 & -99 &  &  & \\
  L0860 & 40969 & 19.82 &  & 0.3826 & 0.52 & 2 & 1 & 102 & -99 &  &  & \\
  L0876 & 110535 & 20.29 & 19.12 &  & 0.54 & 2 & 4 & 102 & 0 & $<$ & 41.47 & \\
  L0890 & 206578 & 21.25 & 18.33 & 0.71 & 0.58 & 2 & 4 & 2 & 30.0 &  & 42.99 & -0.53\\
  L0911 & 206453 & 22.93 & 19.29 & 1.013 & 1.11 & 2 & 2 & 1 & 4.3 &  & 42.58 & -0.63\\
  L0927 & -99 &  &  &  &  &  &  &  &  &  &  & \\
  L0933 & 126512 & 21.25 &  & 0.493 & 0.53 & 2 & 4 & 113 & 0 & $<$ & 41.37 & \\
  L0953 & 170521 &  & 16.40 & 0.0731 & 0.29 & 2 & 4 & 101 & -99 &  &  & \\
  L0973 & 206457 & 22.11 & 19.74 & 0.762 & 0.68 & 0 & 4 & 5 & 122.7 &  & 43.69 & -0.59\\
  L0977 & 206416 & 23.88 & 18.59 & 1.45 & 1.30 & 1 & 4 & 18 & 18.9 &  & 43.67 & -0.35\\
  L0997 & 45980 & 22.89 &  &  & 0.96 & 1 & 4 & 102 & -99 &  &  & \\
  L1034 & 66996 & 24.07 & 19.72 &  & 1.33 & 1 & 4 & 115 & 0 & $<$ & 42.56 & \\
  L1052 & 85797 & 20.03 & 18.85 &  & 0.41 & 2 & 4 & 107 & 0 & $<$ & 41.17 & \\
  L1092 & 92439 & 22.18 & 19.09 &  & 0.62 & 2 & 4 & 108 & 0 & $<$ & 41.63 & \\
  L1115 & 206406 & 20.84 & 18.41 & 0.711 & 0.73 & 2 & 4 & 124 & 77.1 &  & 43.40 & -0.65\\
  L1148 & -1 &  &  &  &  &  &  &  &  &  &  & \\
  L1165 & 117627 & 21.37 & 20.75 &  & 0.33 & 2 & 4 & 123 & 0 & $<$ & 40.91 & \\
  L1200 & 84577 & 25.61 & 21.32 &  & 2.55 & 1 & 4 & 121 & 0 & $<$ & 43.38 & \\
  L1225 & 78963 & 23.09 & 19.37 &  & 0.91 & 2 & 1 & 108 & 0 & $<$ & 42.09 & \\
  L1235 & 151191 & 21.53 &  &  & 0.73 & 2 & 4 & 102 & -99 &  &  & \\
  L1286 & 175609 & 25.29 & 21.27 &  & 1.91 & 1 & 4 & 128 & 0 & $<$ & 43.02 & \\
  L1302 & 206717 & 26.04 & 21.54 &  & 1.49 & 1 & 4 & 124 & 31.5 &  & 43.92 & -0.28\\
  L1306 & 114090 & 25.13 &  &  & 2.23 & 1 & 4 & 128 & 0 & $<$ & 43.21 & \\
  L1364 & 0 &  &  &  &  &  &  &  &  &  &  & \\
  L1374 & 128598 & 19.11 & 17.55 &  & 0.33 & 4 & 3 & 123 & -99 &  &  & \\
  L1404 & -99 &  &  &  &  &  &  &  &  &  &  & \\
\hline\end{tabular}
\label{tab:redshifts}
\end{table*}

\begin{figure*}
\center
\includegraphics[height=4.5cm]{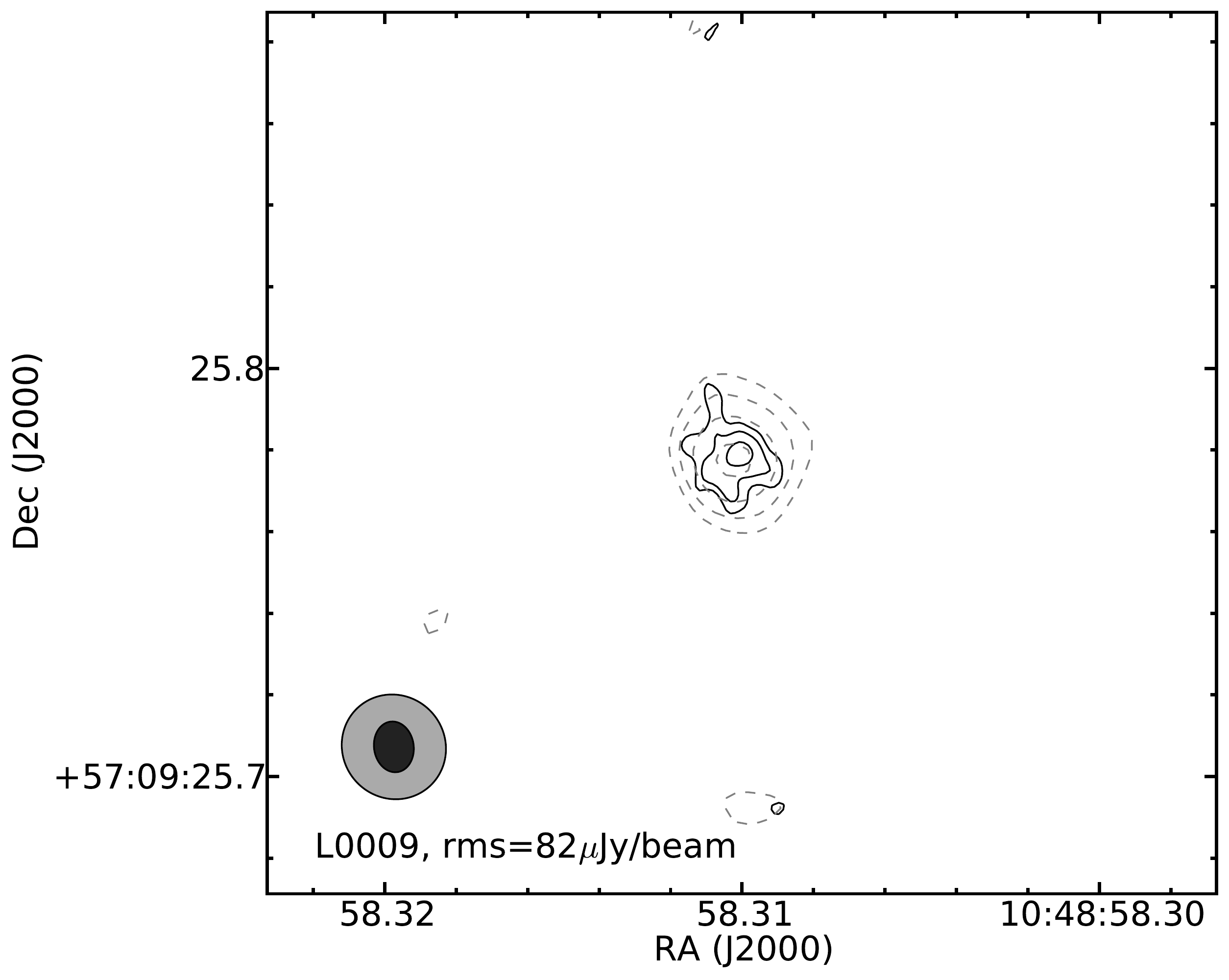}
\includegraphics[height=4.5cm]{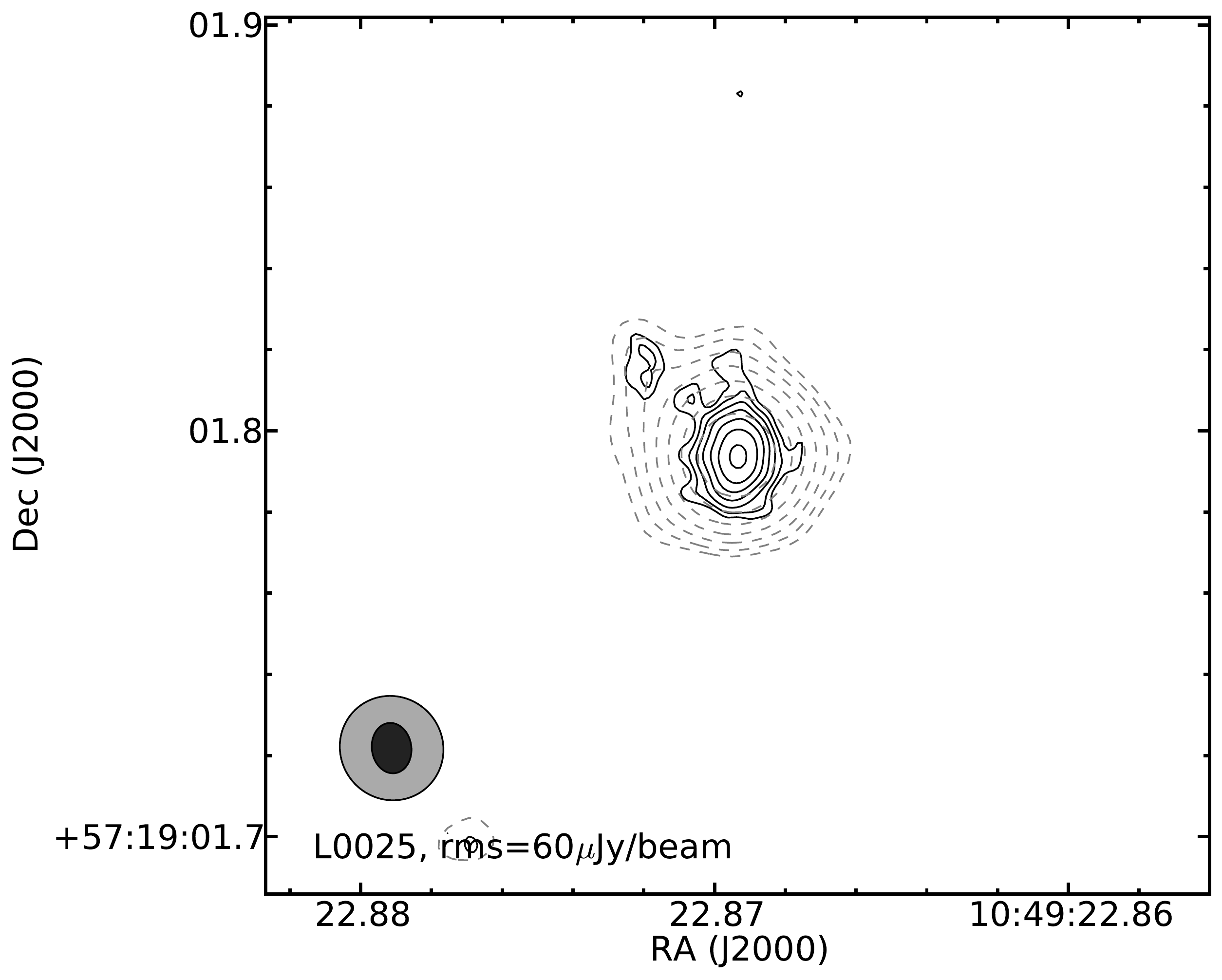}
\includegraphics[height=4.5cm]{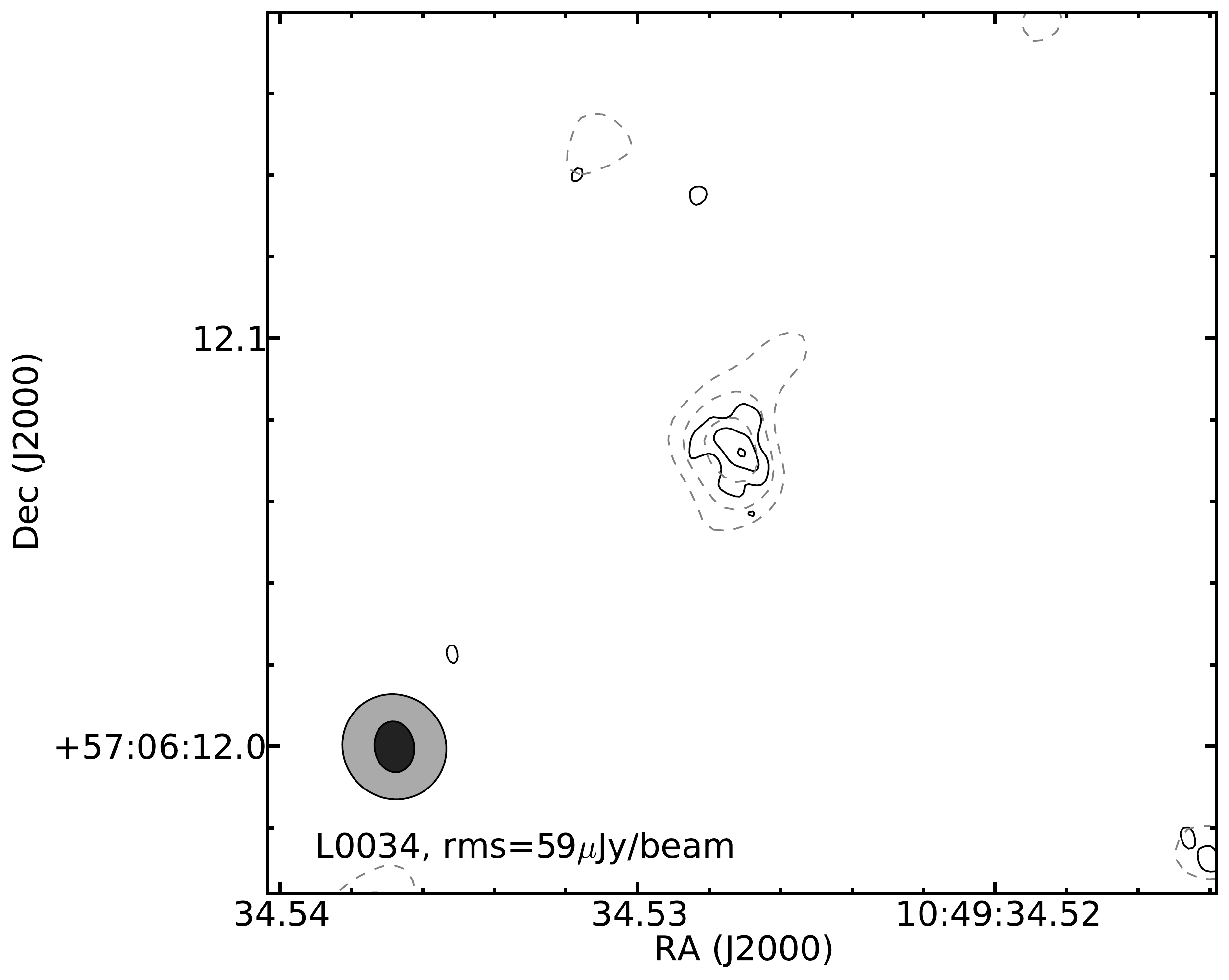}\\
\includegraphics[height=4.5cm]{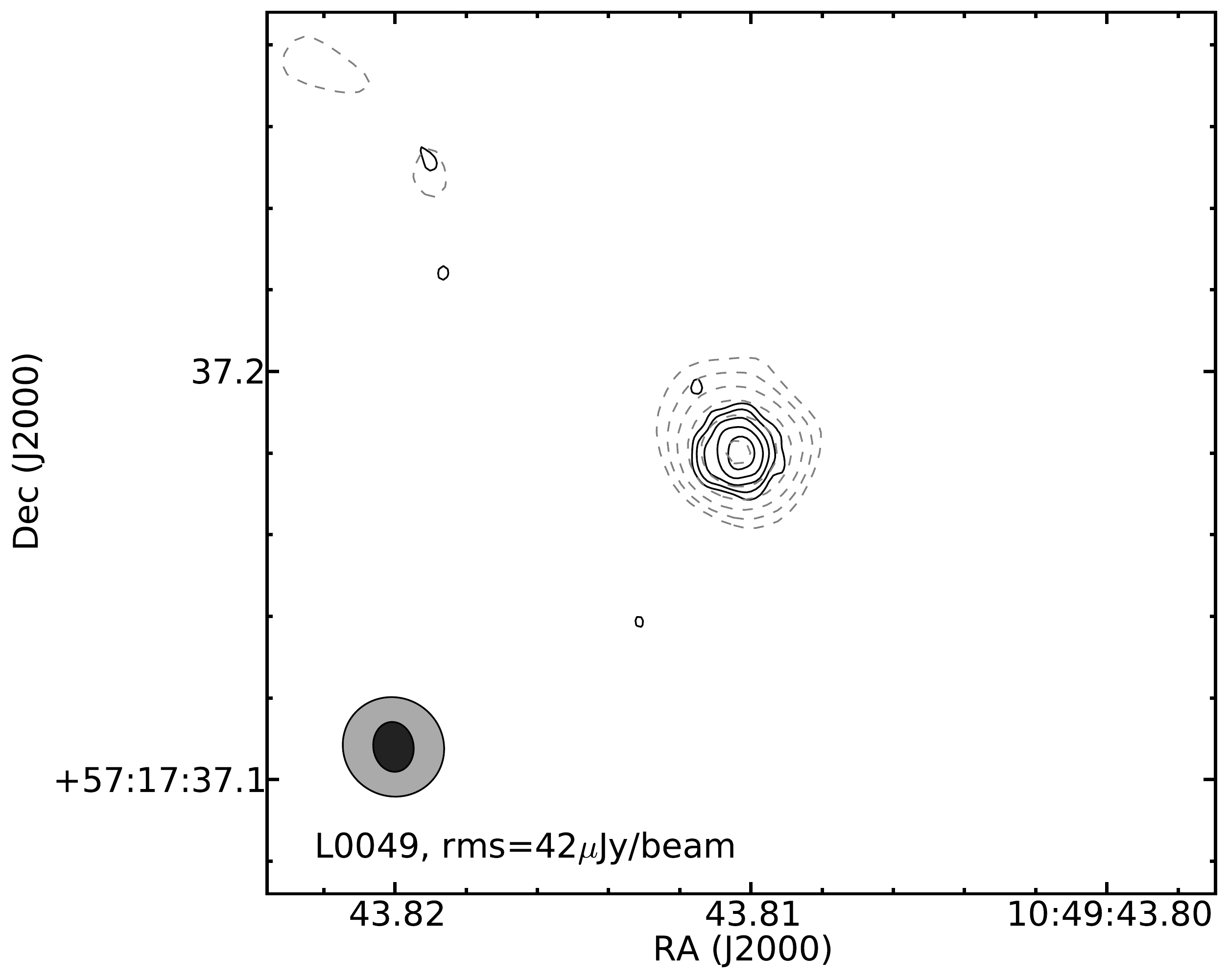}
\includegraphics[height=4.5cm]{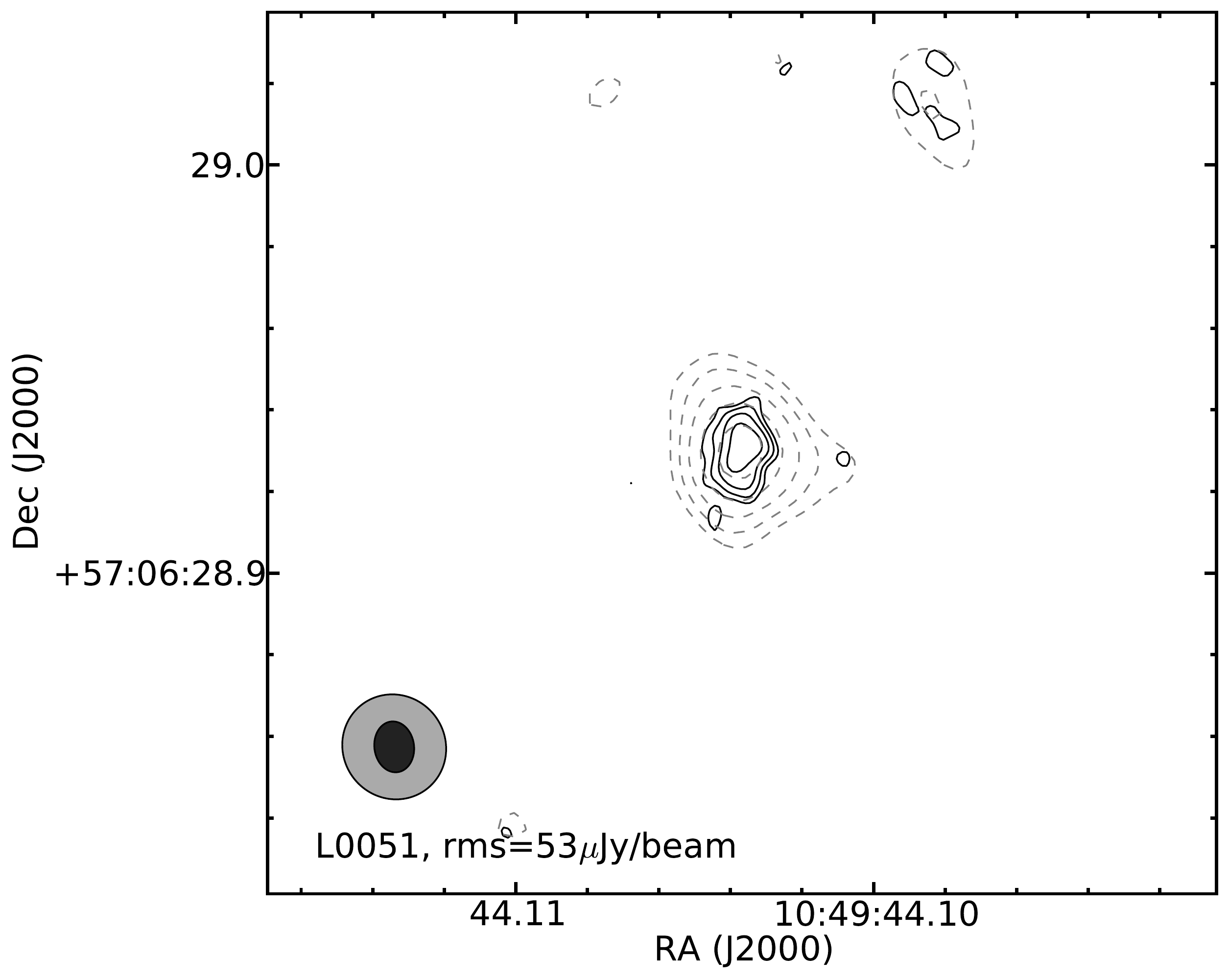}
\includegraphics[height=4.5cm]{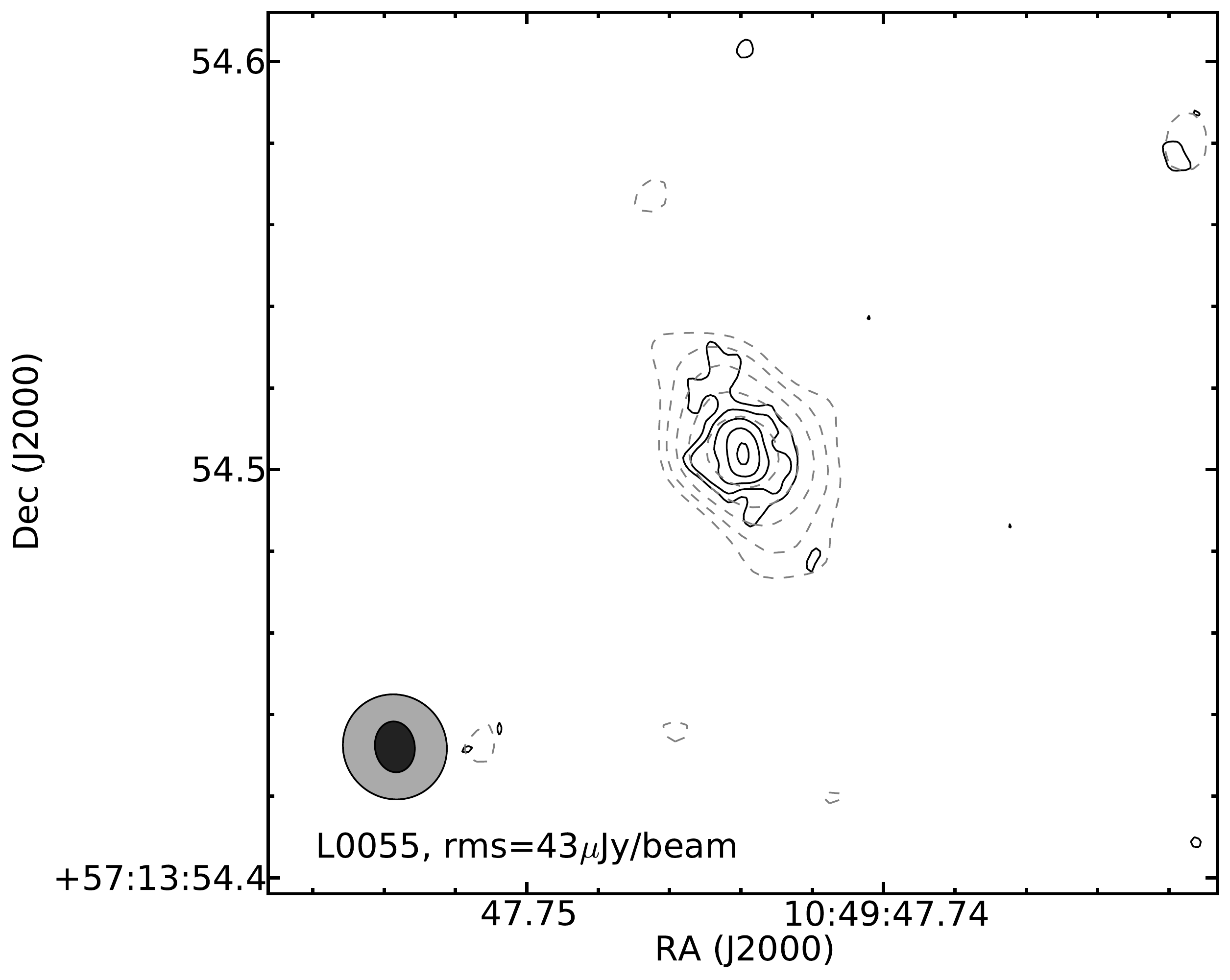}\\
\includegraphics[height=4.5cm]{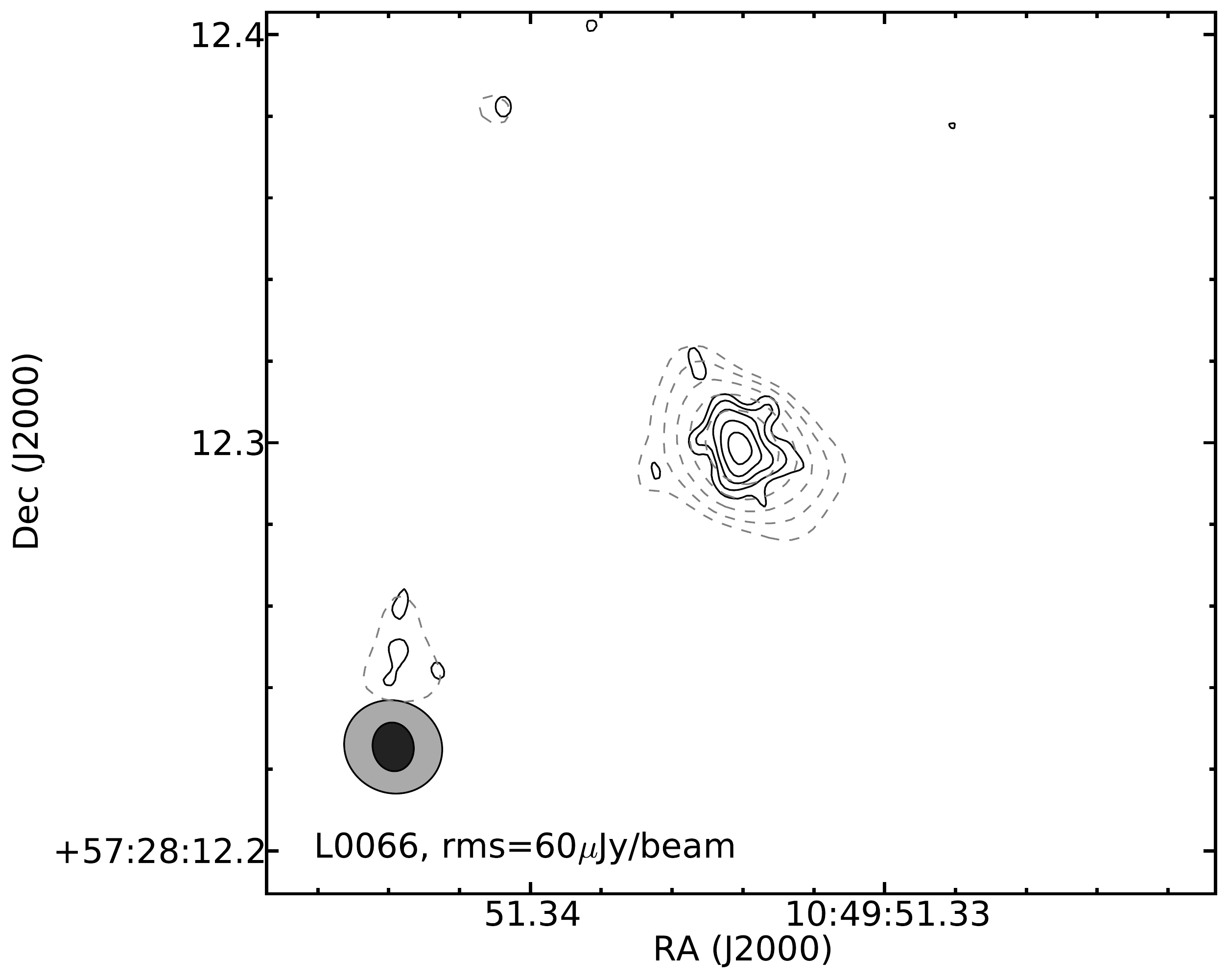}
\includegraphics[height=4.5cm]{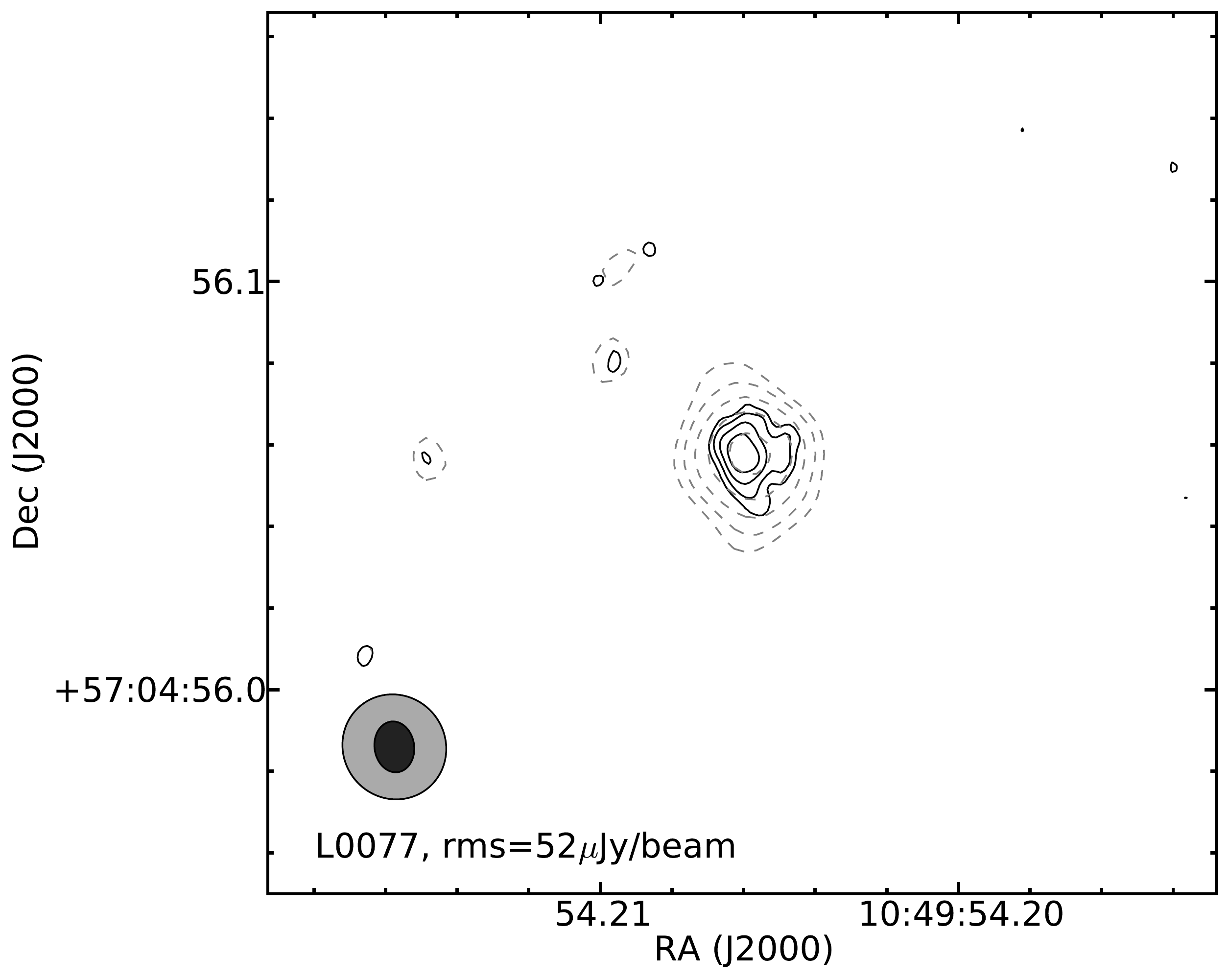}
\includegraphics[height=4.5cm]{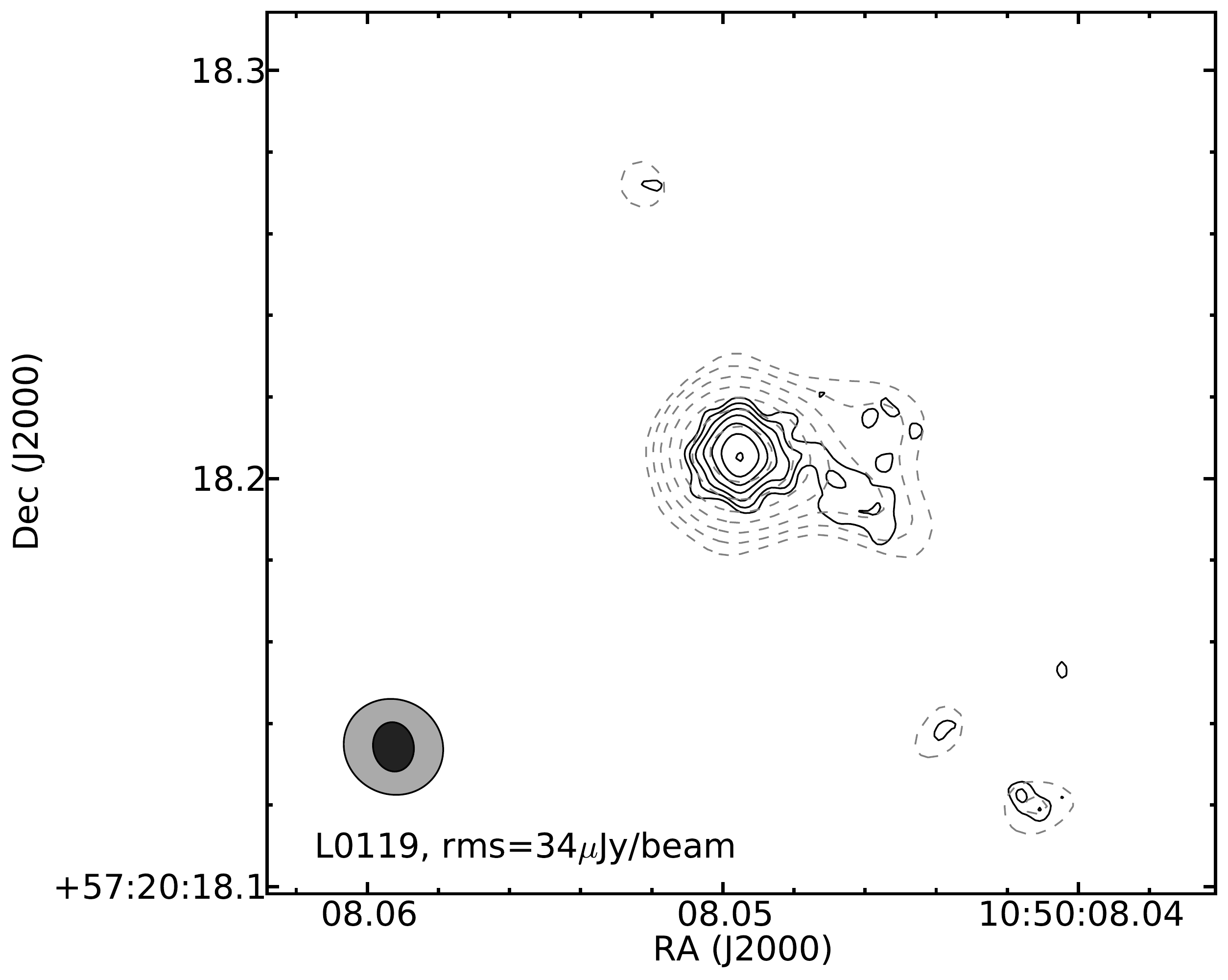}\\
\includegraphics[height=4.5cm]{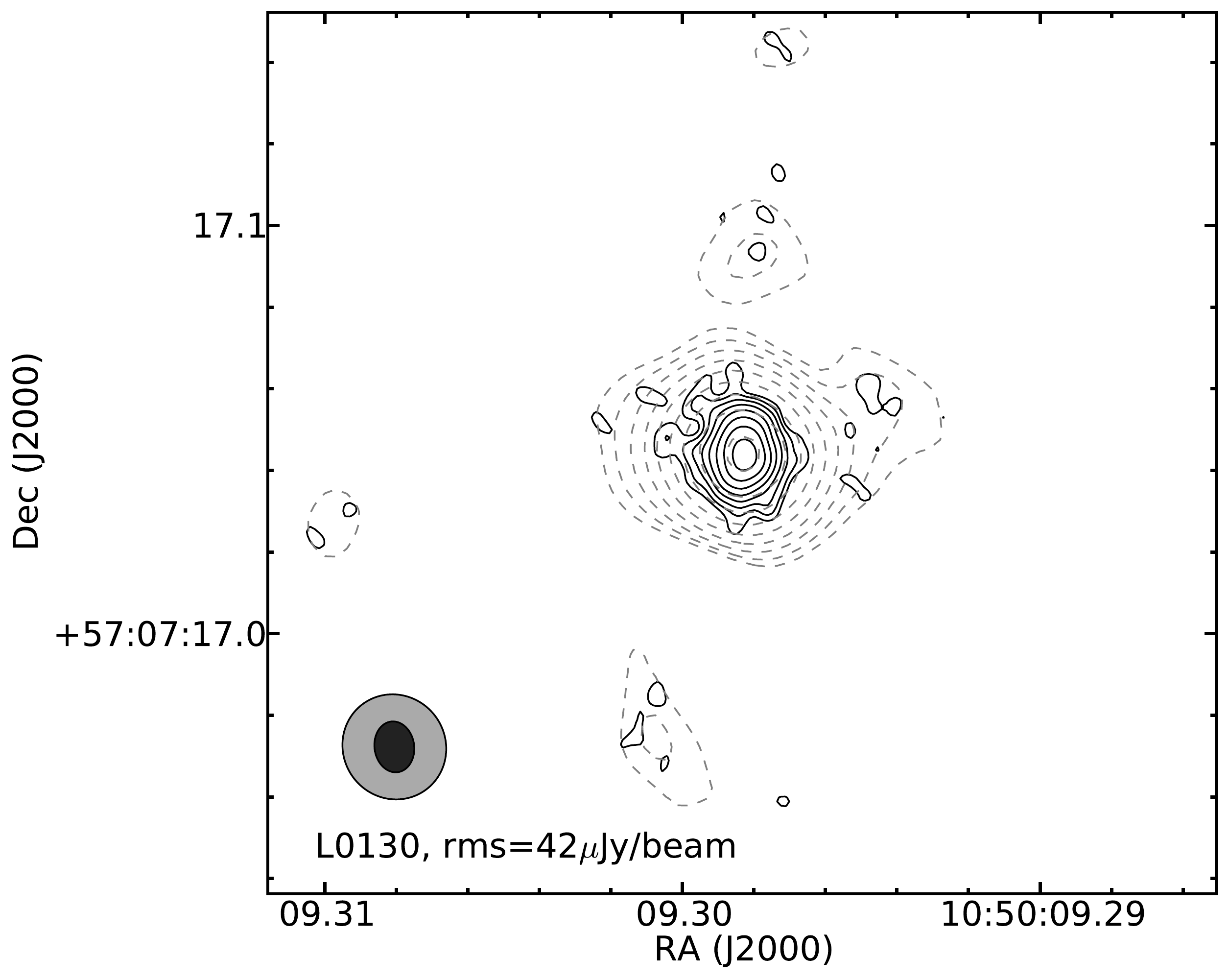}
\includegraphics[height=4.5cm]{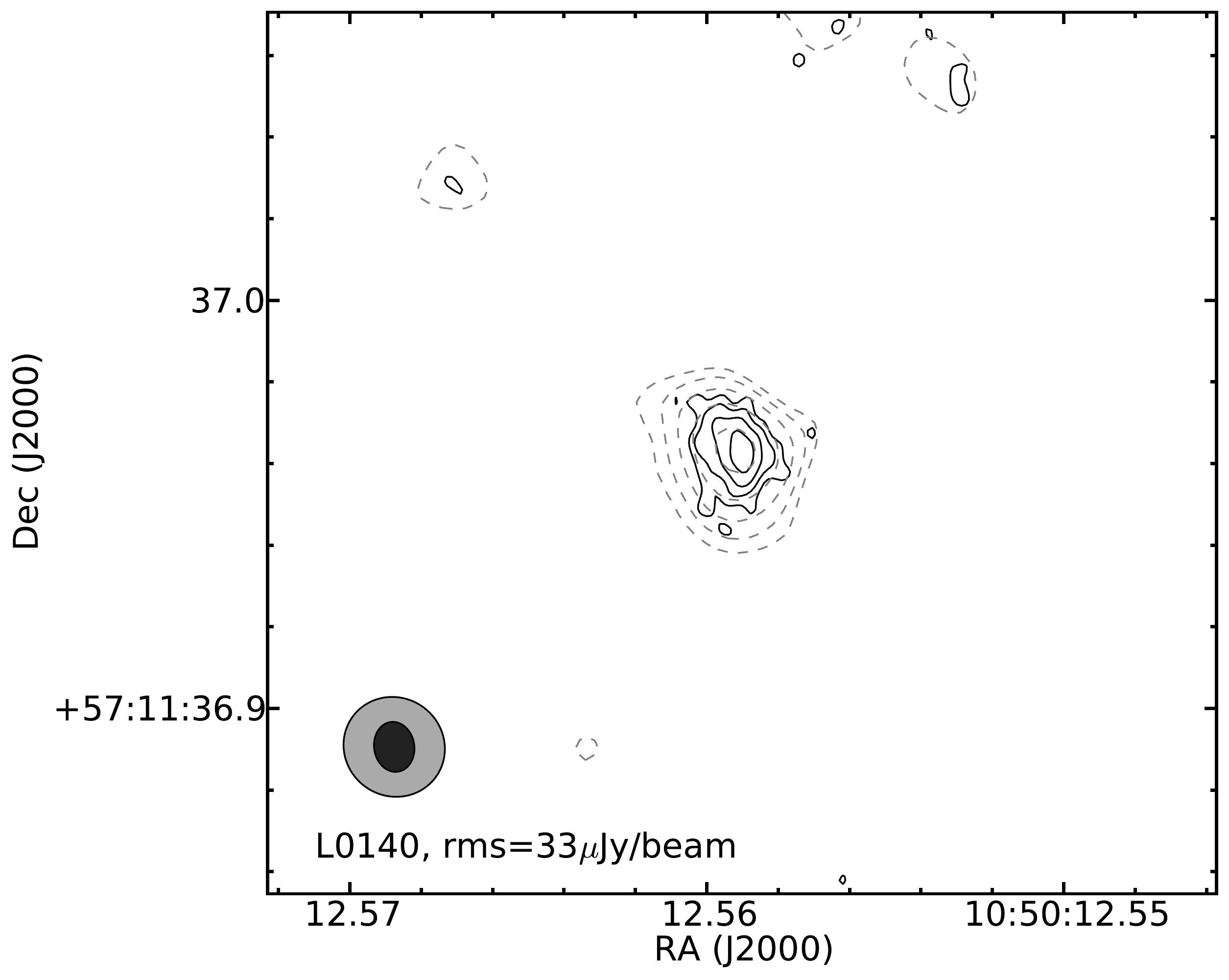}
\includegraphics[height=4.5cm]{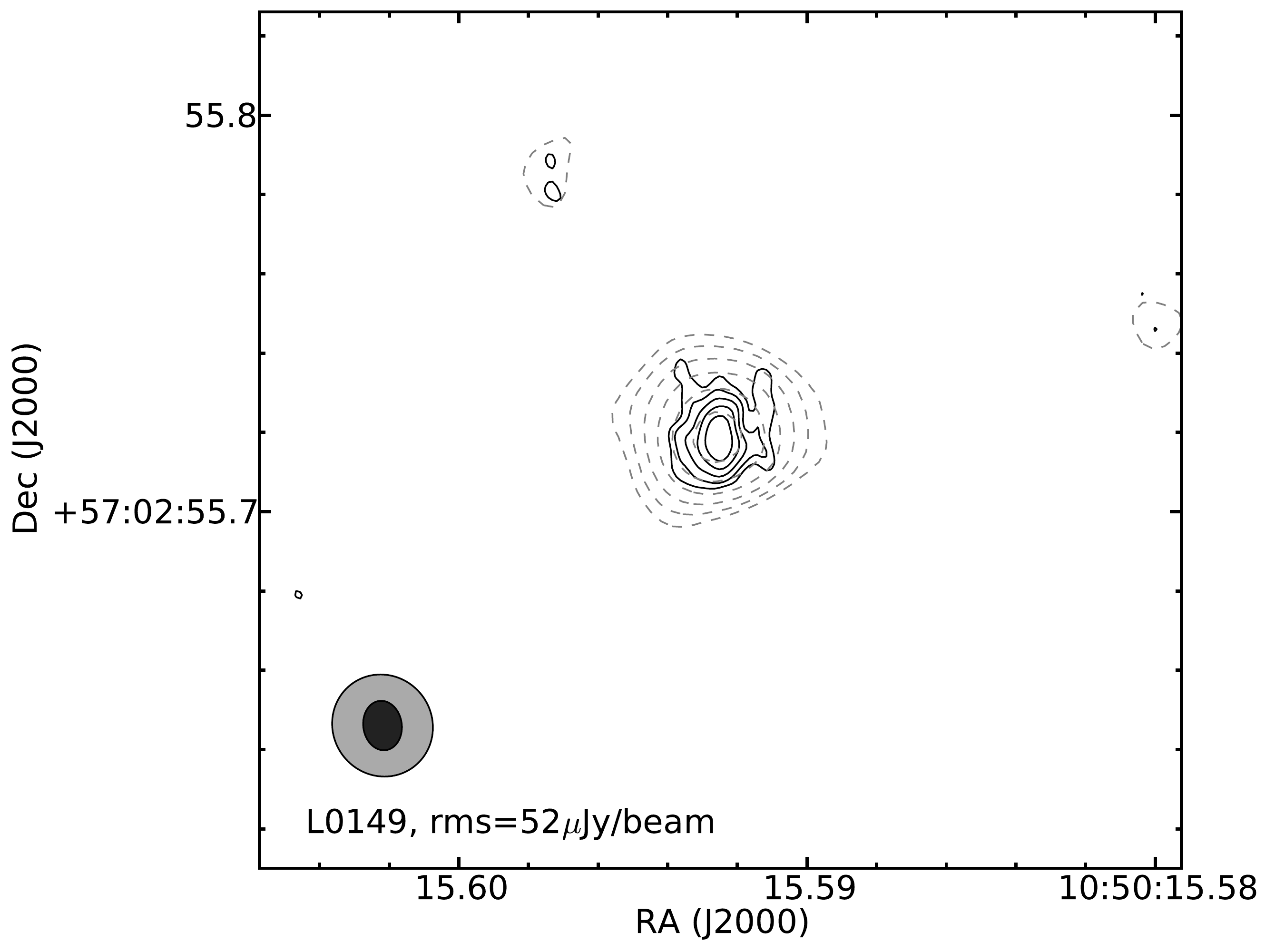}\\
\includegraphics[height=4.5cm]{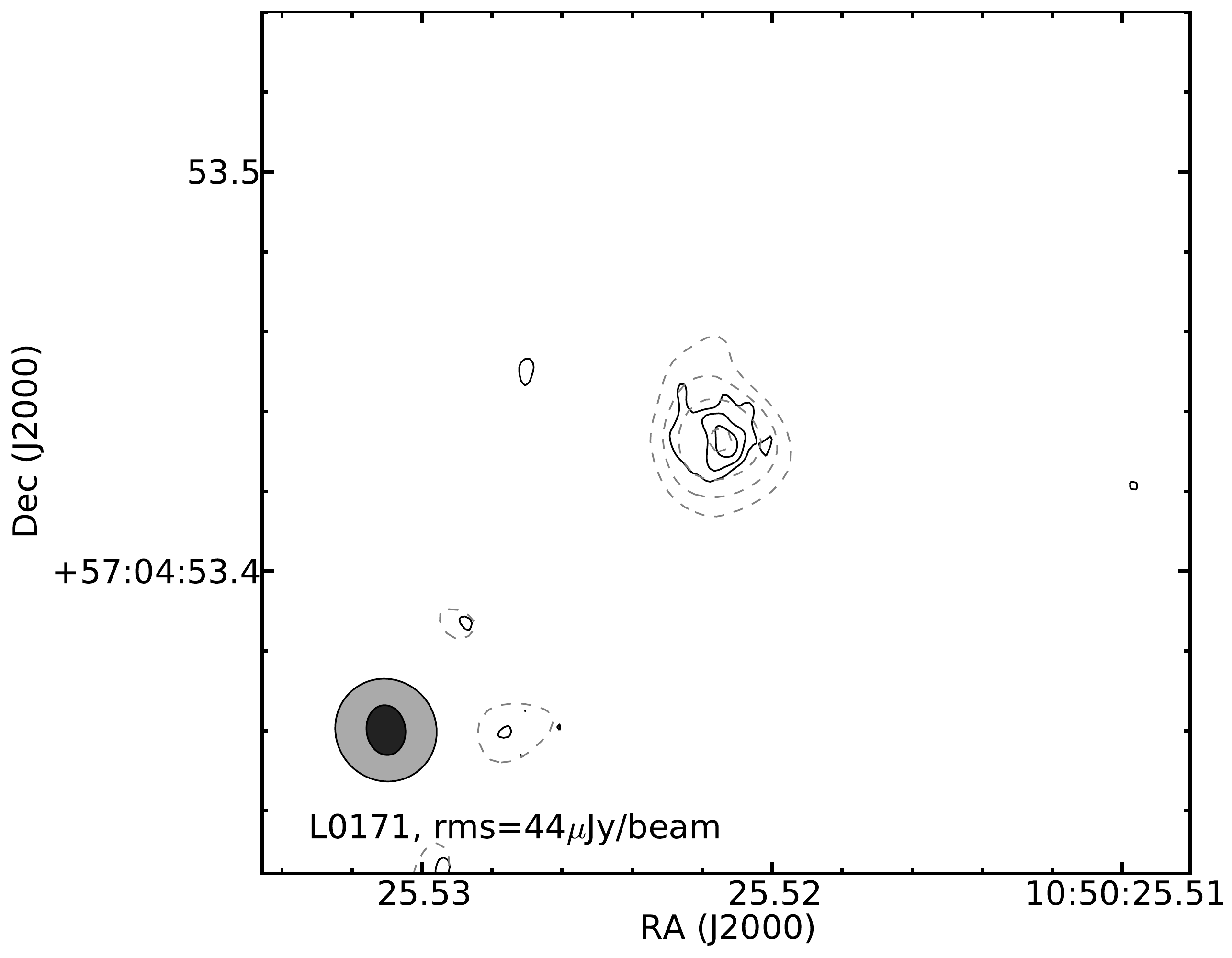}
\includegraphics[height=4.5cm]{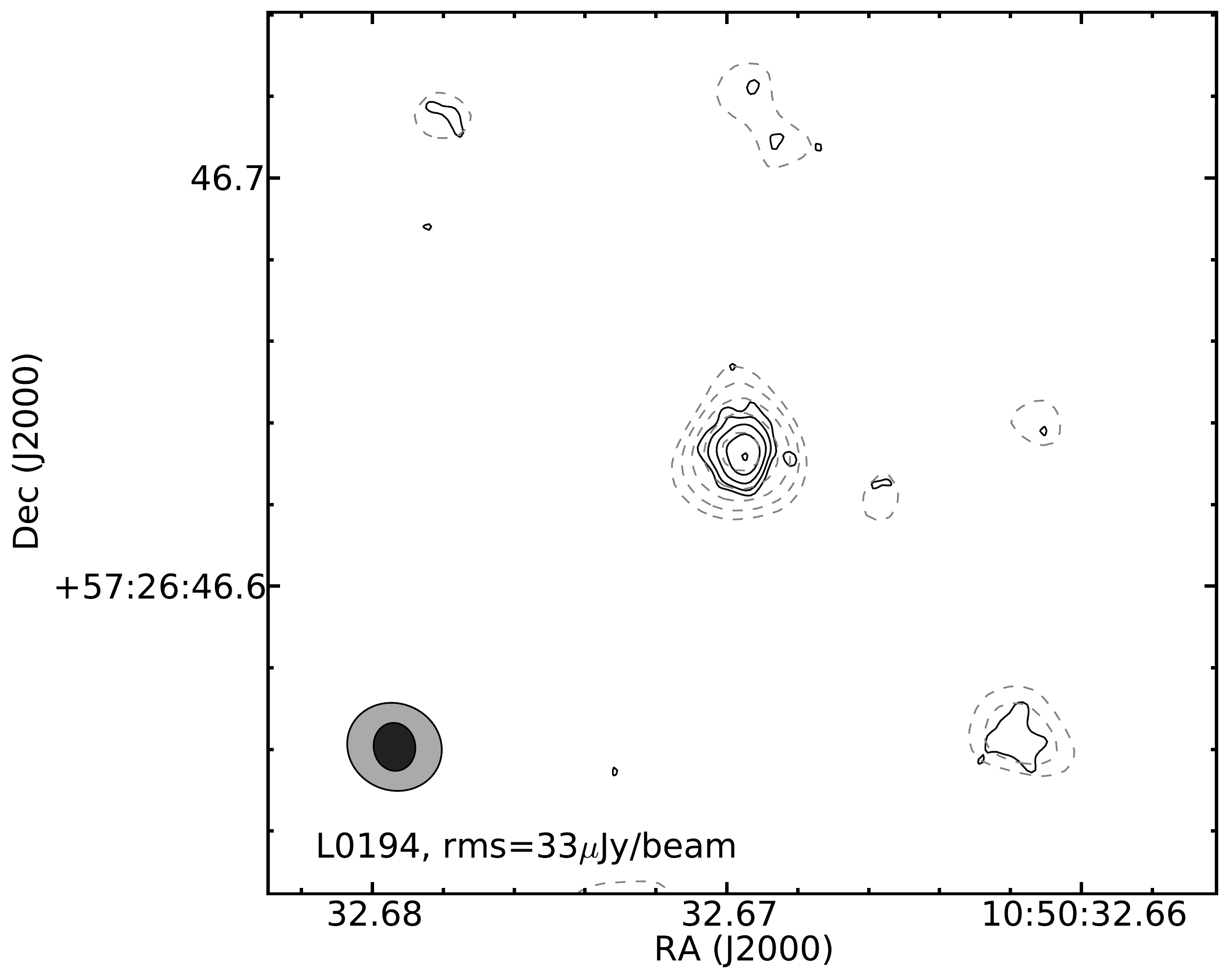}
\includegraphics[height=4.5cm]{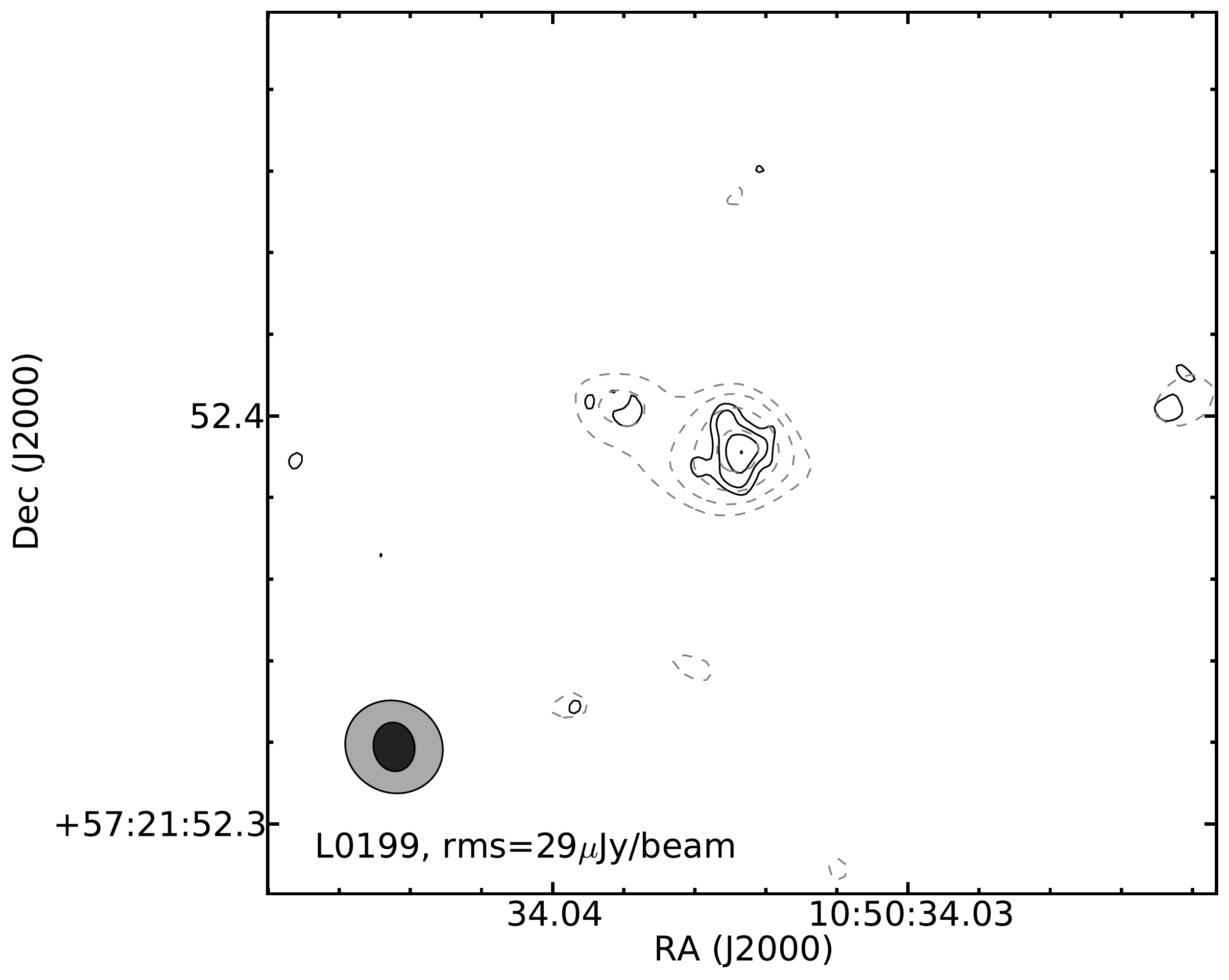}\\
\caption{Contour plots of the VLBA images of the 65 detected
  sources. The plots show regions of 216\,mas$\times$216\,mas, centred
  on the sources, except for the plot of L0703, which shows a region
  of 360\,mas$\times$360\,mas. Black contours indicate the
  full-resolution, naturally-weighted images, whereas grey, dashed
  contours indicate images made with a 10\,M$\lambda$ taper. The black
  and grey ellipses in the lower left corners indicate the size and
  orientation of the restoring beams used with full resolution and a
  10\,M$\lambda$ taper, respectively. In both cases contours start at
  the 3 times the naturally-weighted image noise and increase by
  factors of $\sqrt{2}$. The noise levels of the naturally weighted
  images have been added to the plots. Note that the flux densities
  listed in Tab.~\ref{tab:results1} have been extracted from
  uniformly-weighted images (not shown), to reduce the effect of the
  point spread function (see Sect.~\ref{sec:imaging} and
  Fig.~\ref{fig:beams}).}
\label{fig:all_contour_plots}
\end{figure*}

\begin{figure*}
\ContinuedFloat
\center
\includegraphics[height=4.5cm]{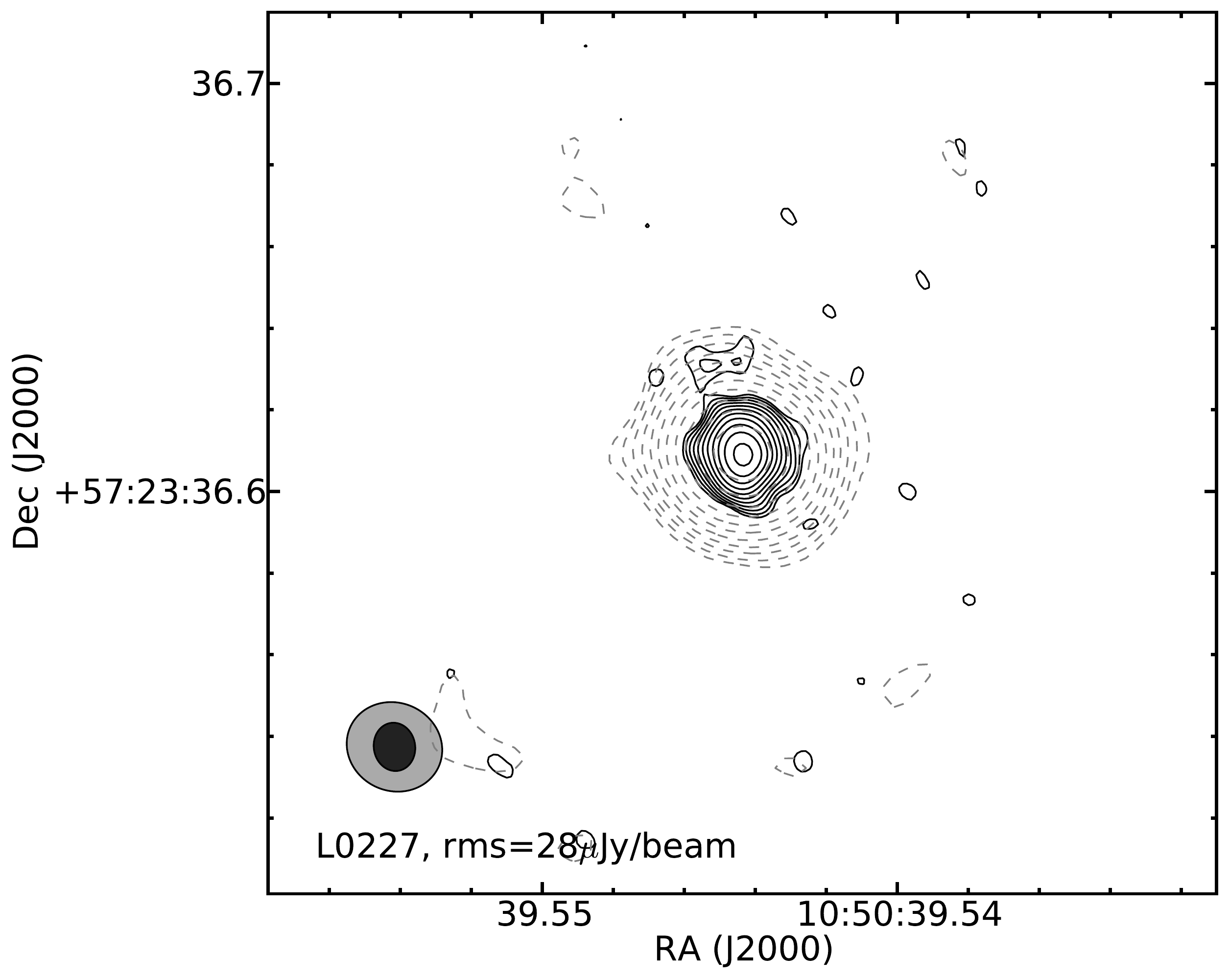}
\includegraphics[height=4.5cm]{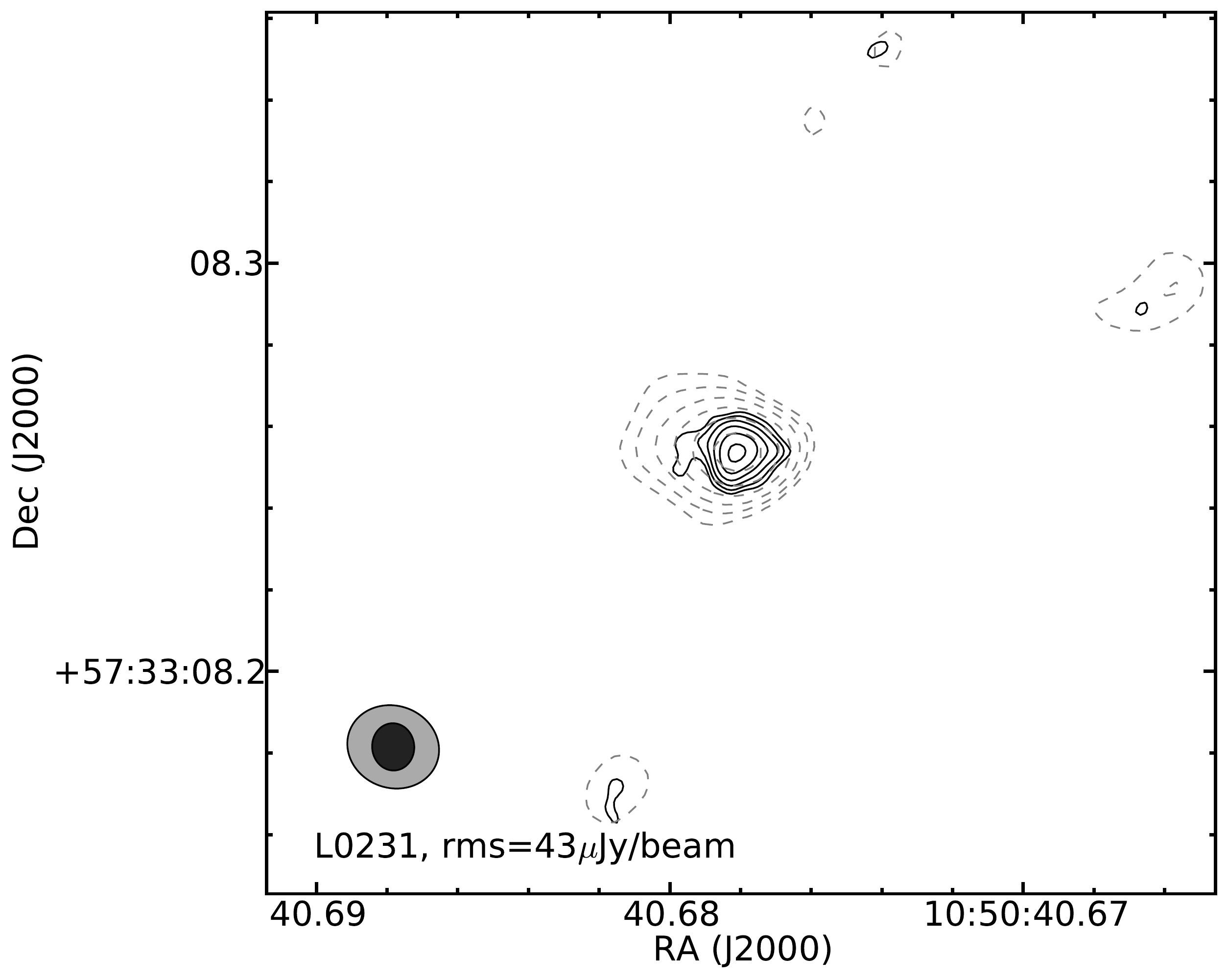}
\includegraphics[height=4.5cm]{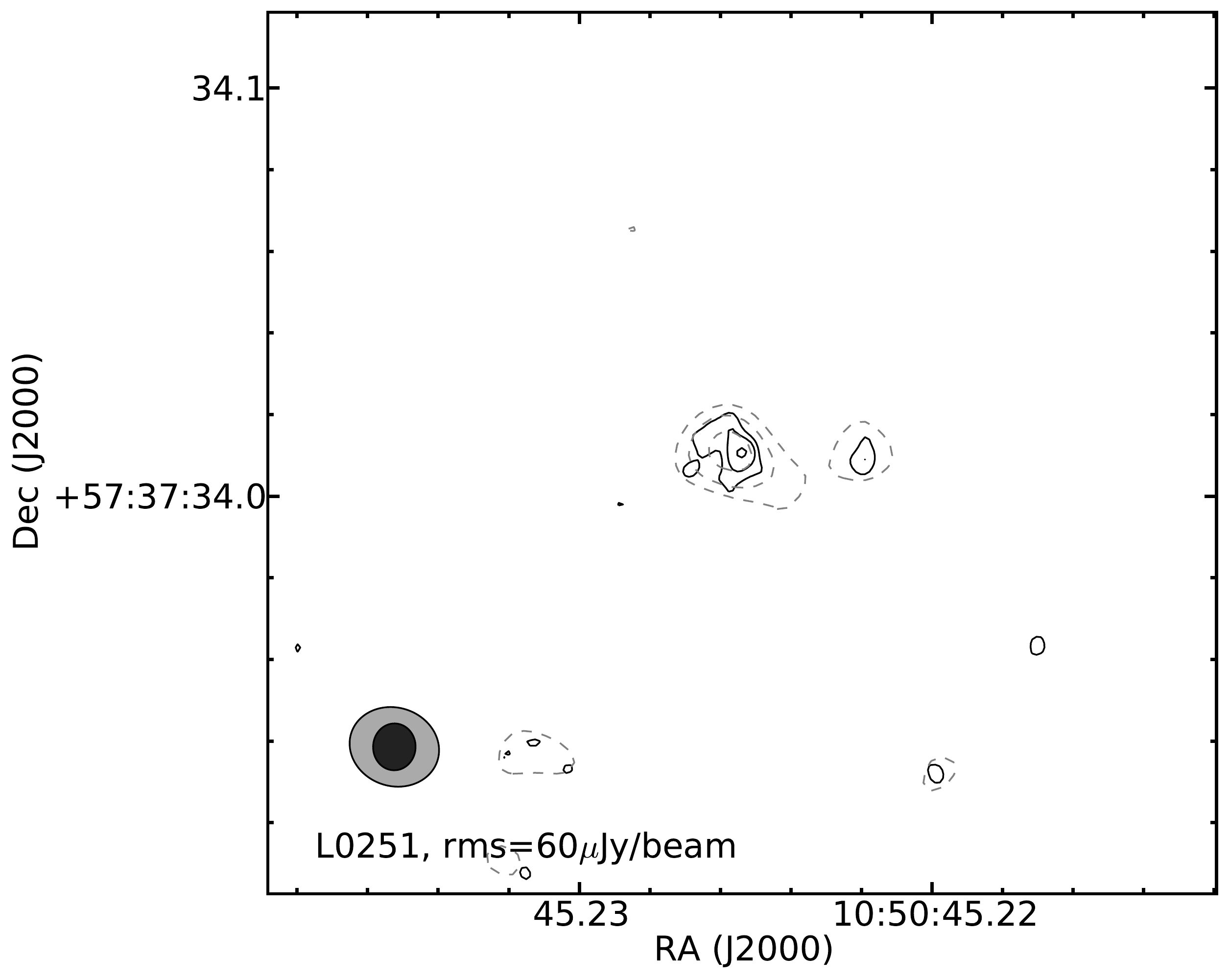}\\
\includegraphics[height=4.5cm]{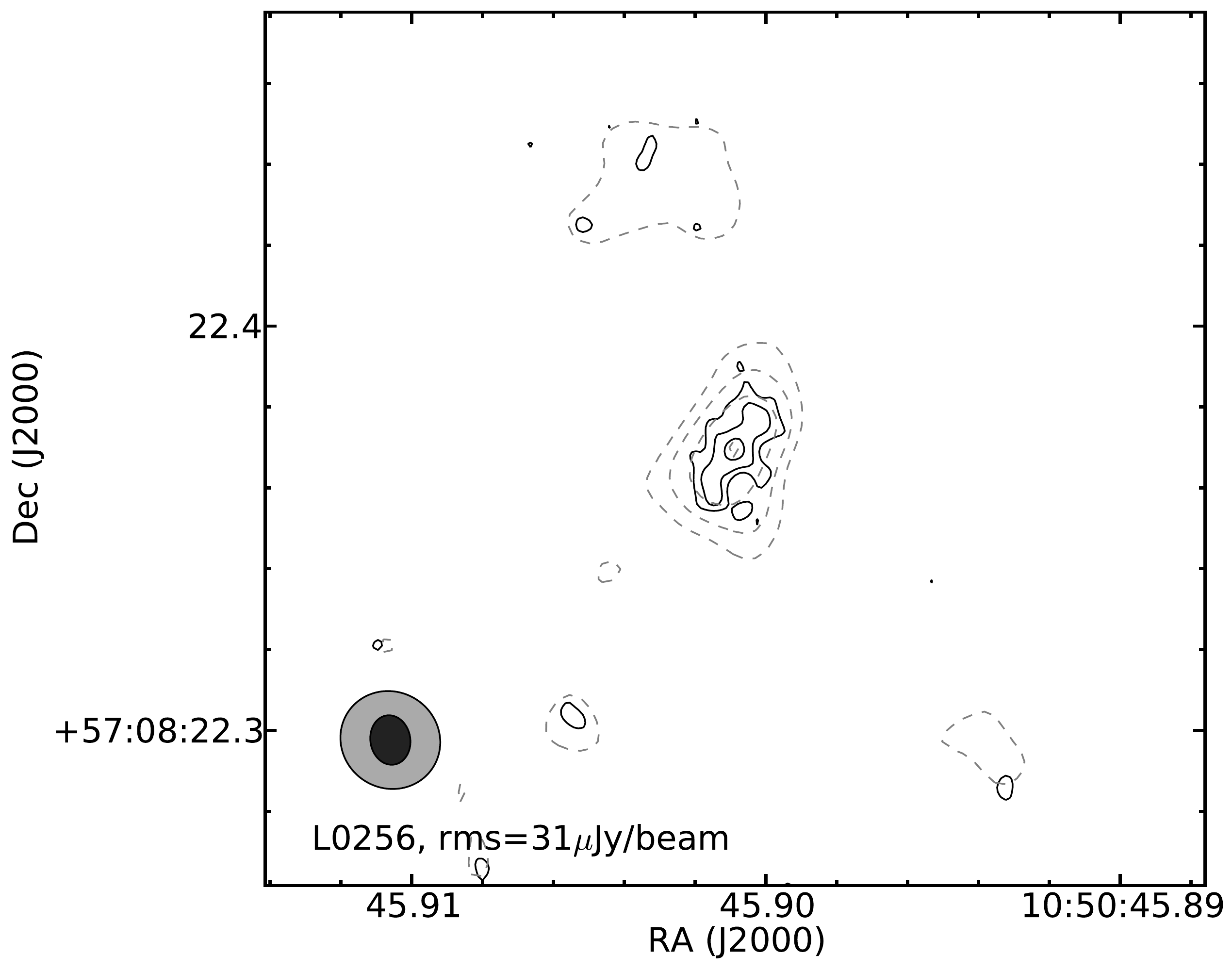}
\includegraphics[height=4.5cm]{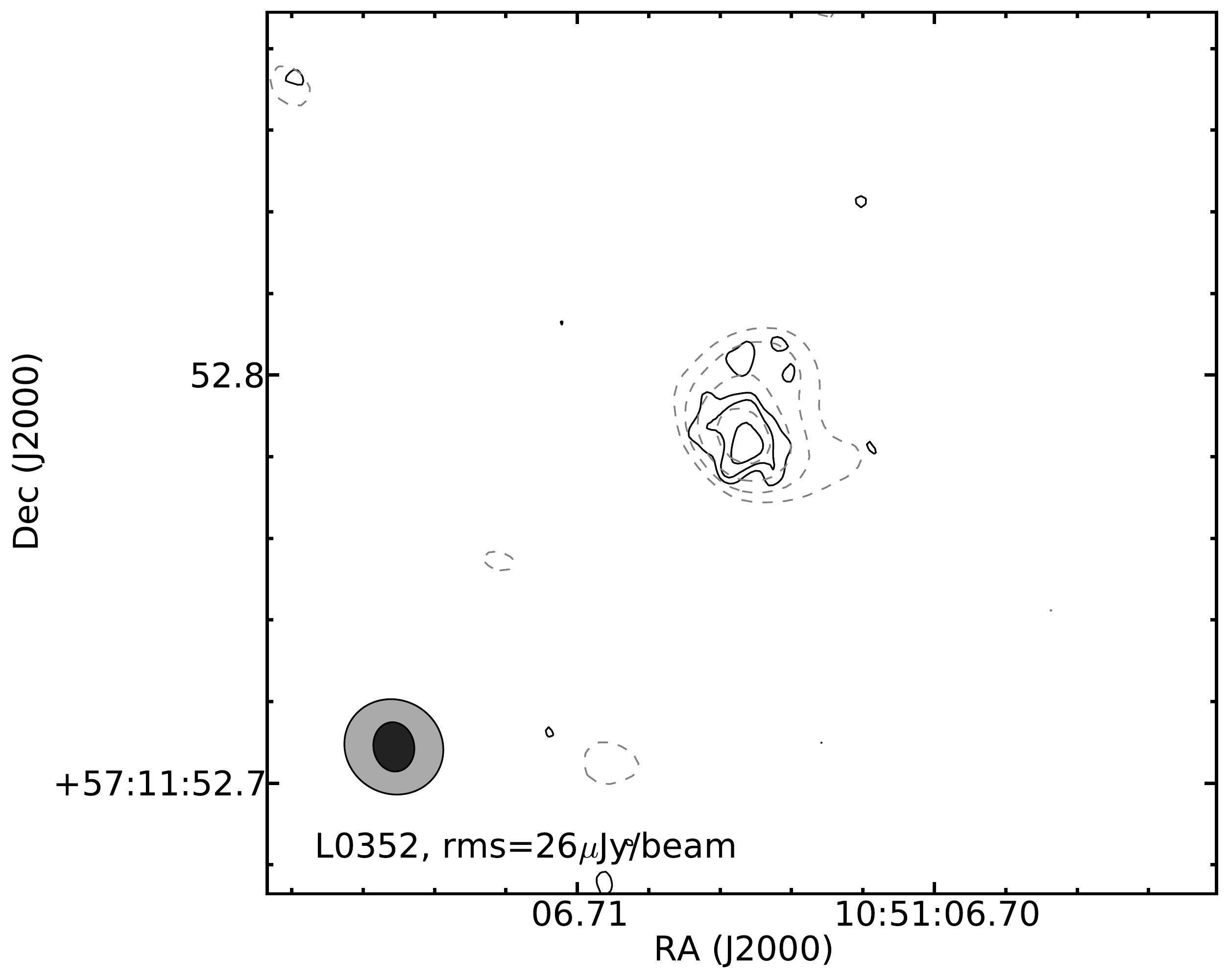}
\includegraphics[height=4.5cm]{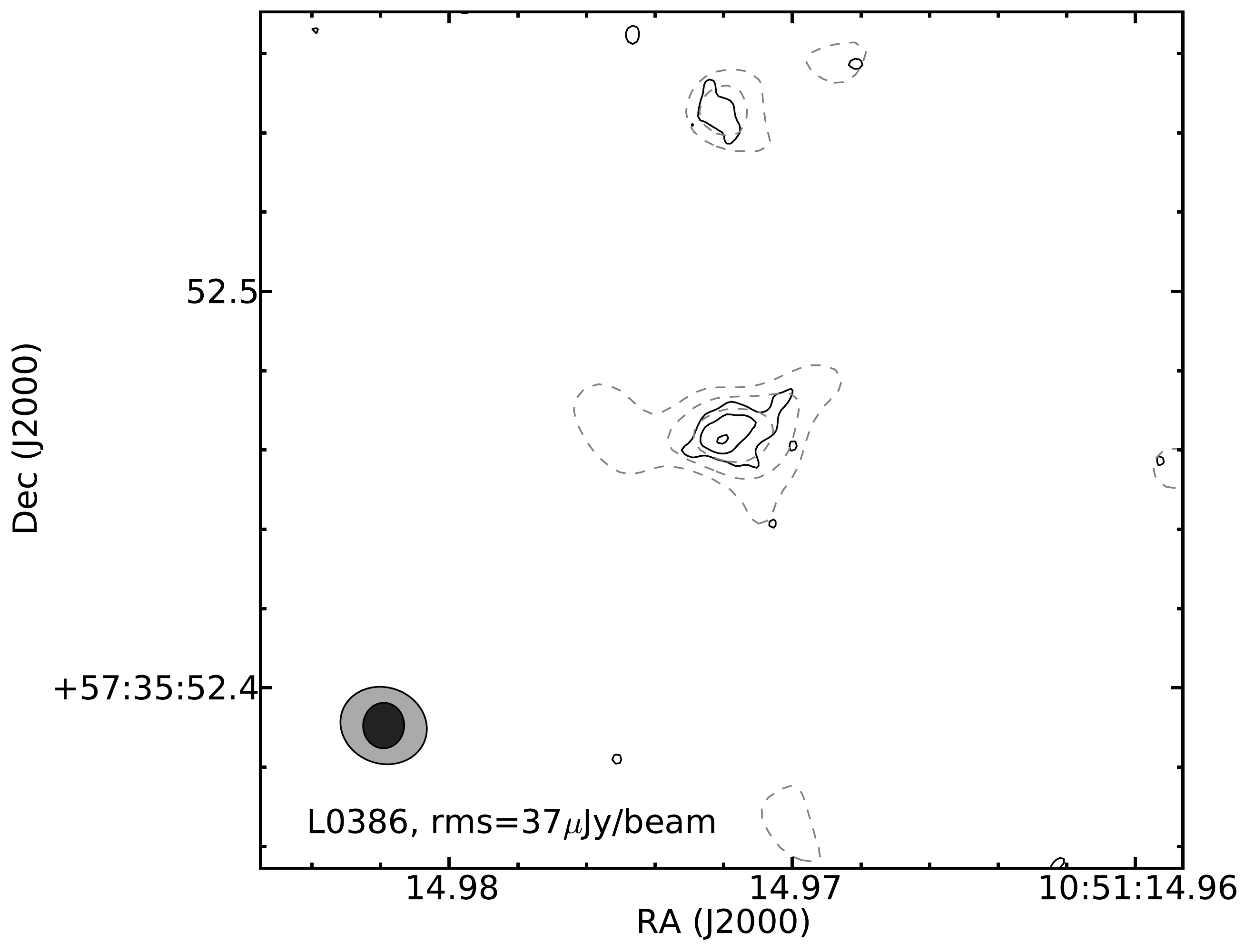}\\
\includegraphics[height=4.5cm]{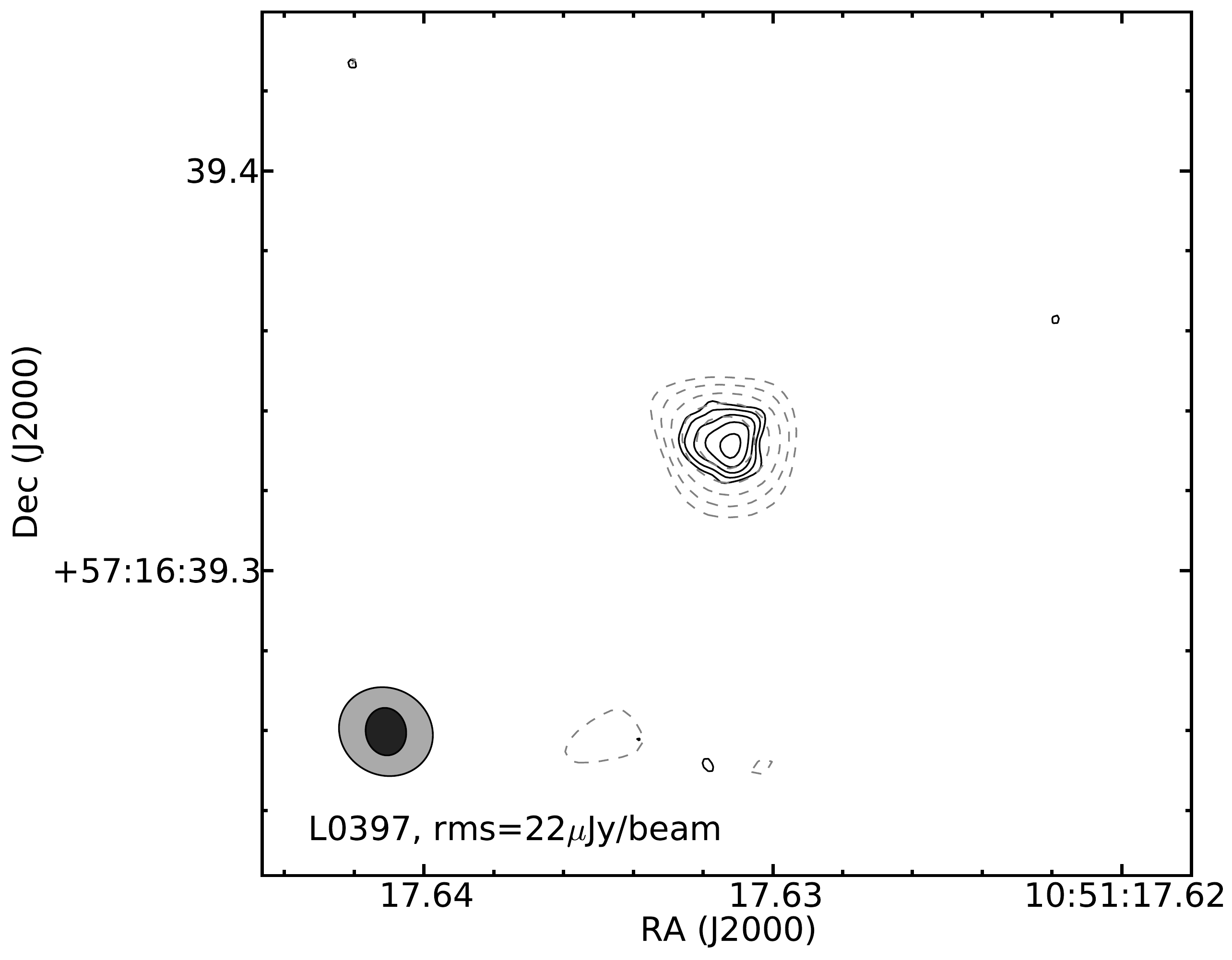}
\includegraphics[height=4.5cm]{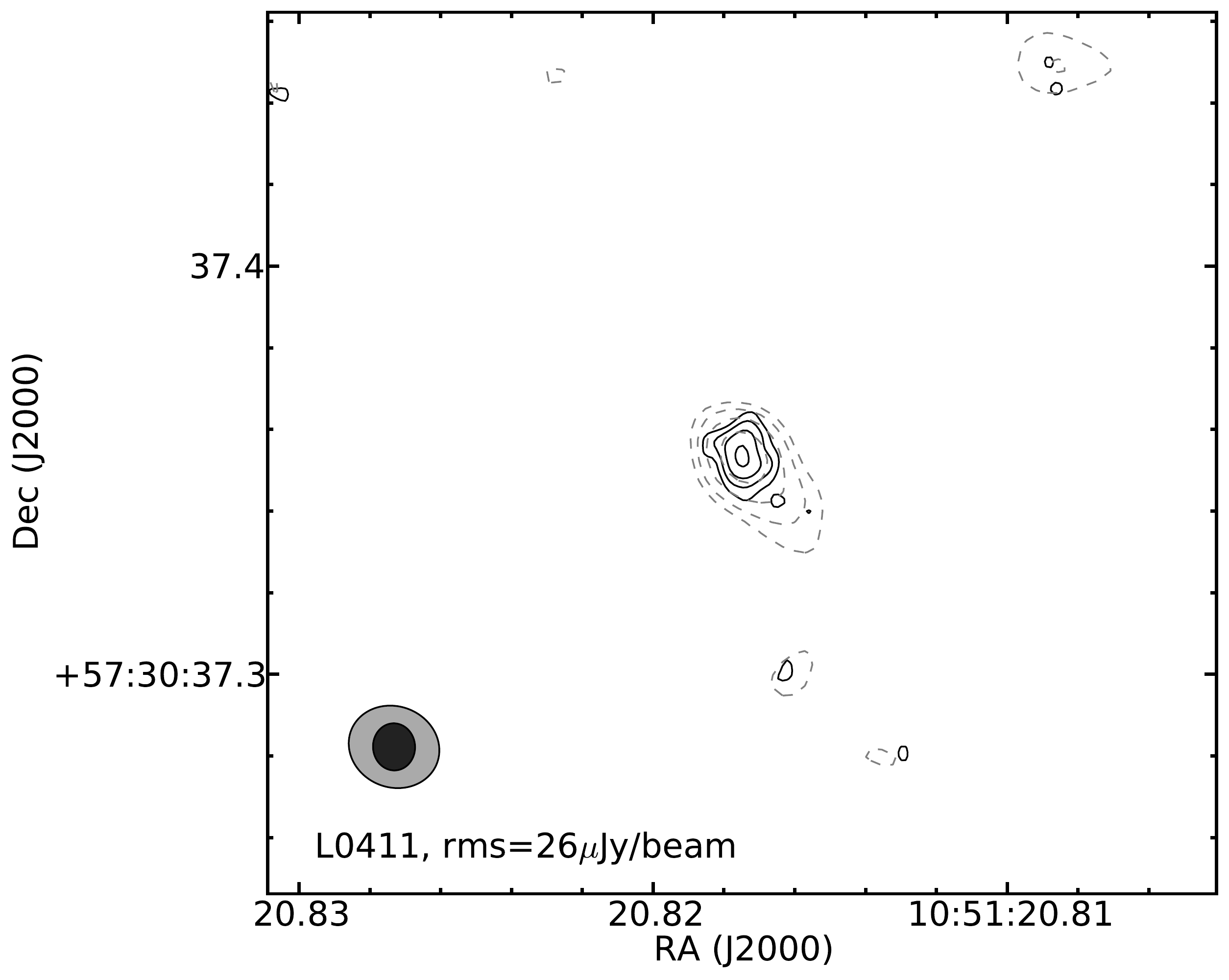}
\includegraphics[height=4.5cm]{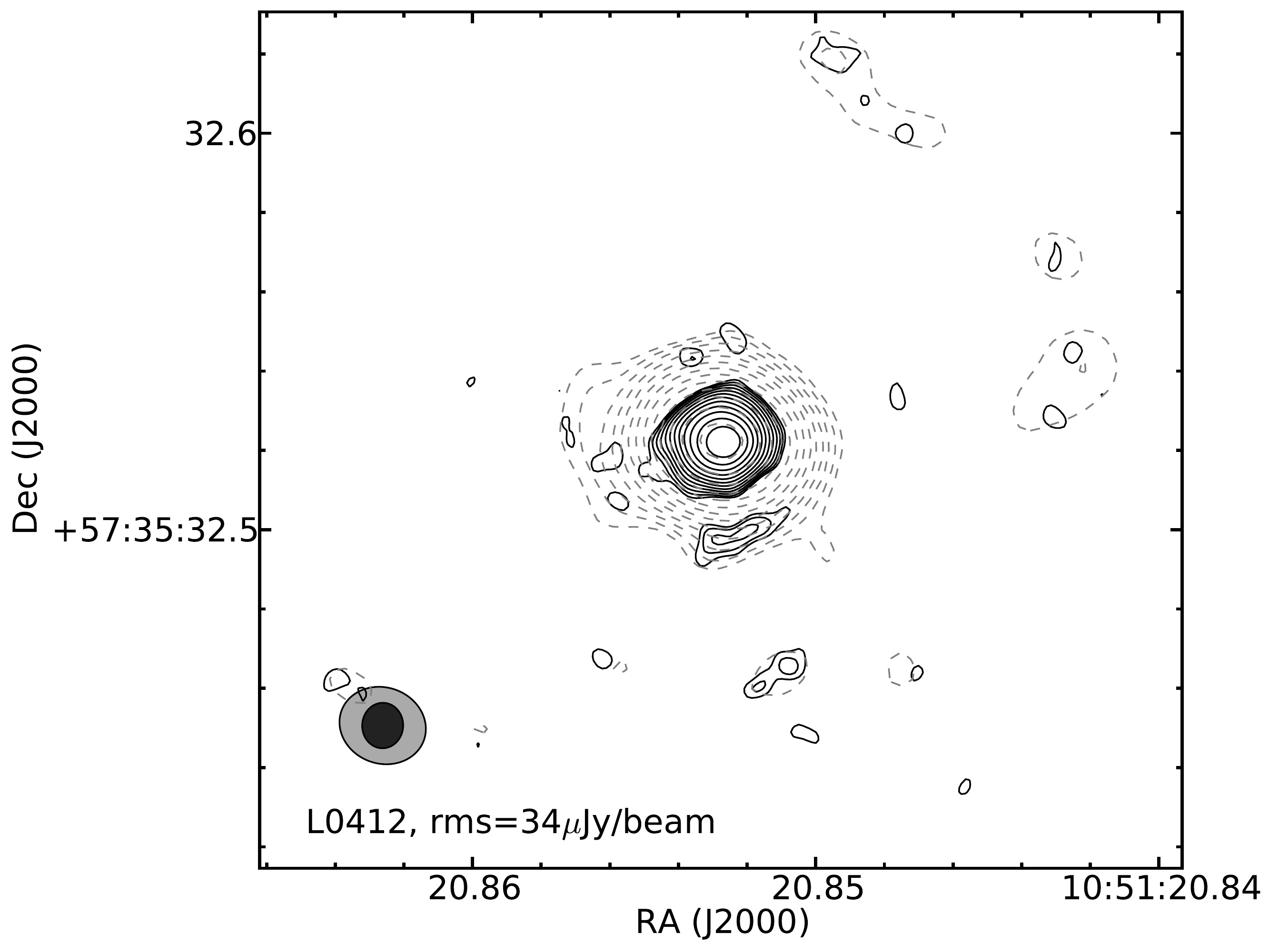}\\
\includegraphics[height=4.5cm]{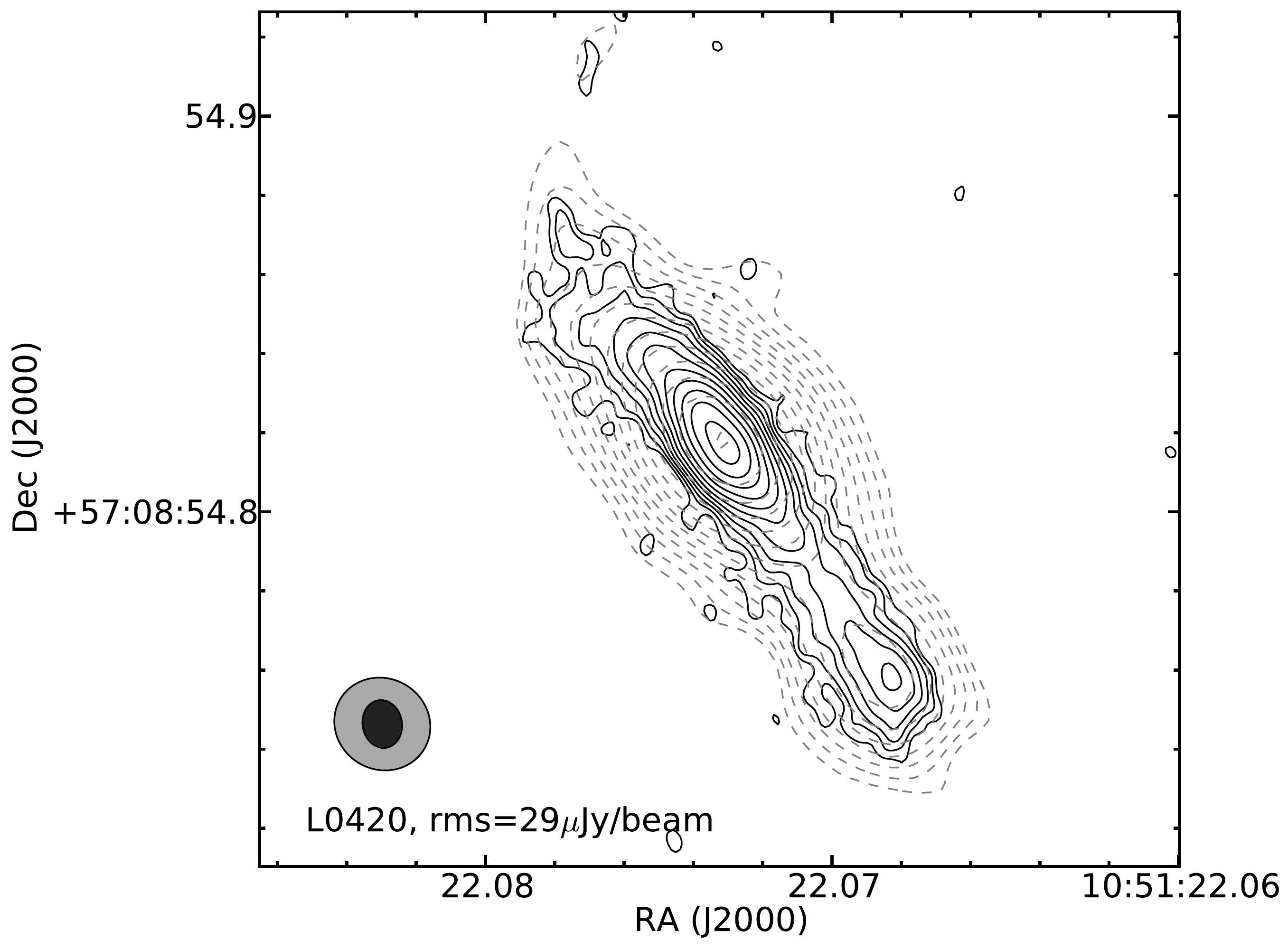}
\includegraphics[height=4.5cm]{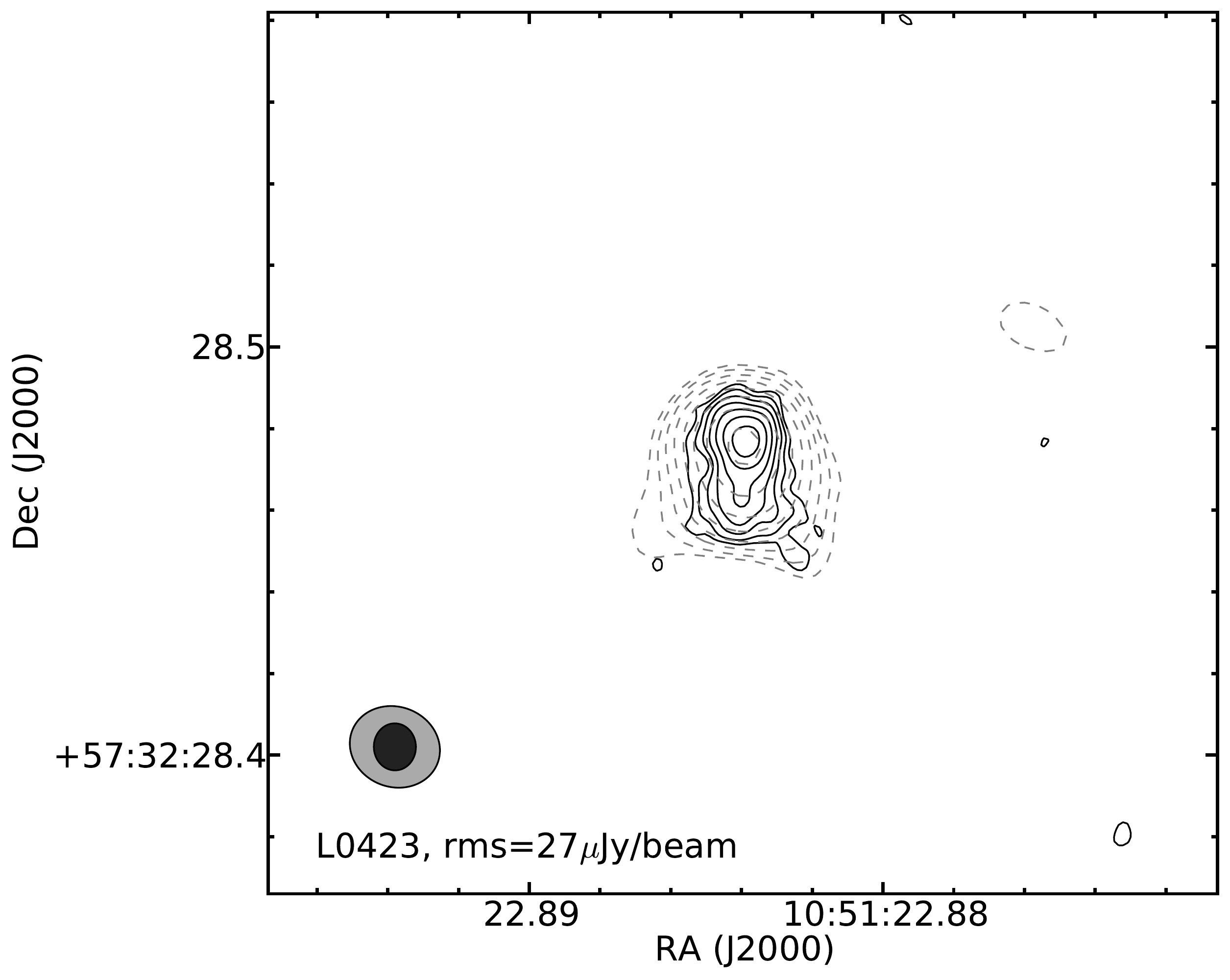}
\includegraphics[height=4.5cm]{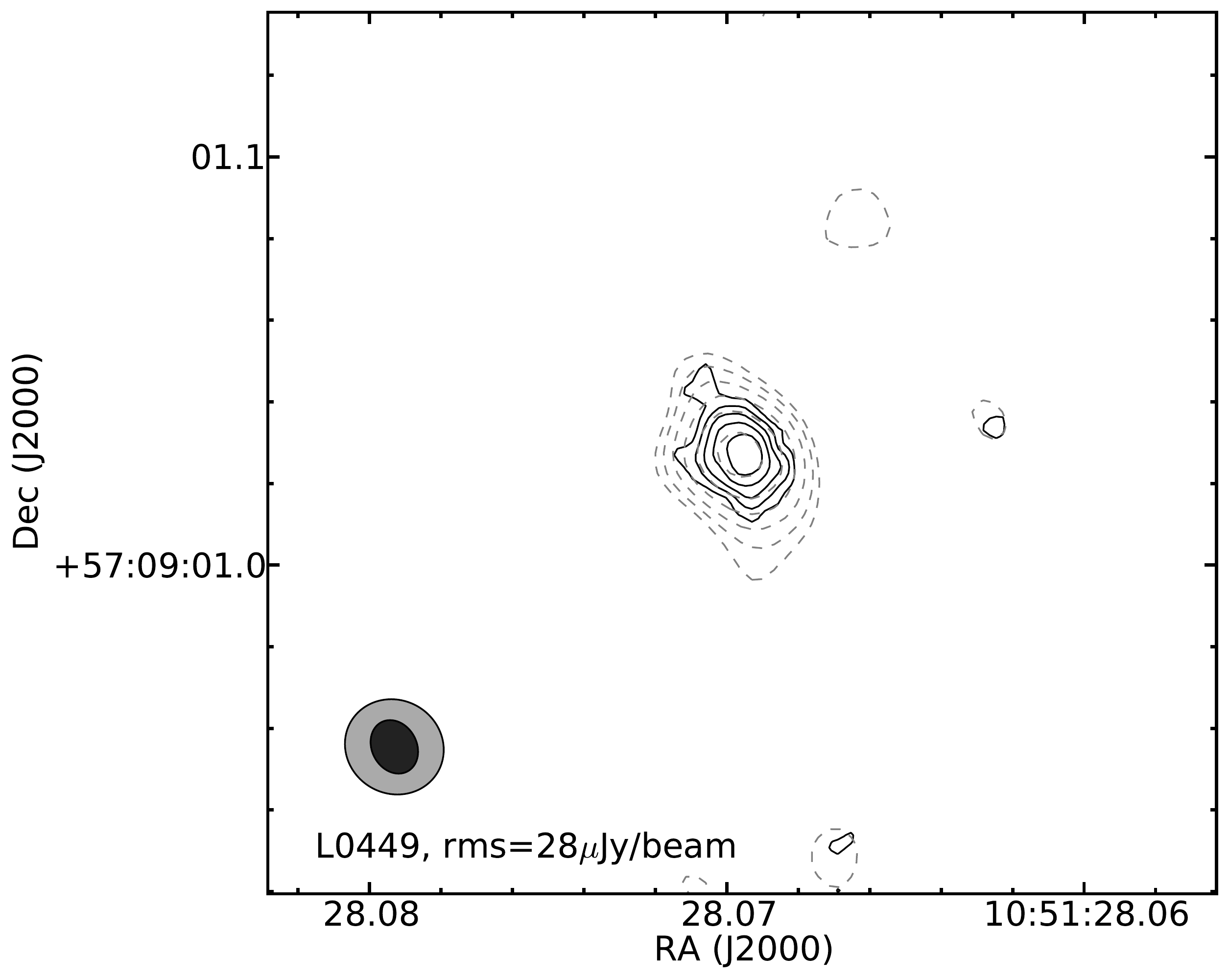}\\
\includegraphics[height=4.5cm]{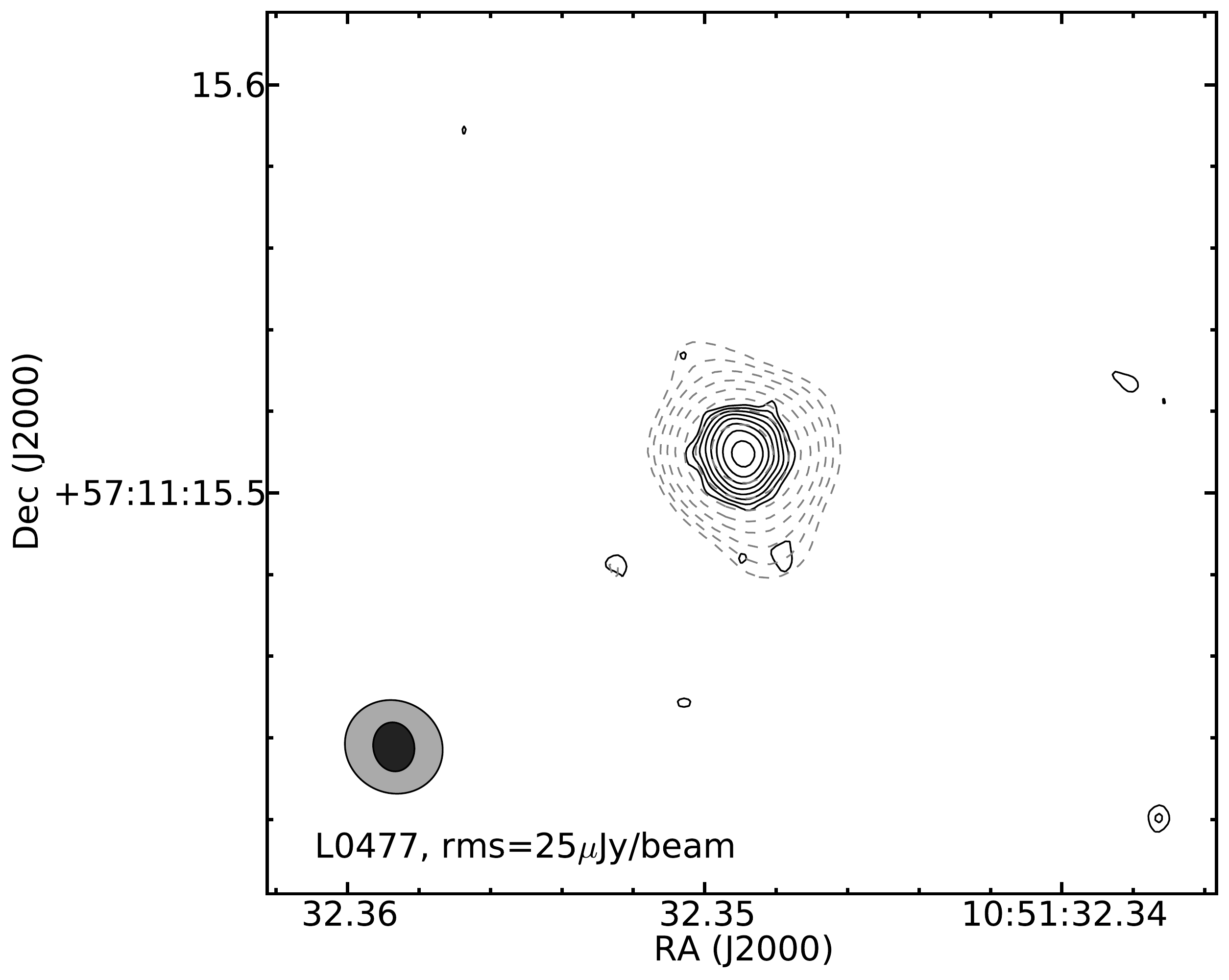}
\includegraphics[height=4.5cm]{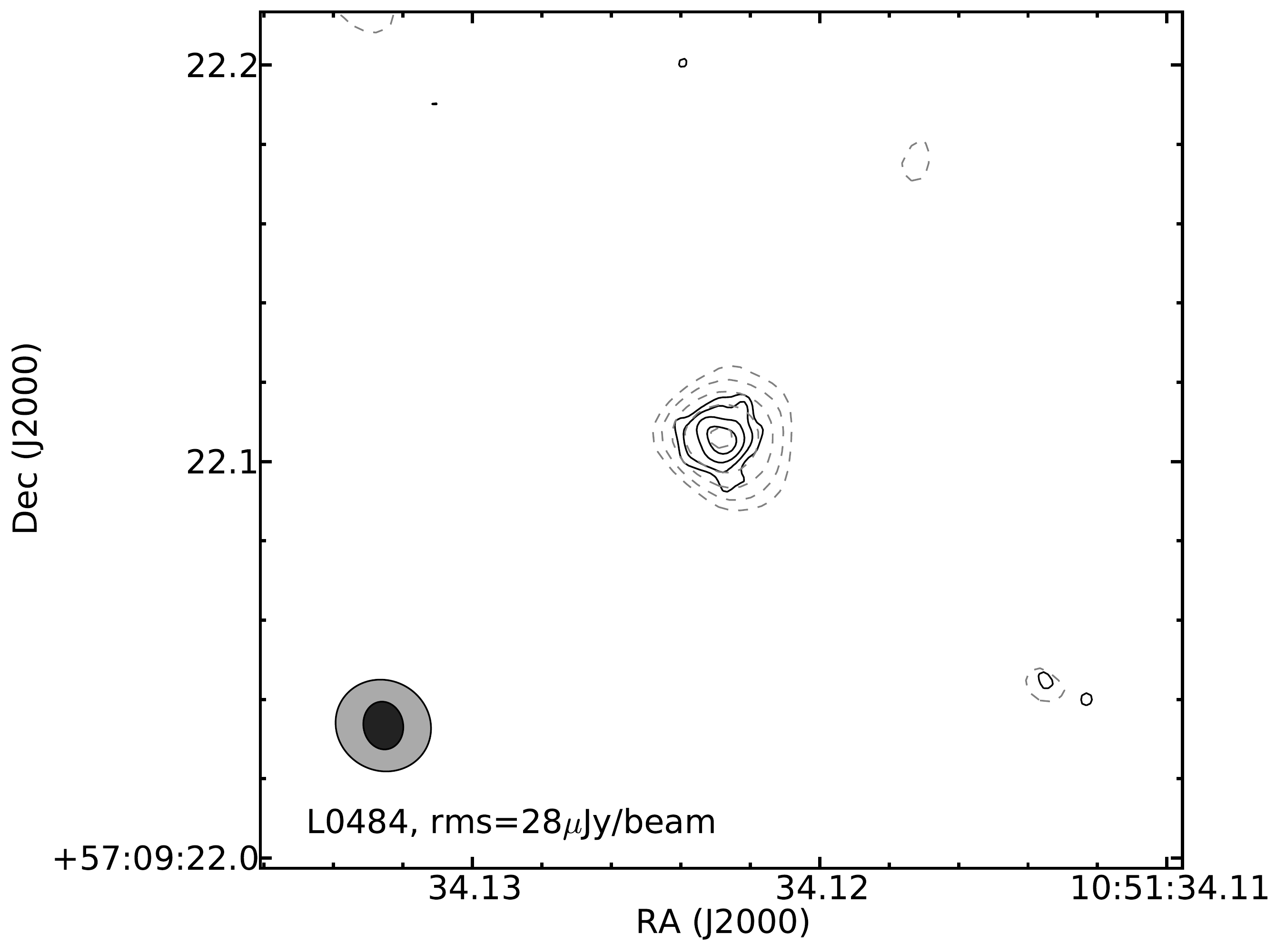}
\includegraphics[height=4.5cm]{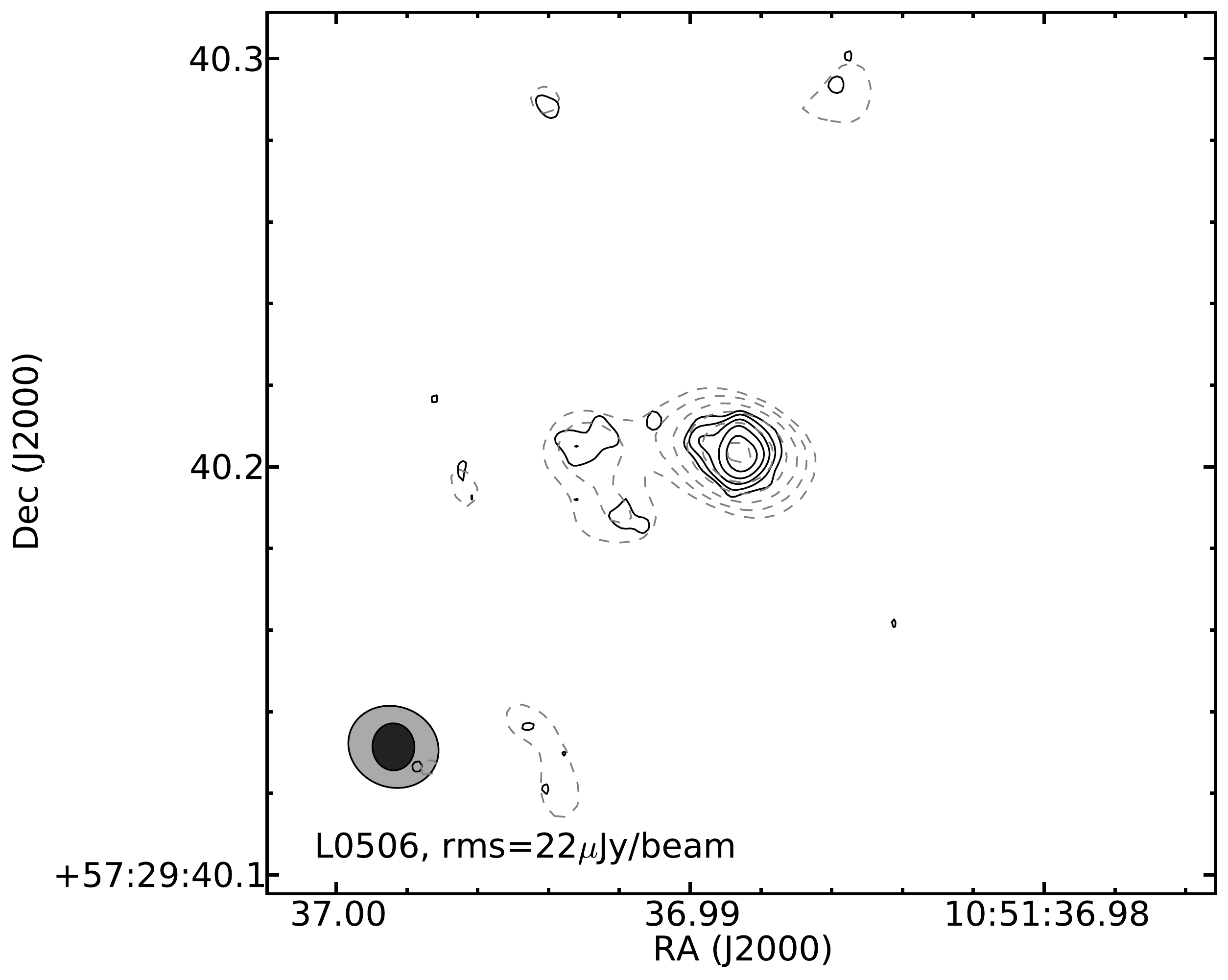}\\
\caption{(Continued)}
\end{figure*}

\begin{figure*}
\ContinuedFloat
\center
\includegraphics[height=4.5cm]{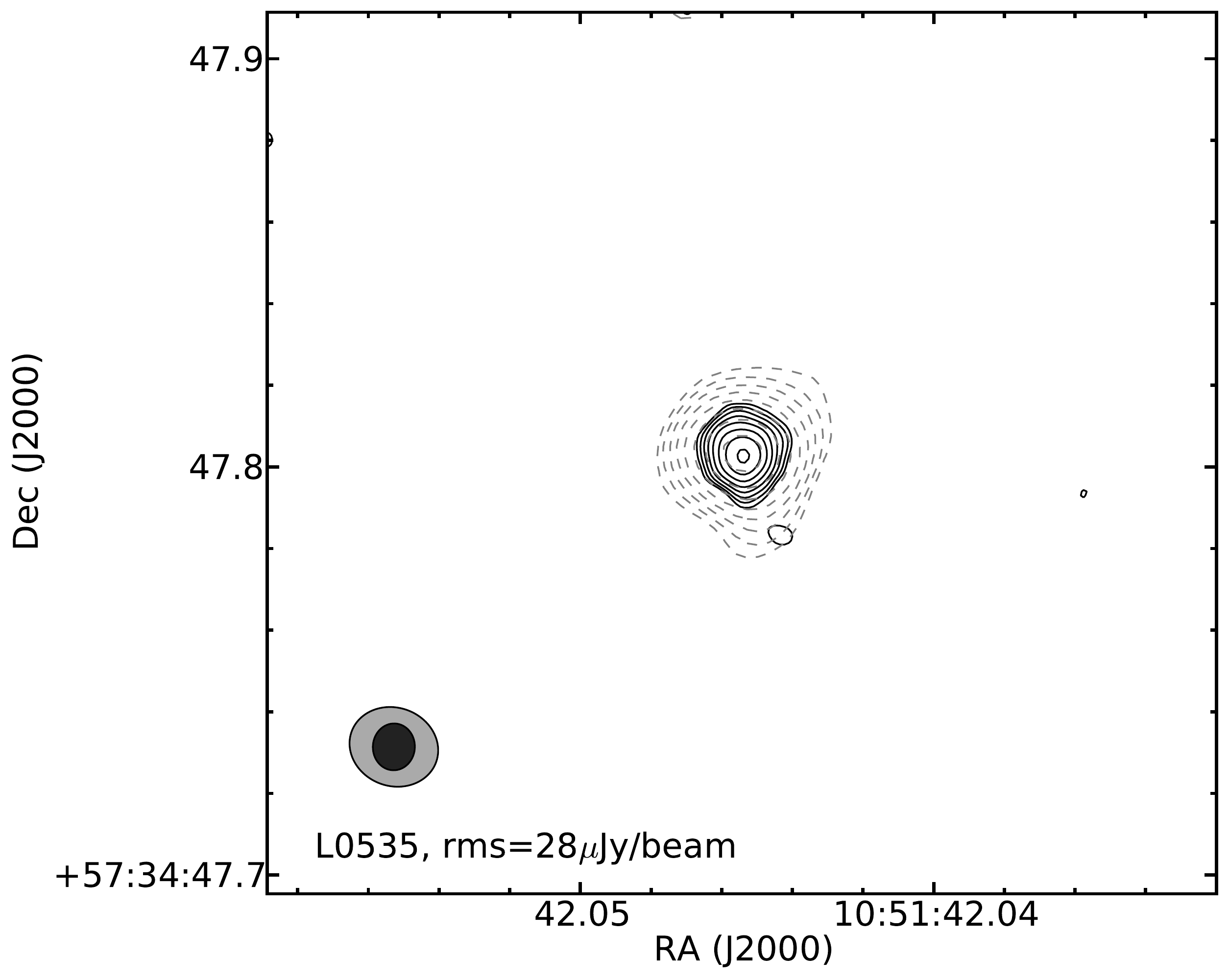}
\includegraphics[height=4.5cm]{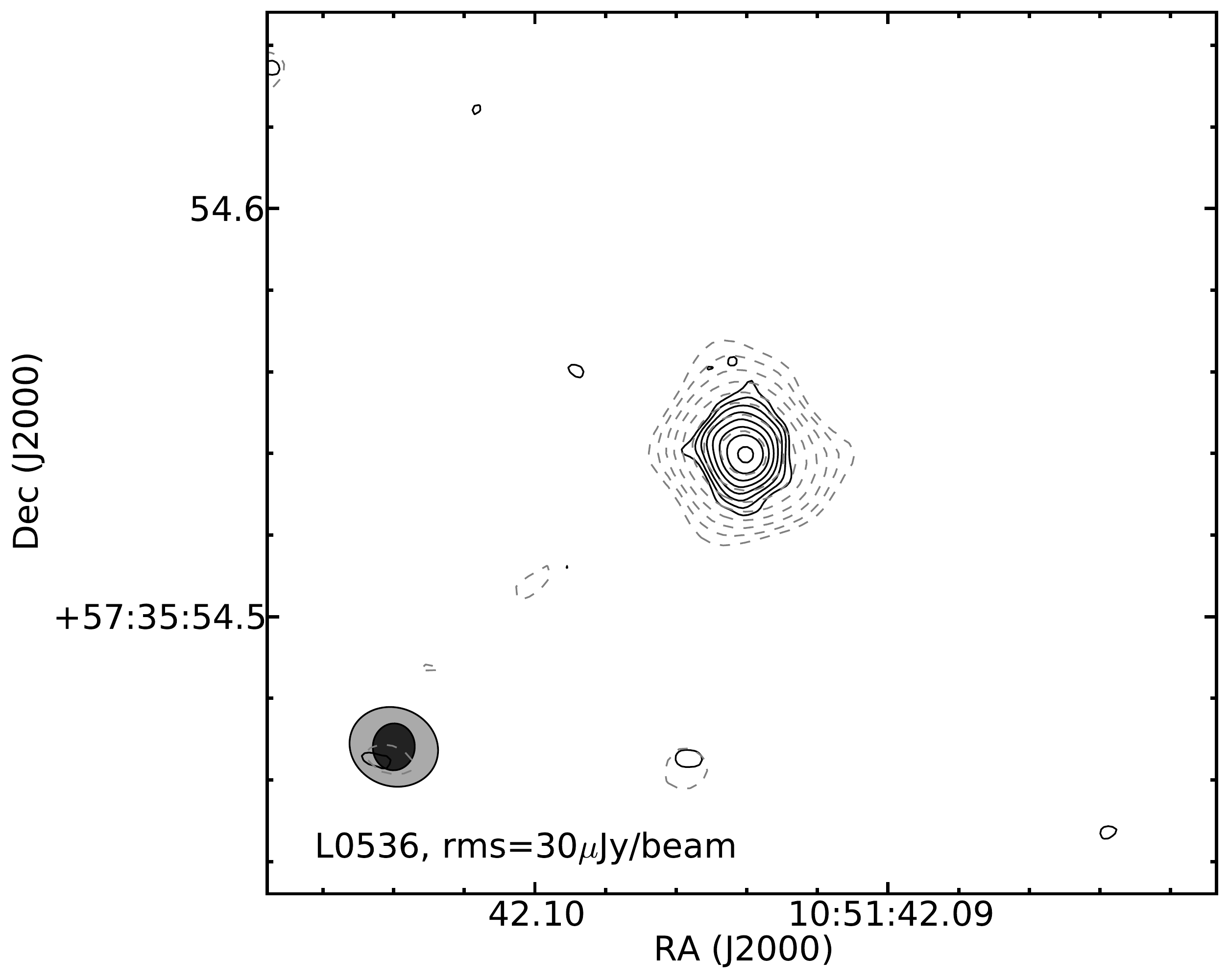}
\includegraphics[height=4.5cm]{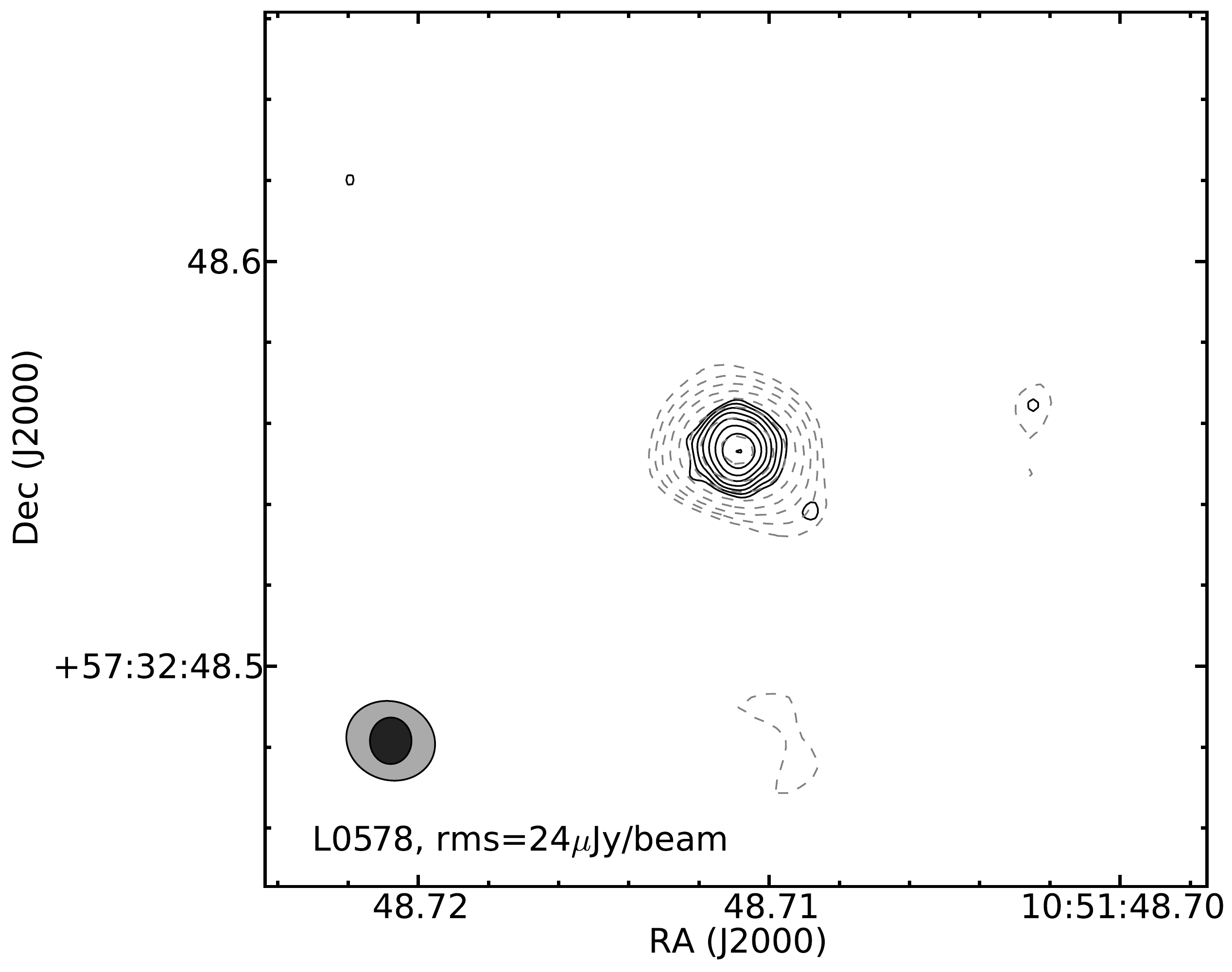}\\
\includegraphics[height=4.5cm]{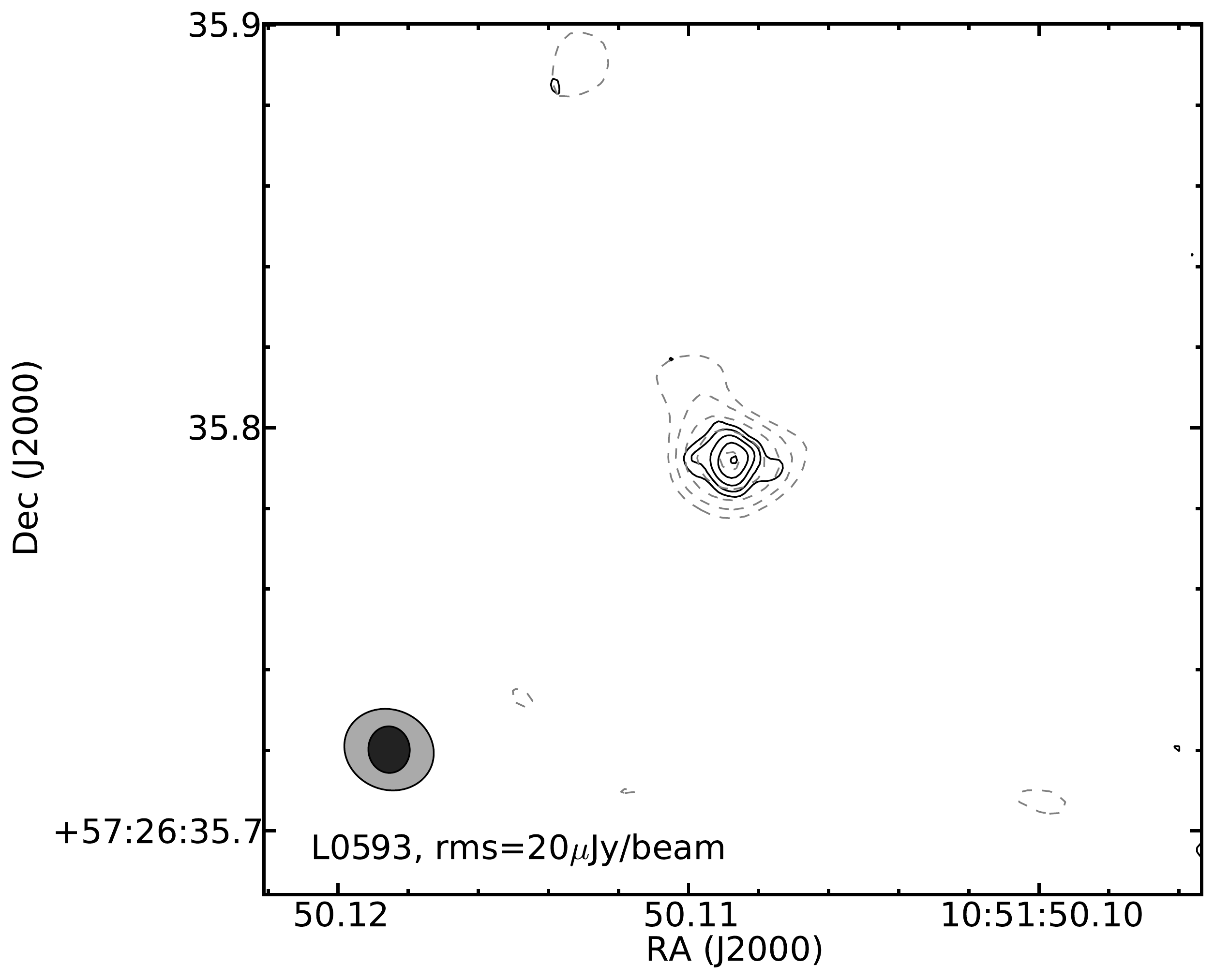}
\includegraphics[height=4.5cm]{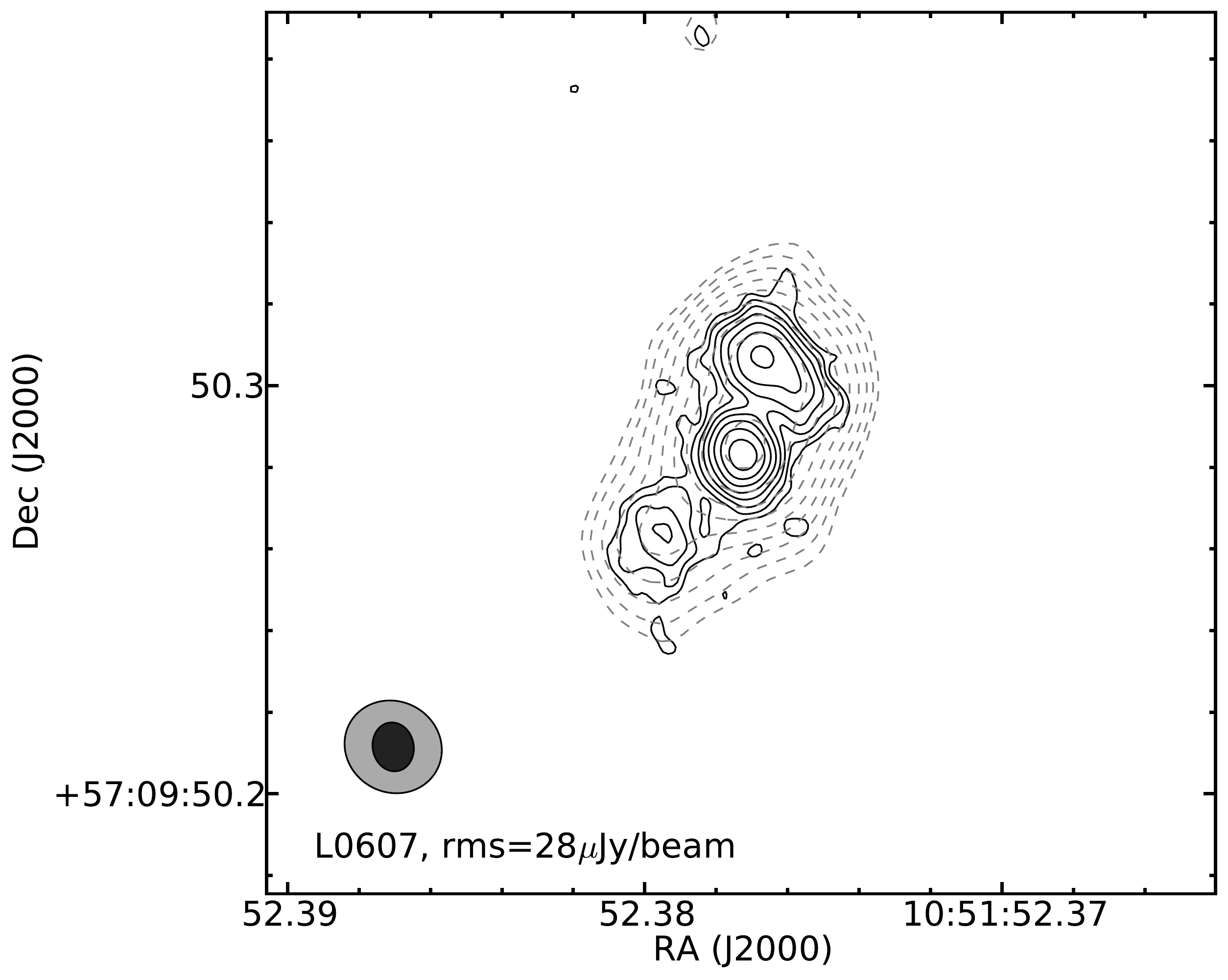}
\includegraphics[height=4.5cm]{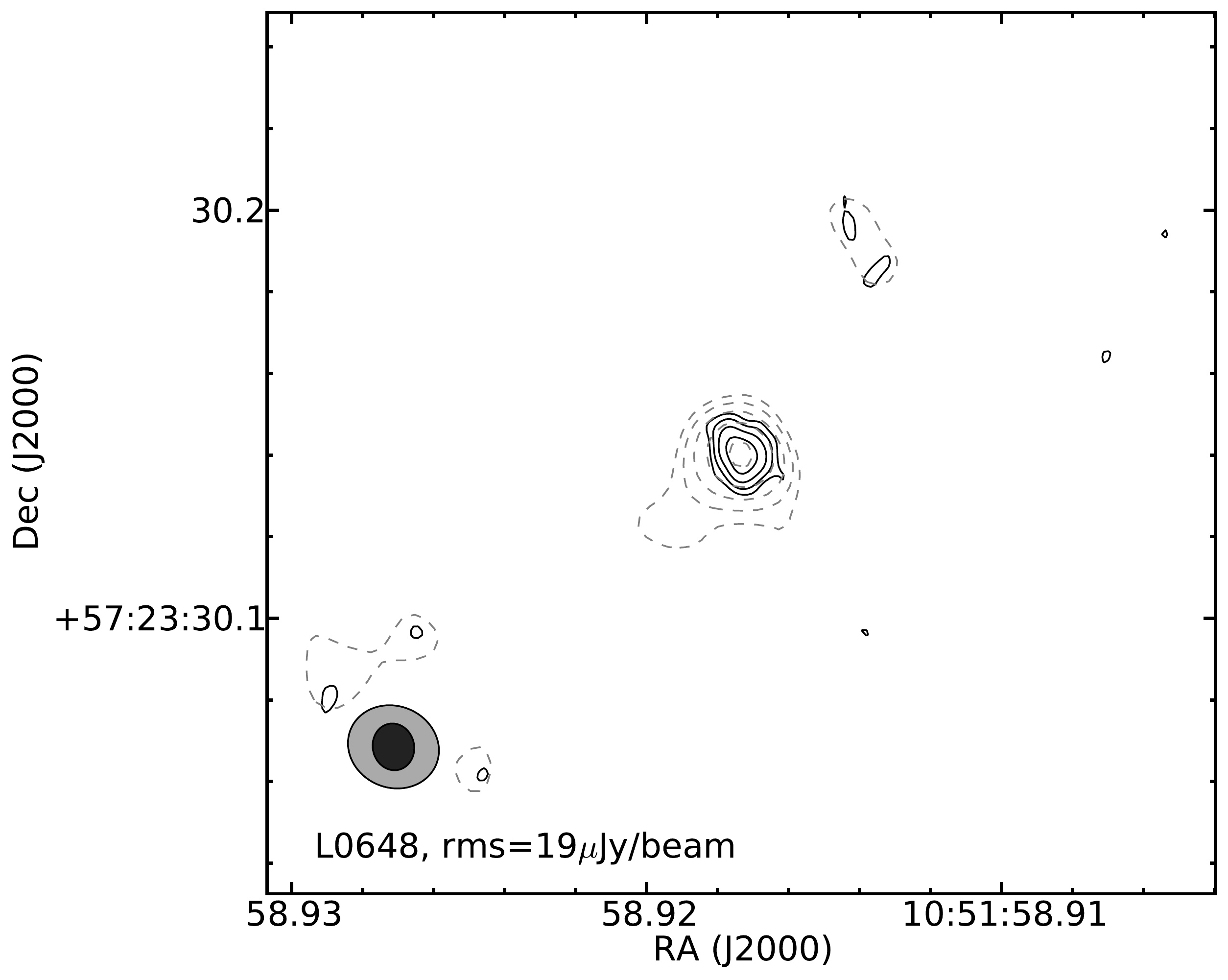}\\
\includegraphics[height=4.5cm]{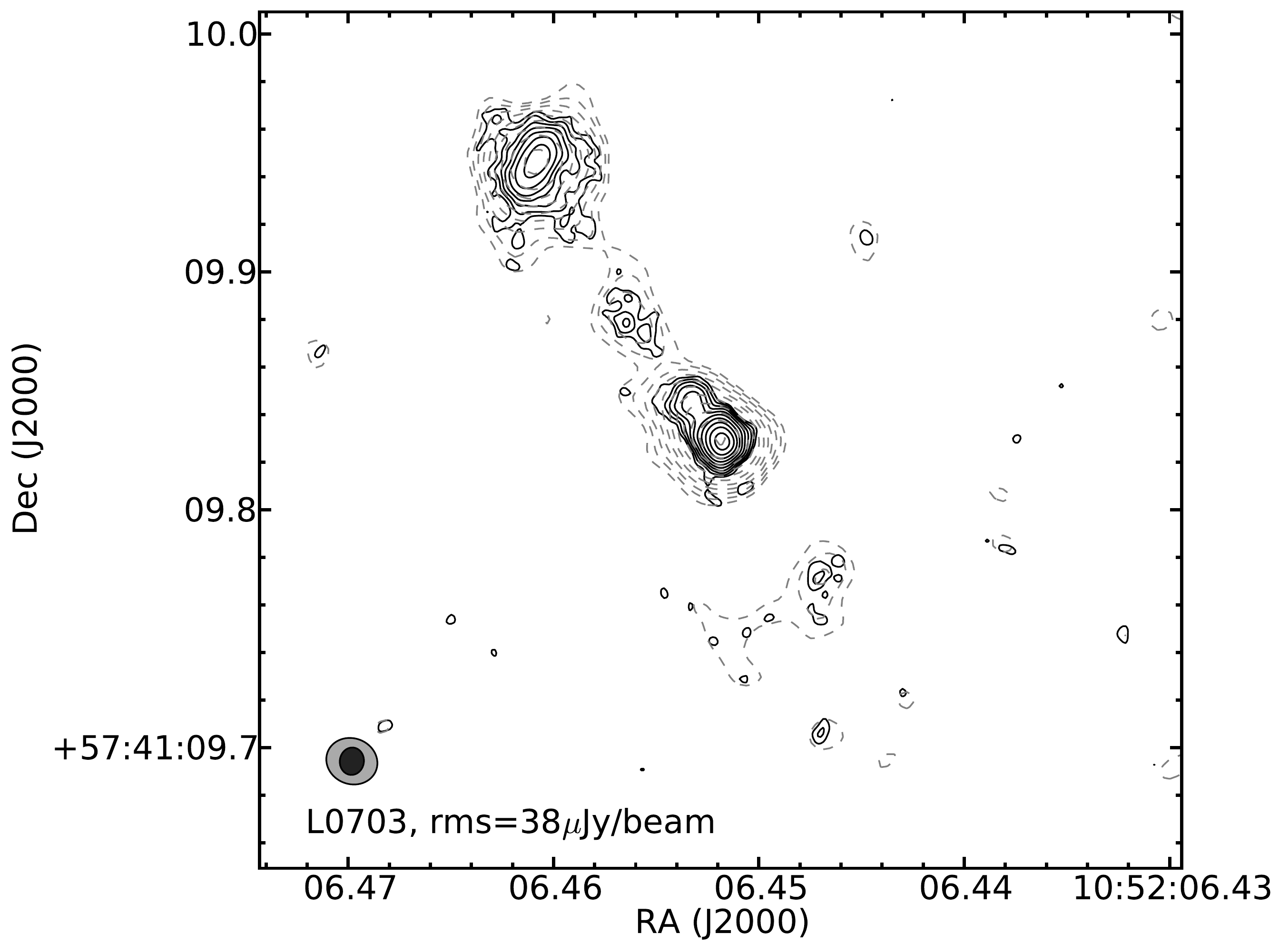}
\includegraphics[height=4.5cm]{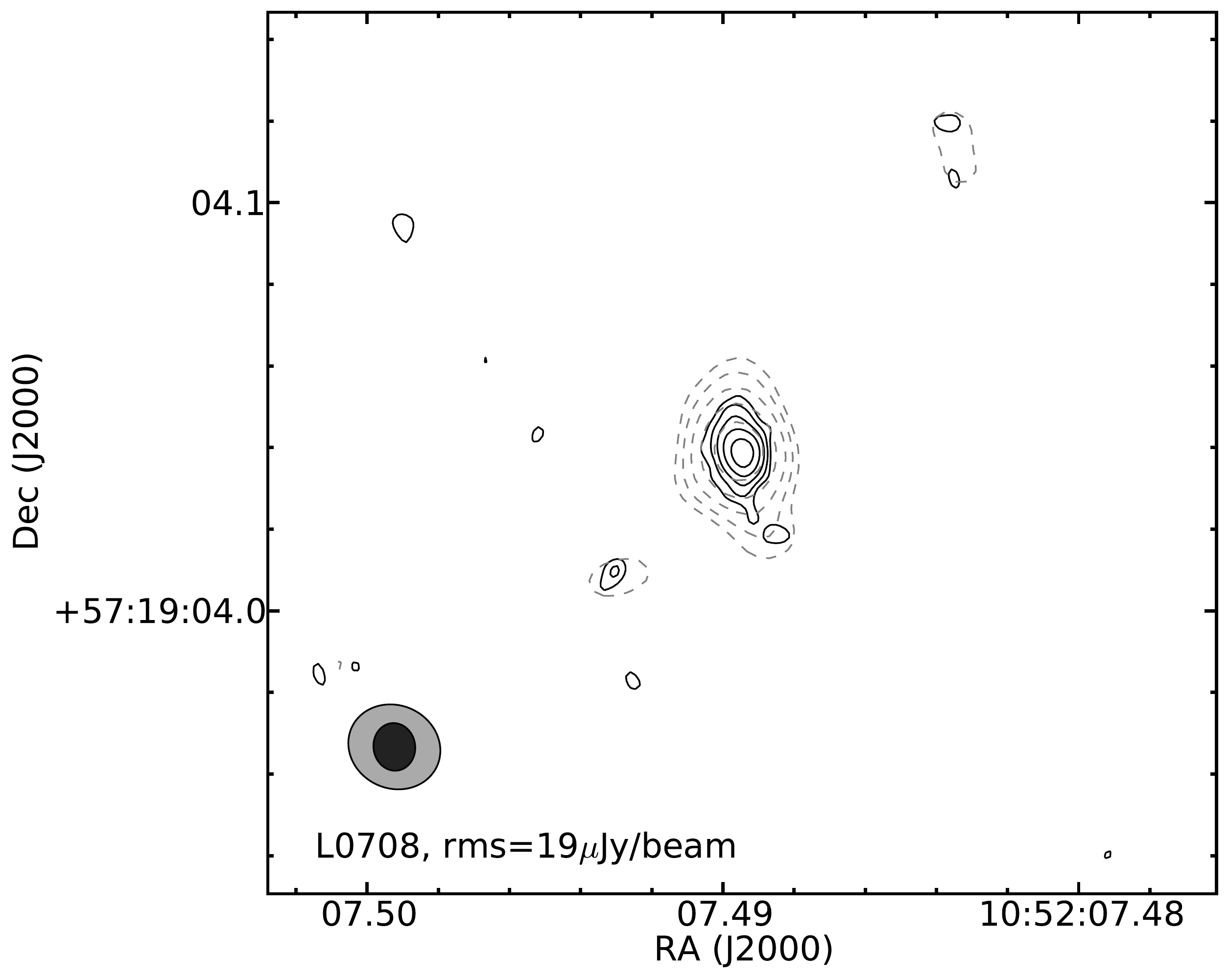}
\includegraphics[height=4.5cm]{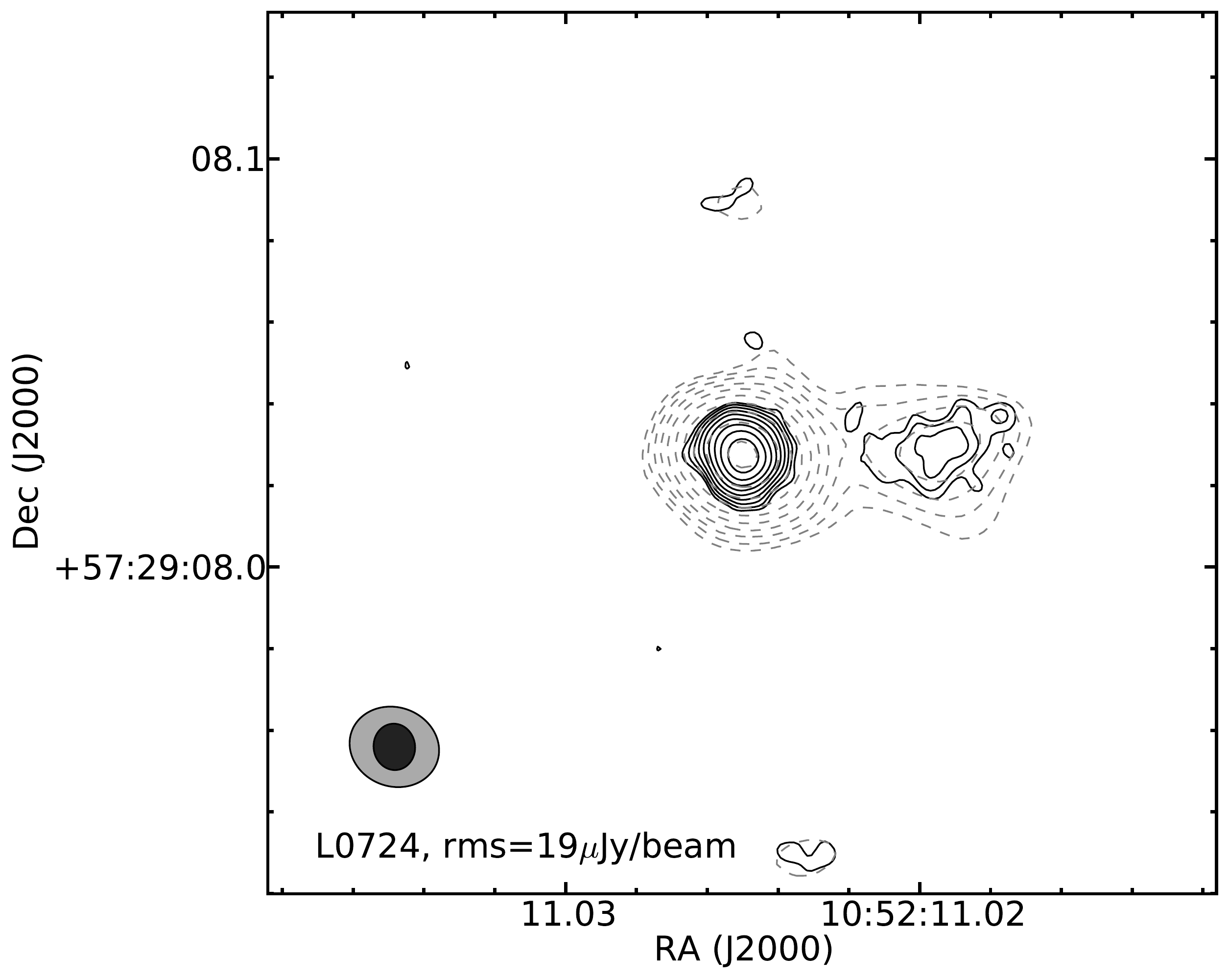}\\
\includegraphics[height=4.5cm]{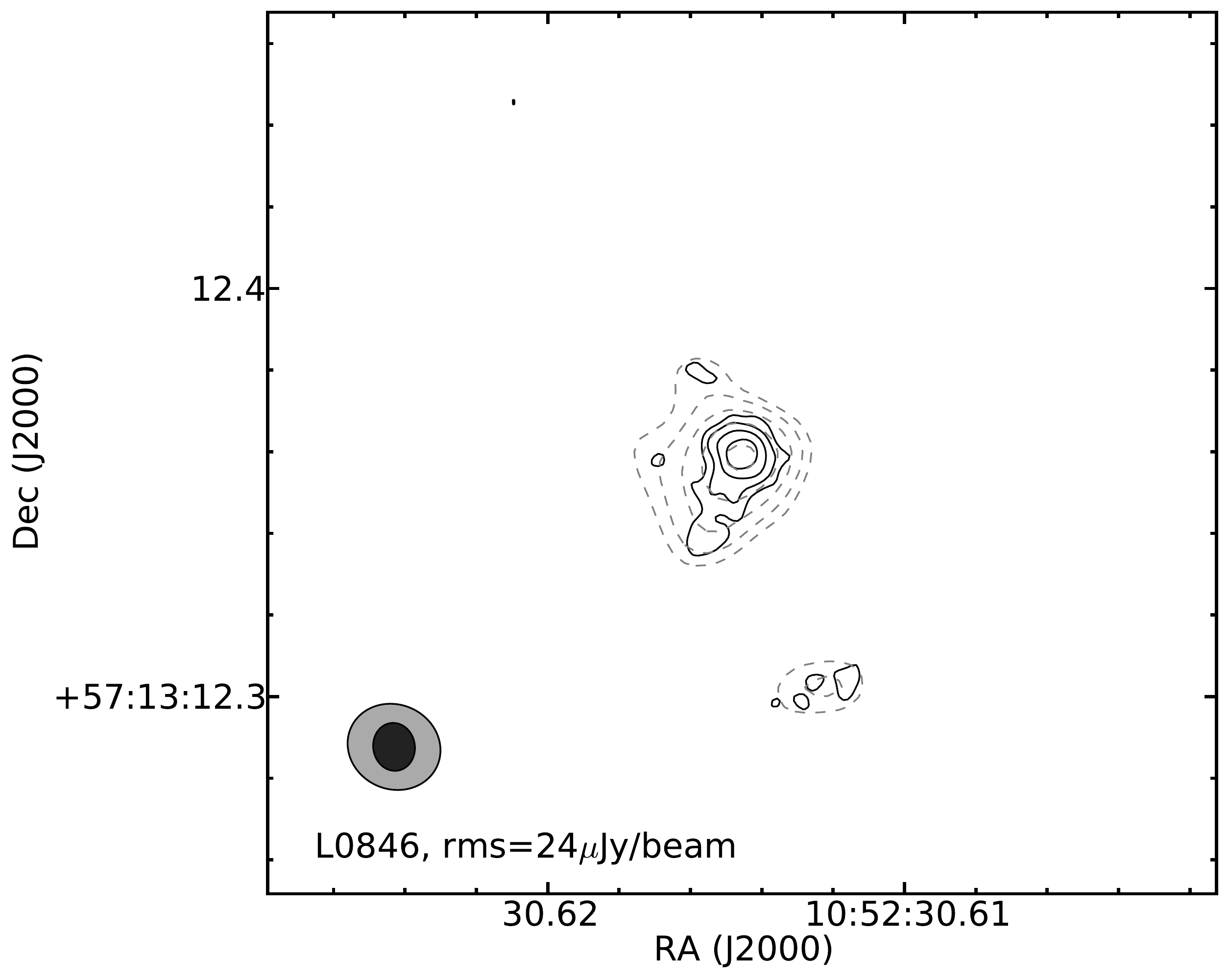}
\includegraphics[height=4.5cm]{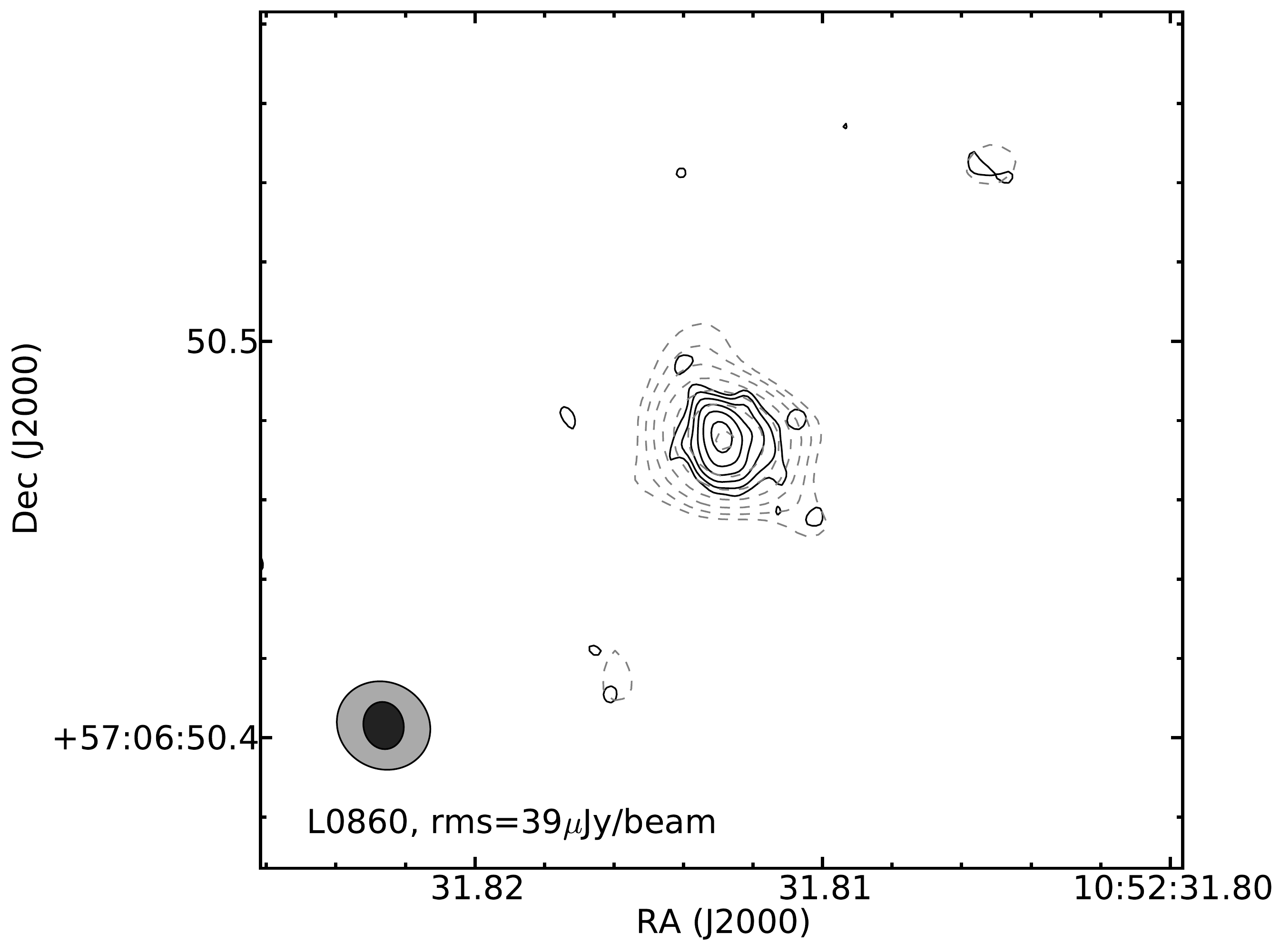}
\includegraphics[height=4.5cm]{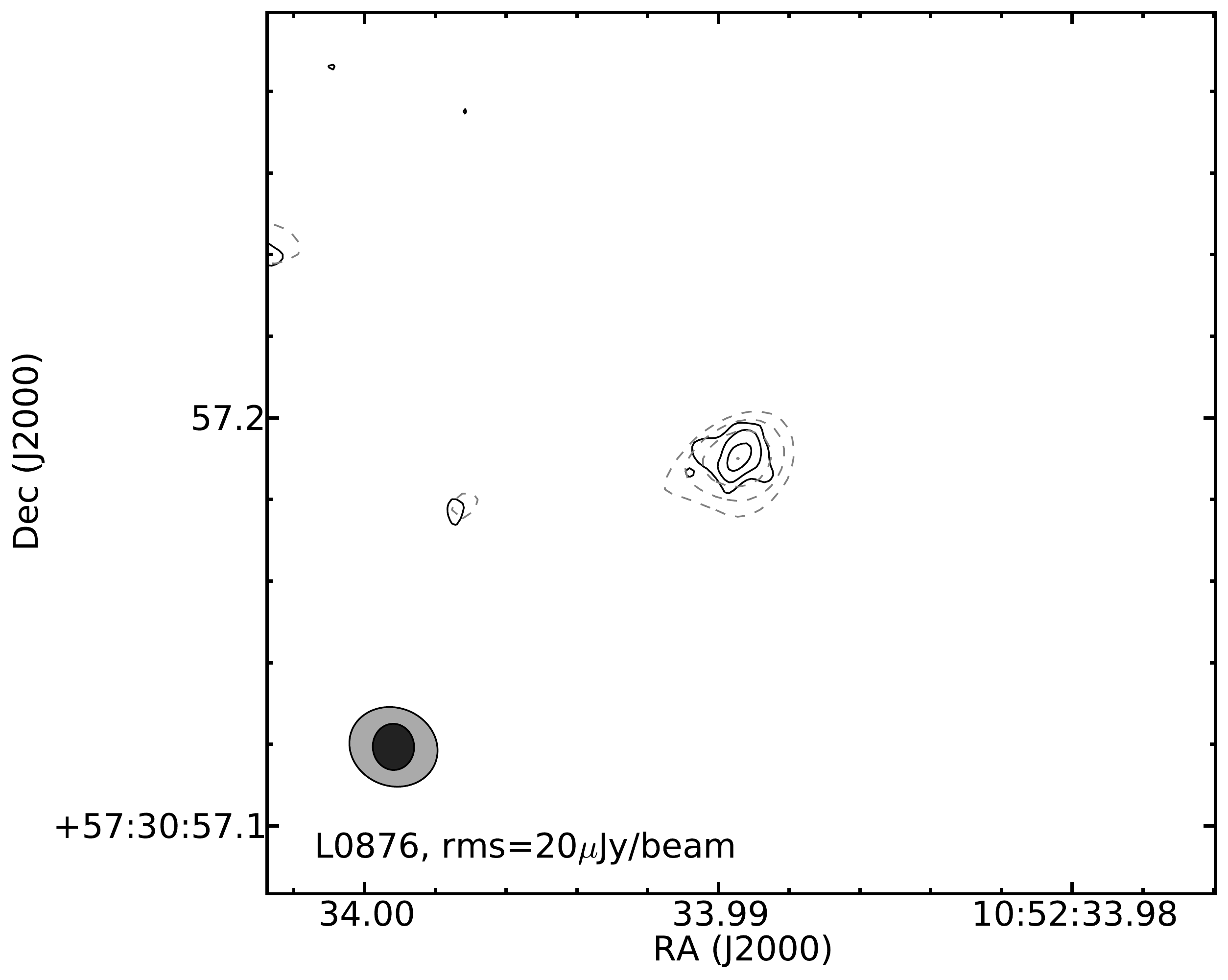}\\
\includegraphics[height=4.5cm]{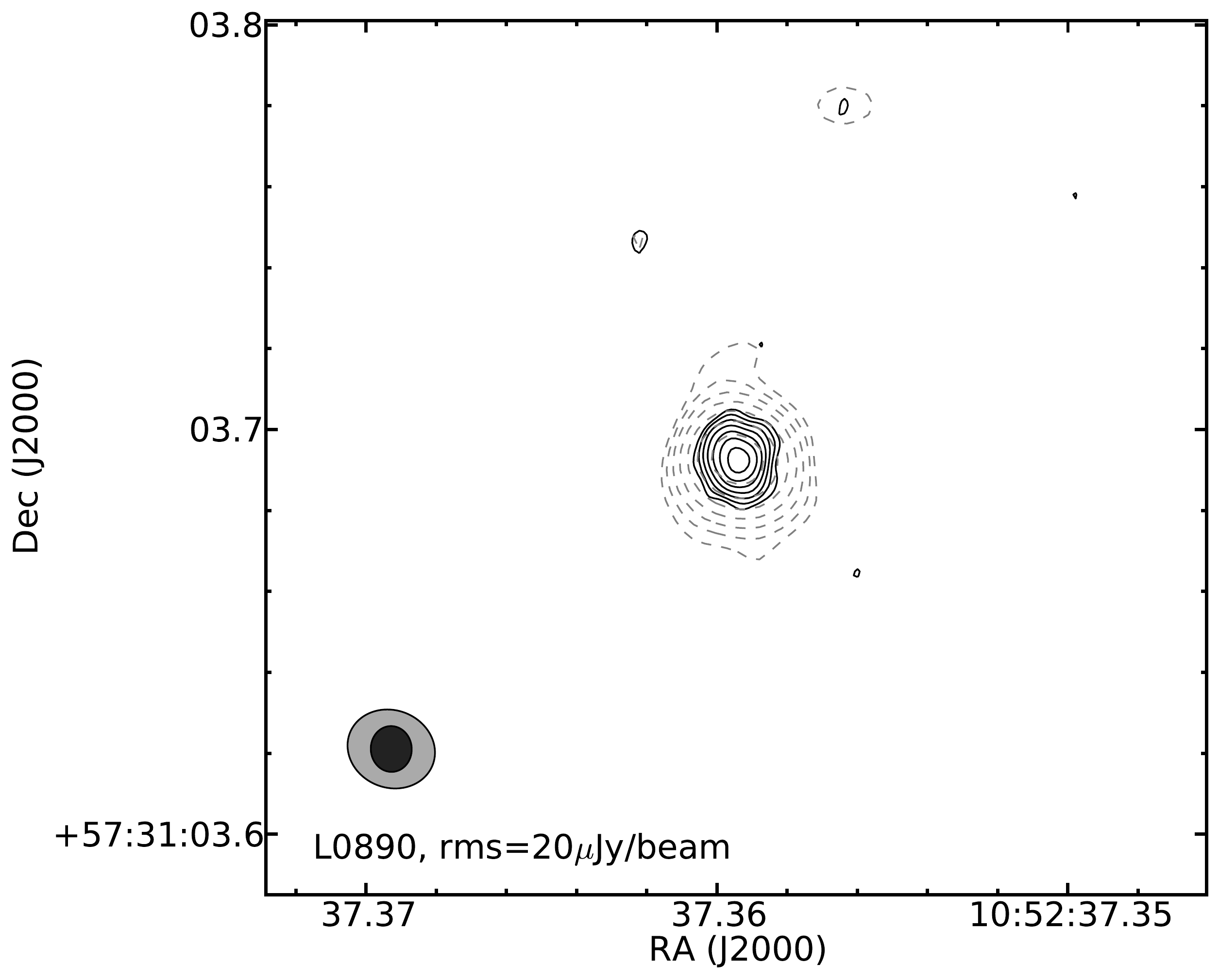}
\includegraphics[height=4.5cm]{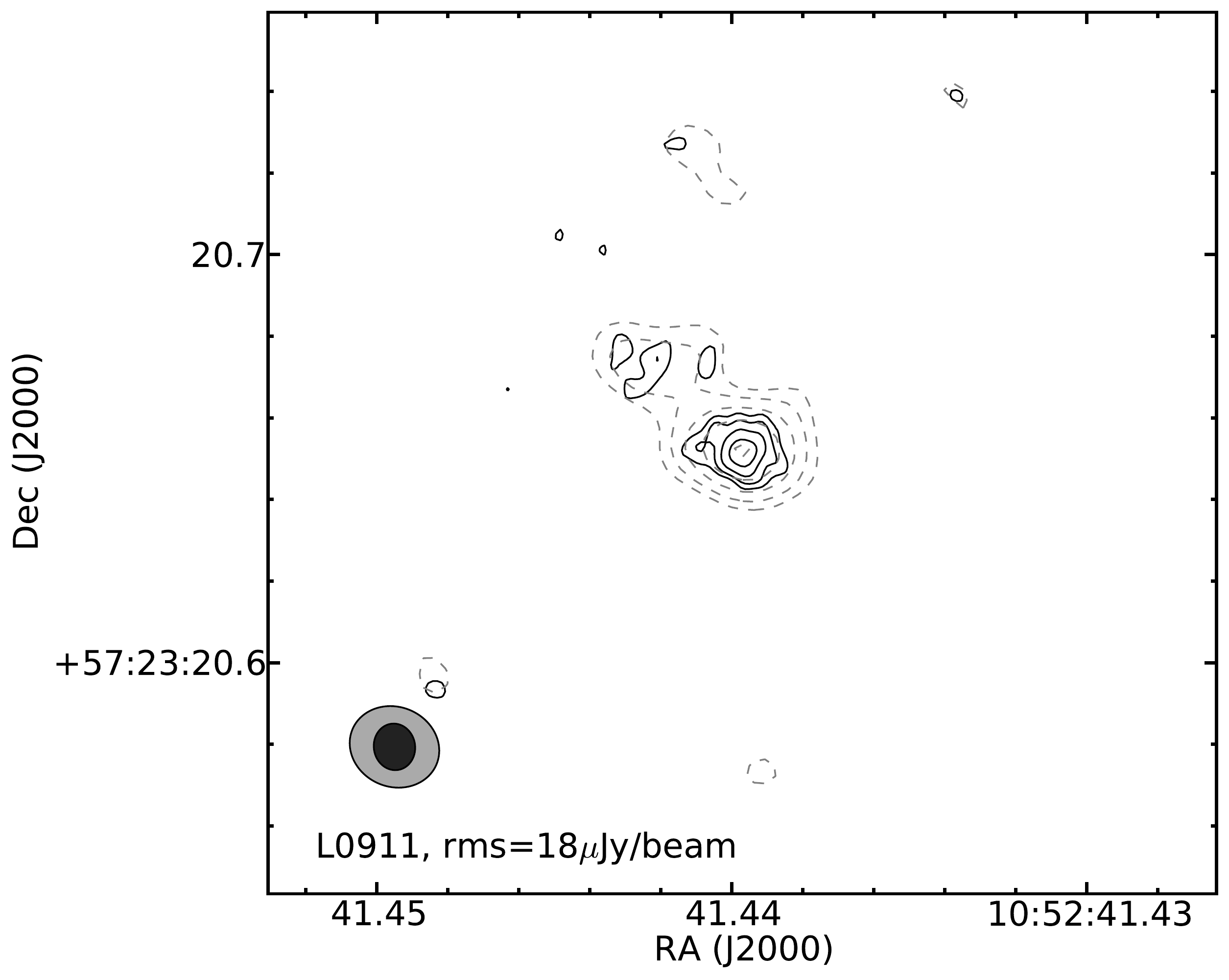}
\includegraphics[height=4.5cm]{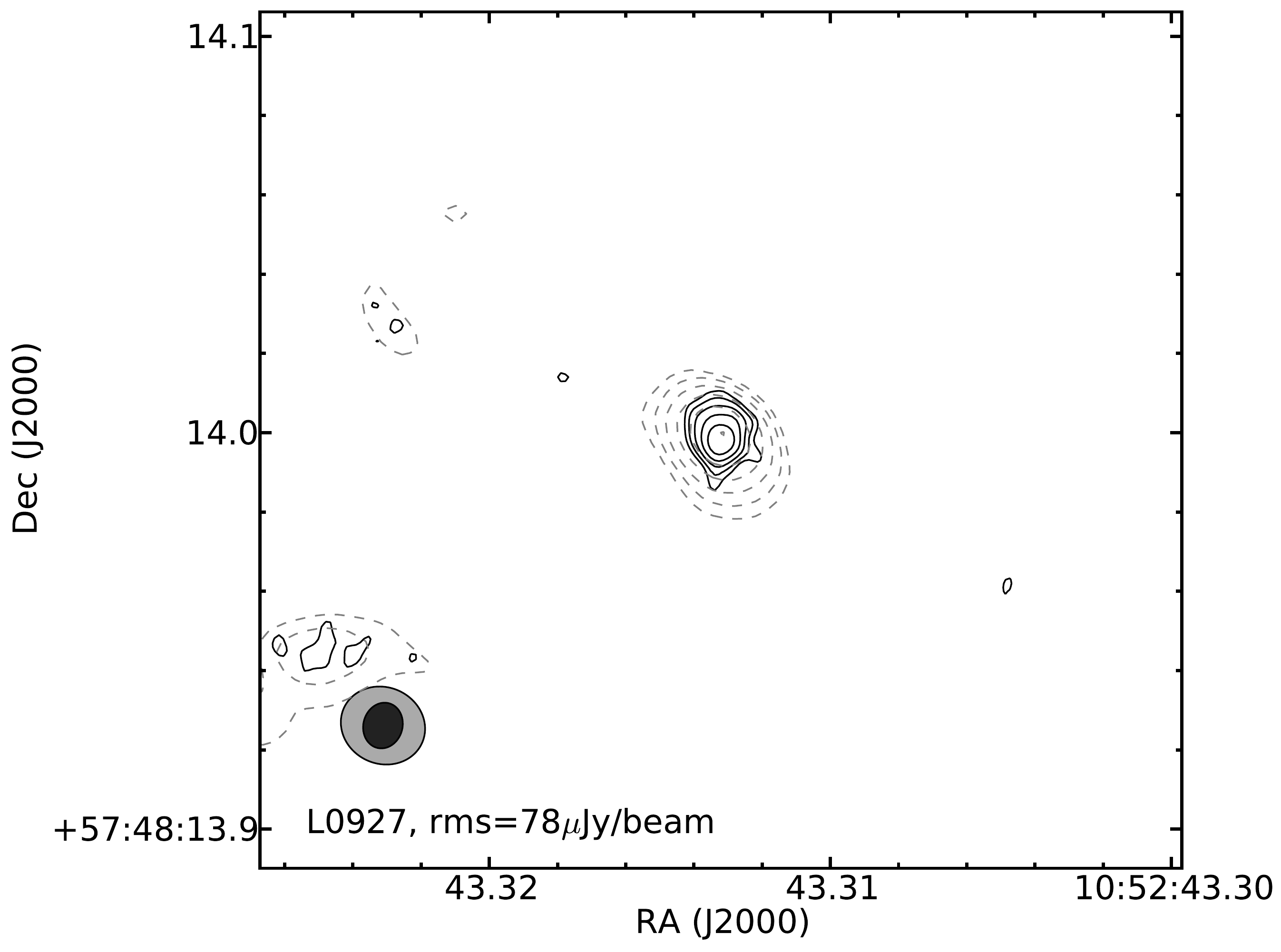}\\
\caption{(Continued)}
\end{figure*}

\begin{figure*}
\ContinuedFloat
\center
\includegraphics[height=4.5cm]{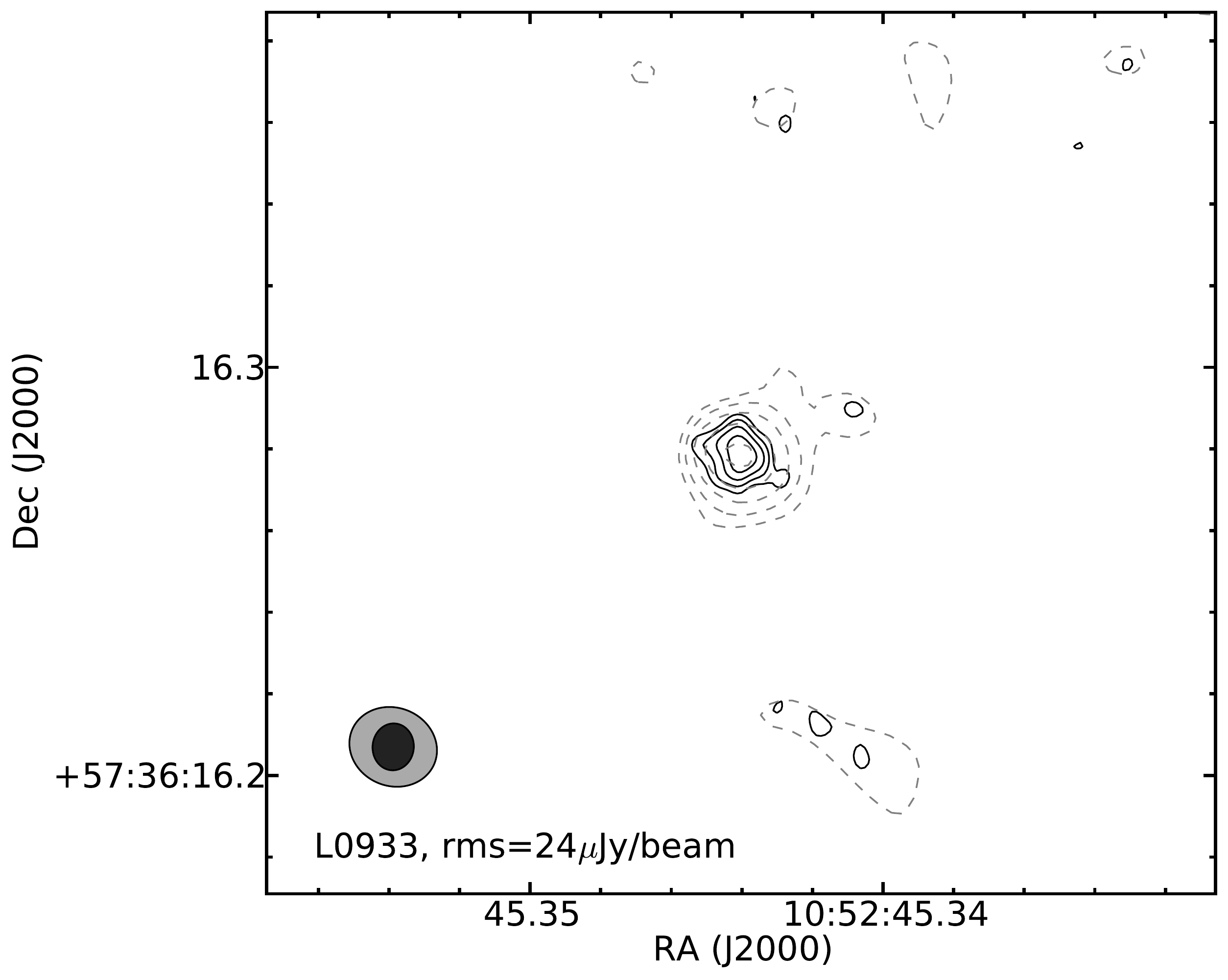}
\includegraphics[height=4.5cm]{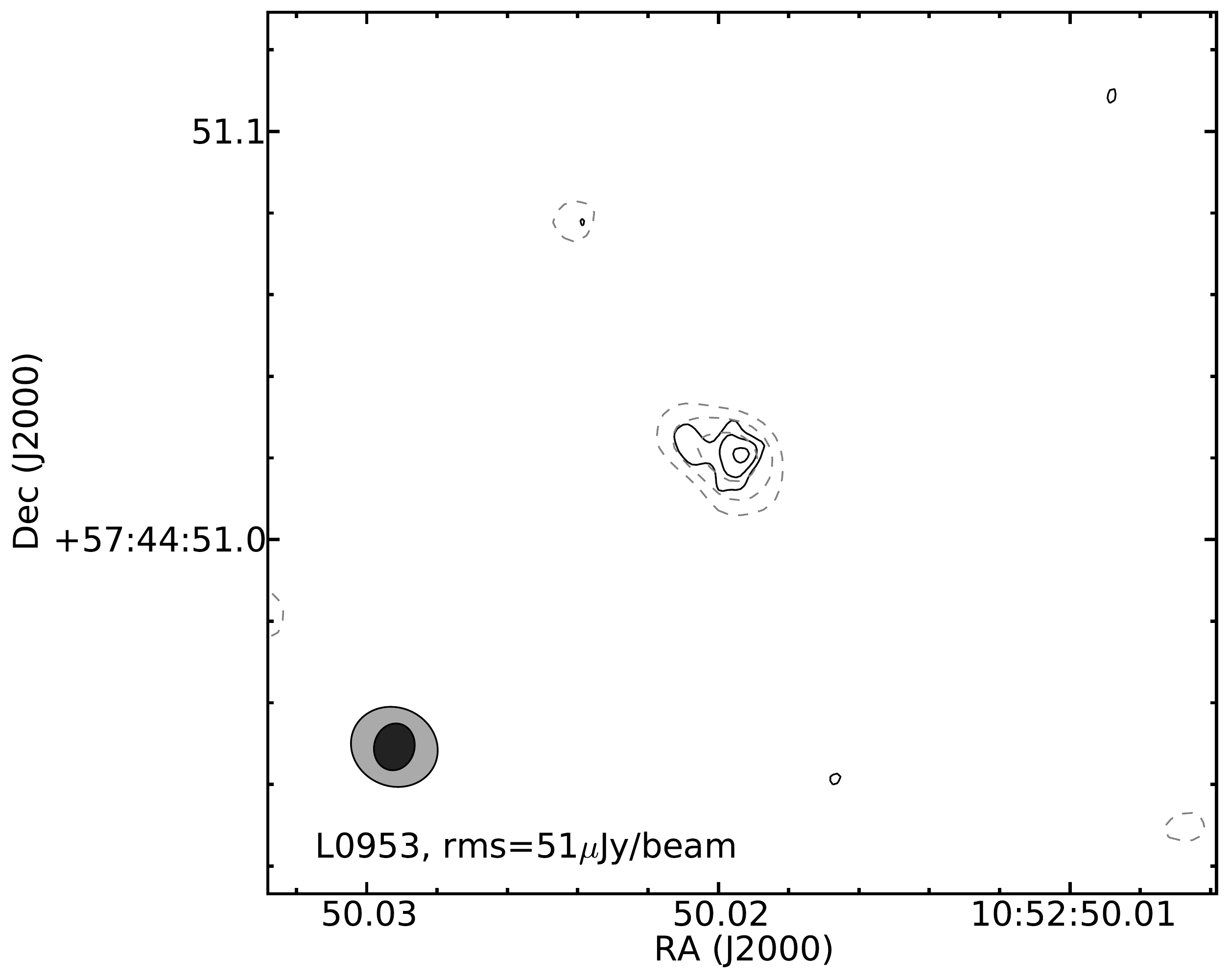}
\includegraphics[height=4.5cm]{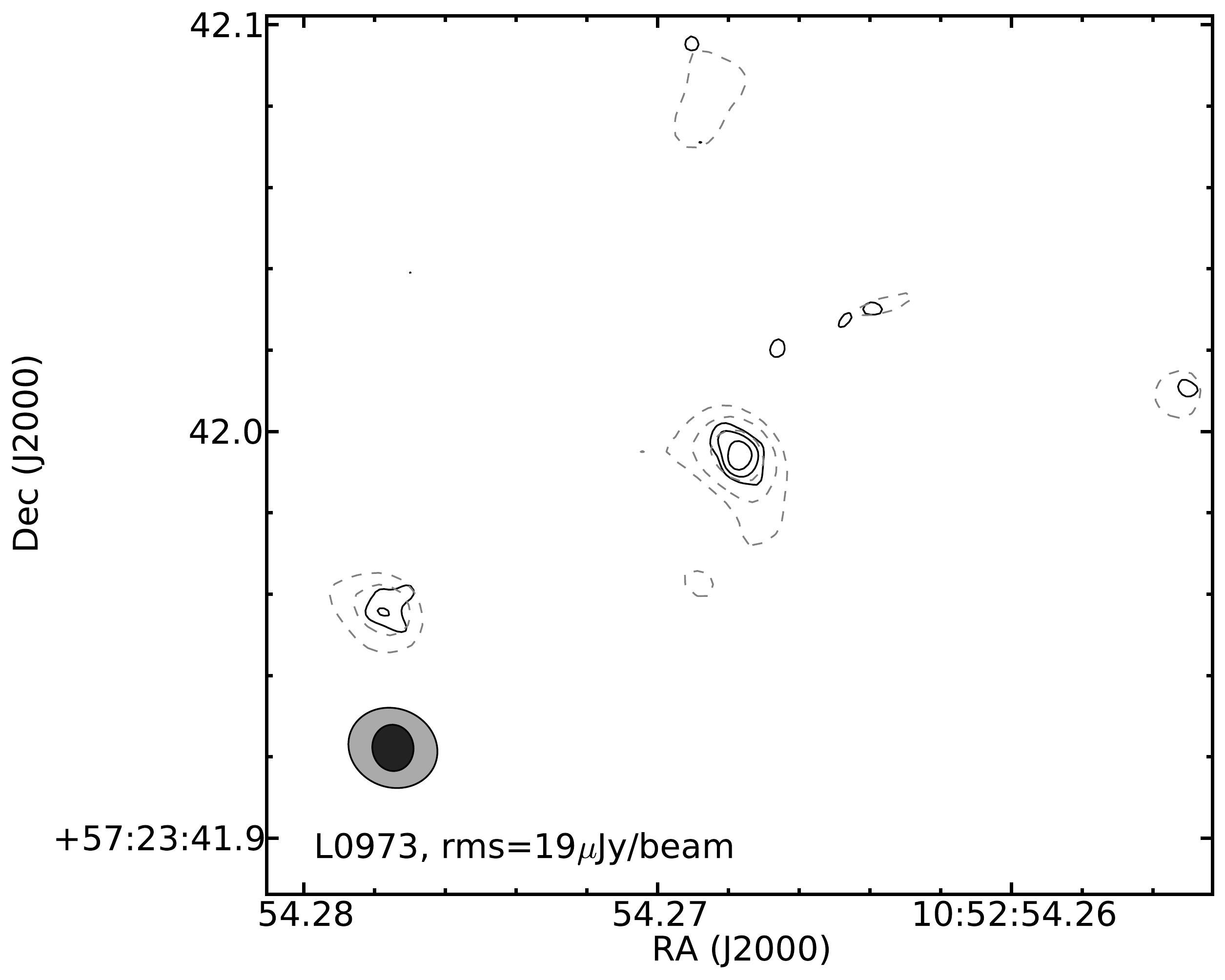}\\
\includegraphics[height=4.5cm]{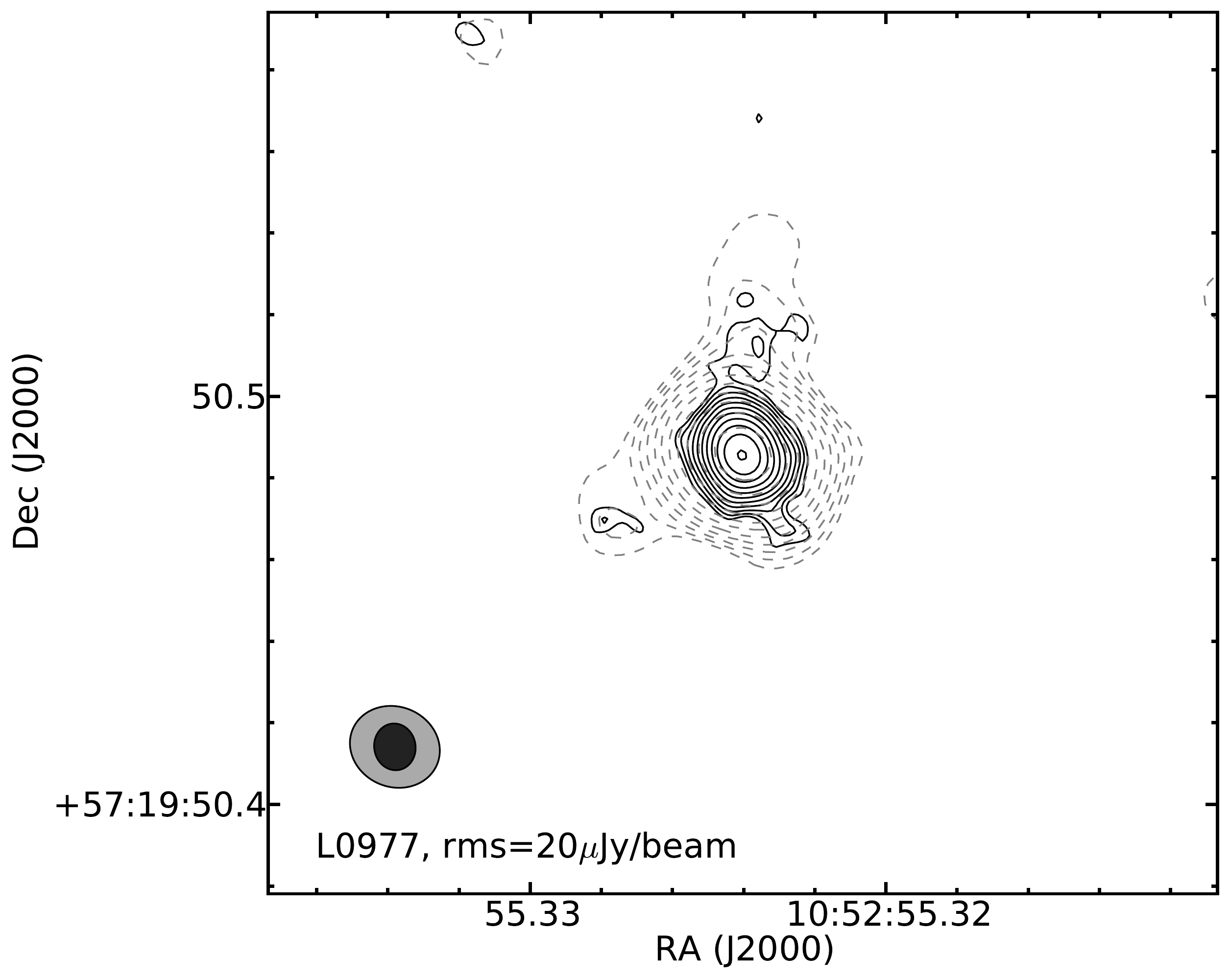}
\includegraphics[height=4.5cm]{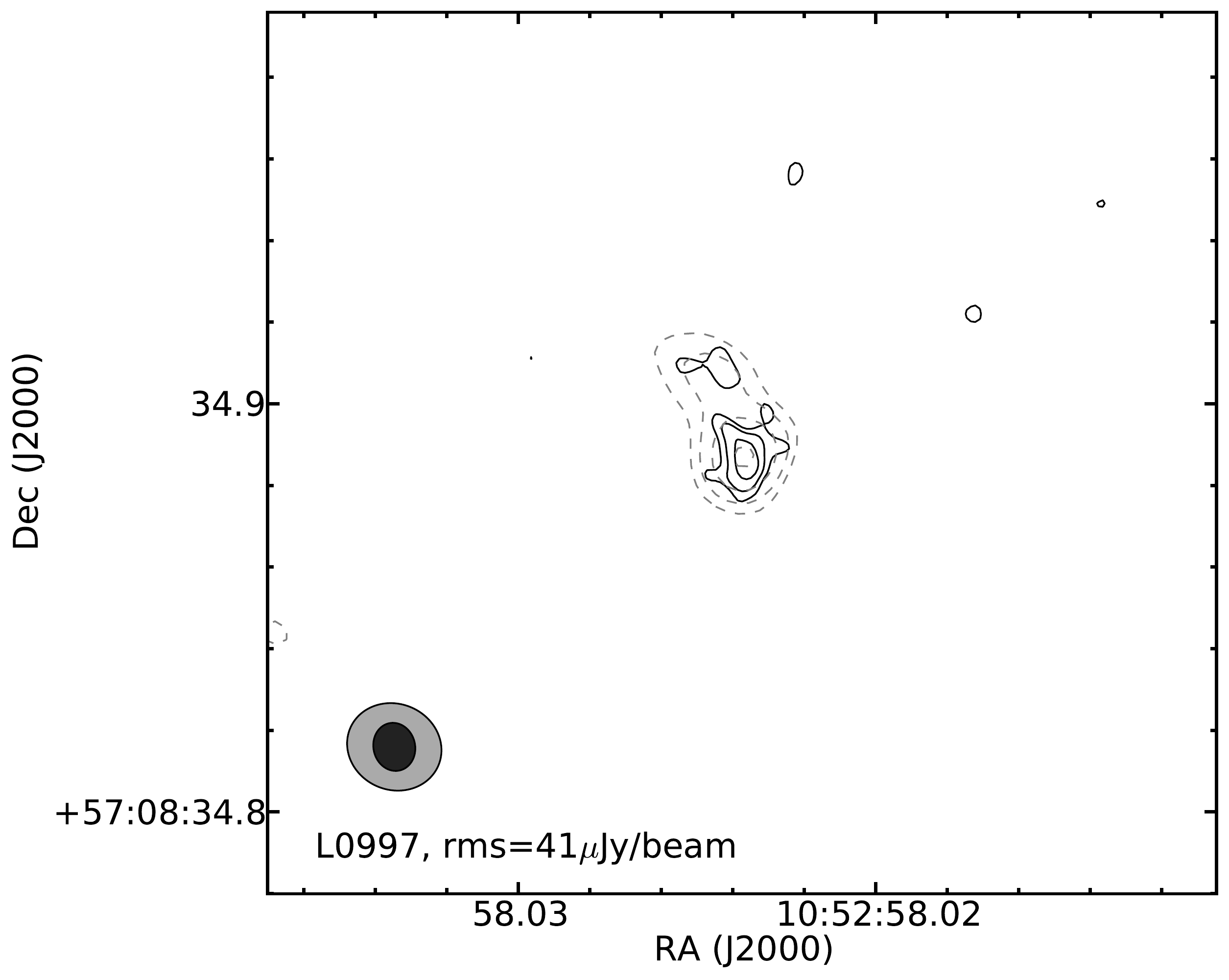}
\includegraphics[height=4.5cm]{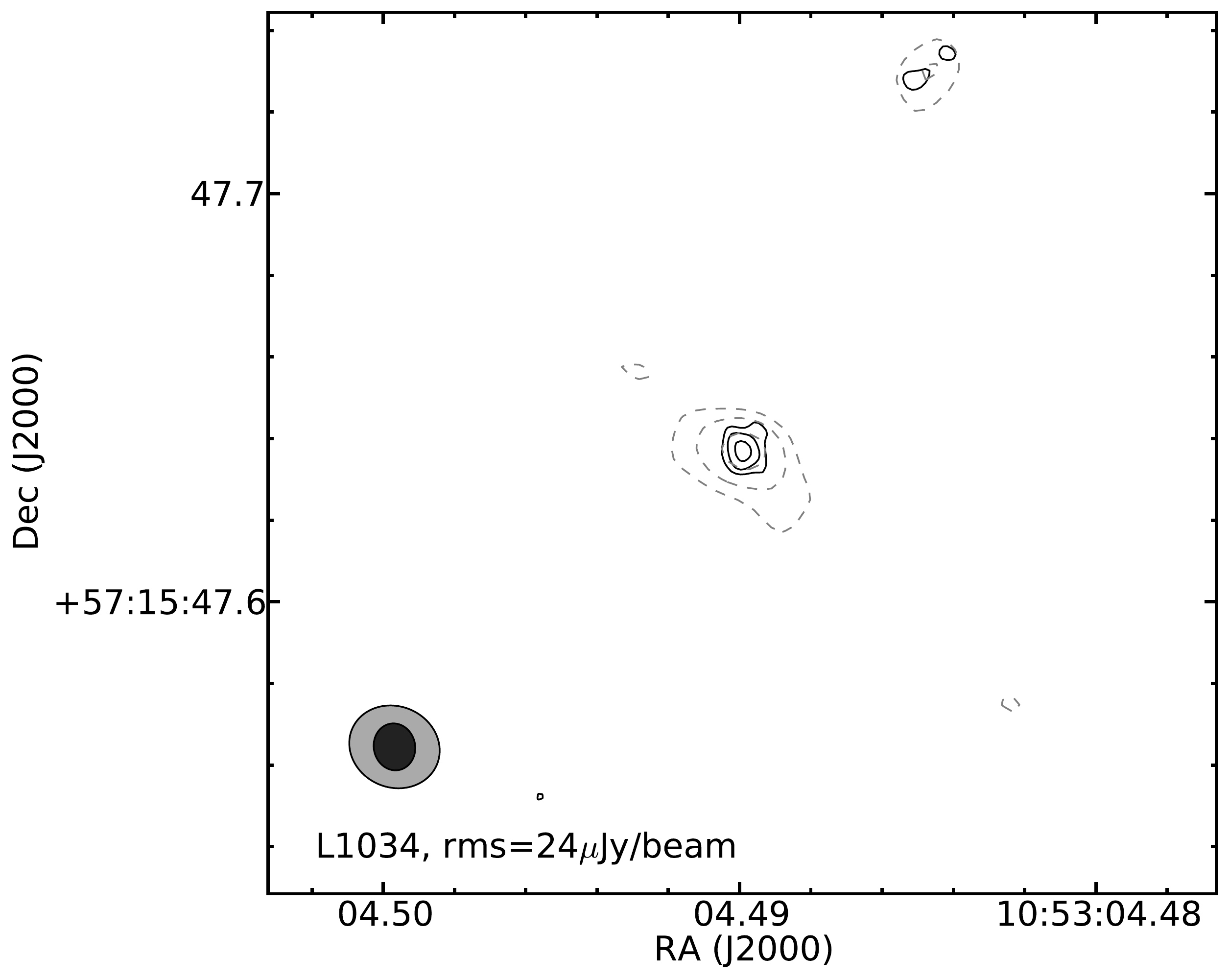}\\
\includegraphics[height=4.5cm]{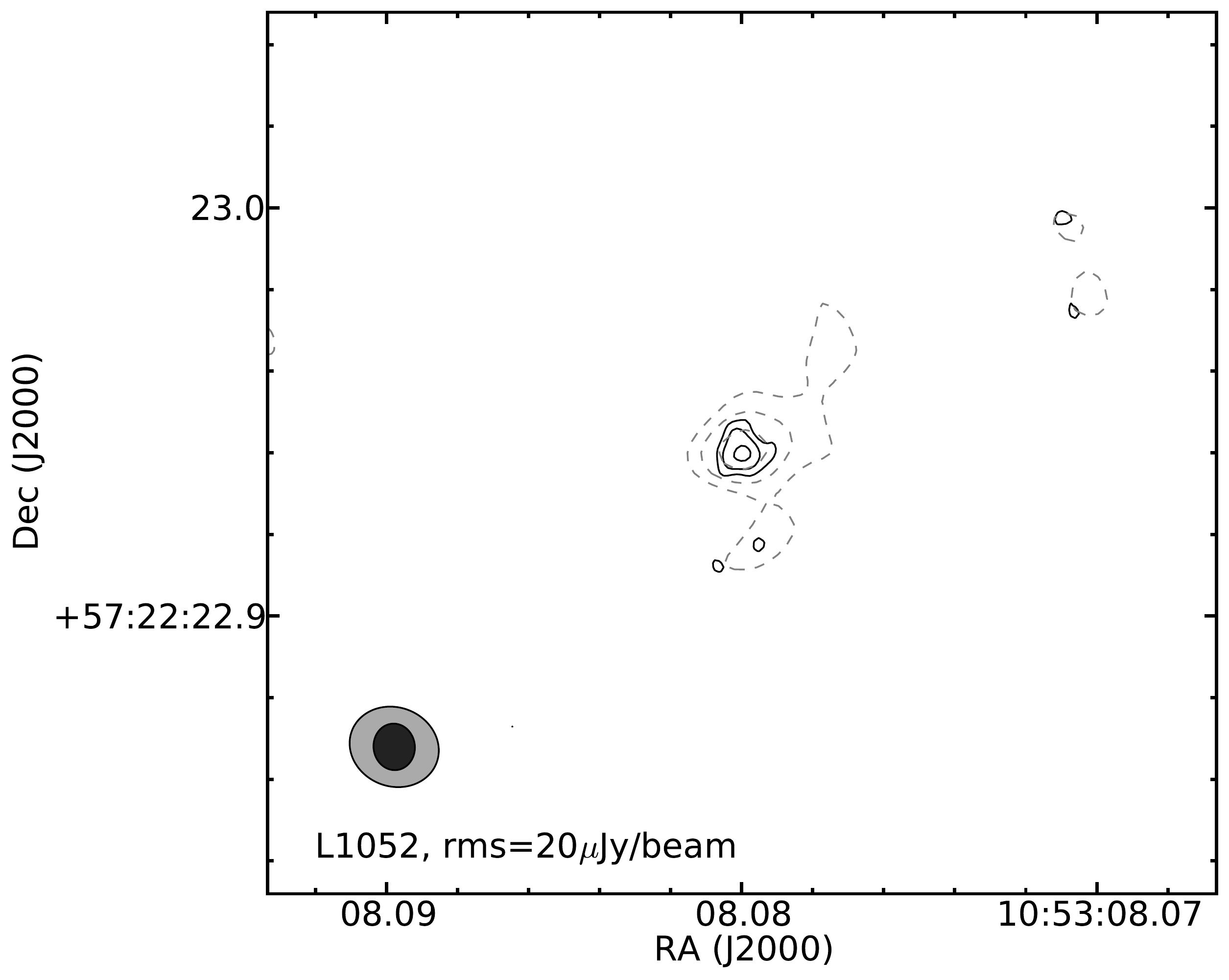}
\includegraphics[height=4.5cm]{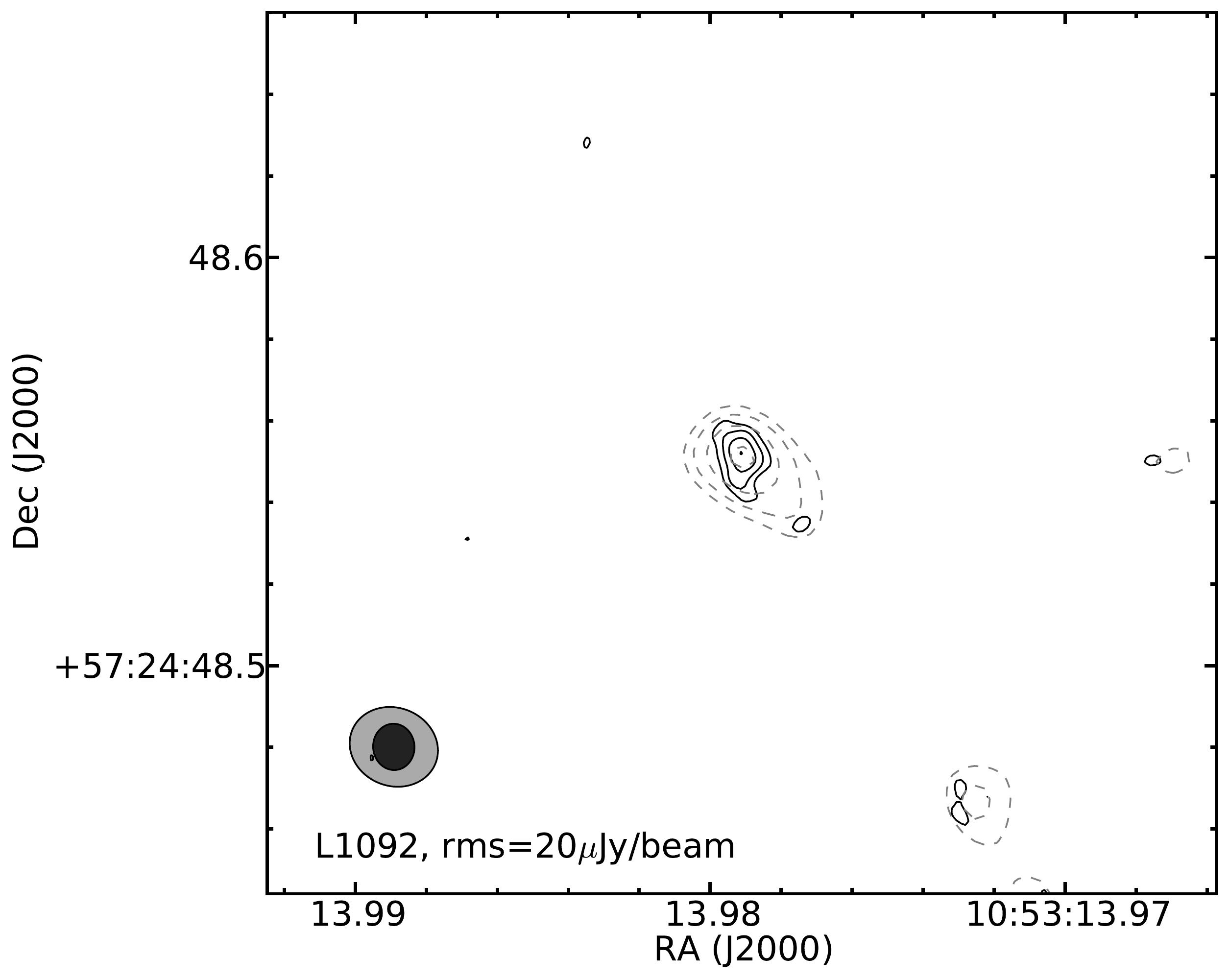}
\includegraphics[height=4.5cm]{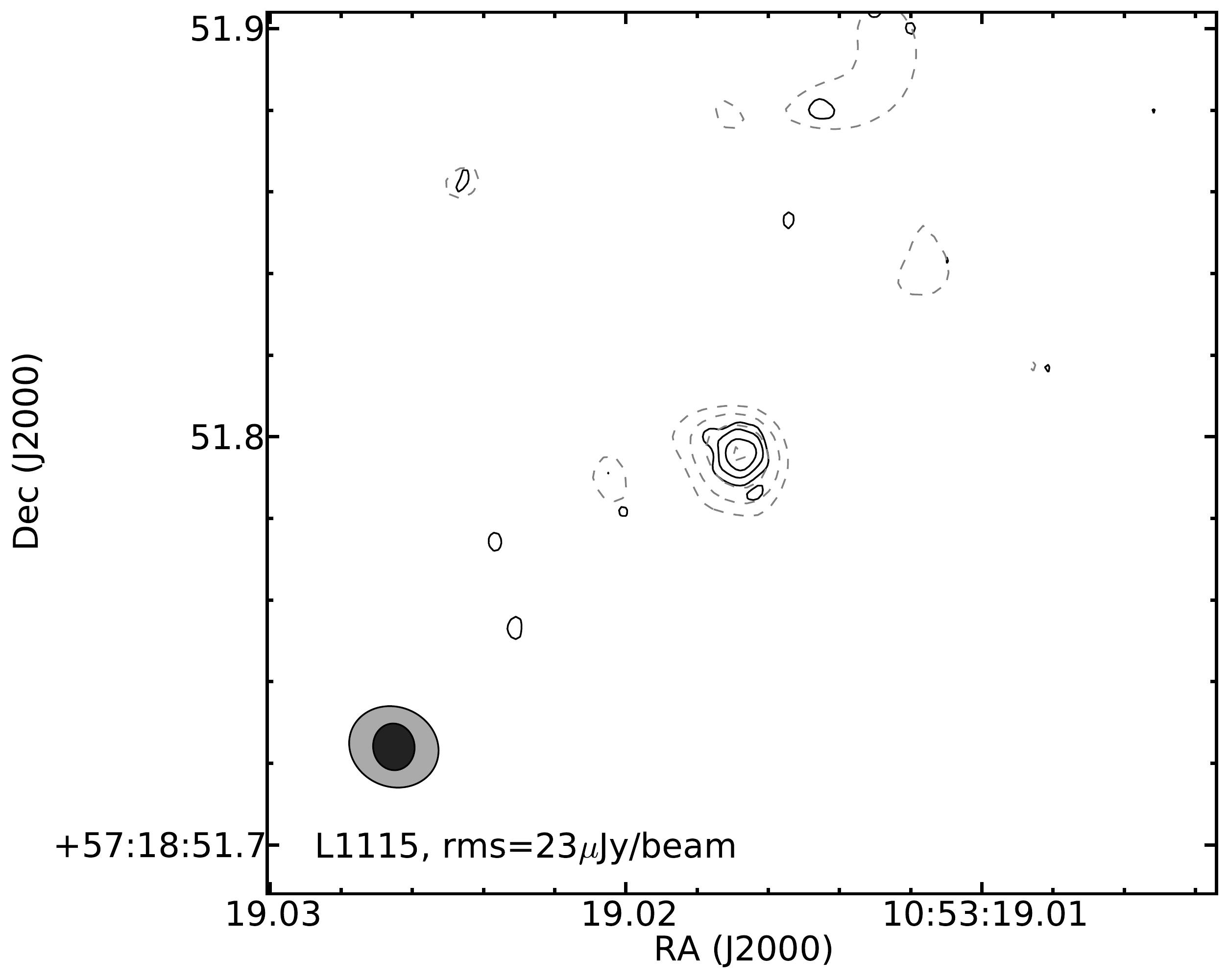}\\
\includegraphics[height=4.5cm]{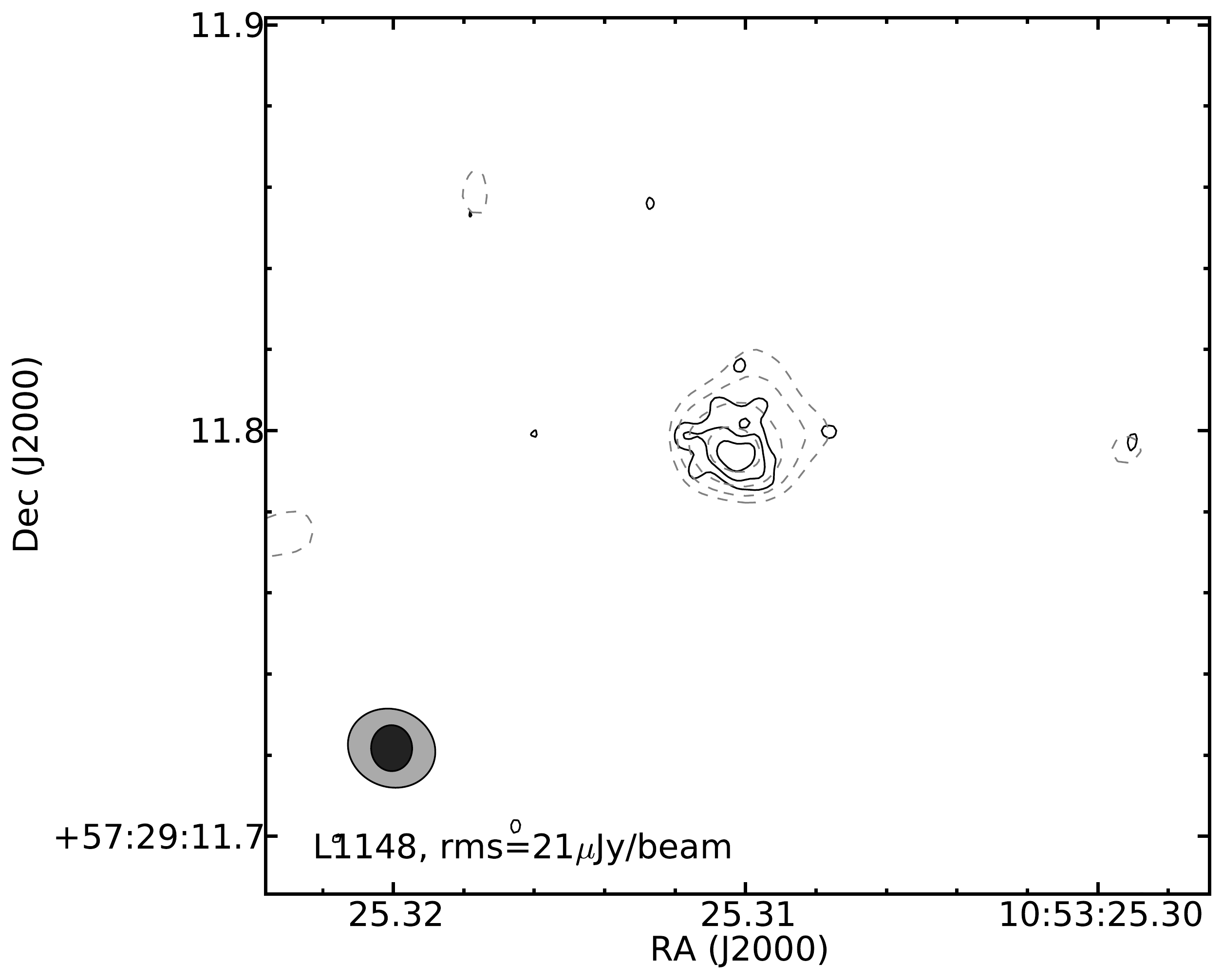}
\includegraphics[height=4.5cm]{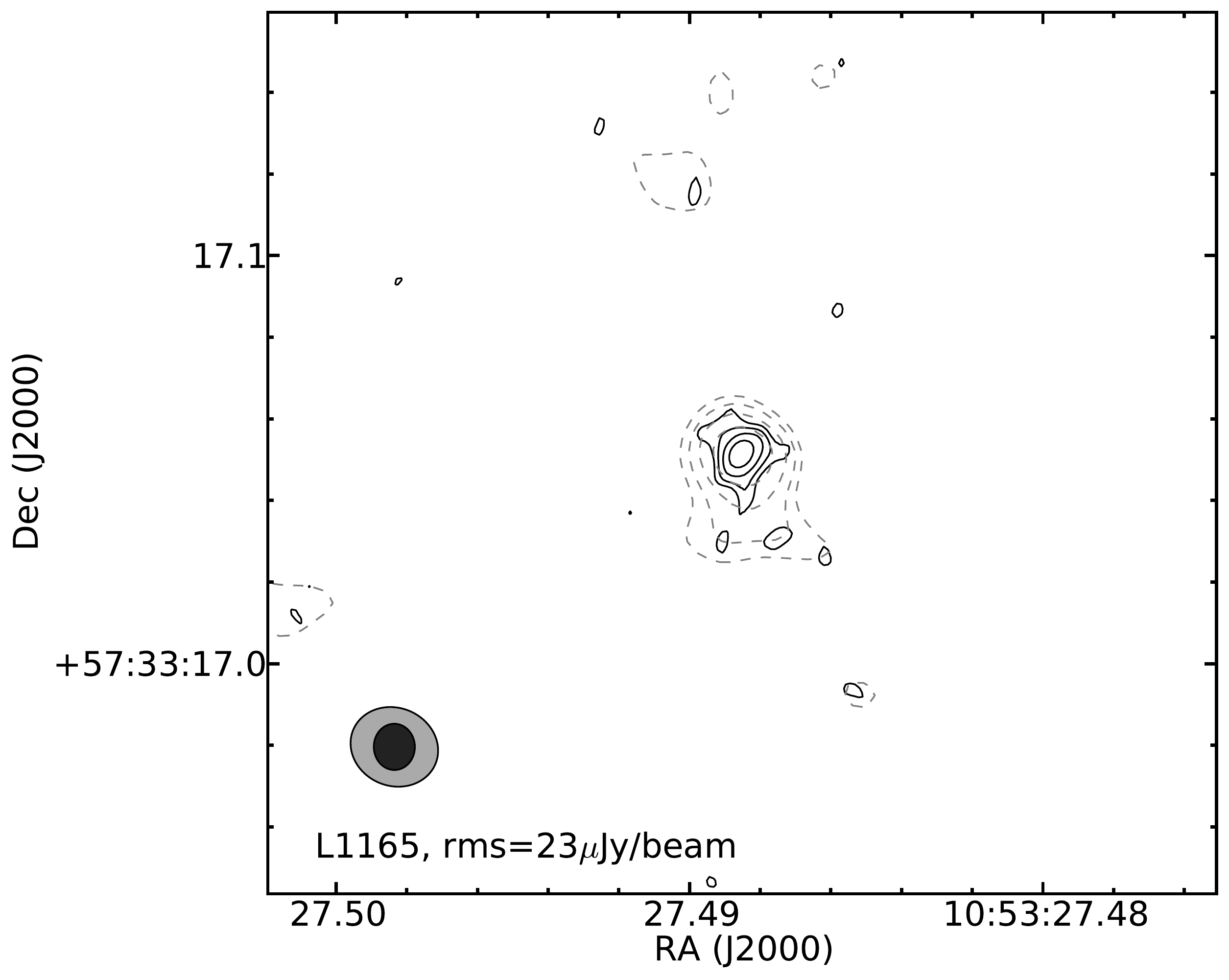}
\includegraphics[height=4.5cm]{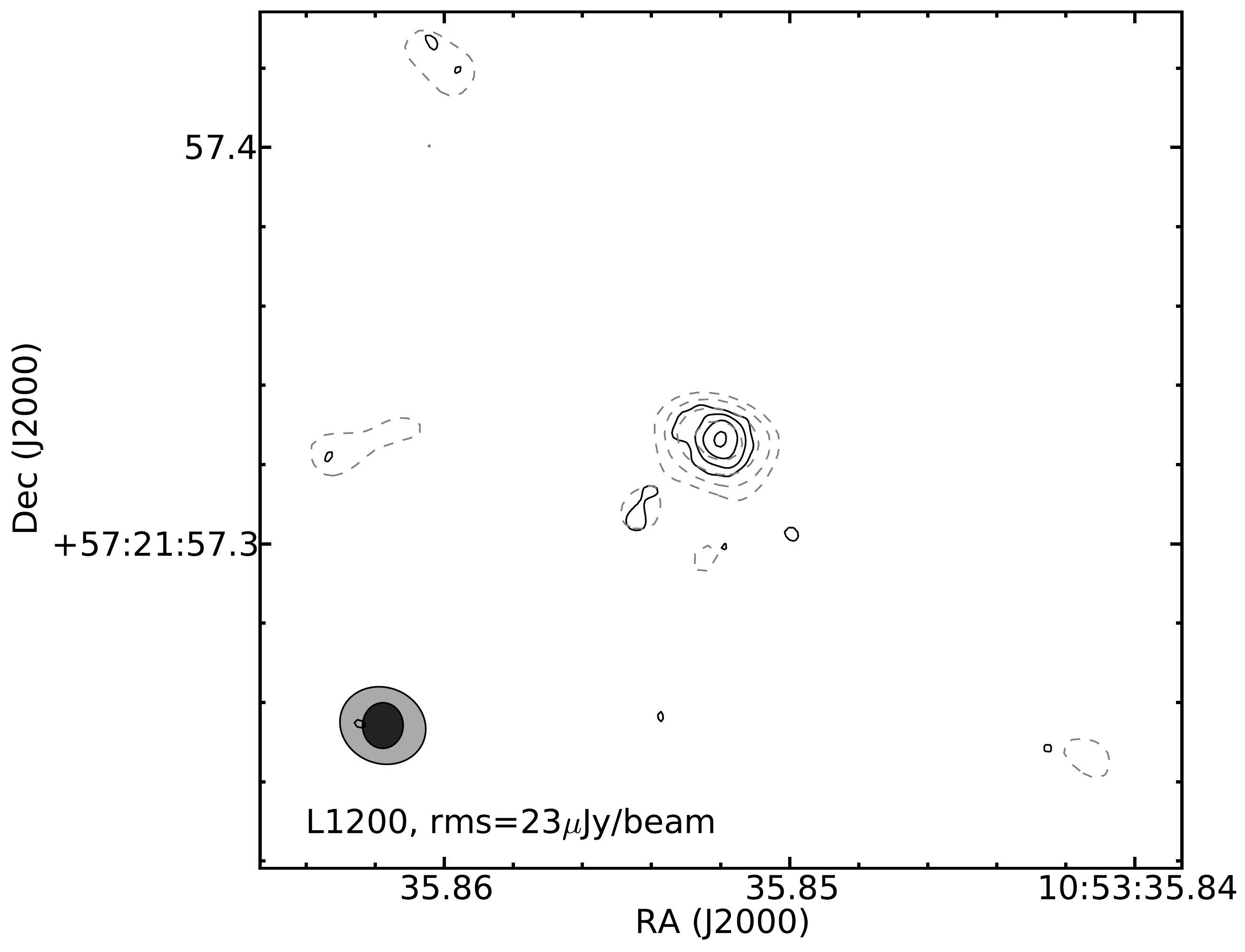}\\
\includegraphics[height=4.5cm]{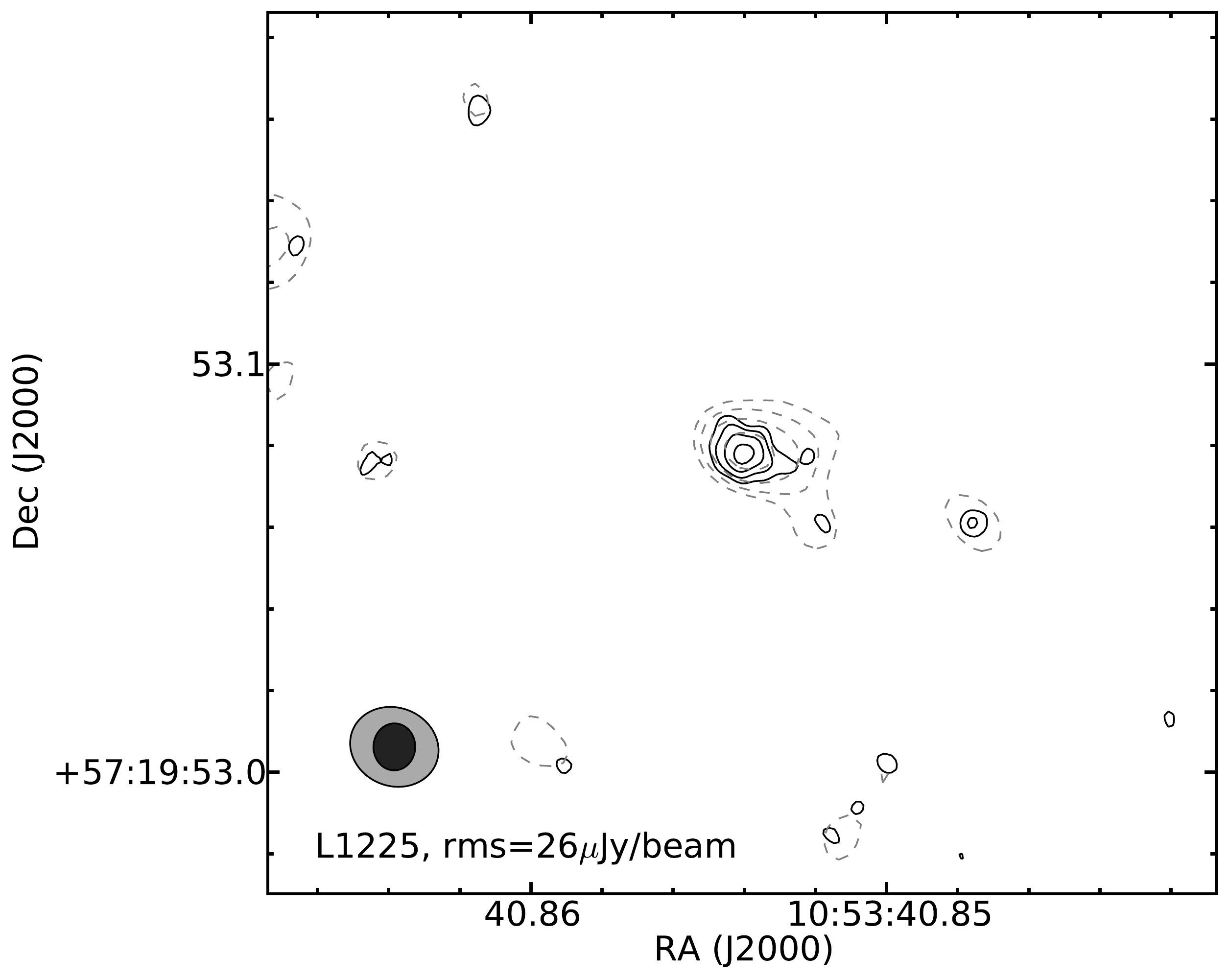}
\includegraphics[height=4.5cm]{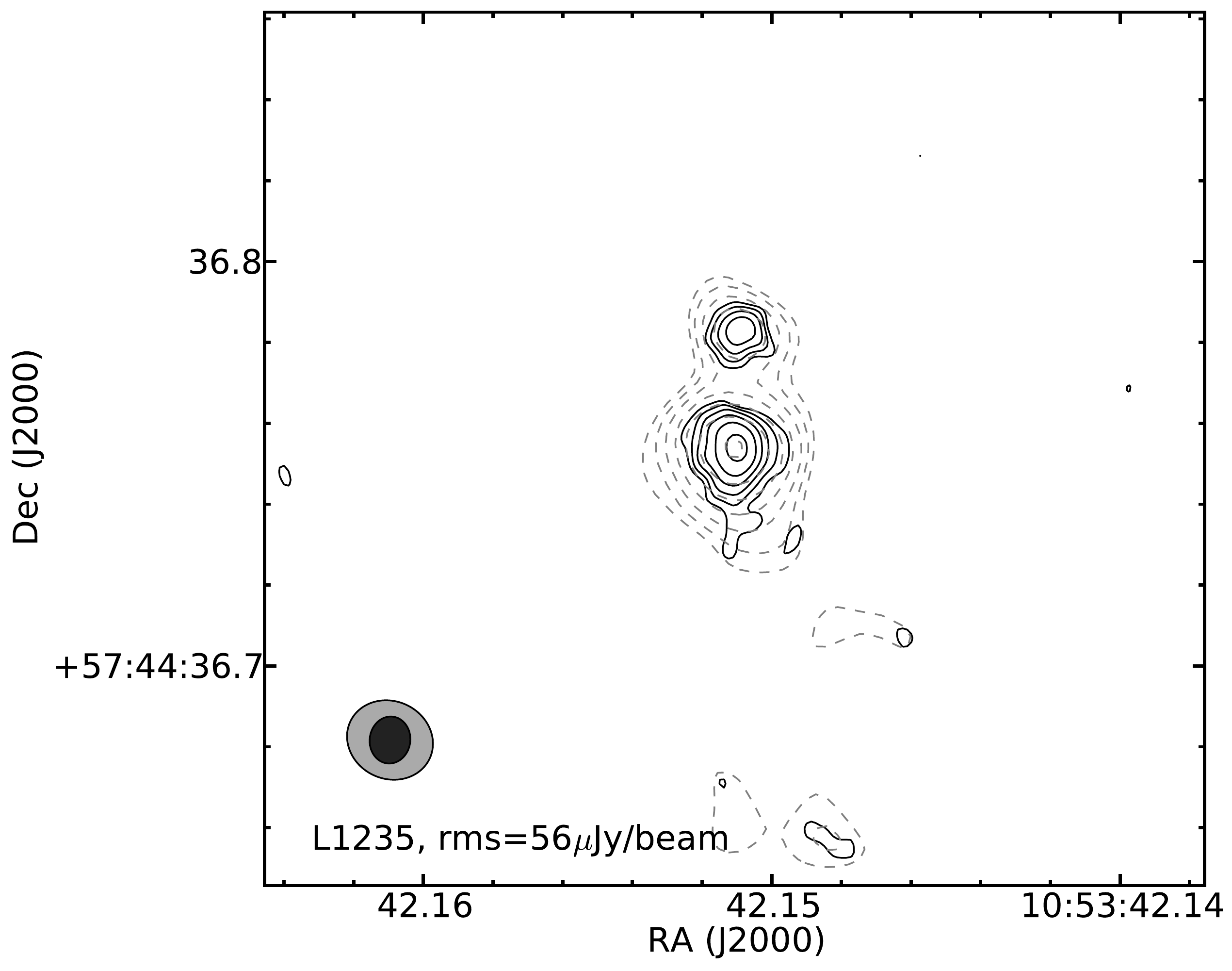}
\includegraphics[height=4.5cm]{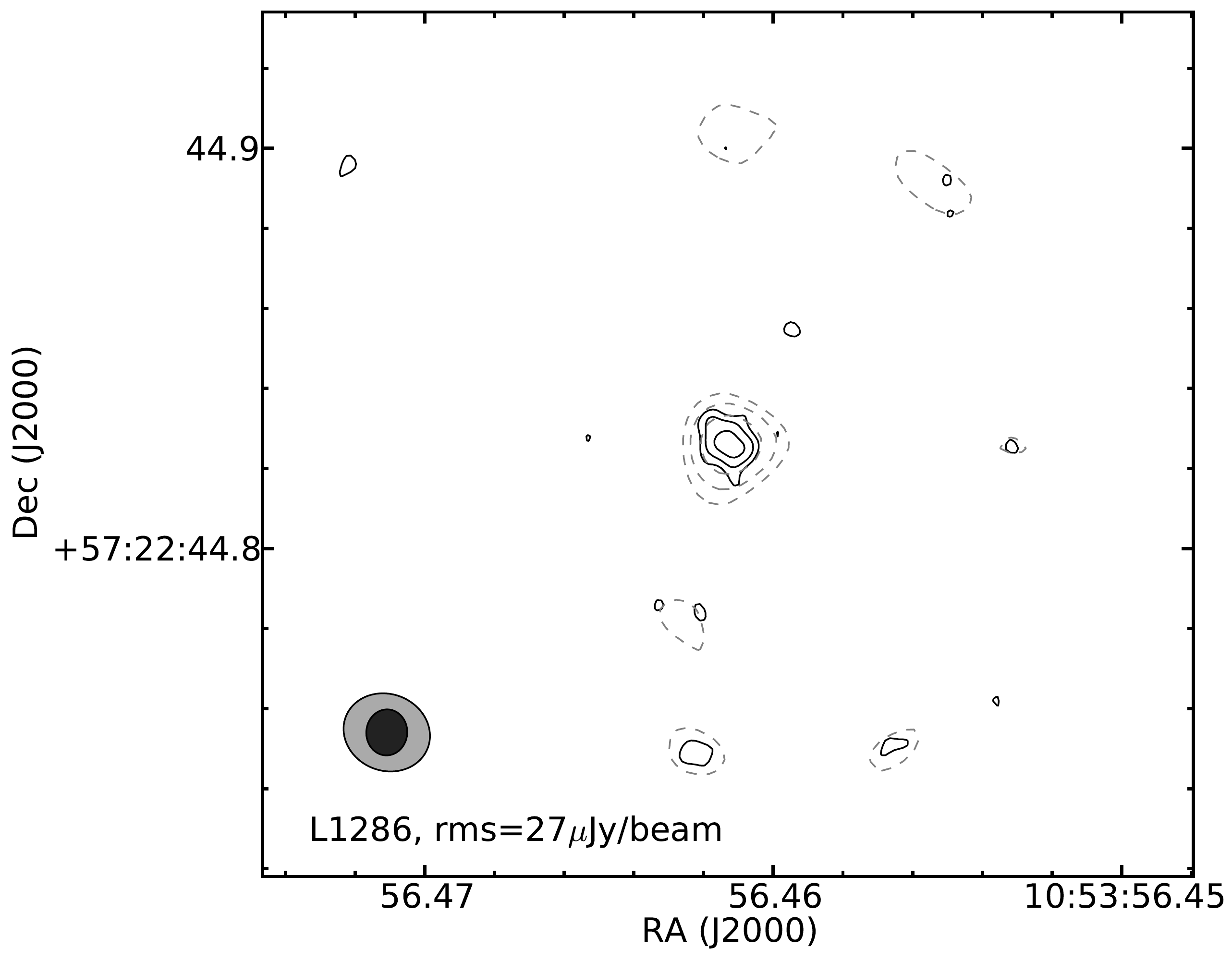}\\
\caption{(Continued)}
\end{figure*}

\begin{figure*}
\ContinuedFloat
\center
\includegraphics[height=4.5cm]{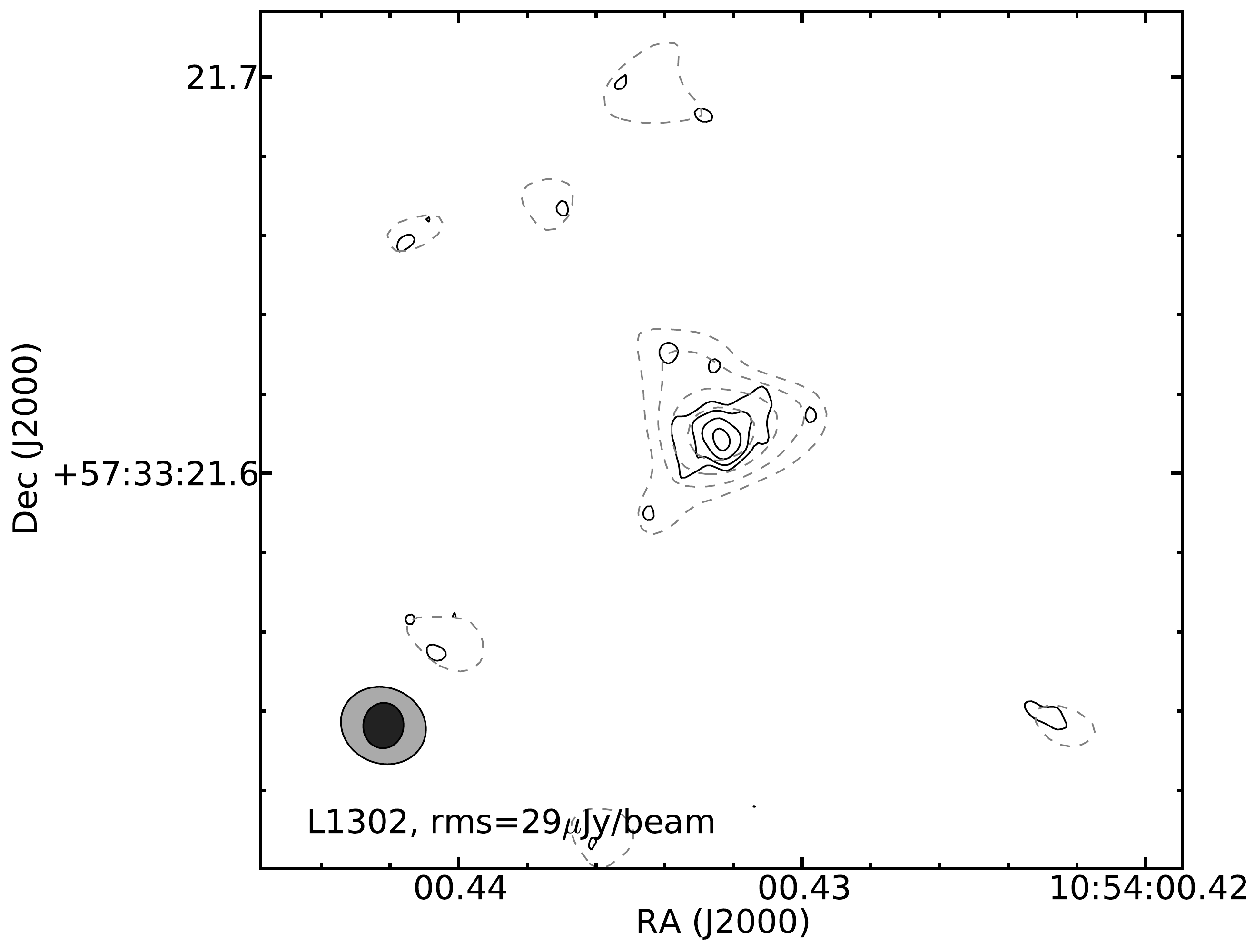}
\includegraphics[height=4.5cm]{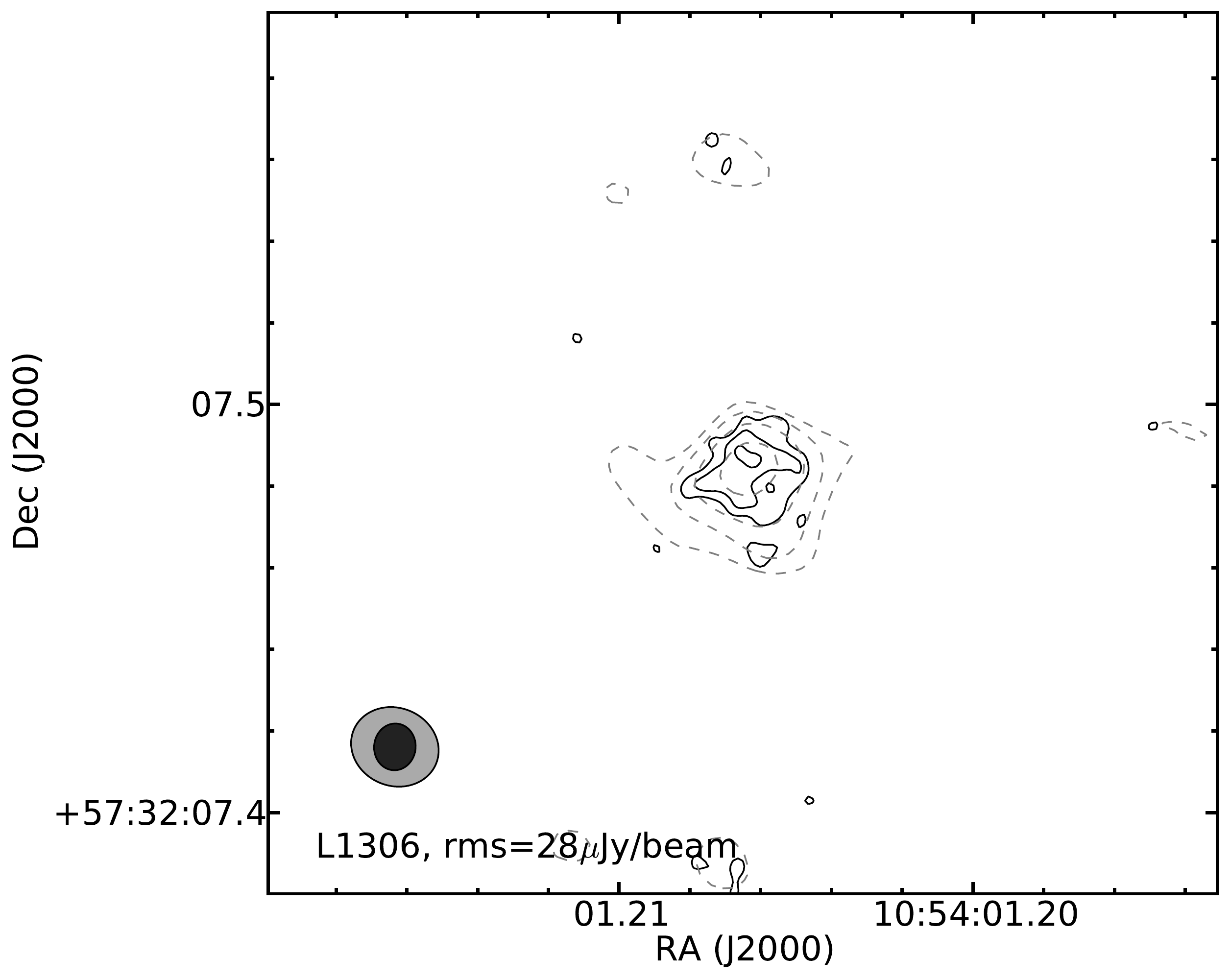}
\includegraphics[height=4.5cm]{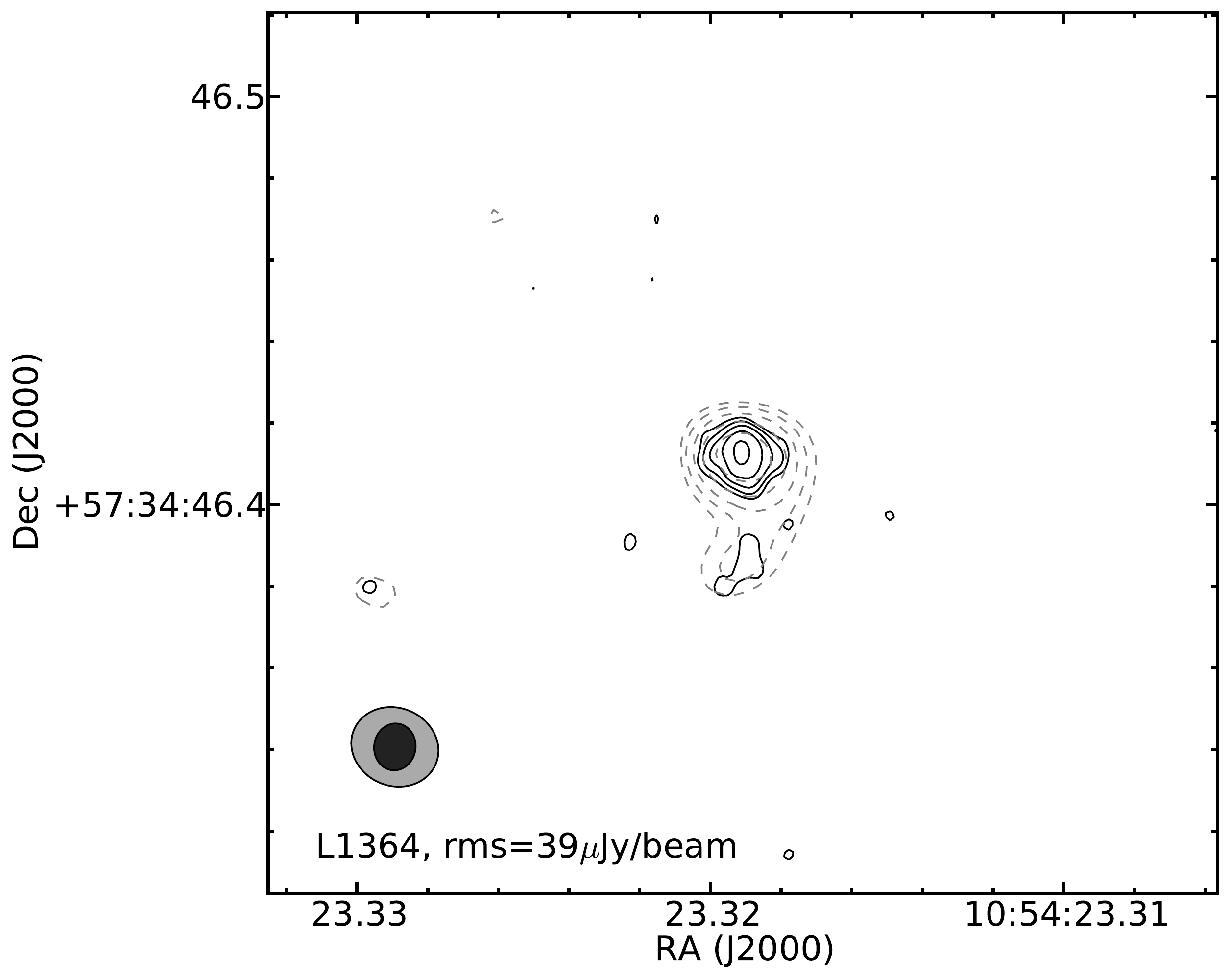}\\
\includegraphics[height=4.5cm]{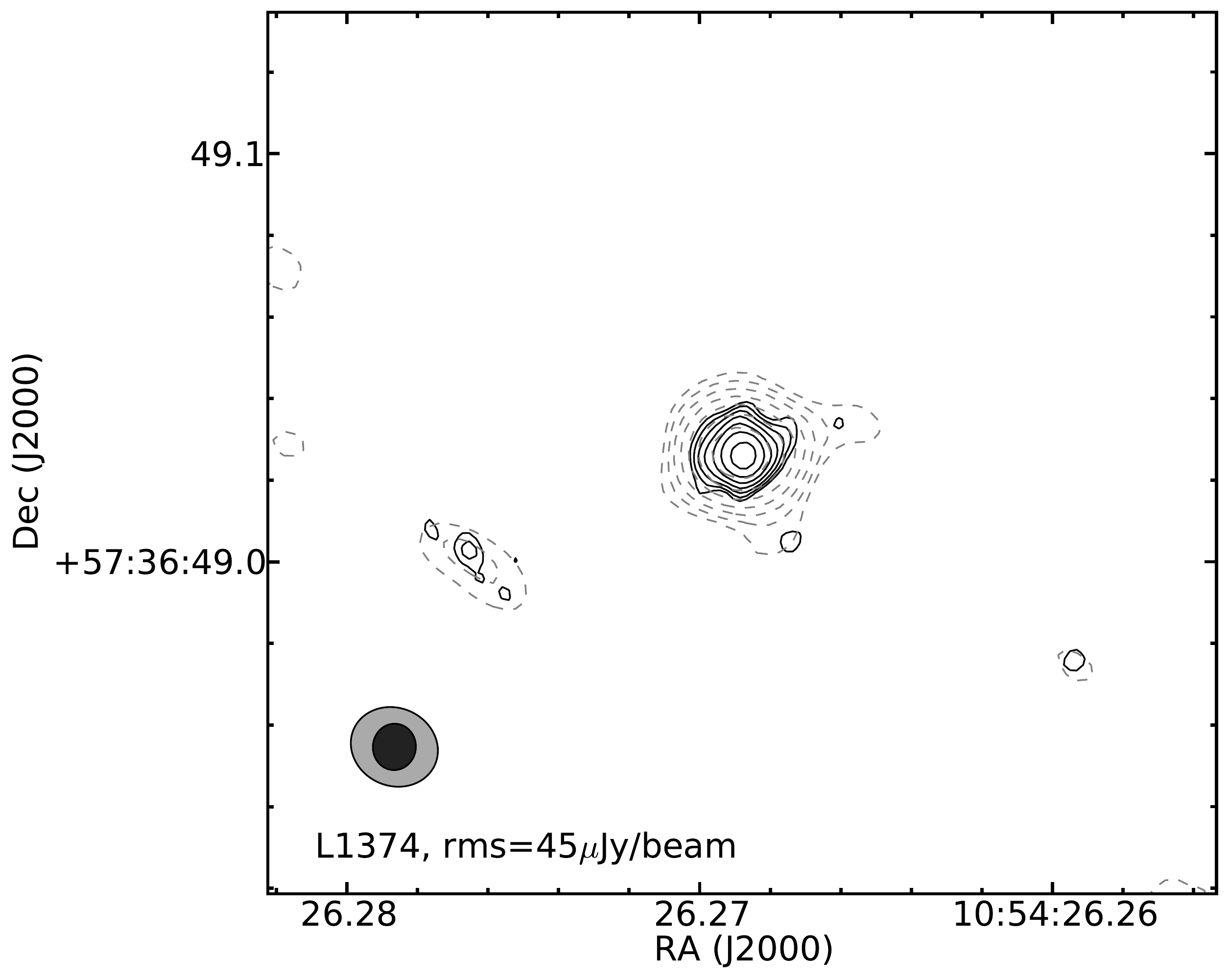}
\includegraphics[height=4.5cm]{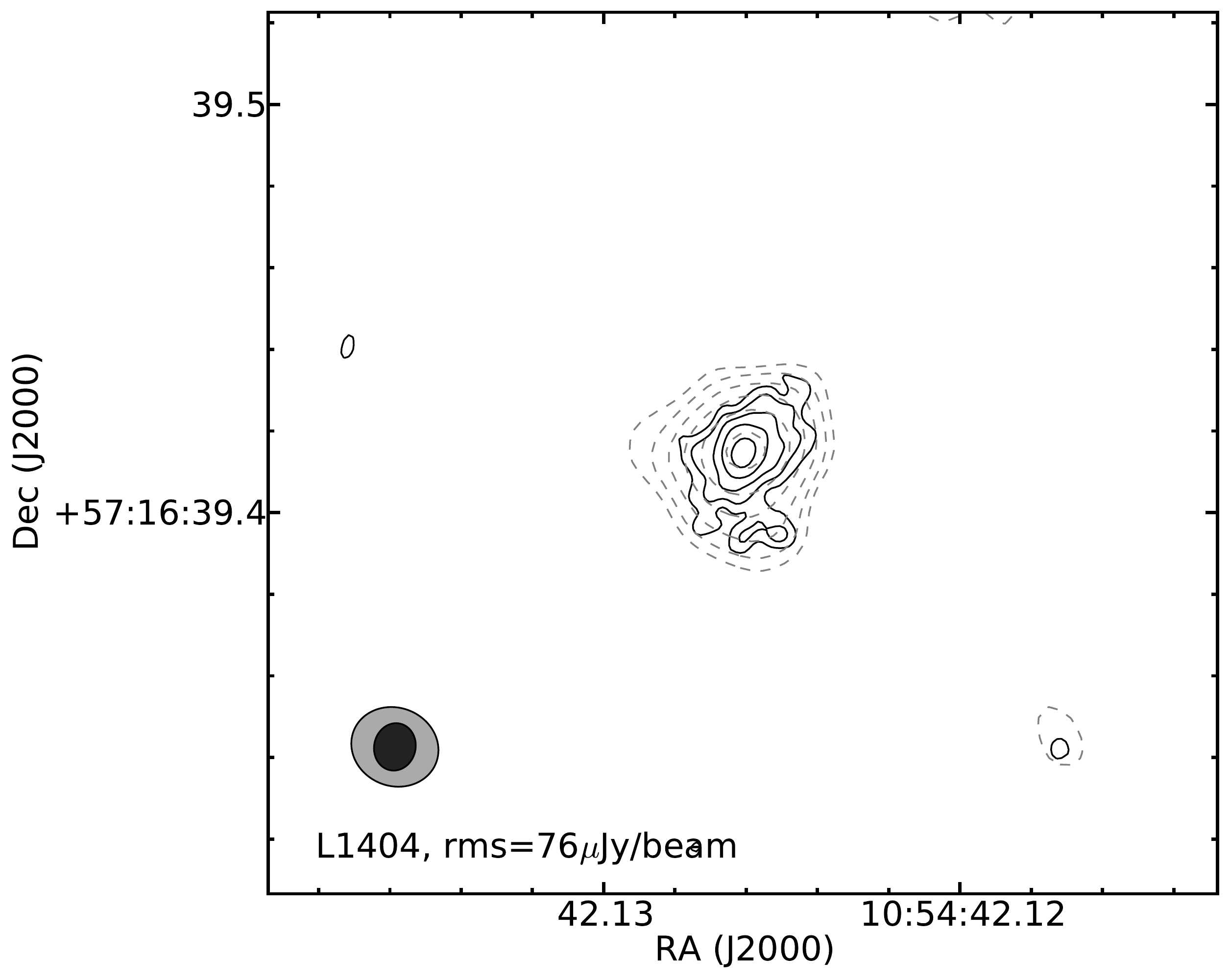}
\caption{(Continued)}
\end{figure*}

\begin{figure*}
\center
\includegraphics[height=4.5cm]{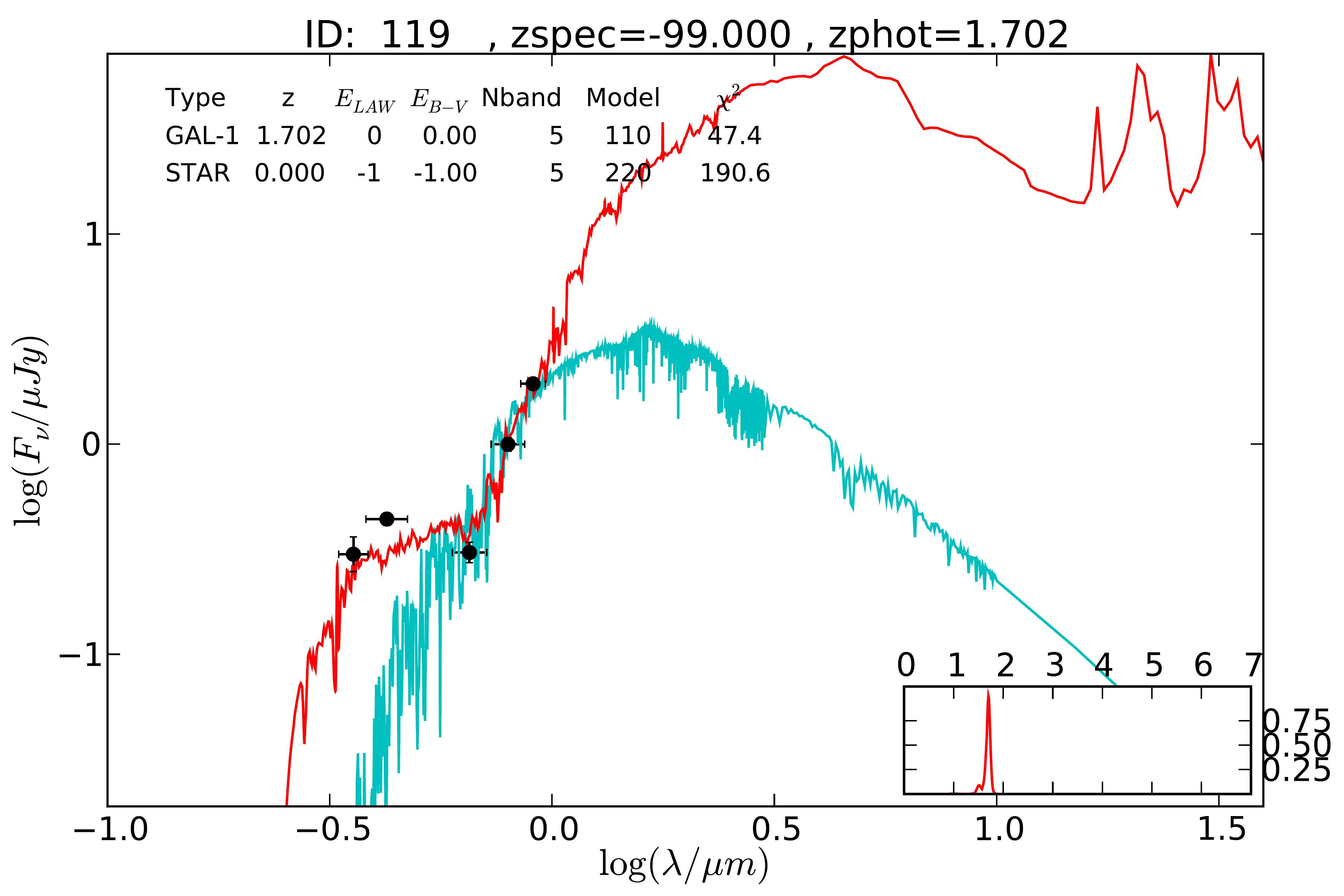}
\includegraphics[height=4.5cm]{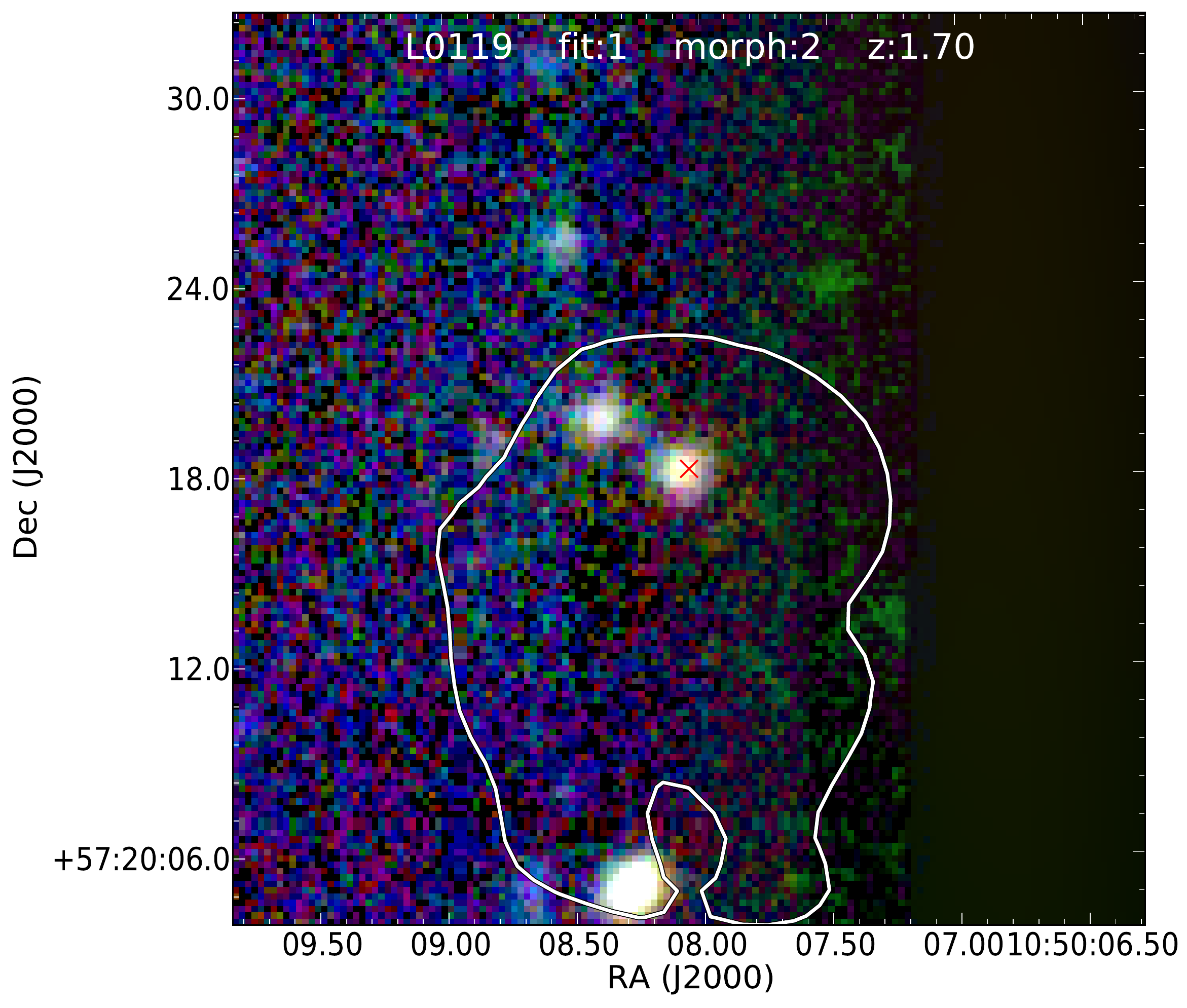}\\
\includegraphics[height=4.5cm]{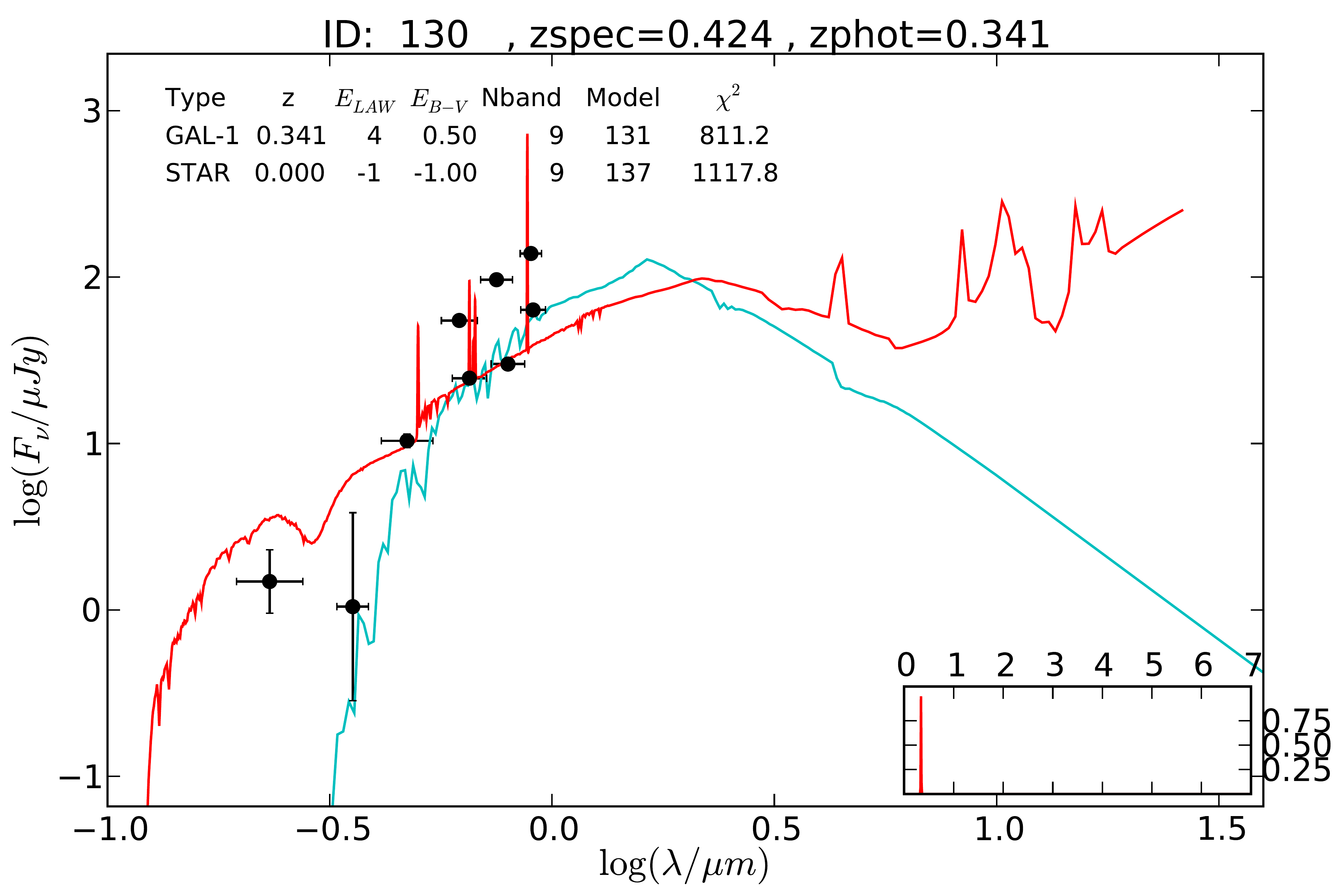}
\includegraphics[height=4.5cm]{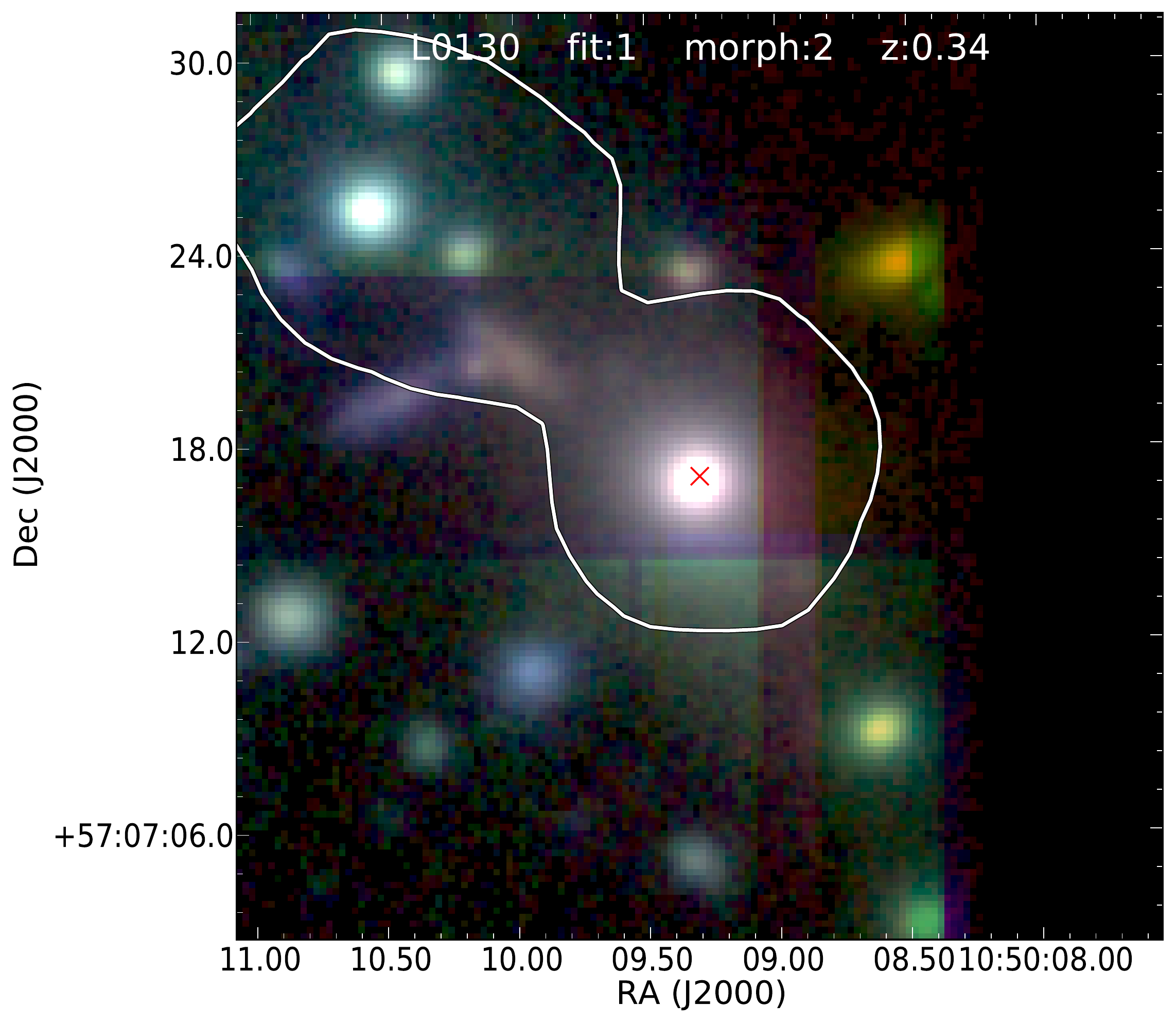}\\
\includegraphics[height=4.5cm]{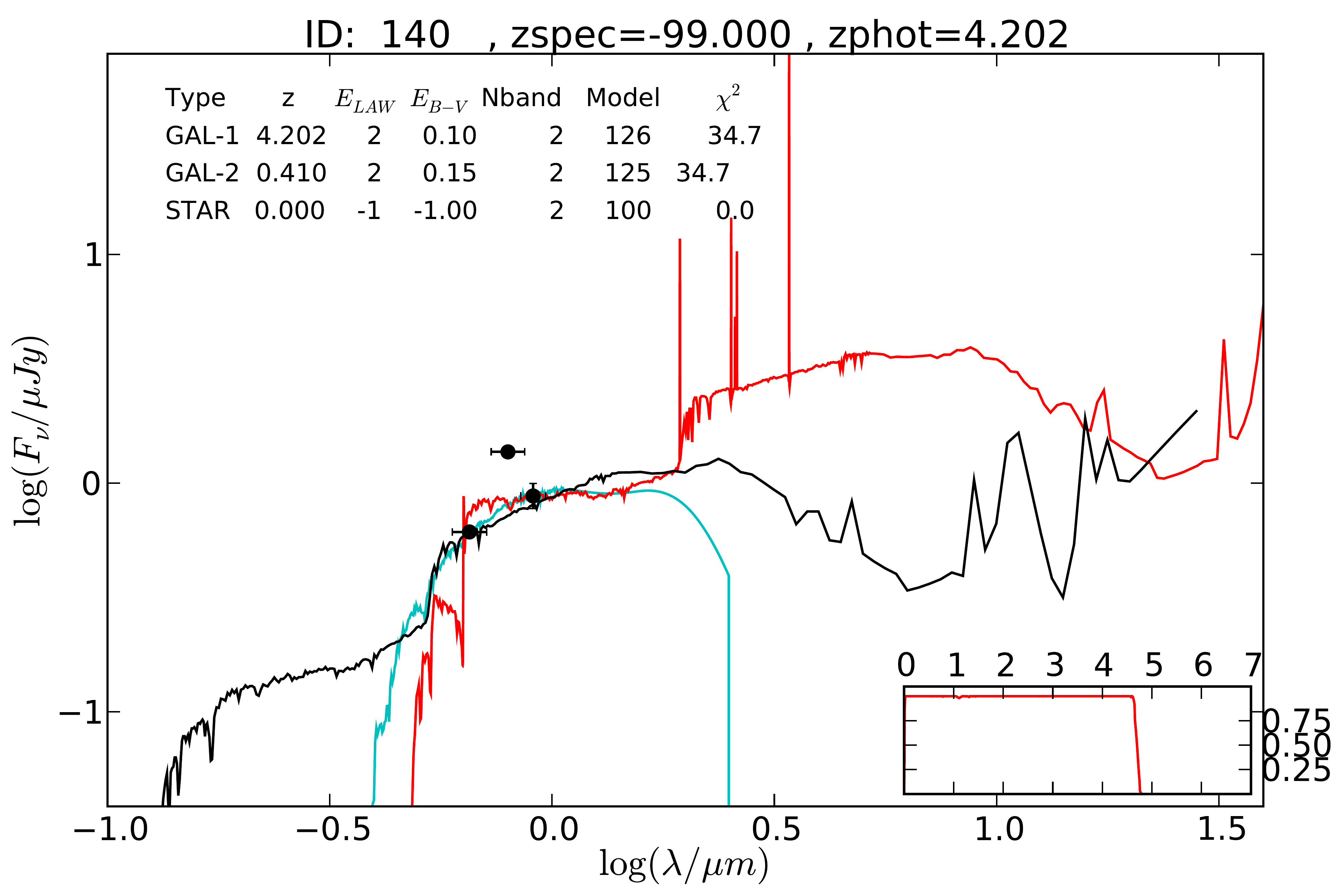}
\includegraphics[height=4.5cm]{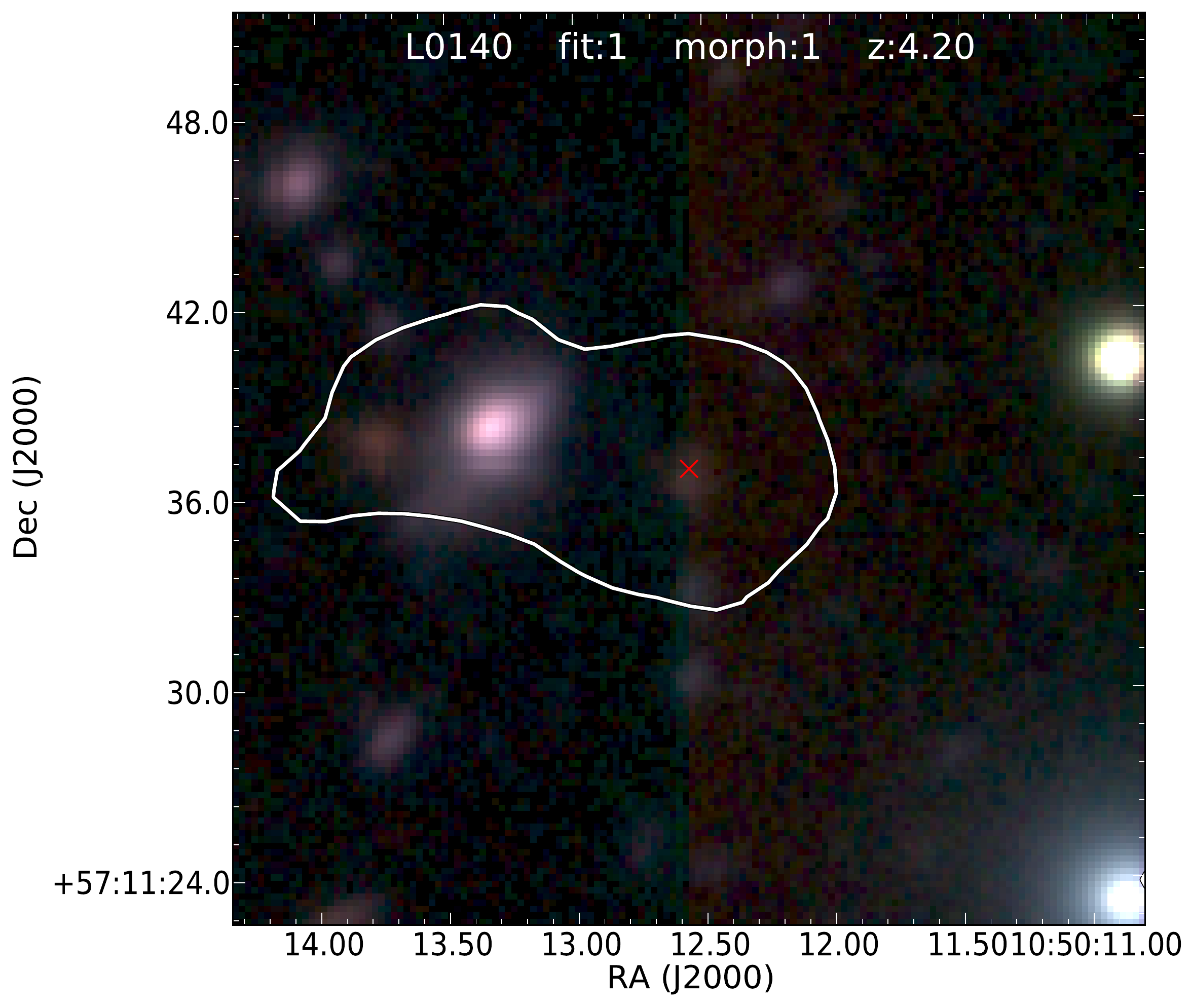}\\
\includegraphics[height=4.5cm]{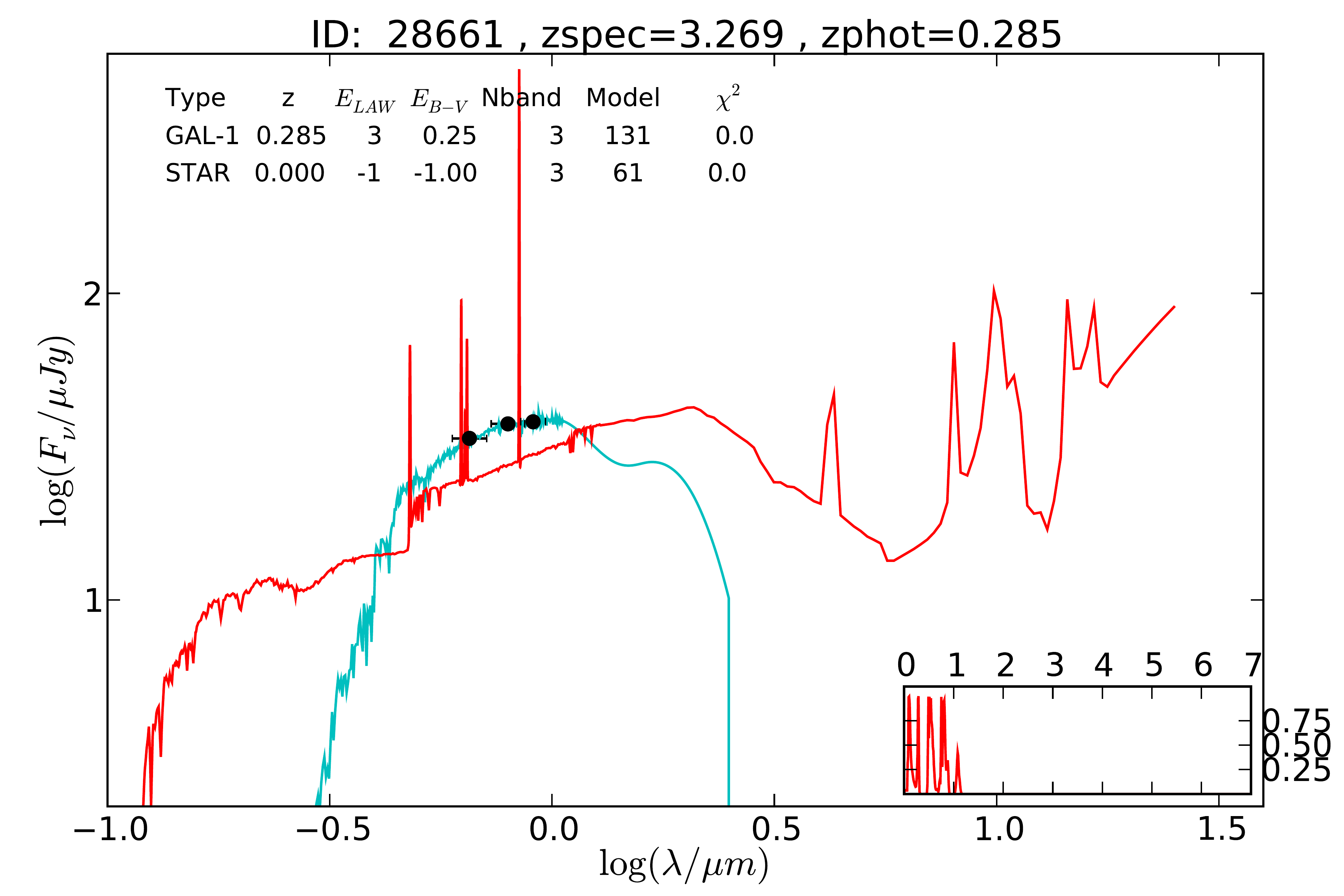}
\includegraphics[height=4.5cm]{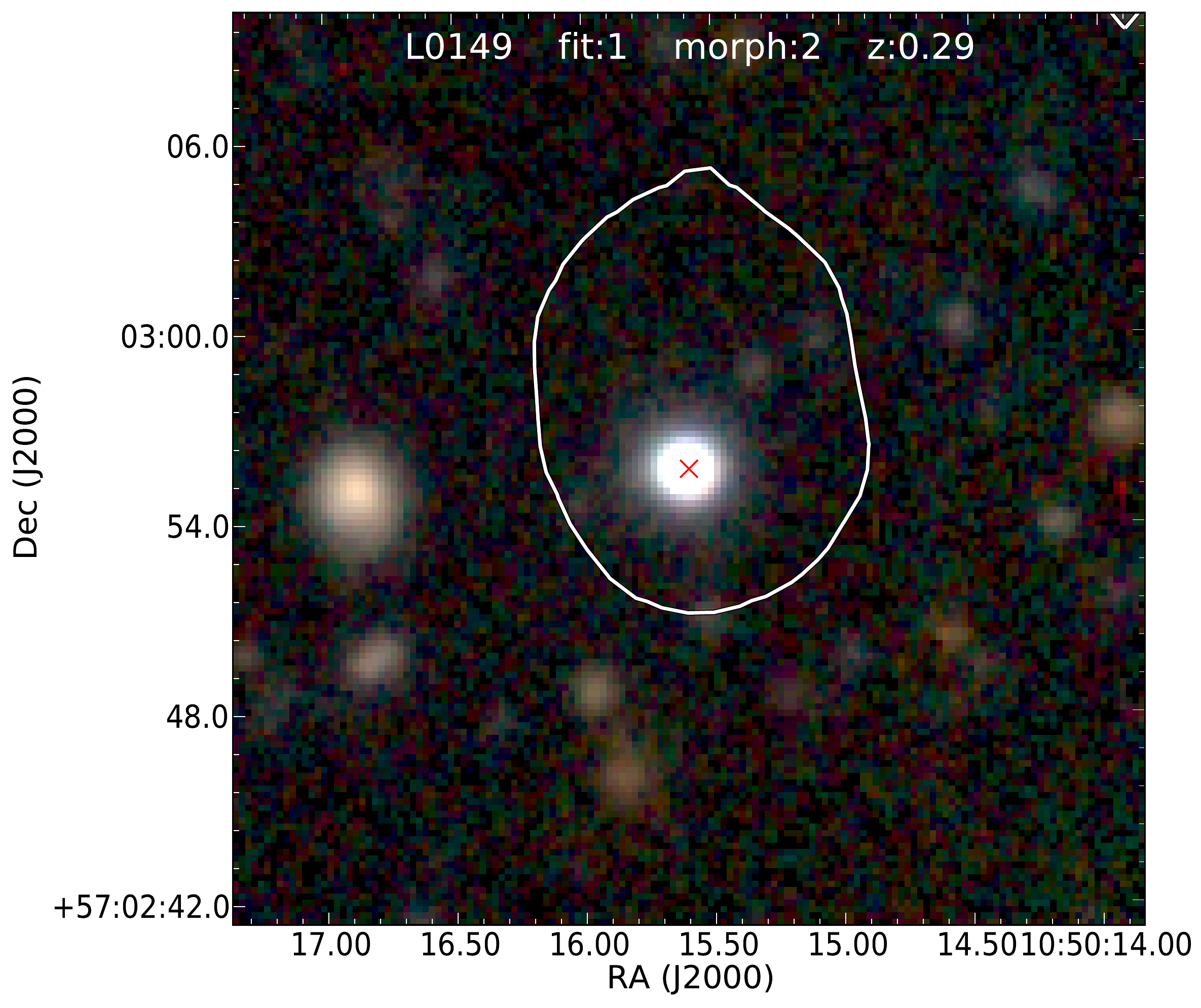}\\
\includegraphics[height=4.5cm]{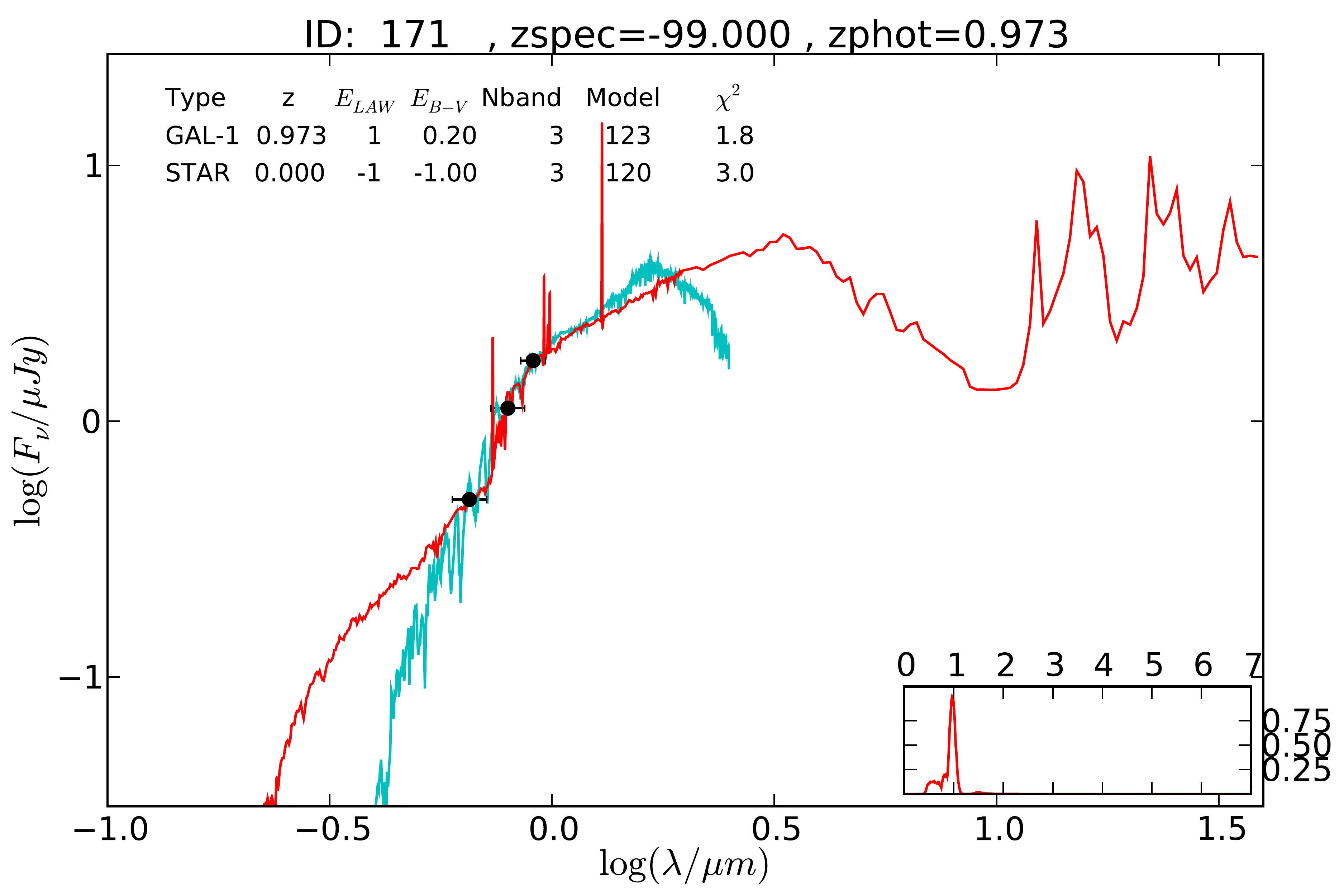}
\includegraphics[height=4.5cm]{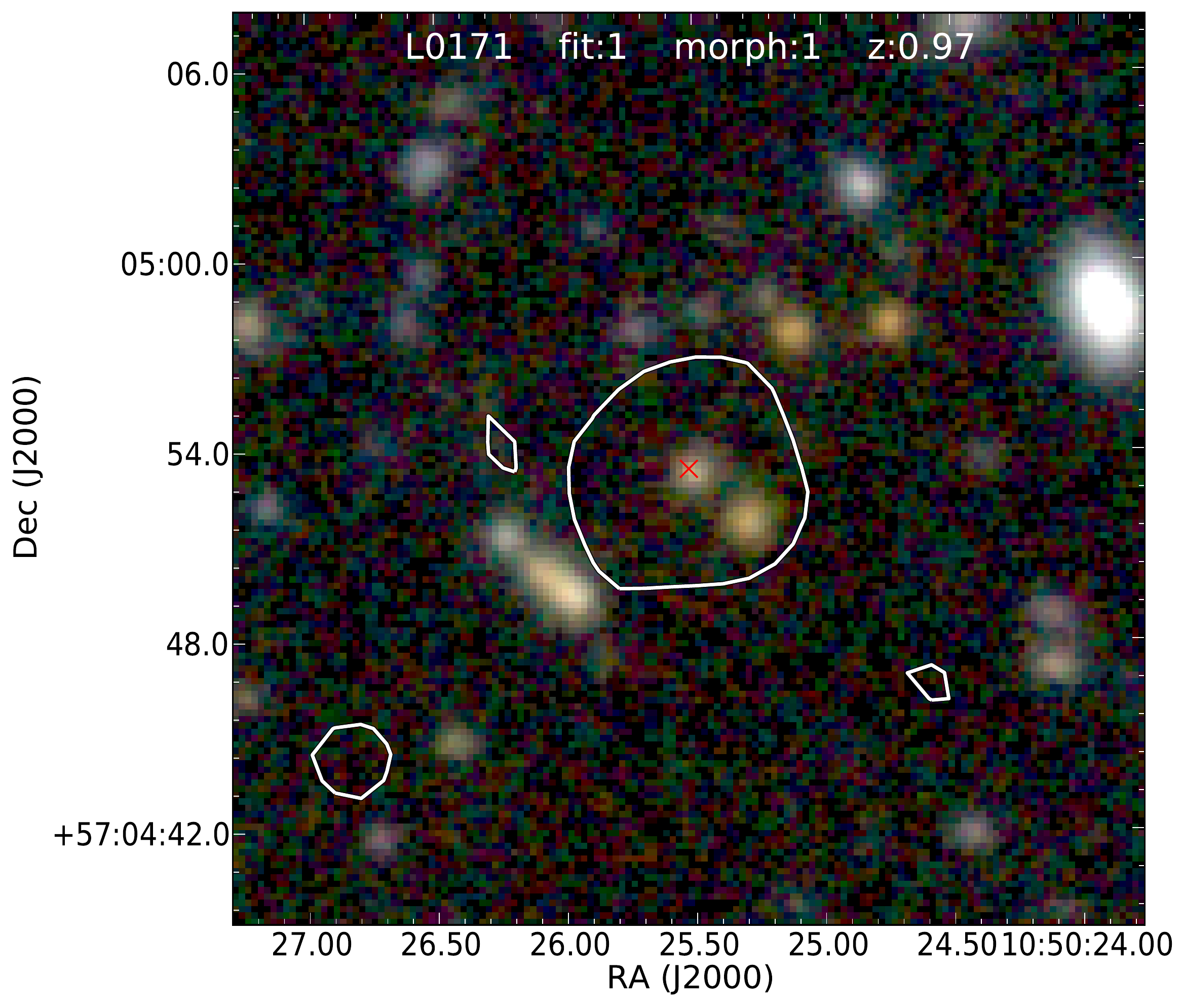}\\
\caption{{\it Left column:} UV to mid-infrared SED fits. Black points:
  photometric observations, red line: best-fitting model, cyan line:
  stellar template, black line (where applicable): galaxy template of
  second photometric redshift solution. The inset shows the
  probability distribution function of the best-fit model. {\it right
    column:} RGB cutouts made from Subaru $R_cI_cz'$ images, with a
  single VLA contour overlaid and the VLBA position indicated with red
  crosses. These images show regions 28.8\,arcsec$\times$28.8\,arcsec
  in size.}
\label{fig:spectra+rgb}
\end{figure*}

\begin{figure*}
\ContinuedFloat
\center
\includegraphics[height=4.5cm]{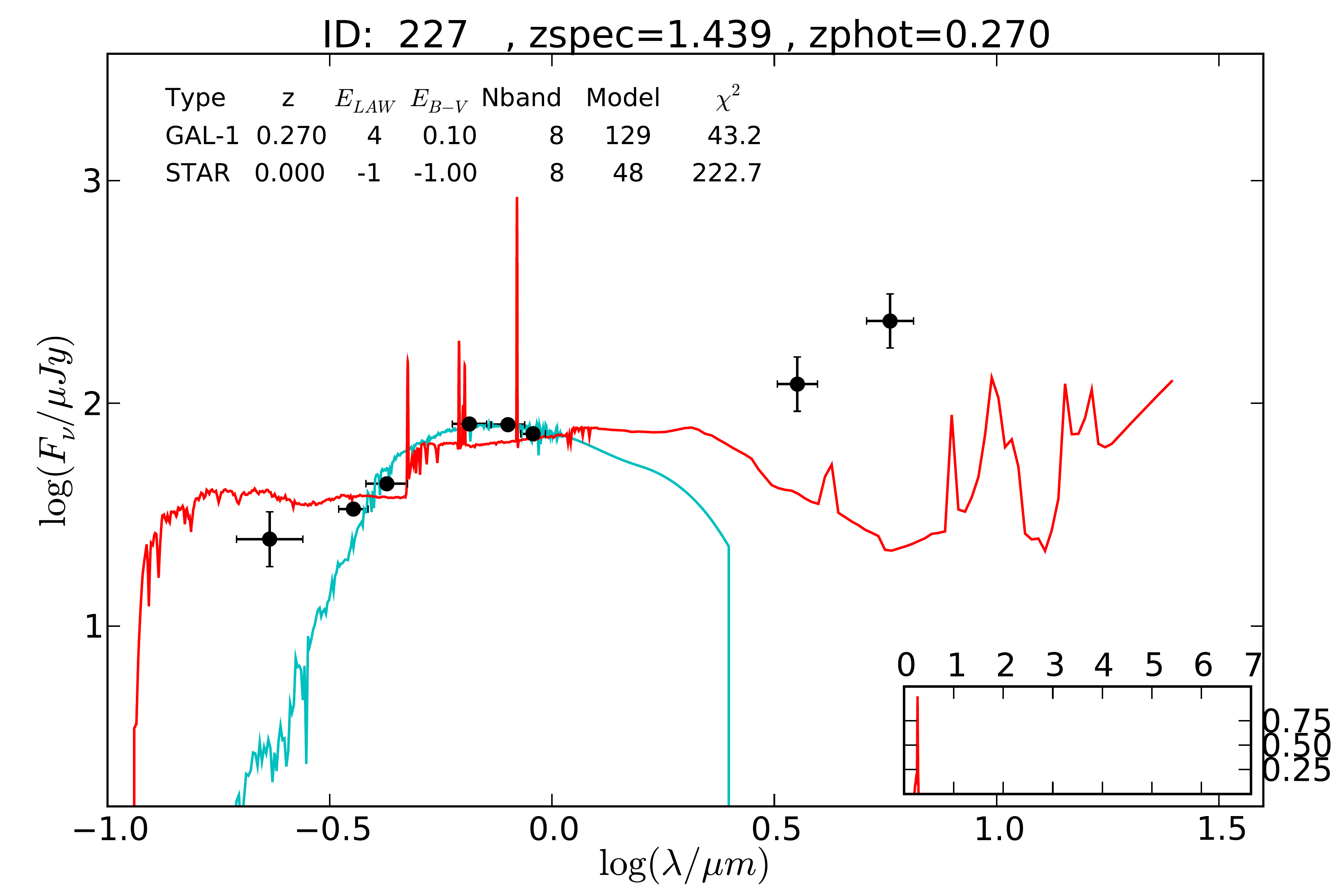}
\includegraphics[height=4.5cm]{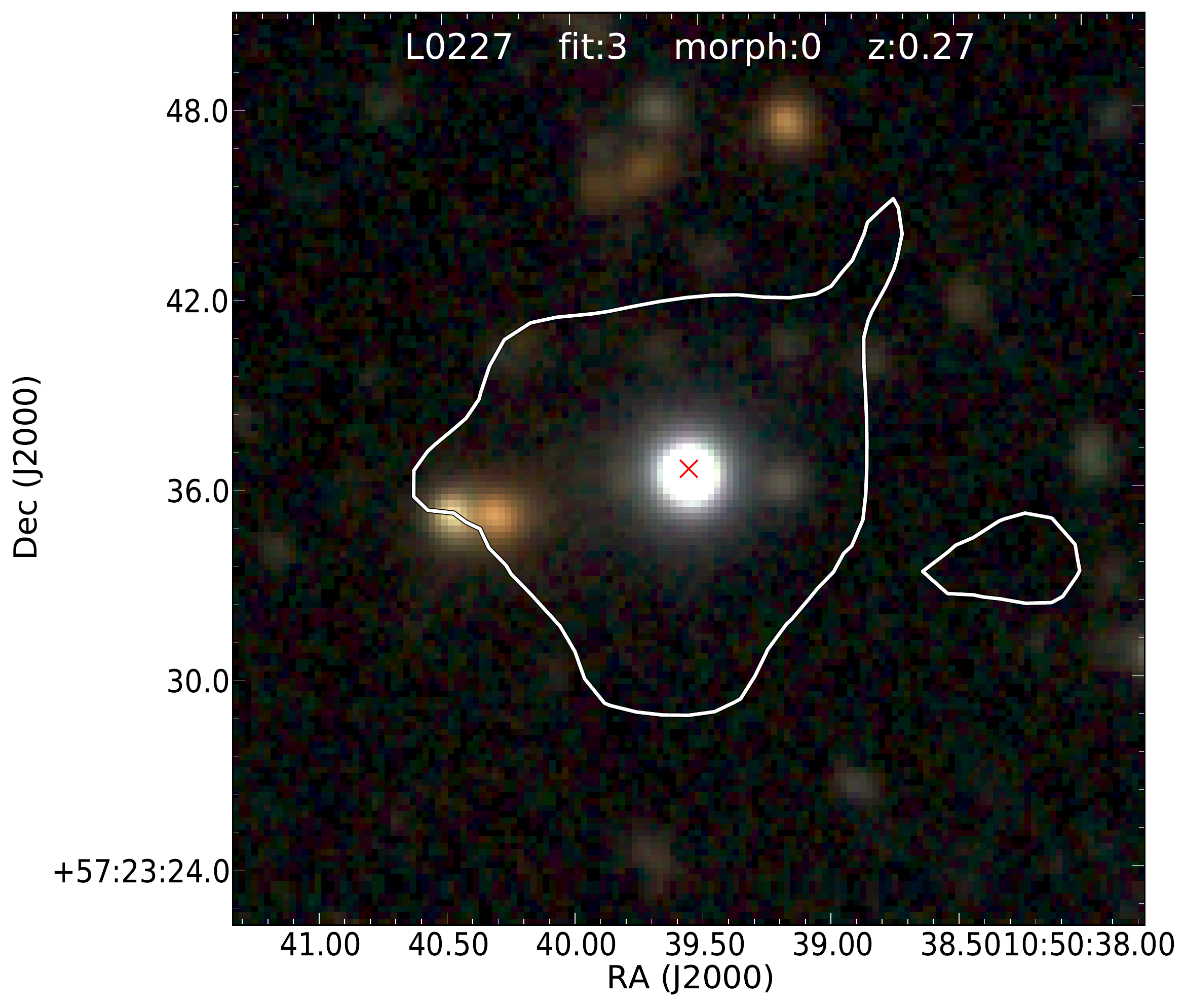}\\
\includegraphics[height=4.5cm]{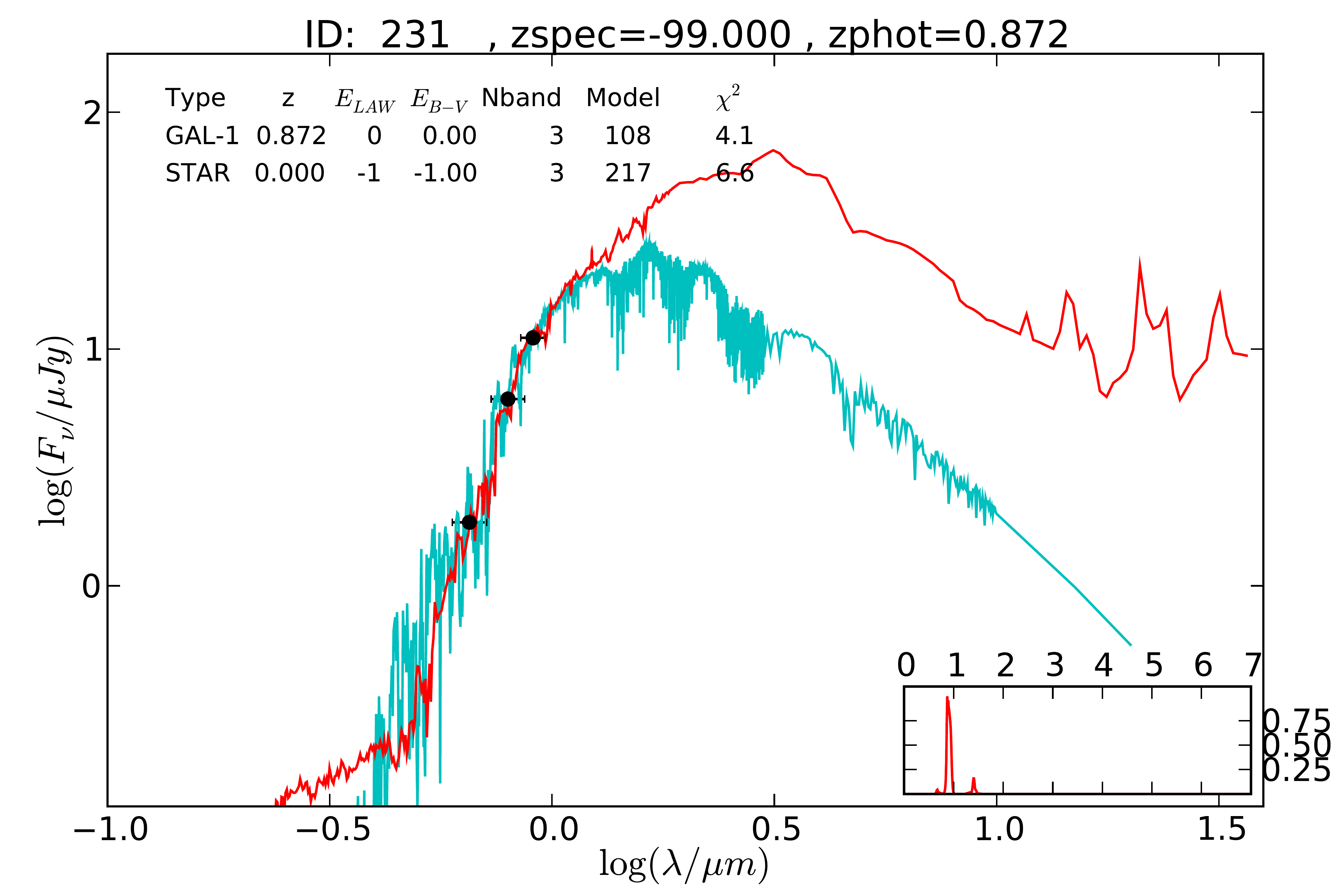}
\includegraphics[height=4.5cm]{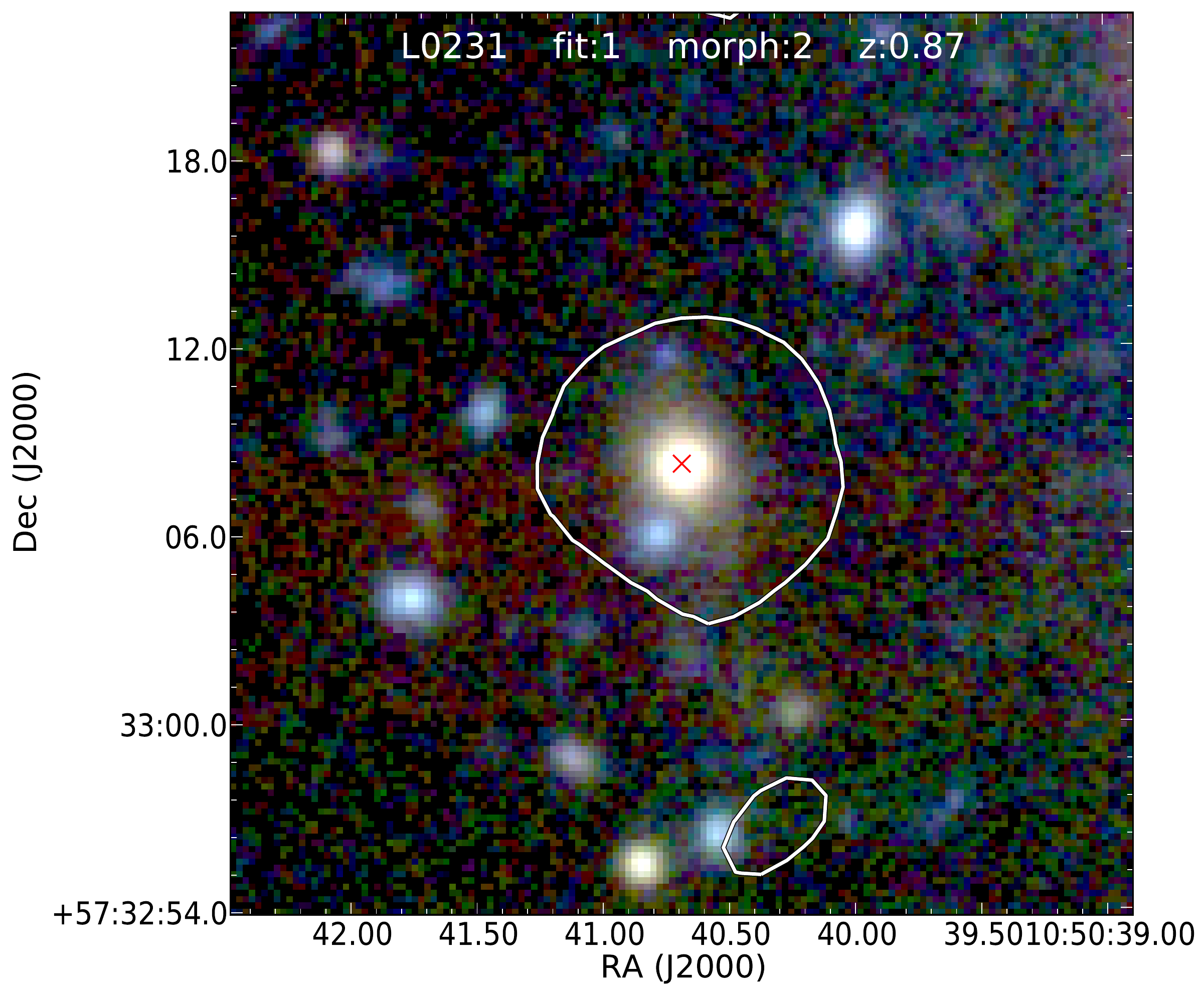}\\
\includegraphics[height=4.5cm]{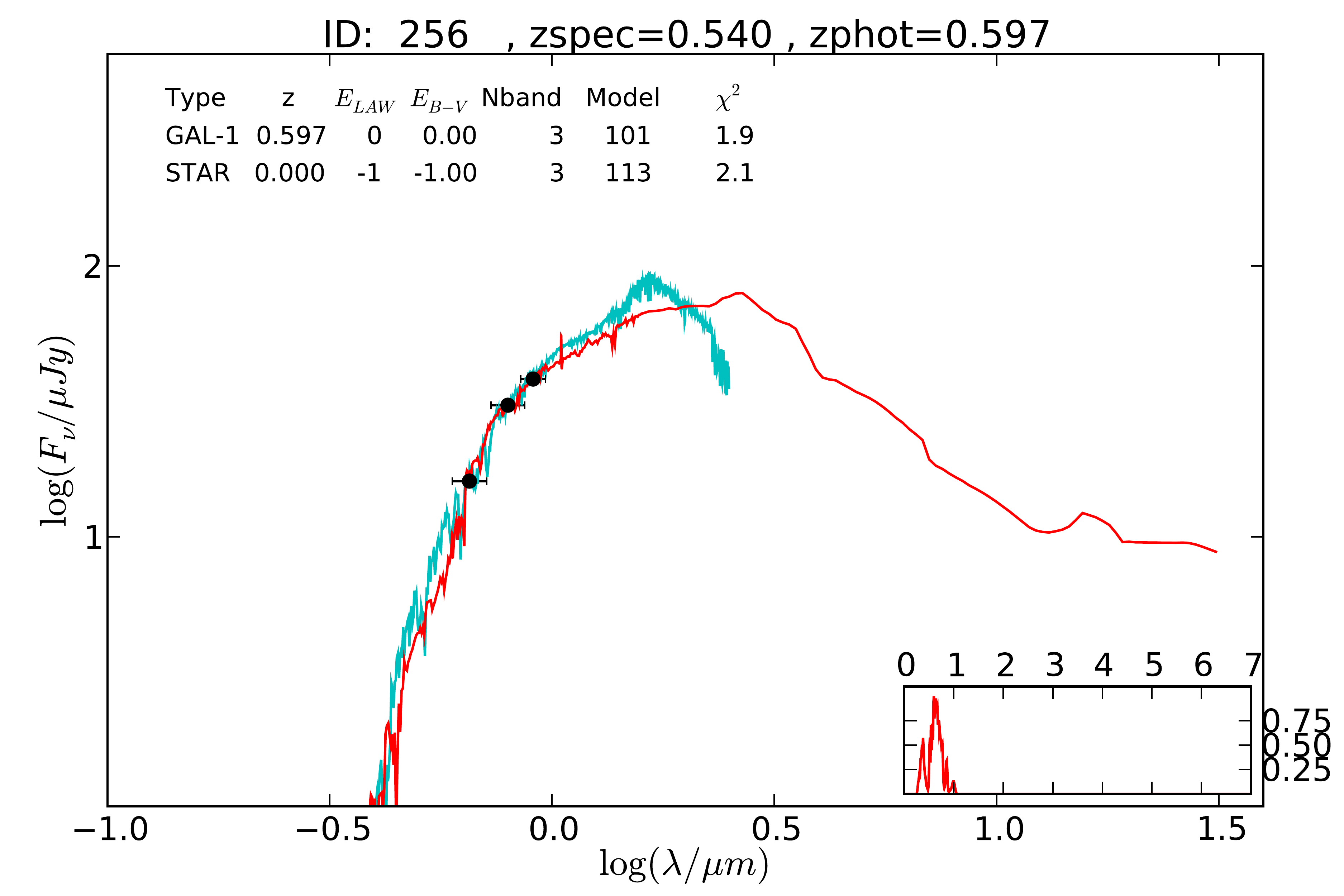}
\includegraphics[height=4.5cm]{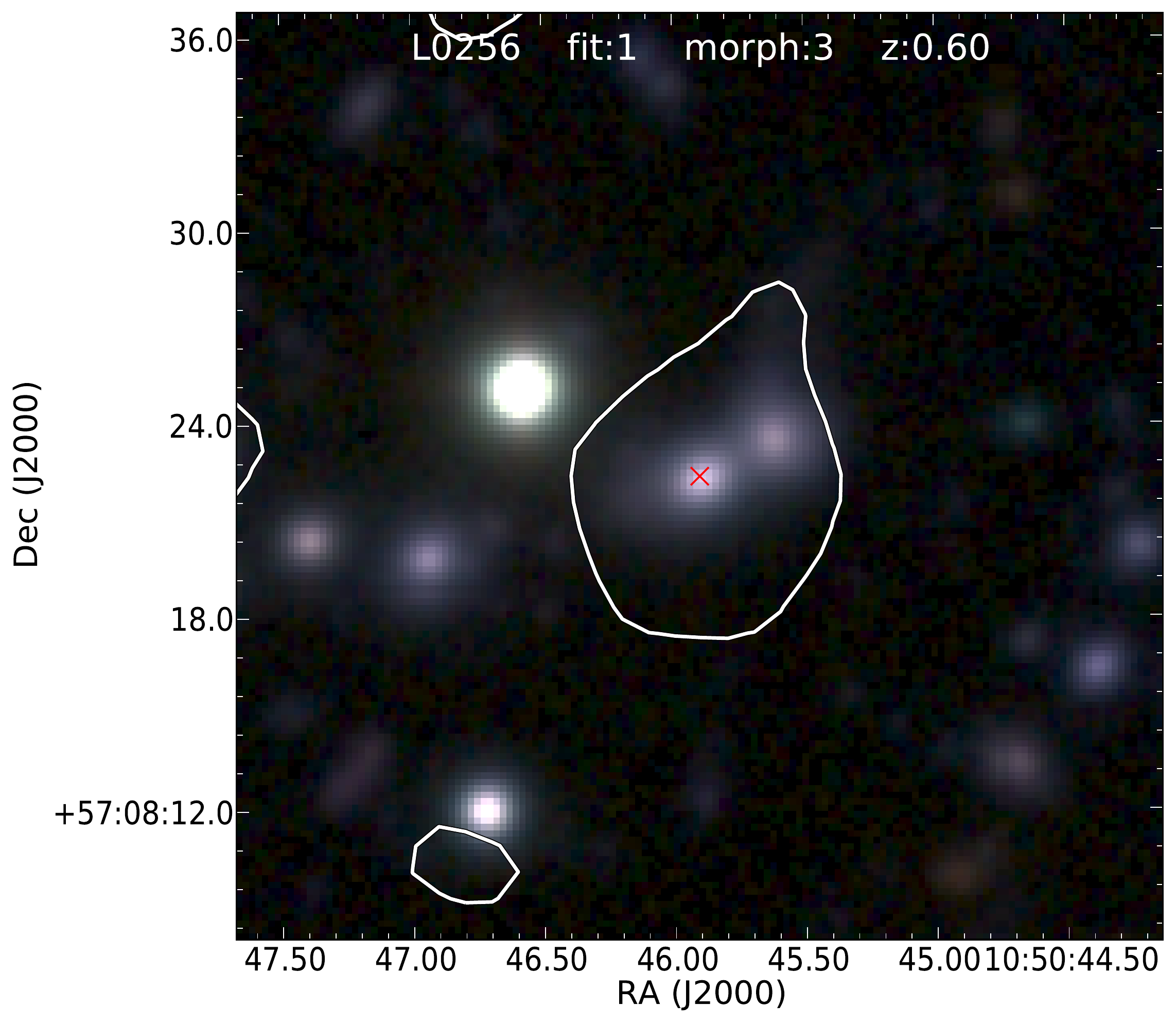}\\
\includegraphics[height=4.5cm]{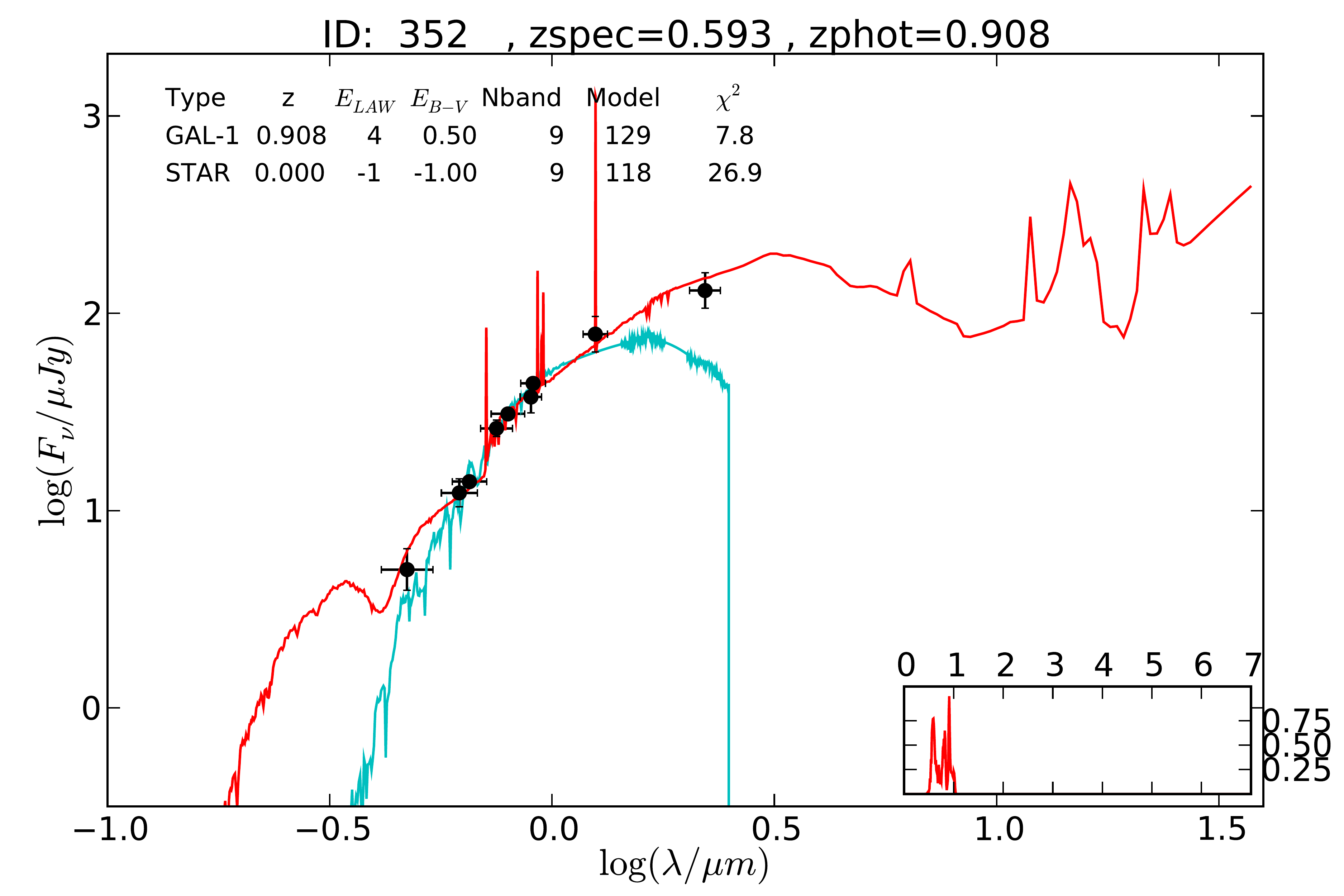}
\includegraphics[height=4.5cm]{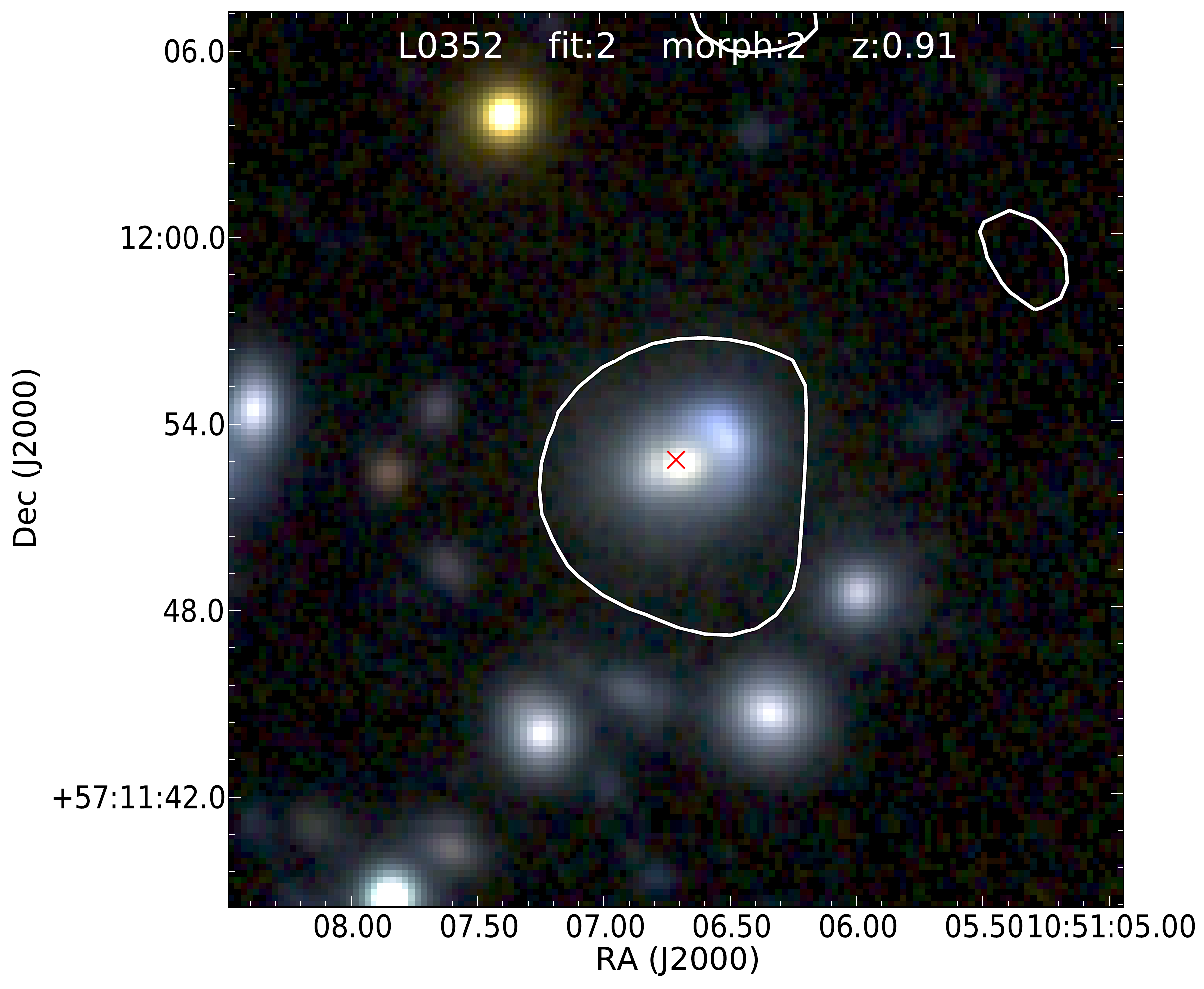}\\
\includegraphics[height=4.5cm]{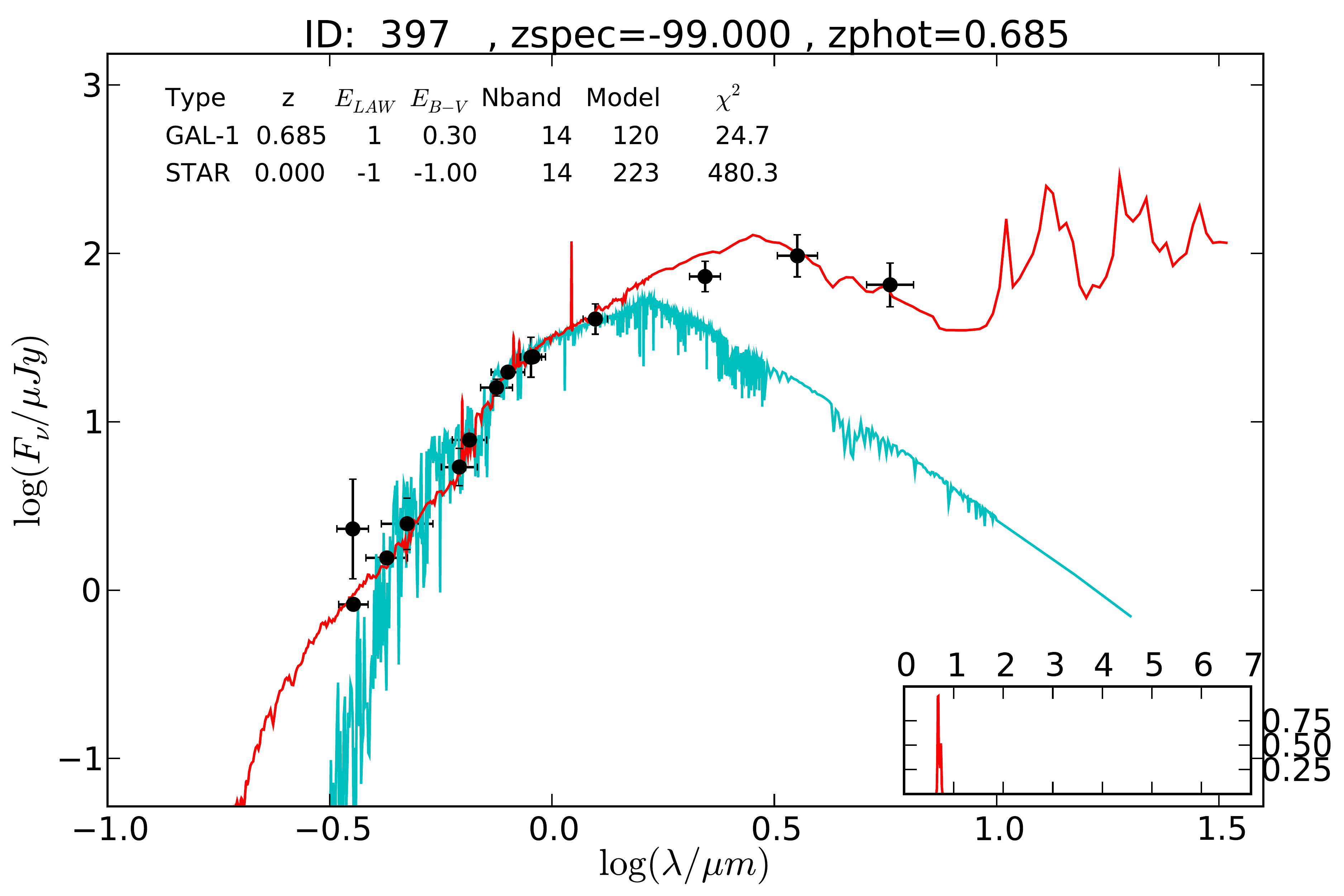}
\includegraphics[height=4.5cm]{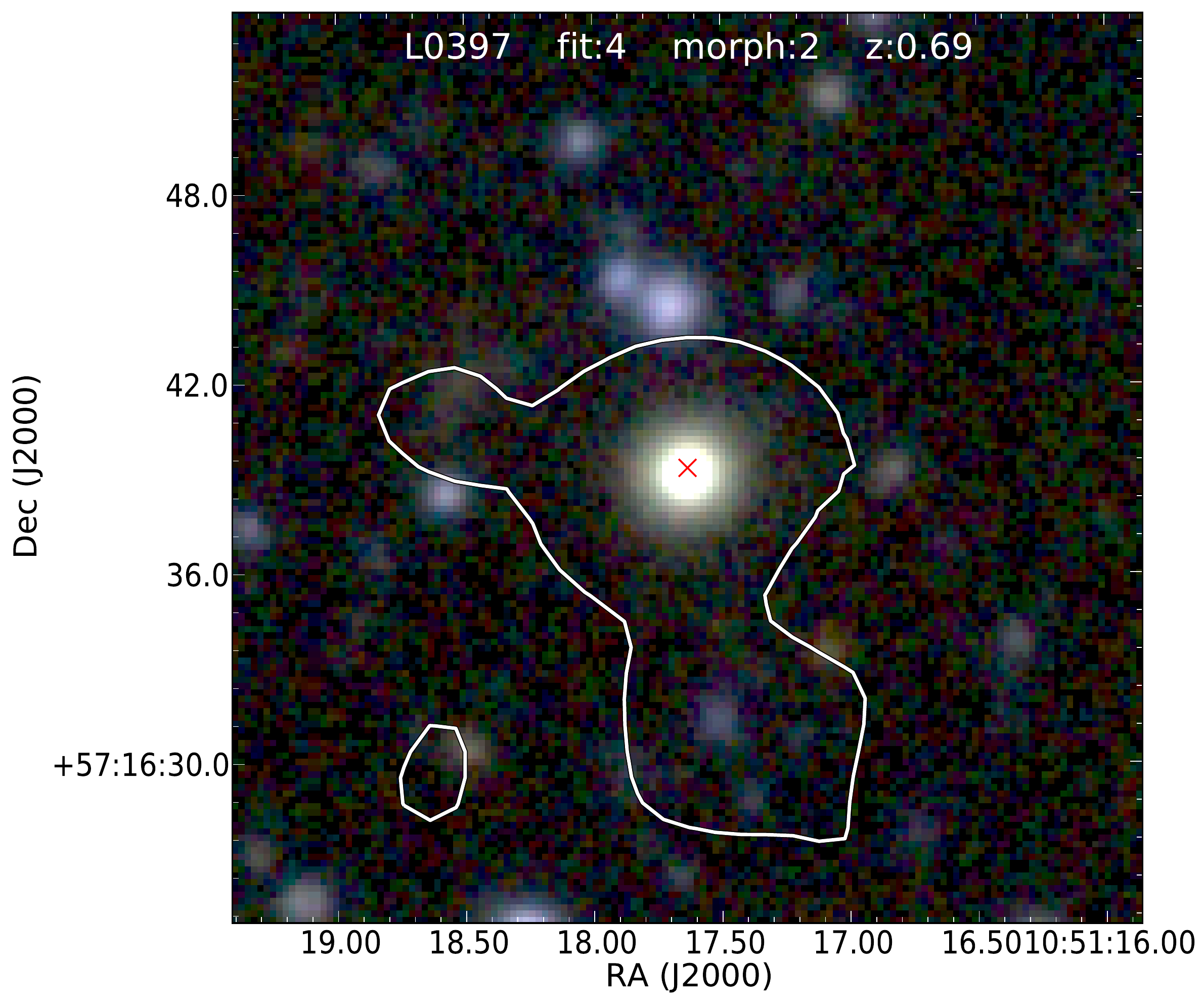}\\
\caption{(Continued)}
\end{figure*}

\begin{figure*}
\ContinuedFloat
\center
\includegraphics[height=4.5cm]{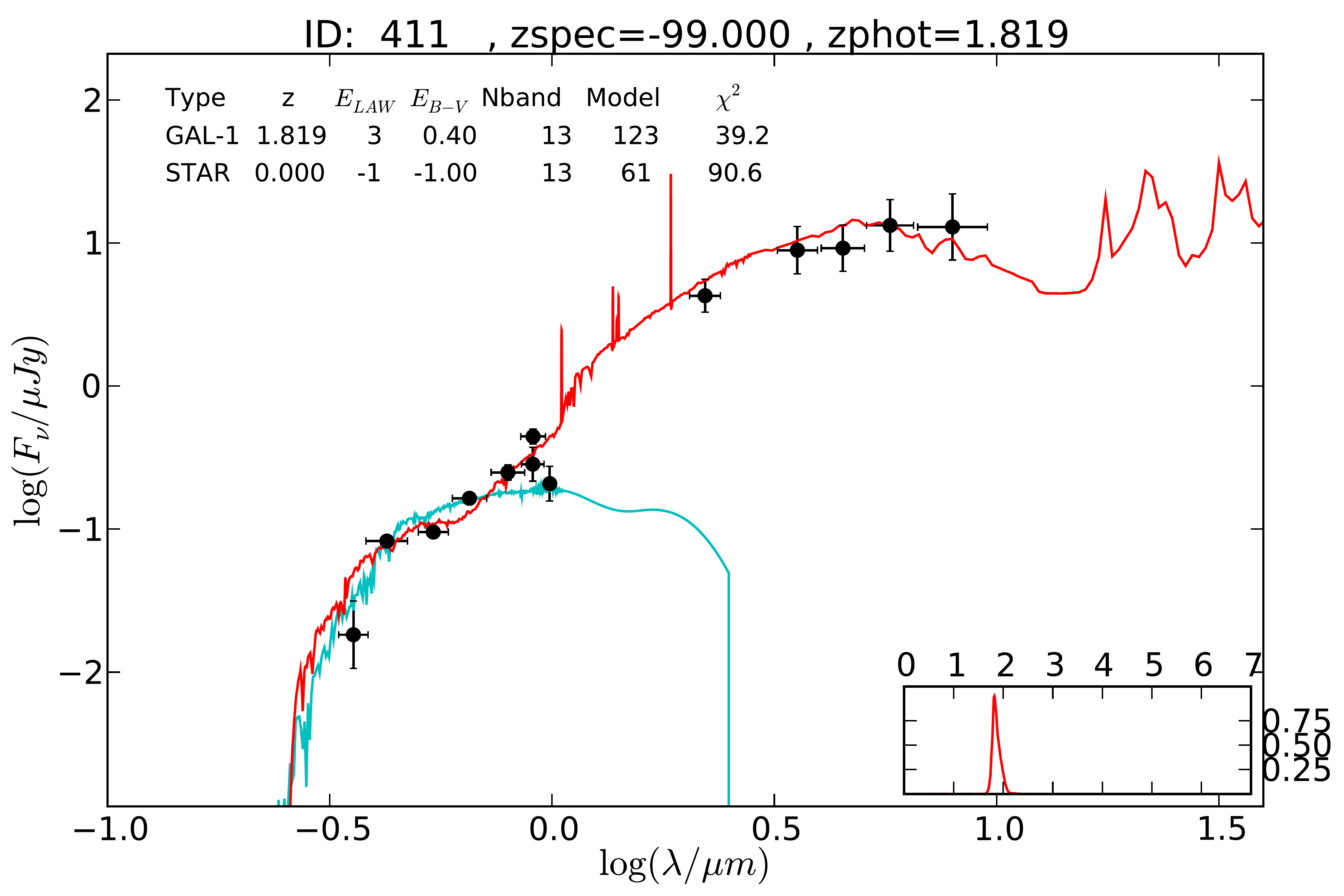}
\includegraphics[height=4.5cm]{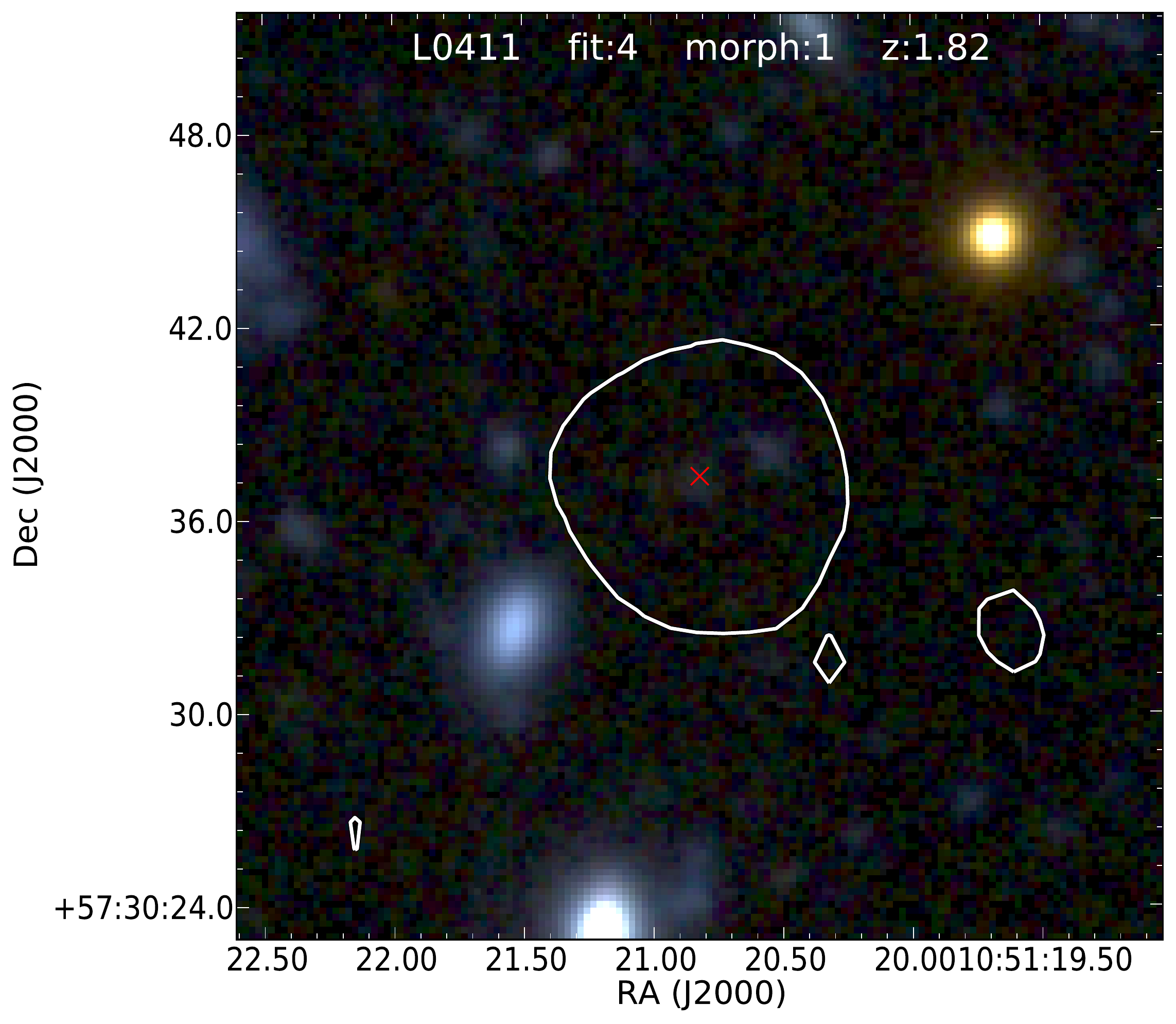}\\
\includegraphics[height=4.5cm]{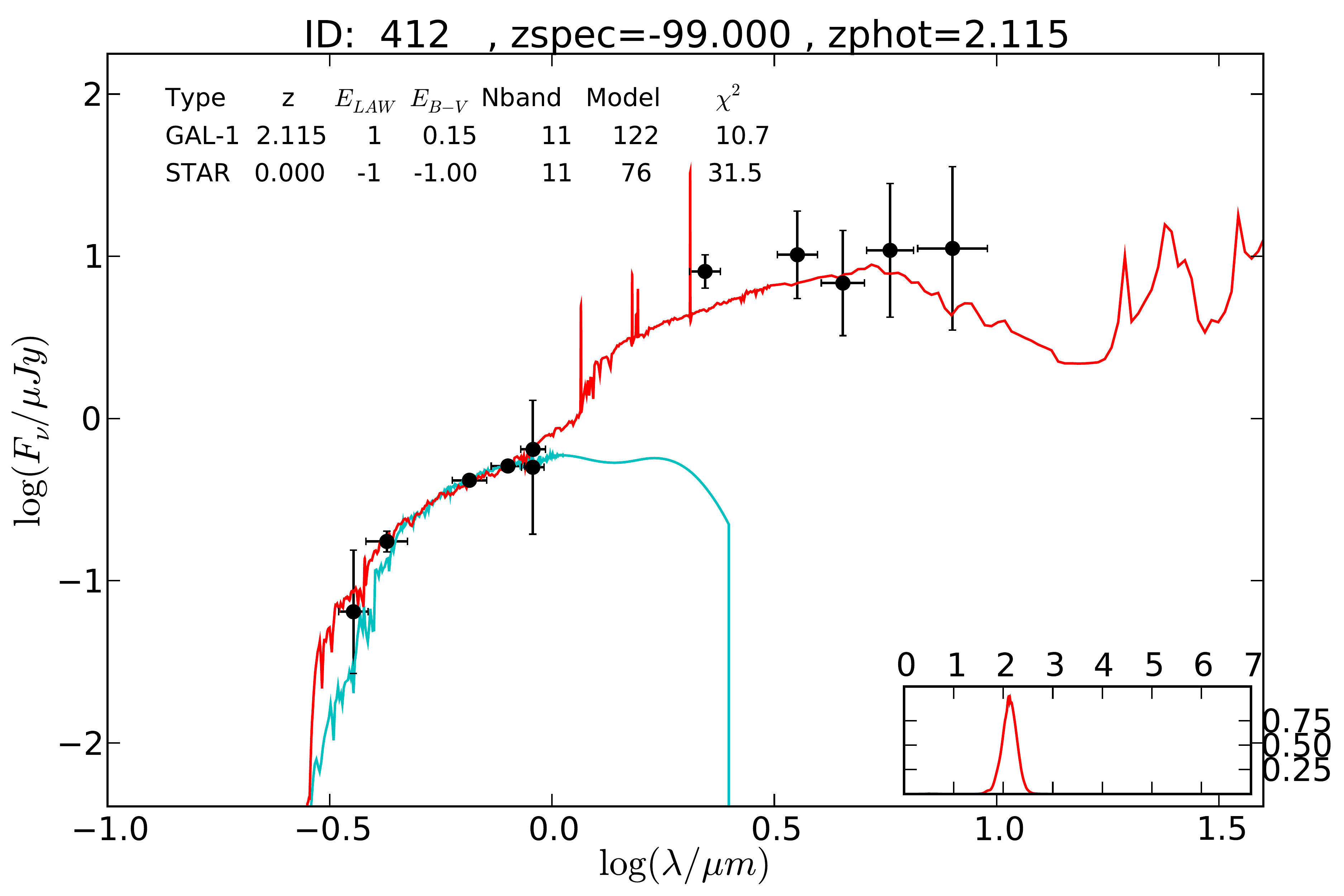}
\includegraphics[height=4.5cm]{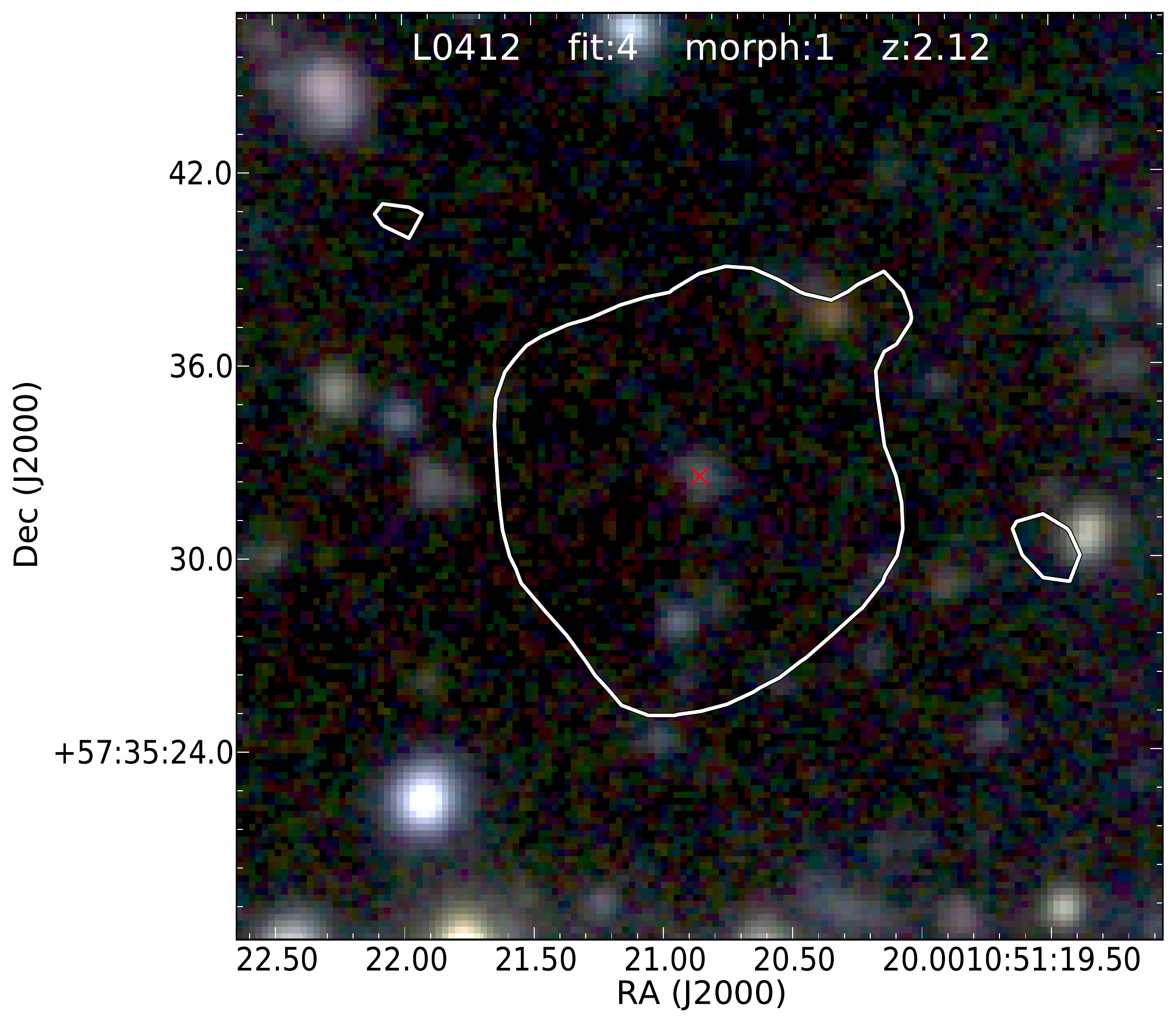}\\
\includegraphics[height=4.5cm]{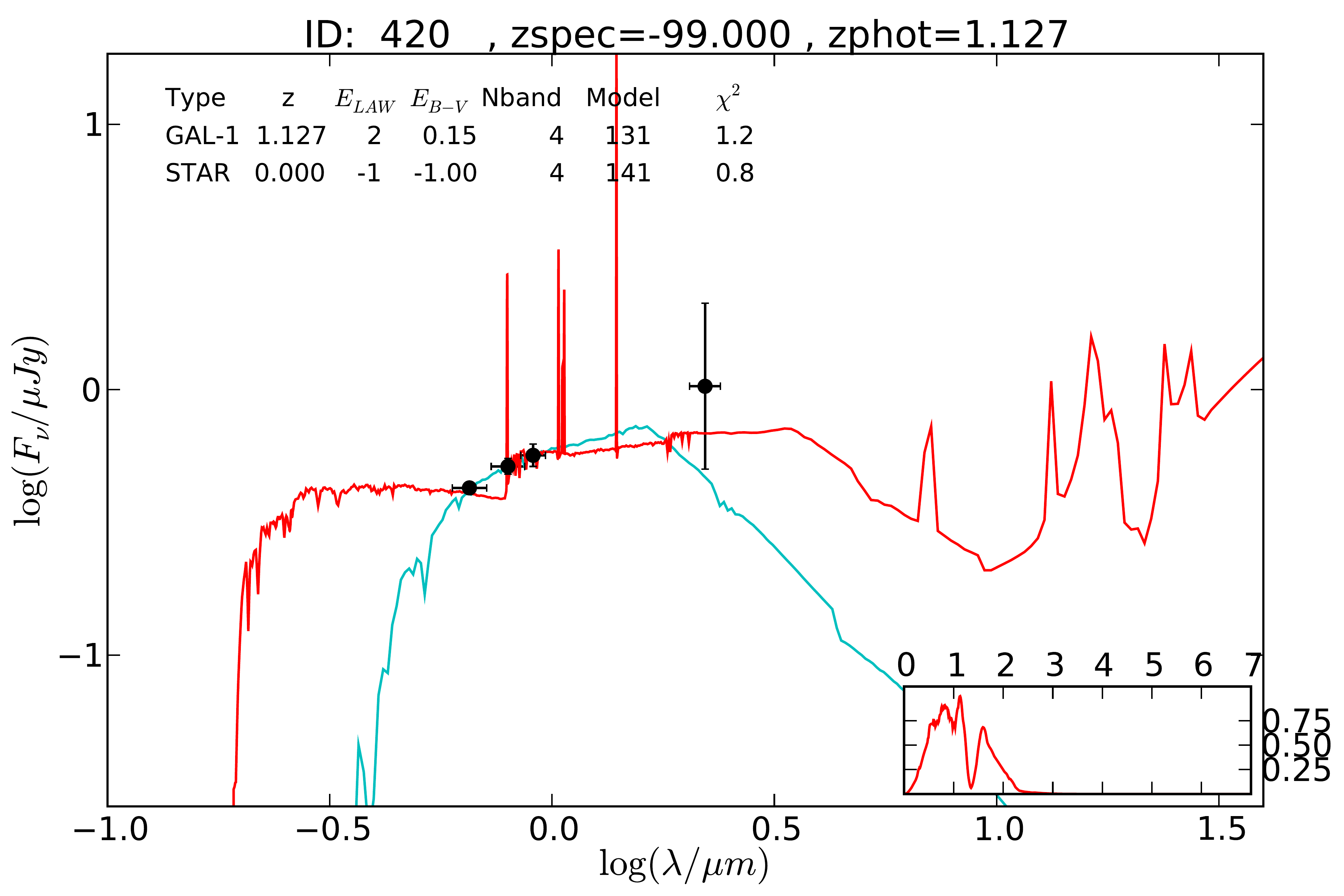}
\includegraphics[height=4.5cm]{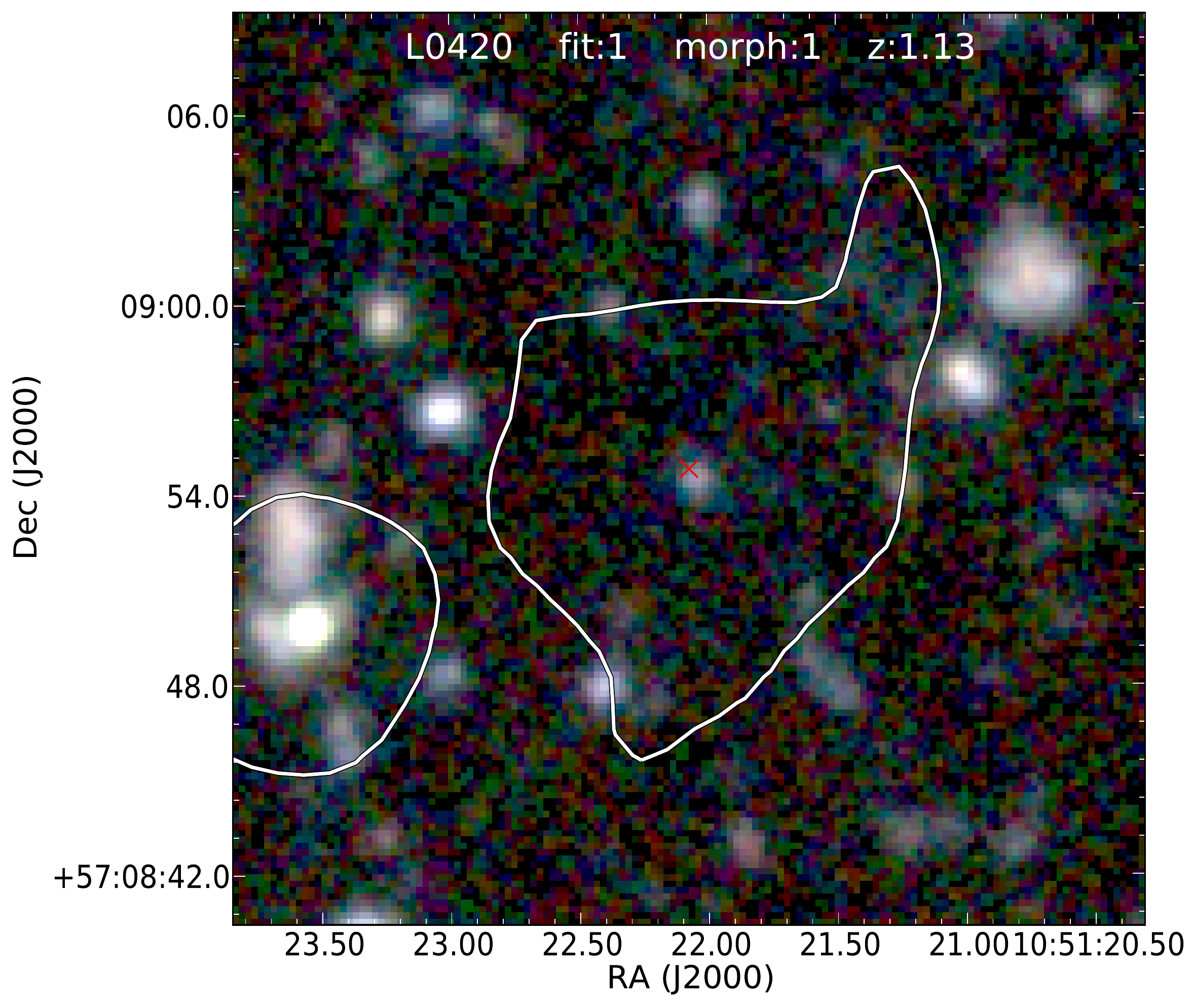}\\
\includegraphics[height=4.5cm]{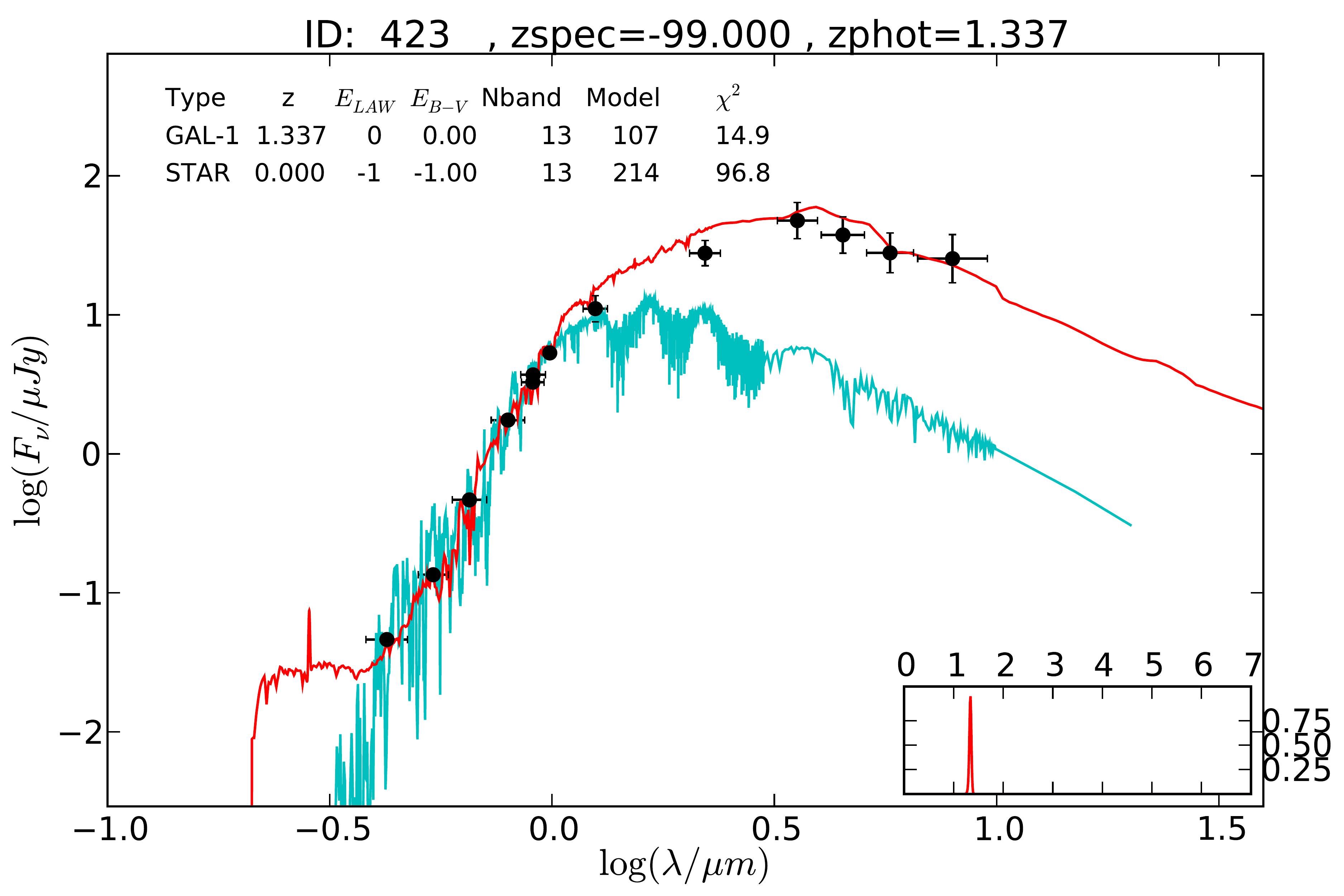}
\includegraphics[height=4.5cm]{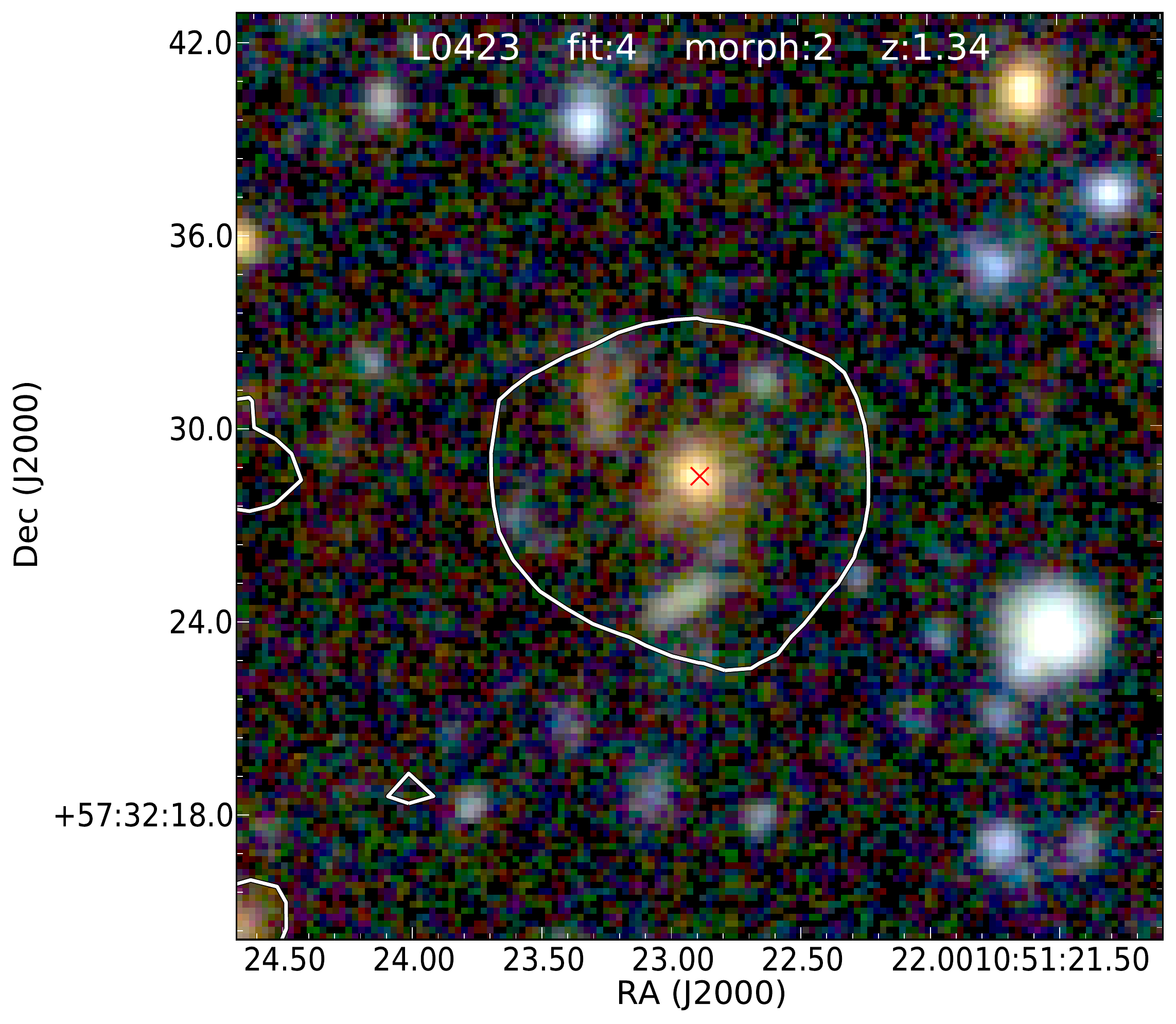}\\
\includegraphics[height=4.5cm]{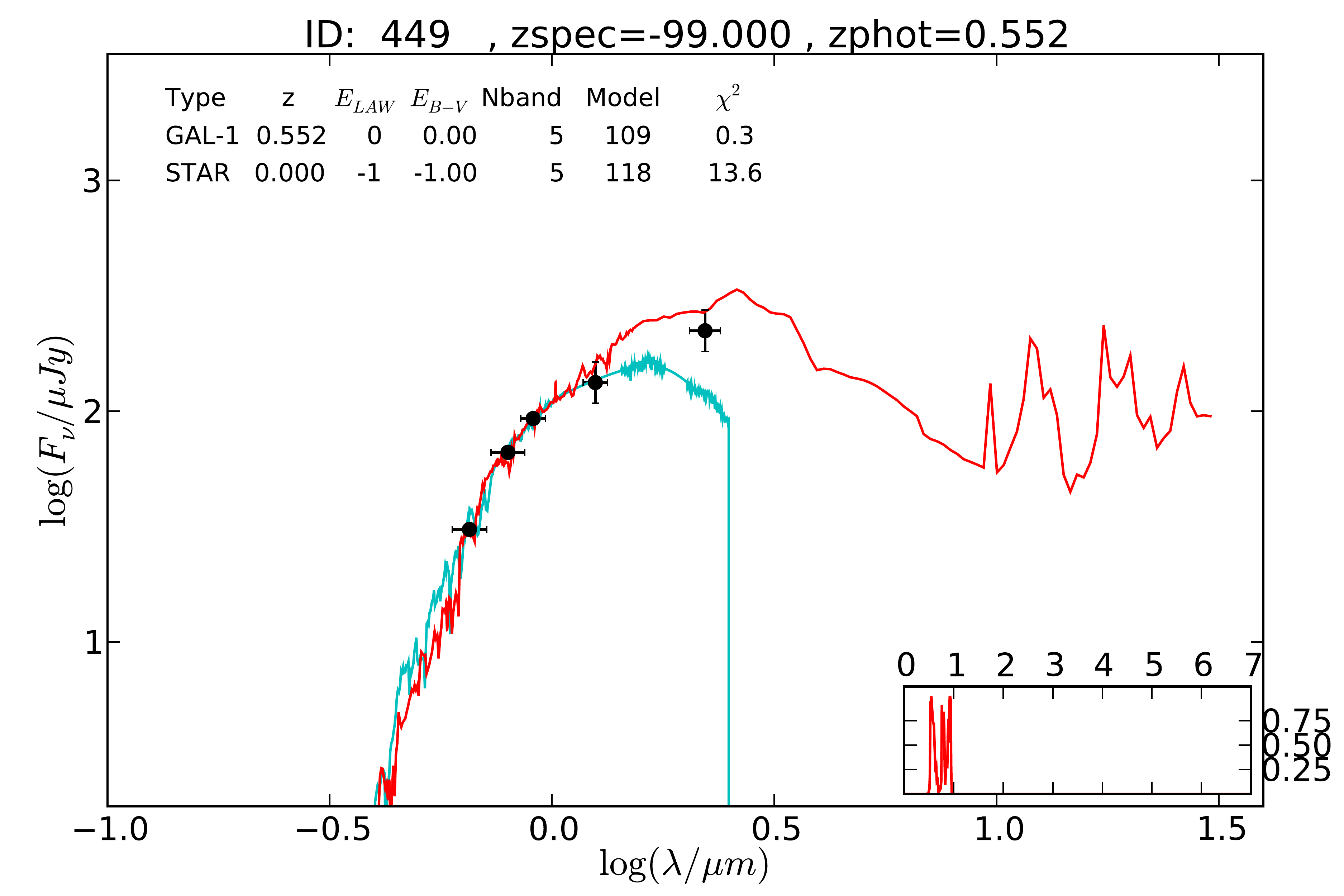}
\includegraphics[height=4.5cm]{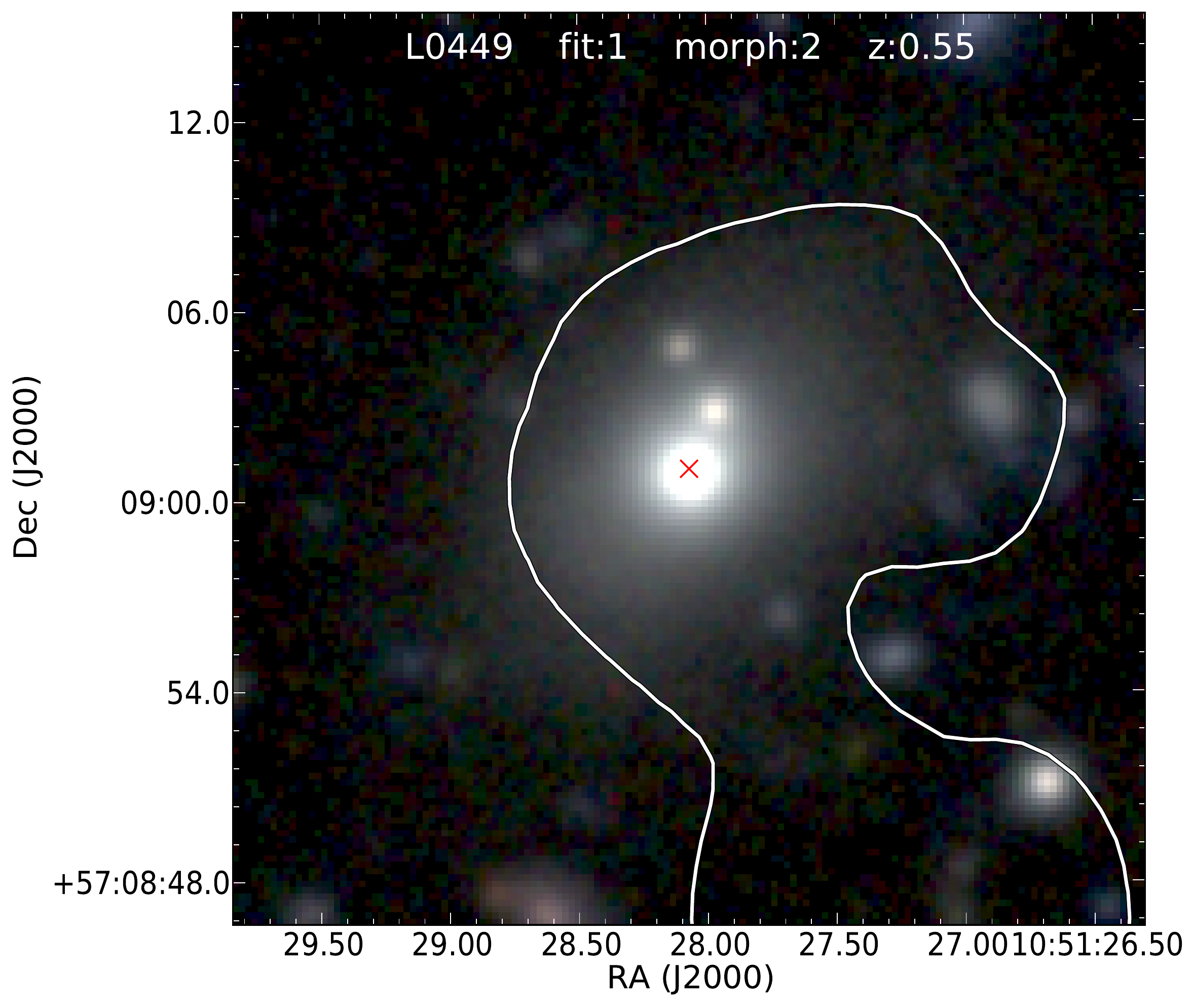}\\
\caption{(Continued)}
\end{figure*}

\begin{figure*}
\ContinuedFloat
\center
\includegraphics[height=4.5cm]{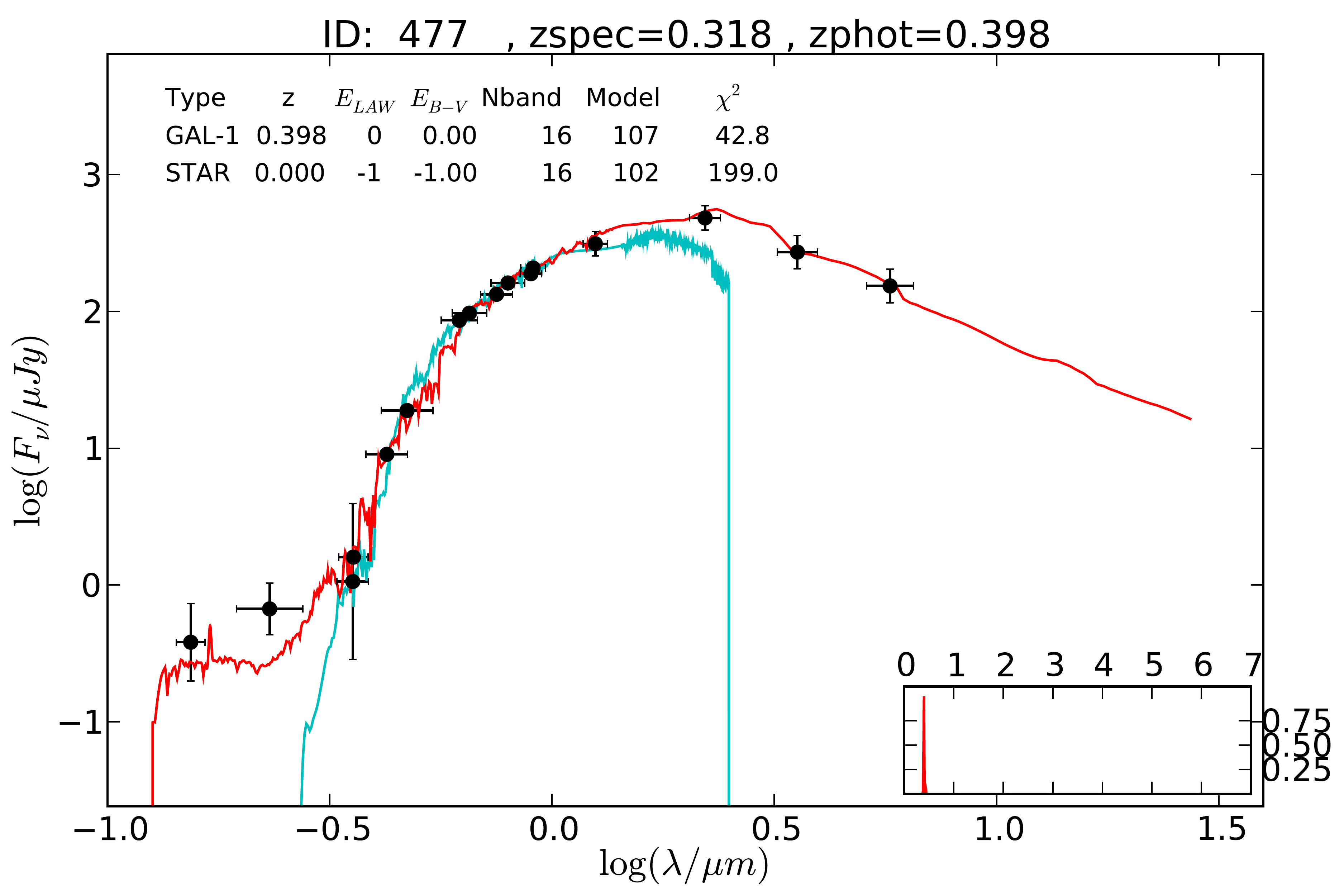}
\includegraphics[height=4.5cm]{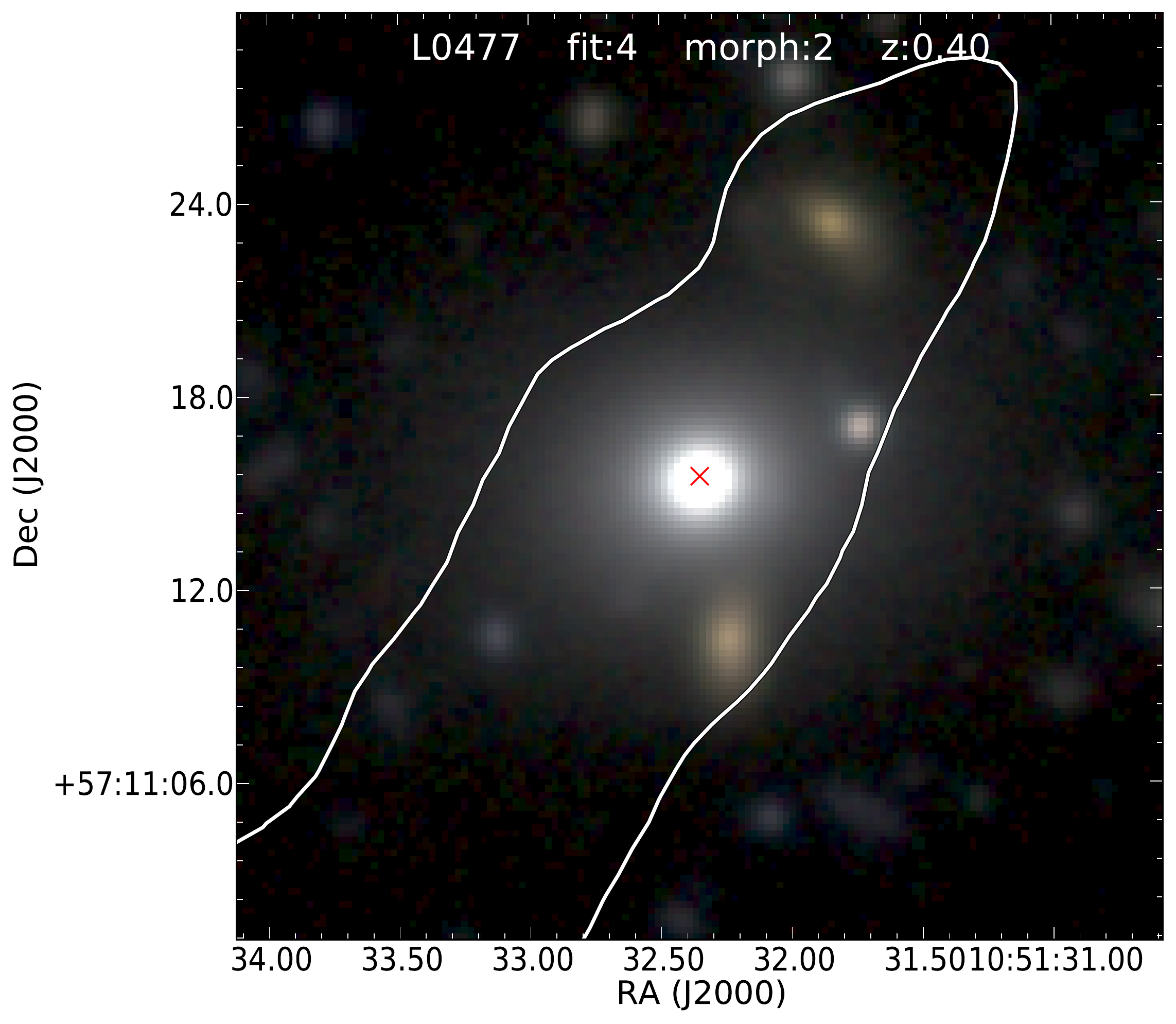}\\
\includegraphics[height=4.5cm]{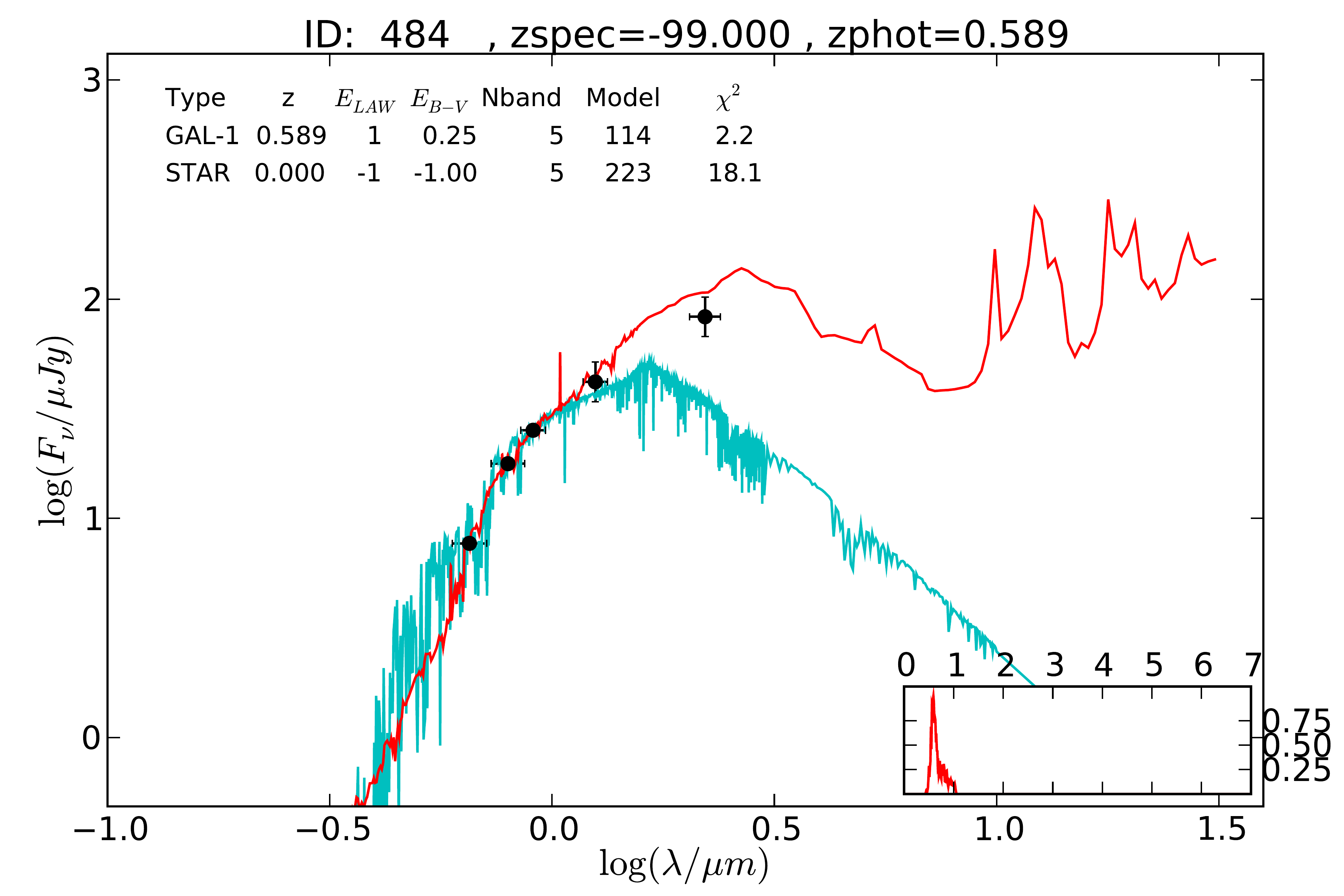}
\includegraphics[height=4.5cm]{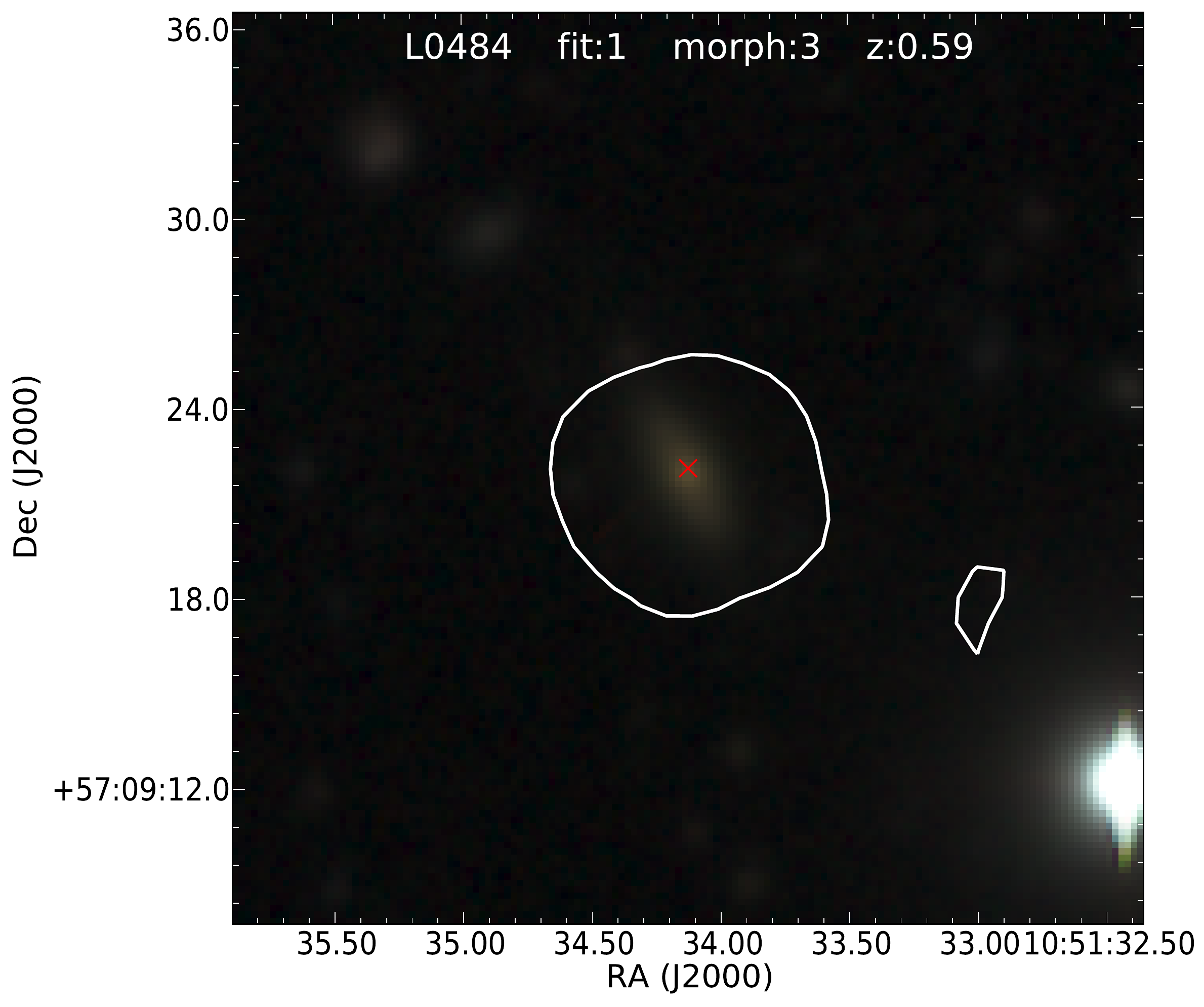}\\
\includegraphics[height=4.5cm]{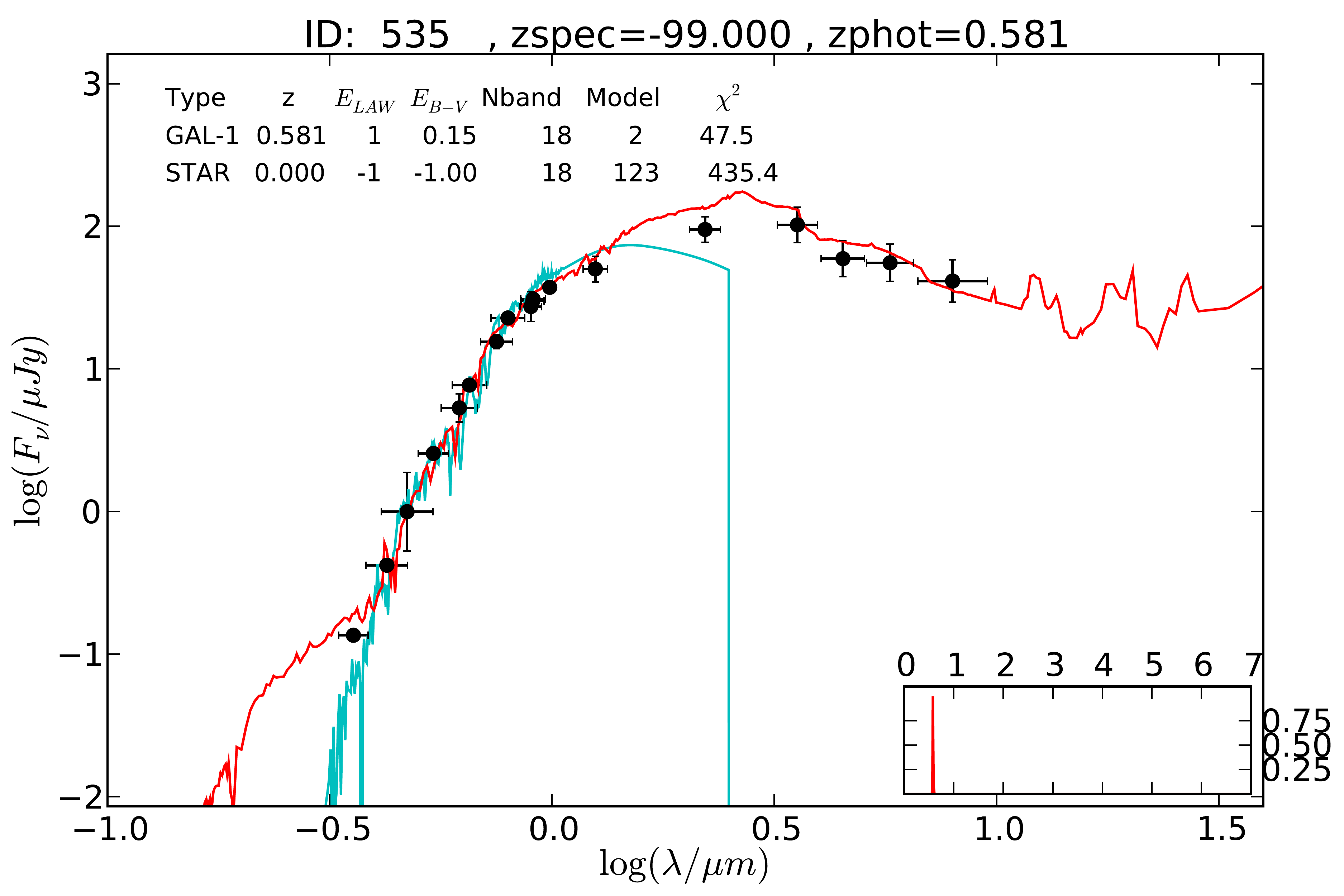}
\includegraphics[height=4.5cm]{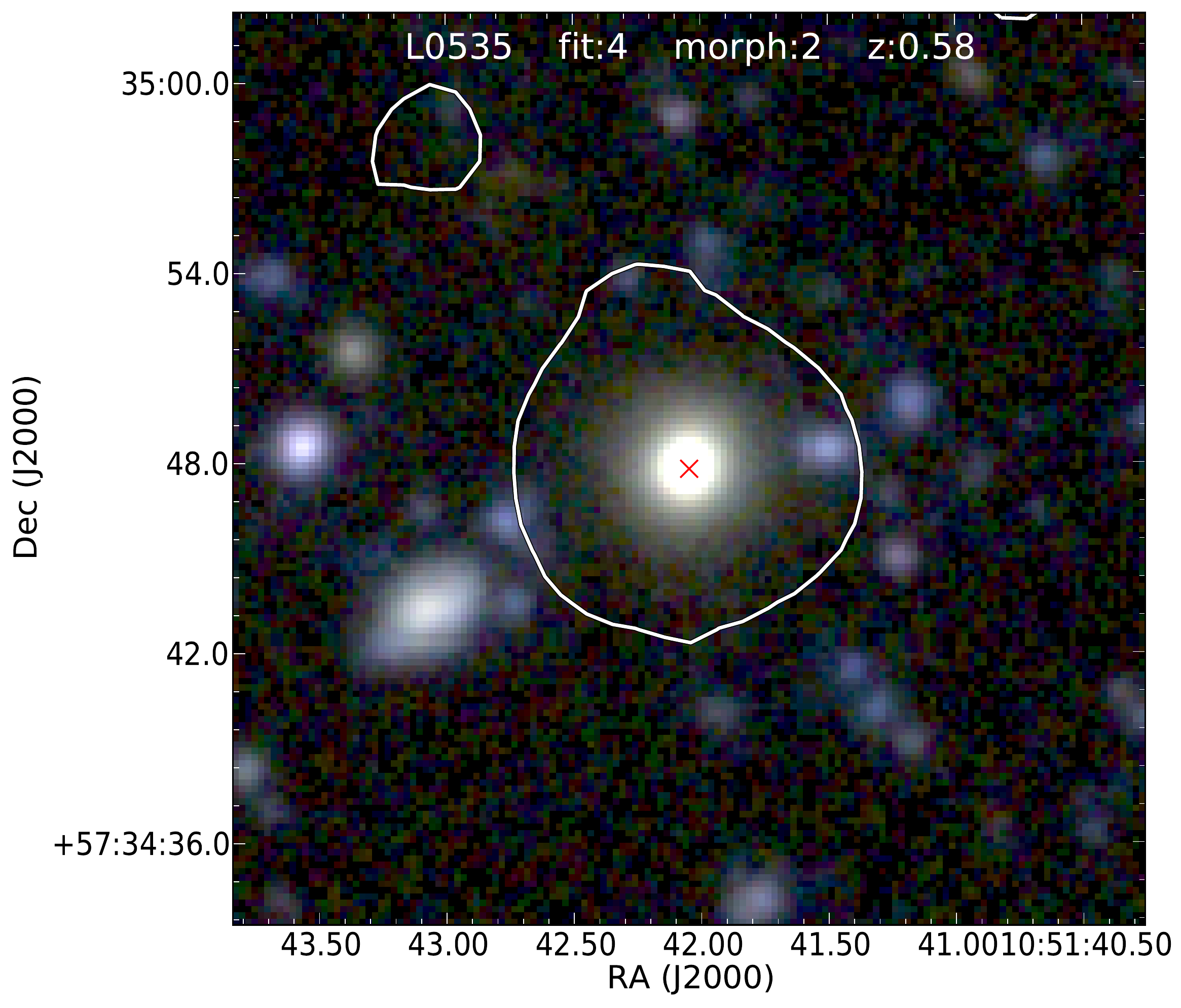}\\
\includegraphics[height=4.5cm]{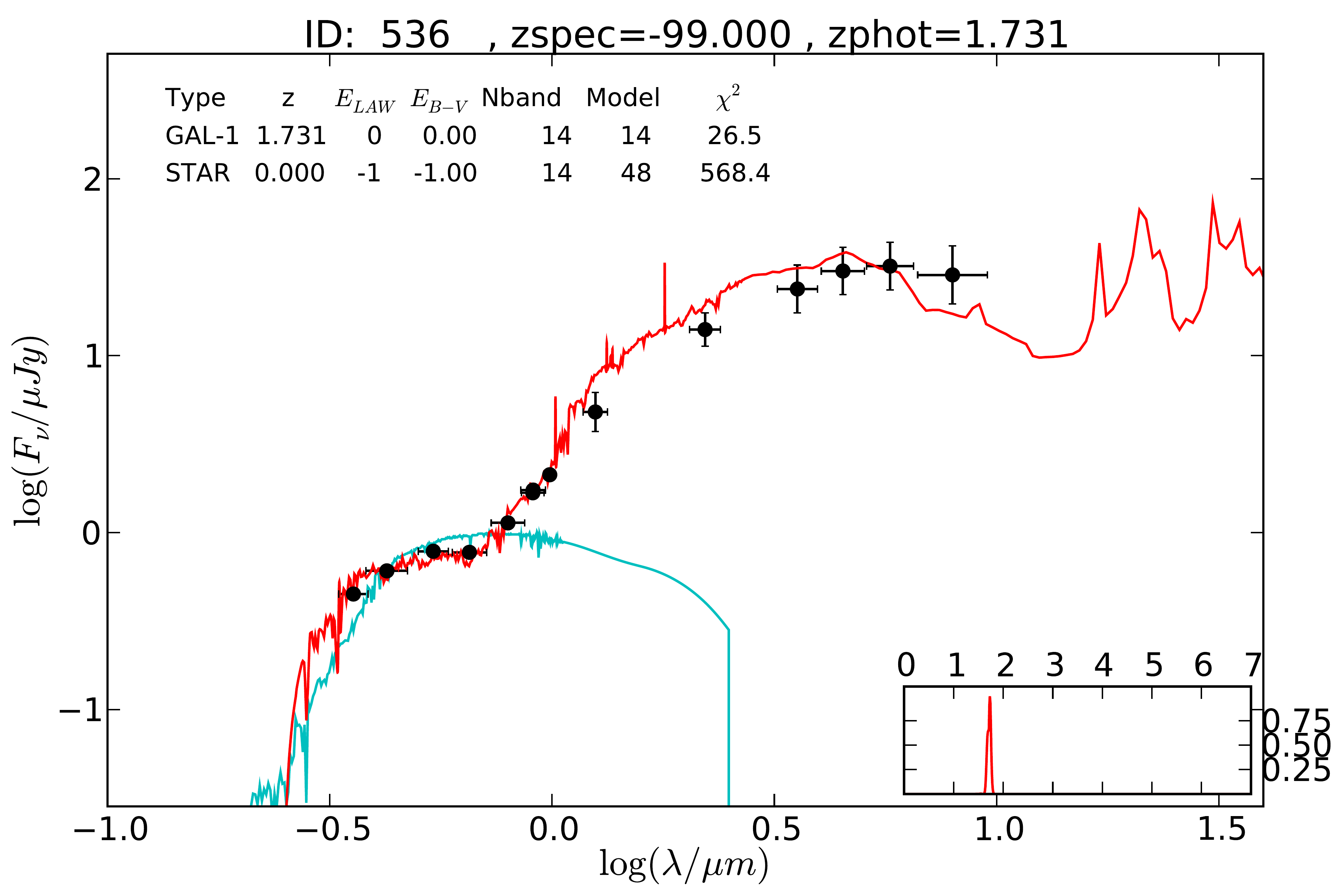}
\includegraphics[height=4.5cm]{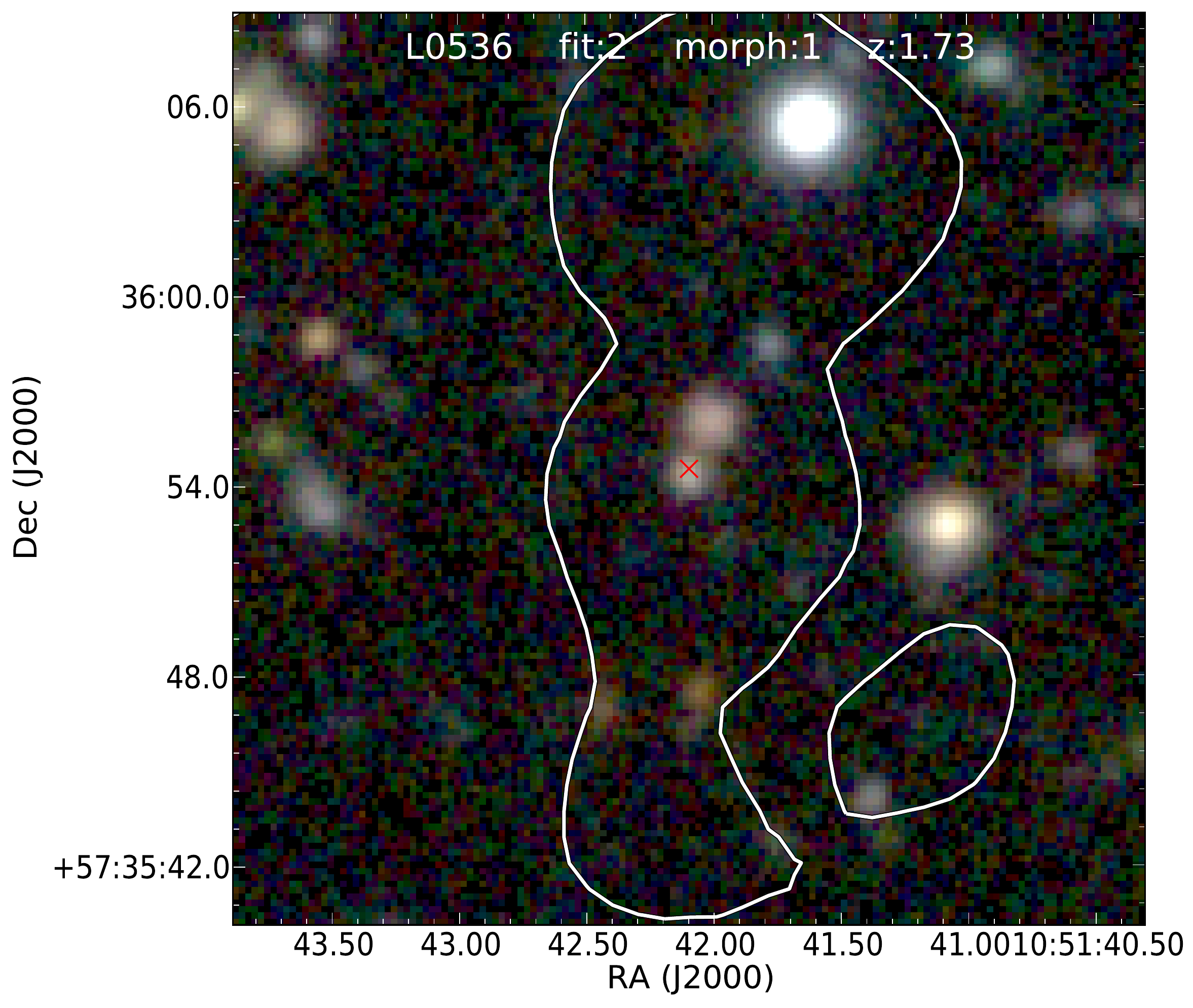}\\
\includegraphics[height=4.5cm]{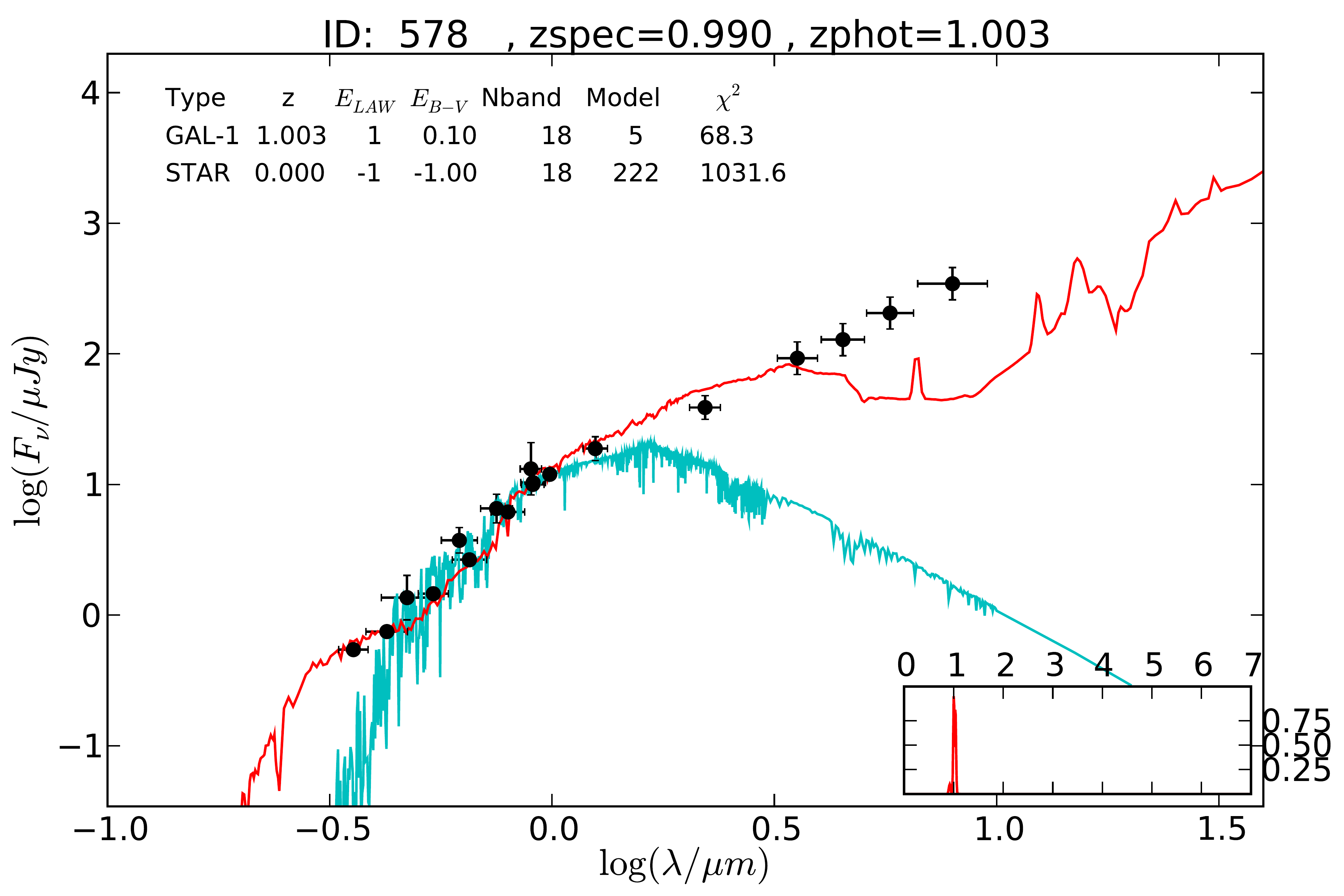}
\includegraphics[height=4.5cm]{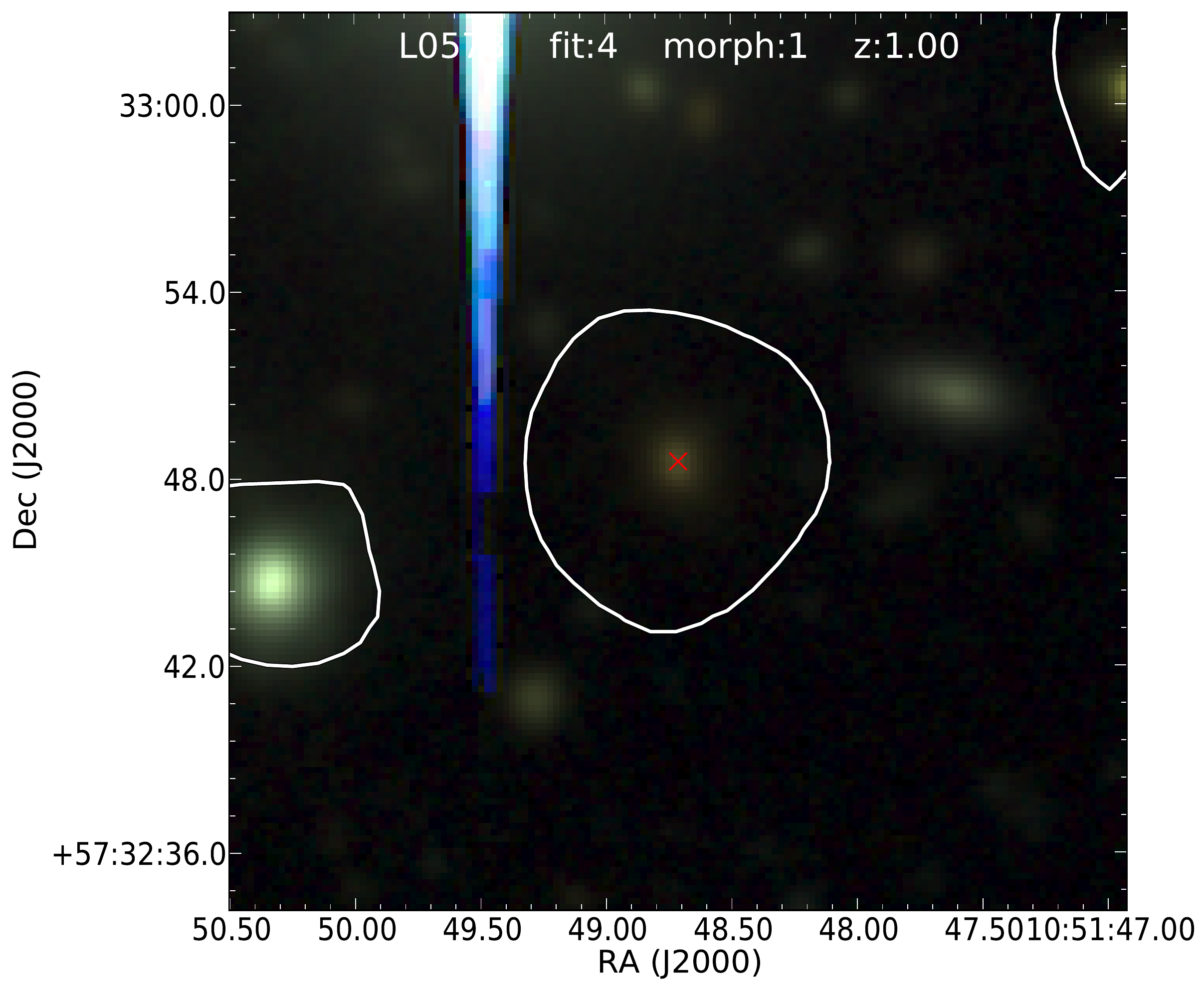}\\
\caption{(Continued)}
\end{figure*}

\begin{figure*}
\ContinuedFloat
\center
\includegraphics[height=4.5cm]{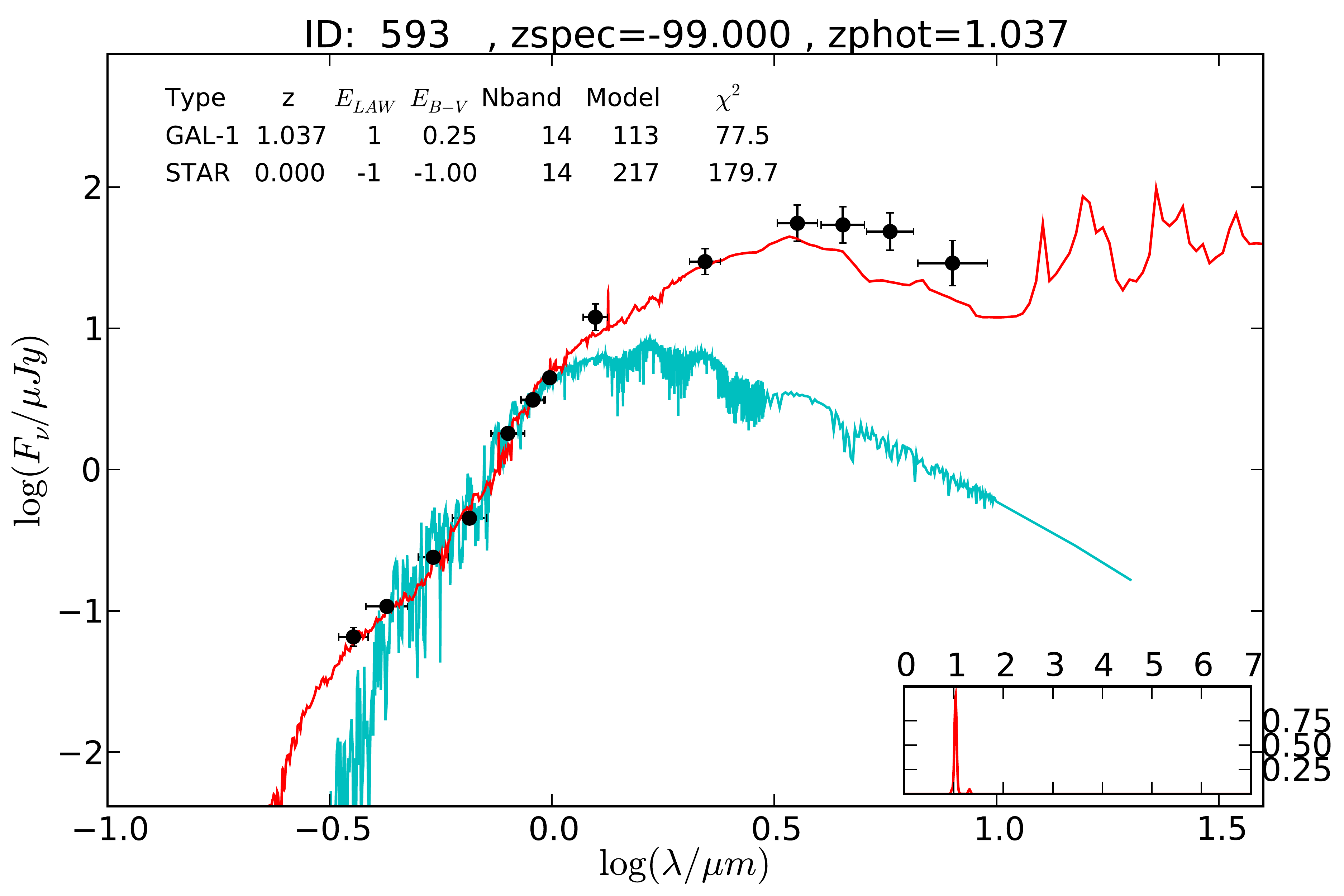}
\includegraphics[height=4.5cm]{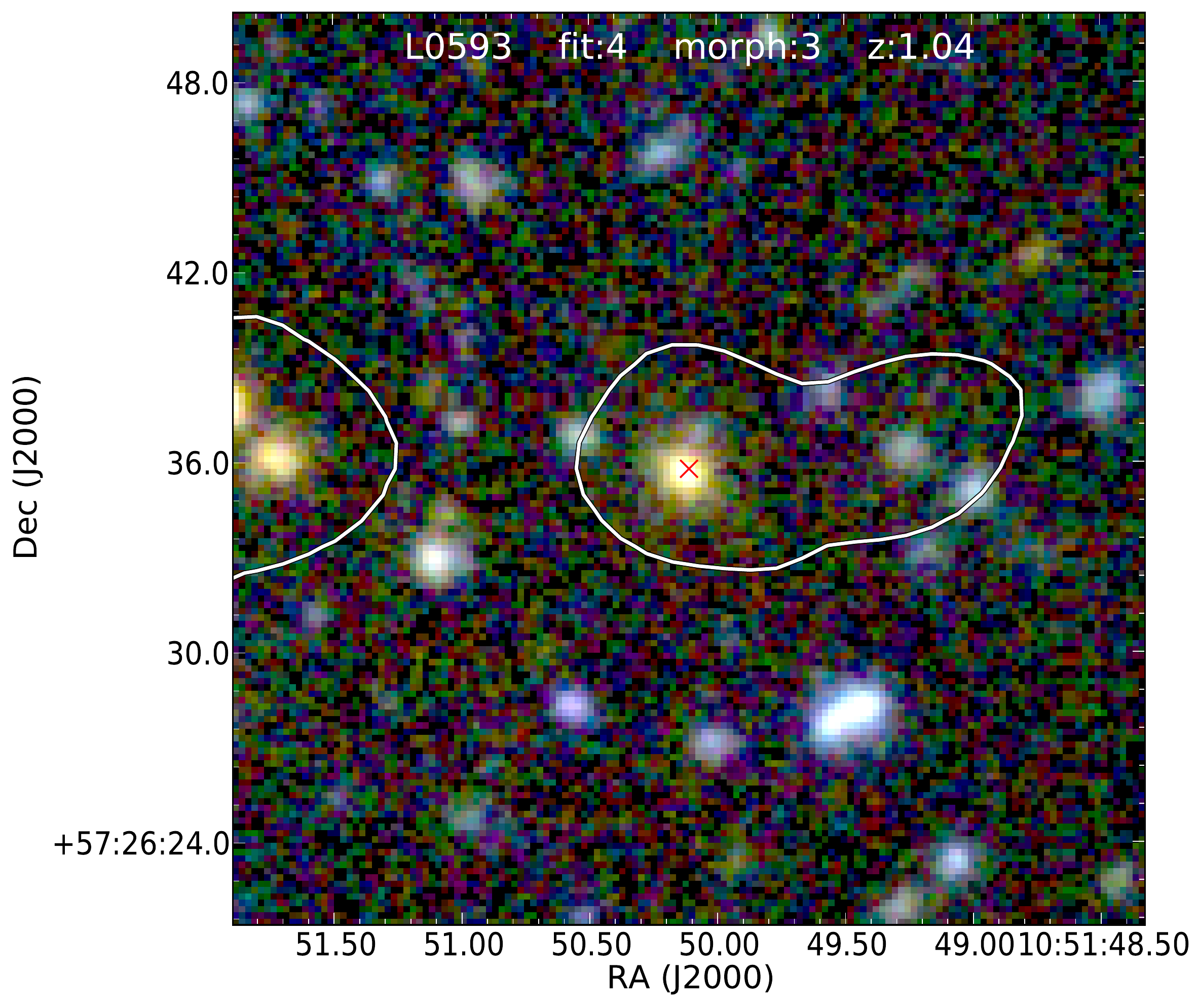}\\
\includegraphics[height=4.5cm]{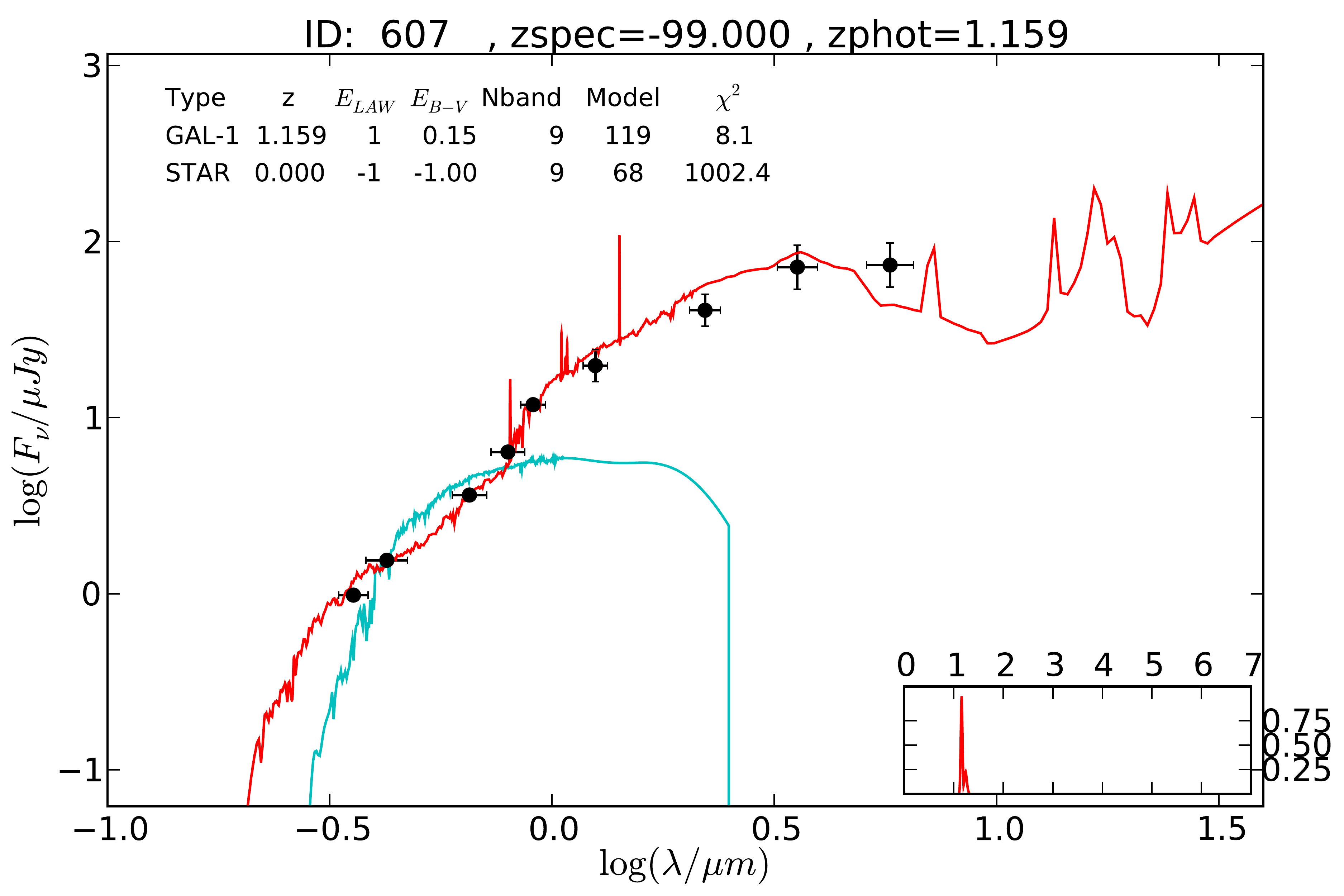}
\includegraphics[height=4.5cm]{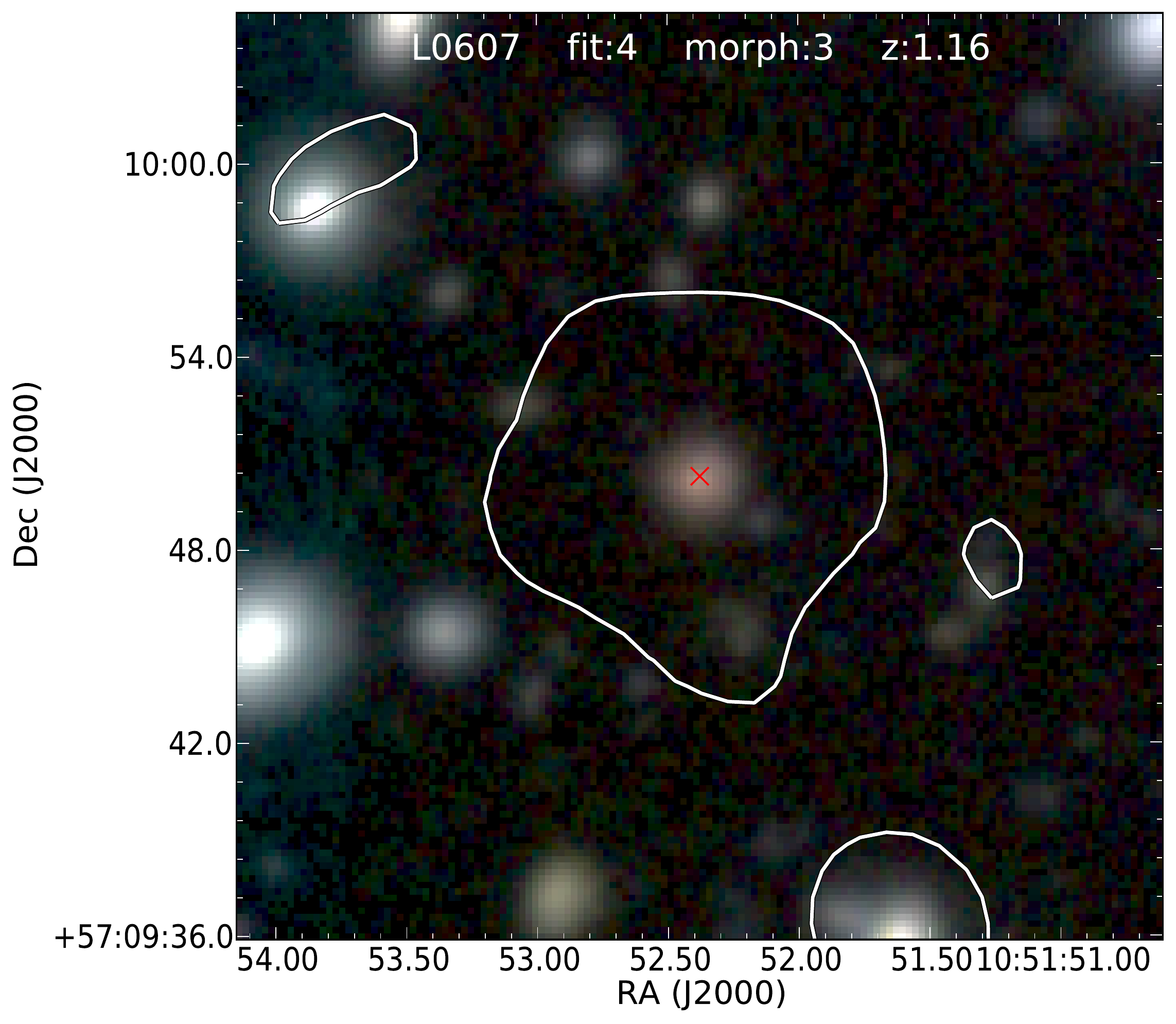}\\
\includegraphics[height=4.5cm]{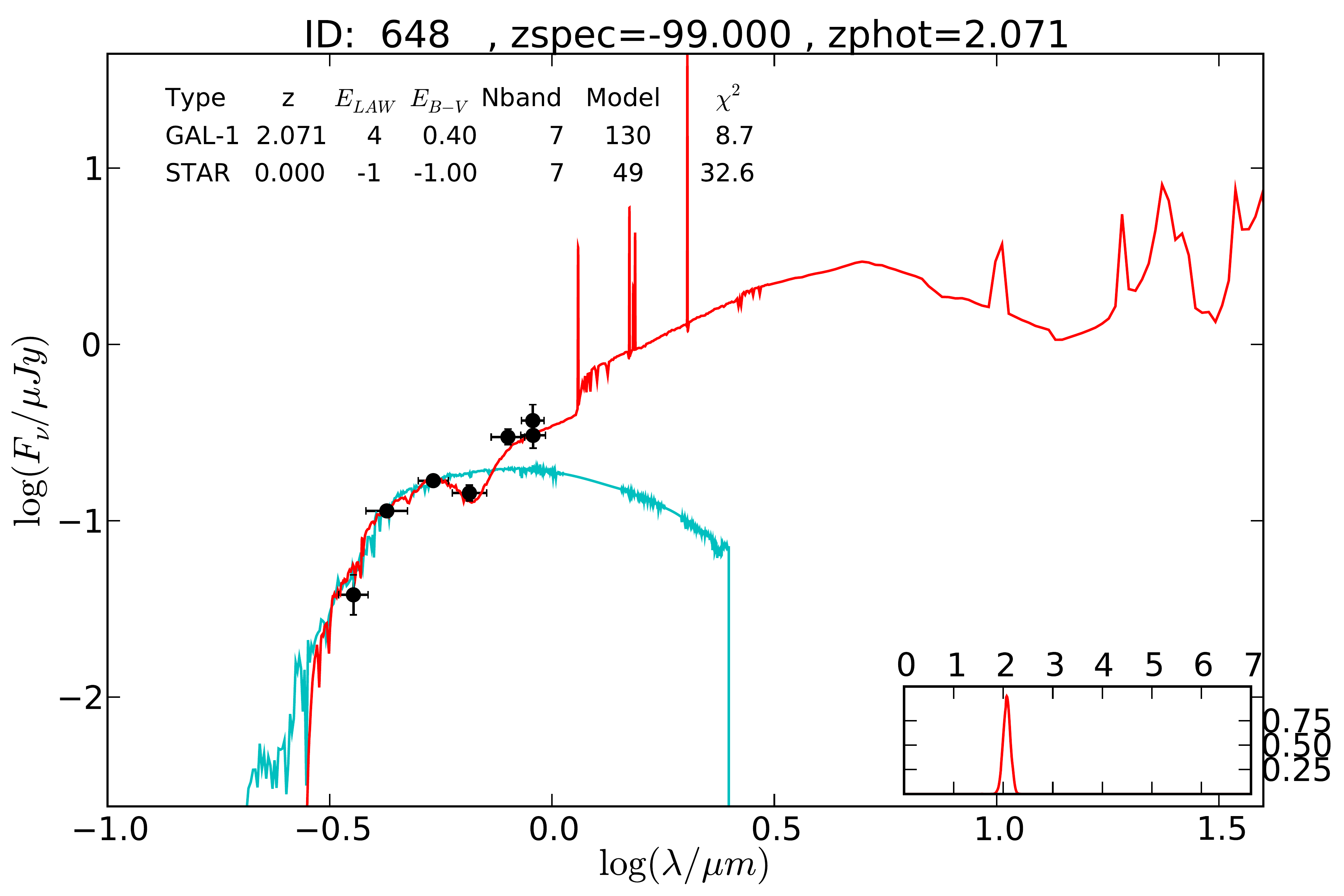}
\includegraphics[height=4.5cm]{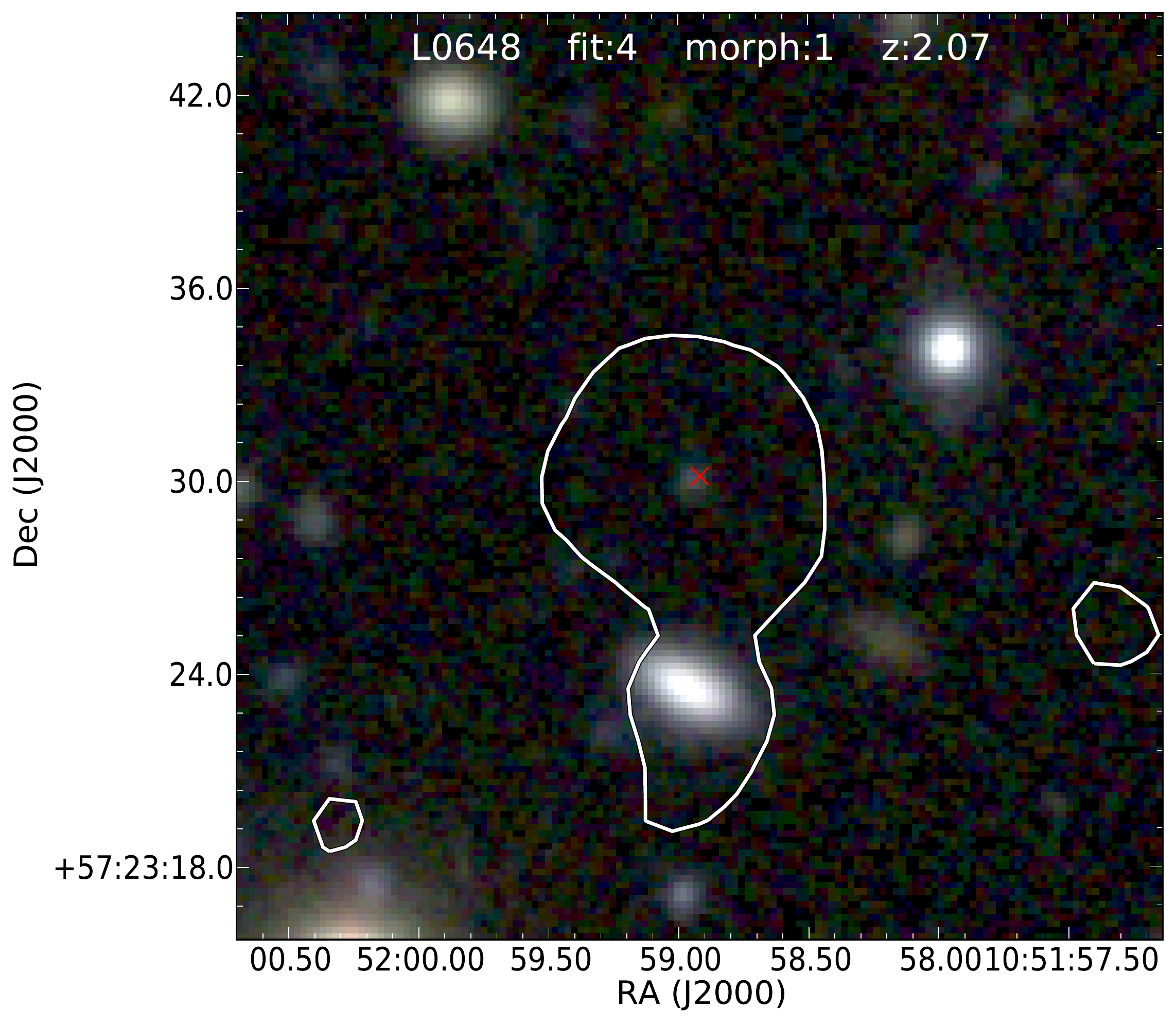}\\
\includegraphics[height=4.5cm]{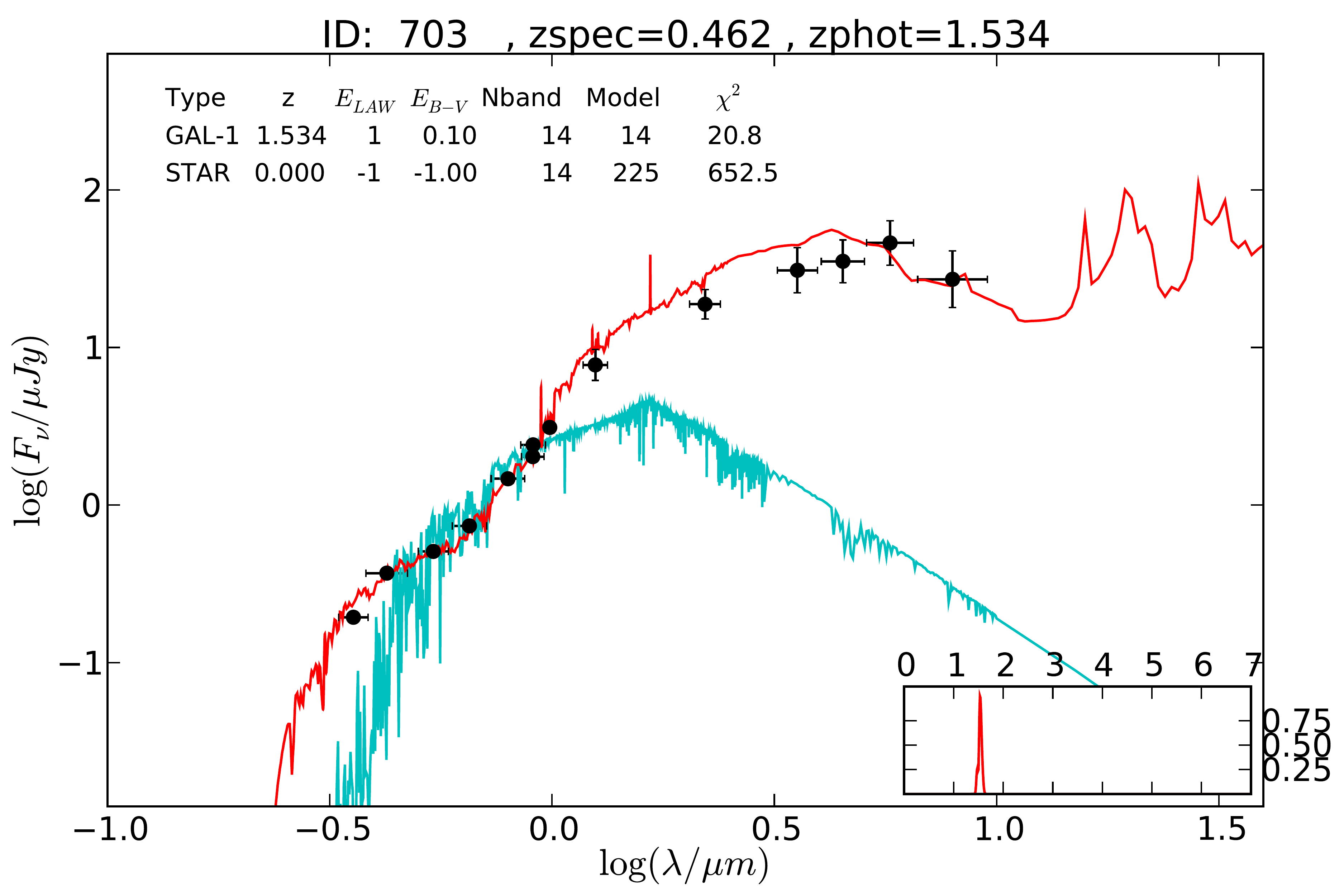}
\includegraphics[height=4.5cm]{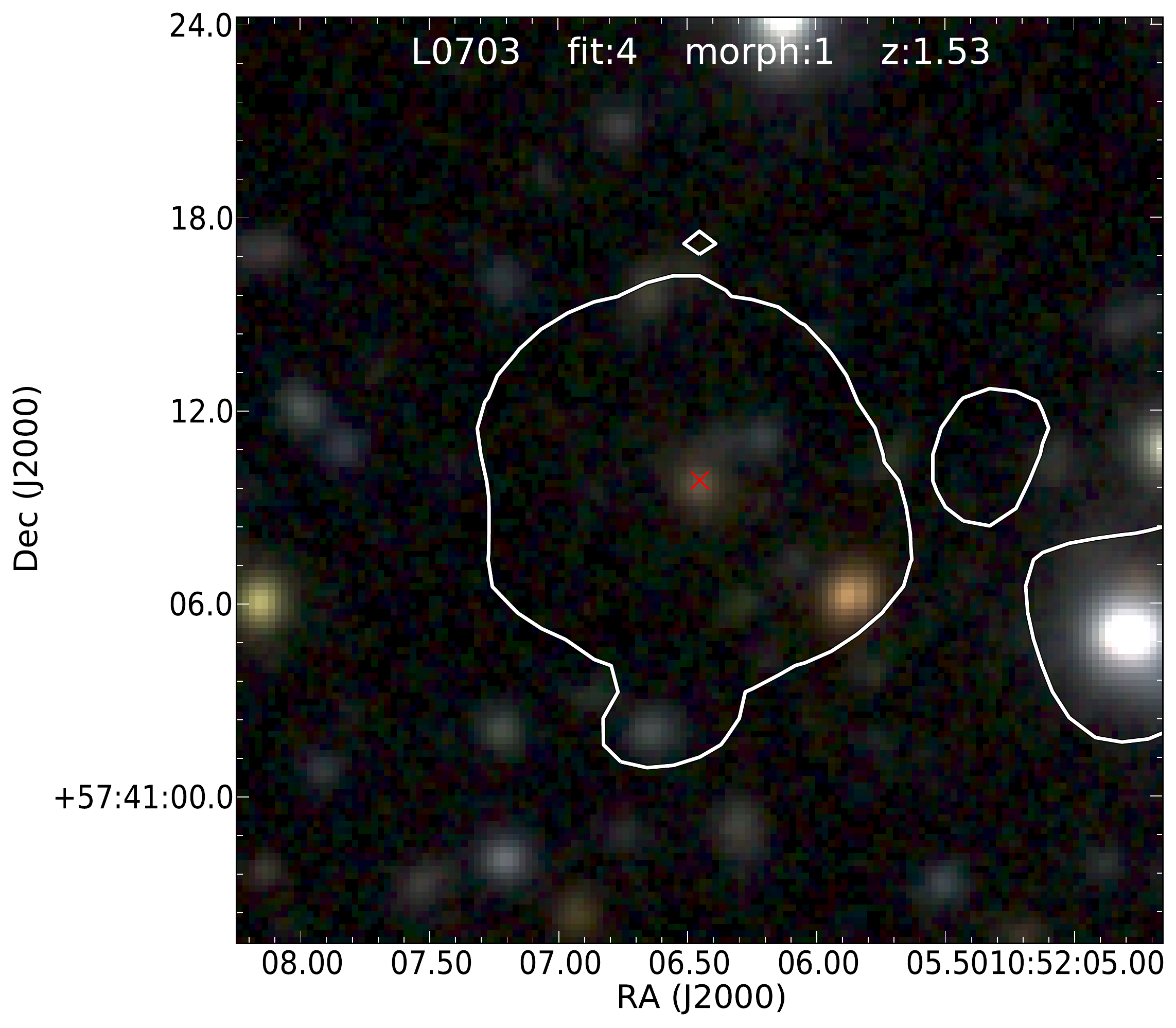}\\
\includegraphics[height=4.5cm]{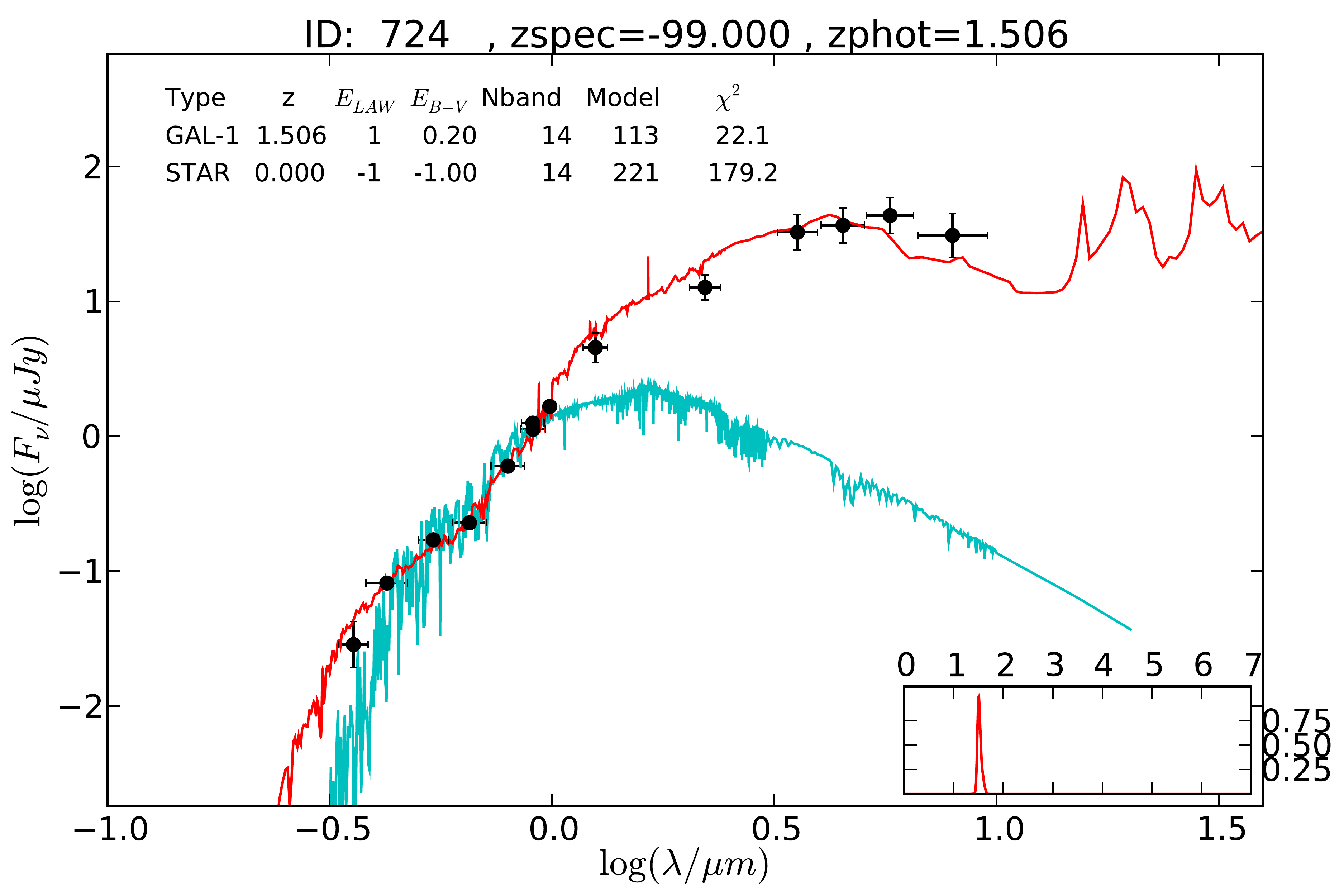}
\includegraphics[height=4.5cm]{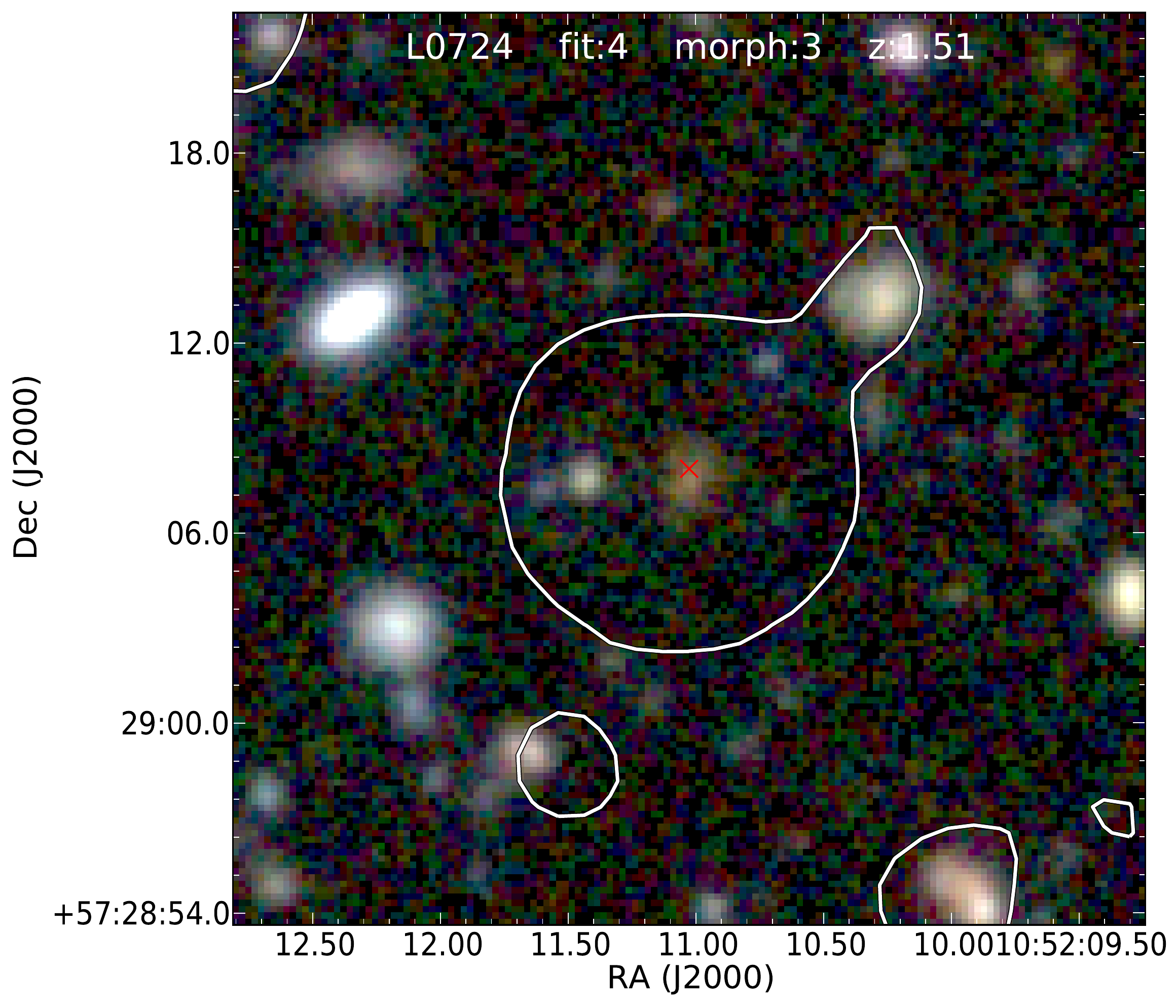}\\
\caption{(Continued)}
\end{figure*}

\begin{figure*}
\ContinuedFloat
\center
\includegraphics[height=4.5cm]{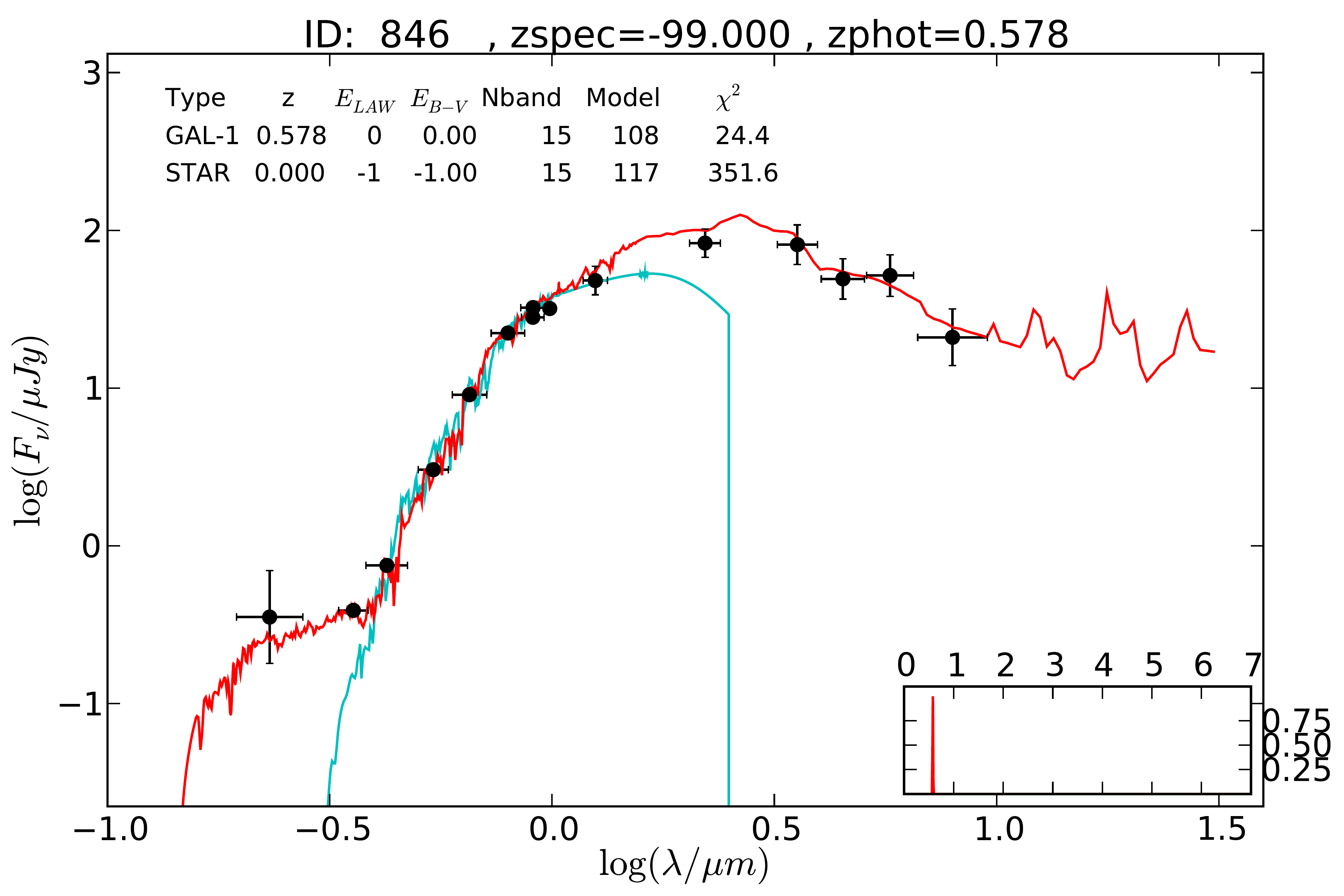}
\includegraphics[height=4.5cm]{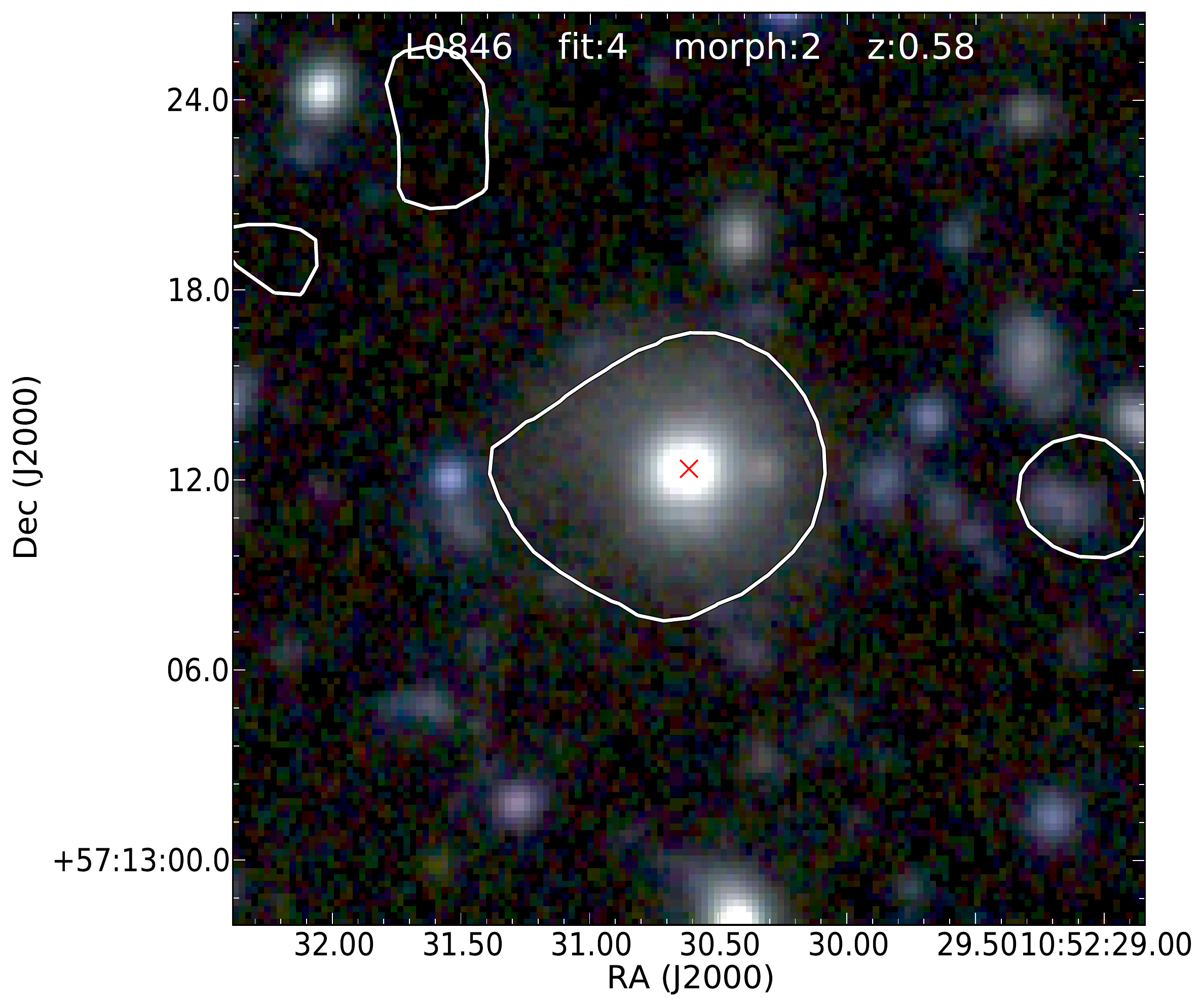}\\
\includegraphics[height=4.5cm]{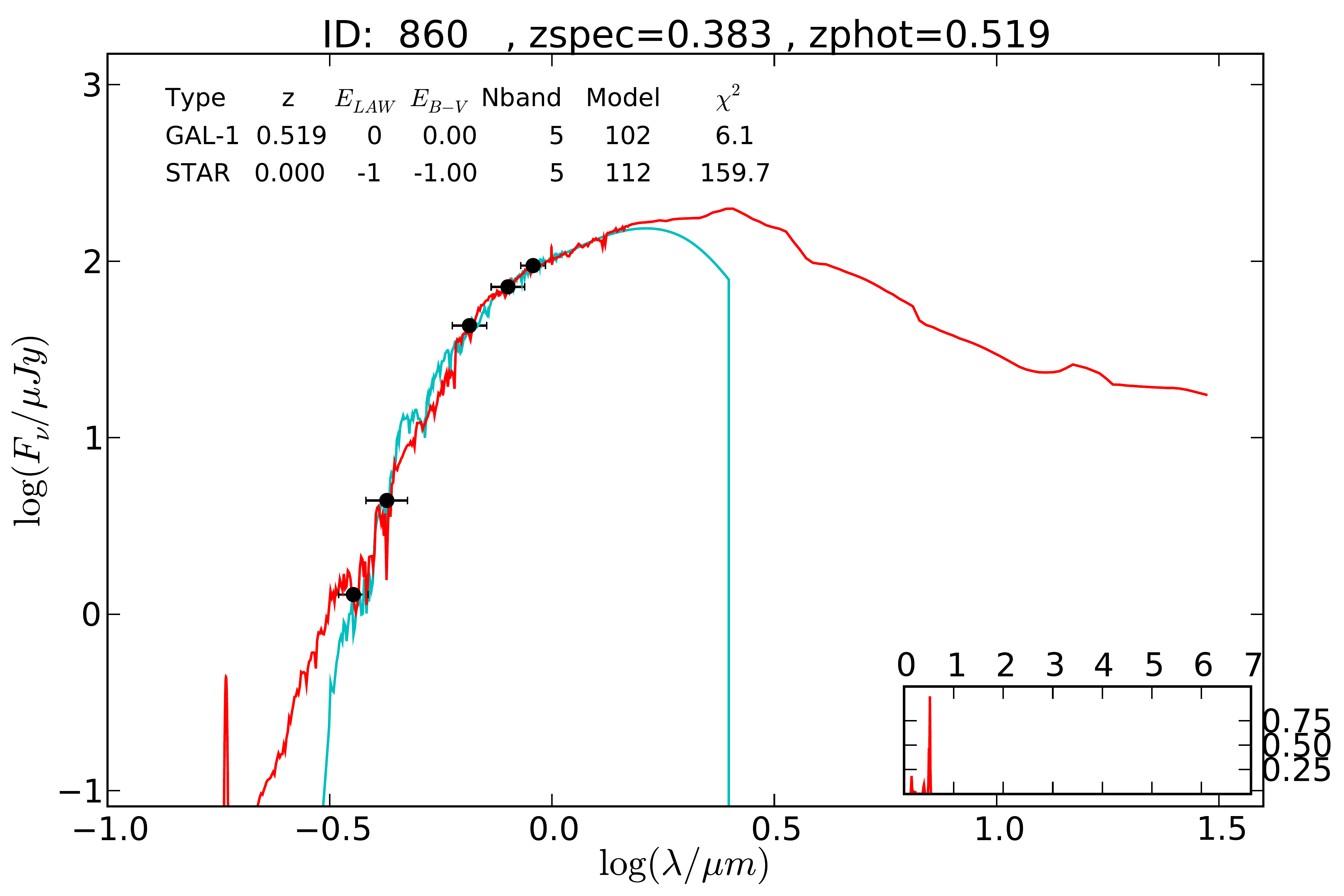}
\includegraphics[height=4.5cm]{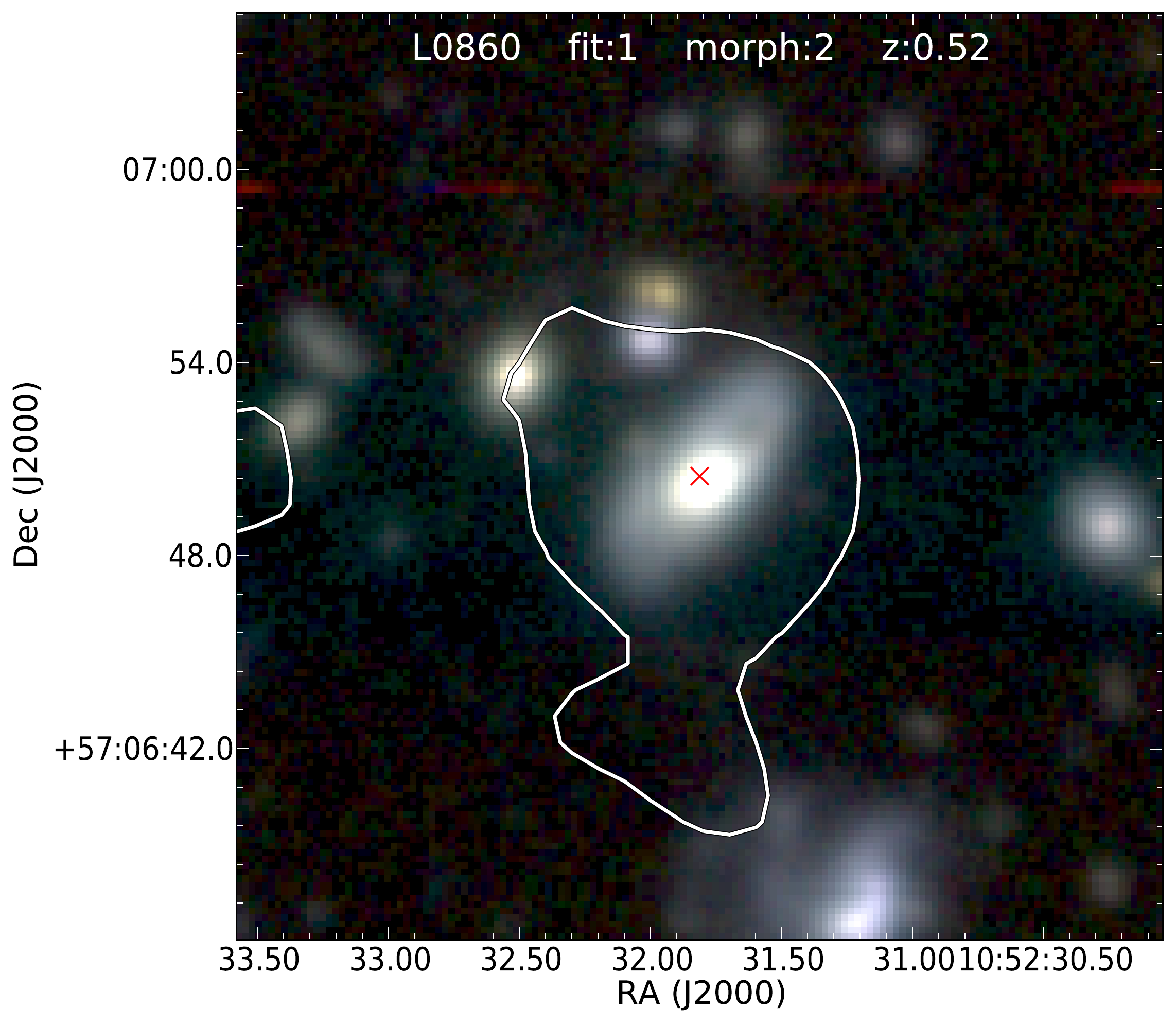}\\
\includegraphics[height=4.5cm]{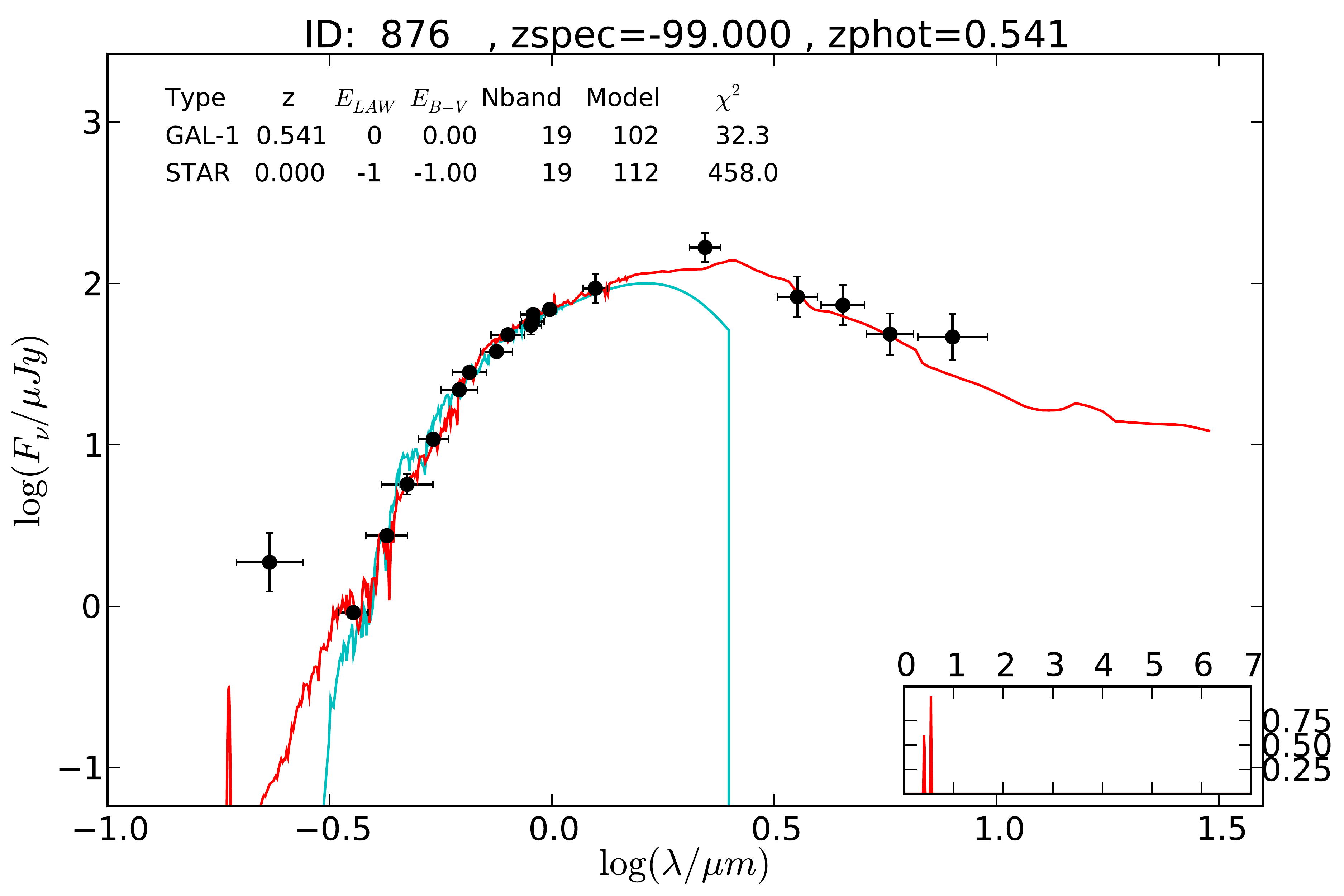}
\includegraphics[height=4.5cm]{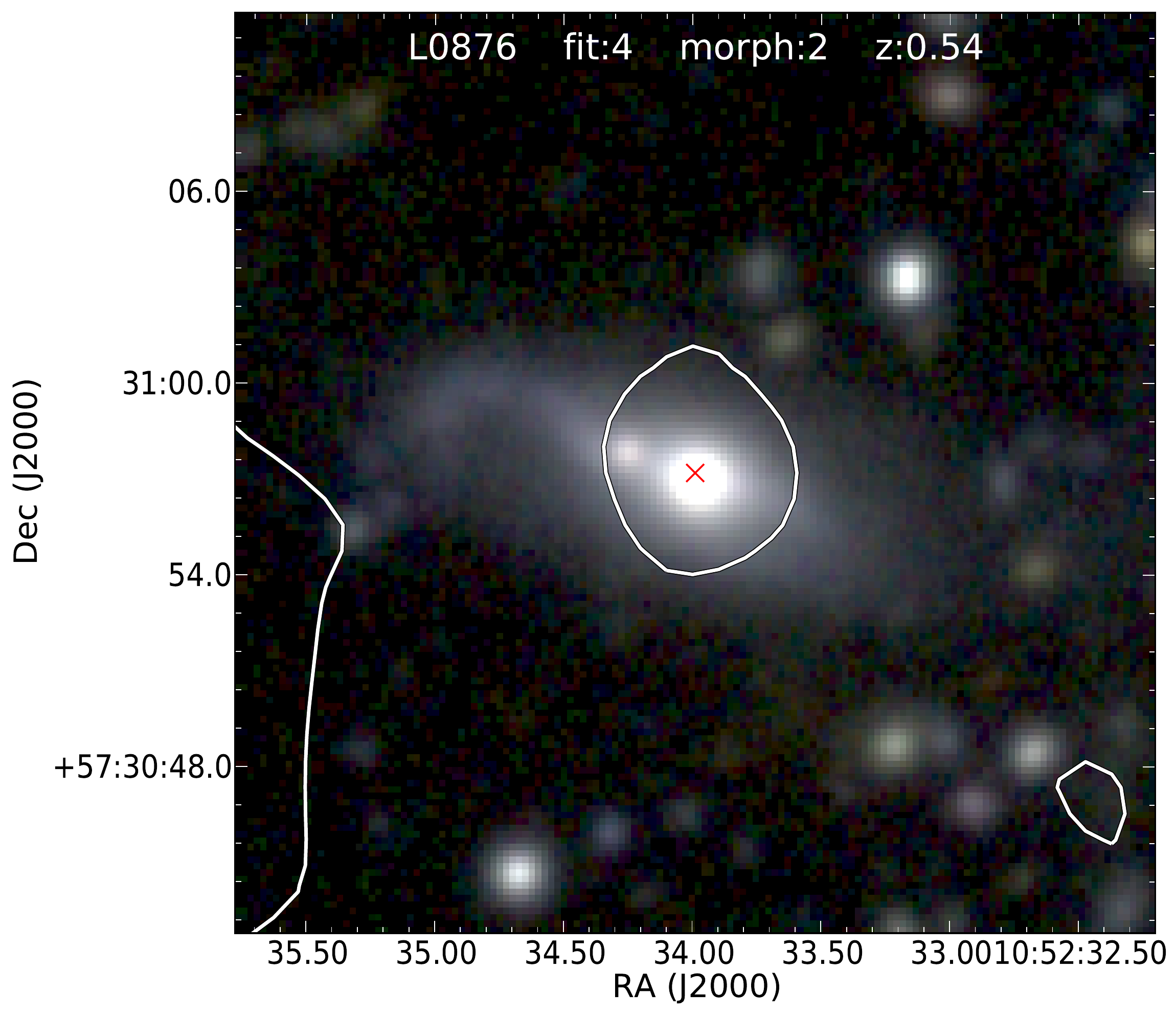}\\
\includegraphics[height=4.5cm]{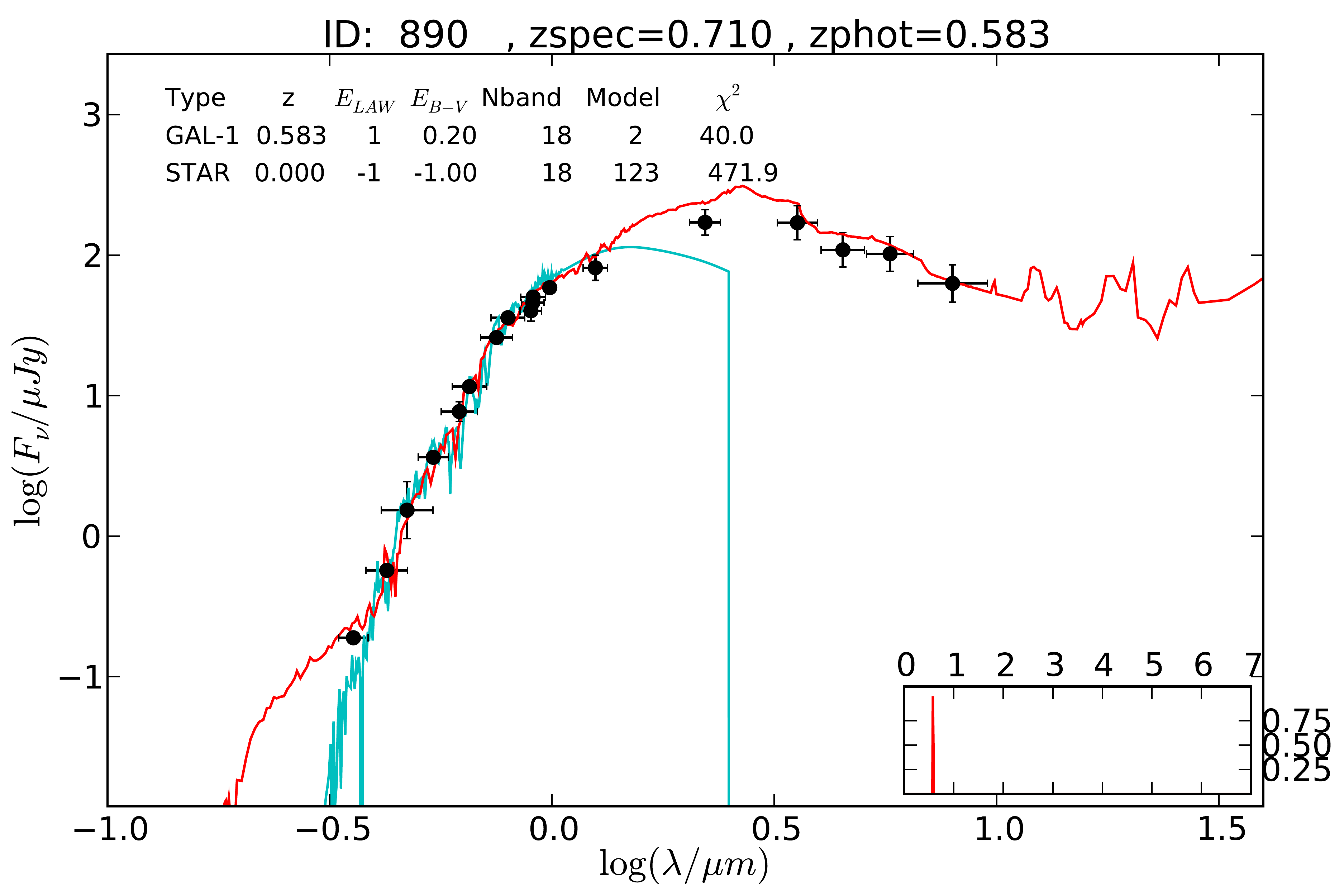}
\includegraphics[height=4.5cm]{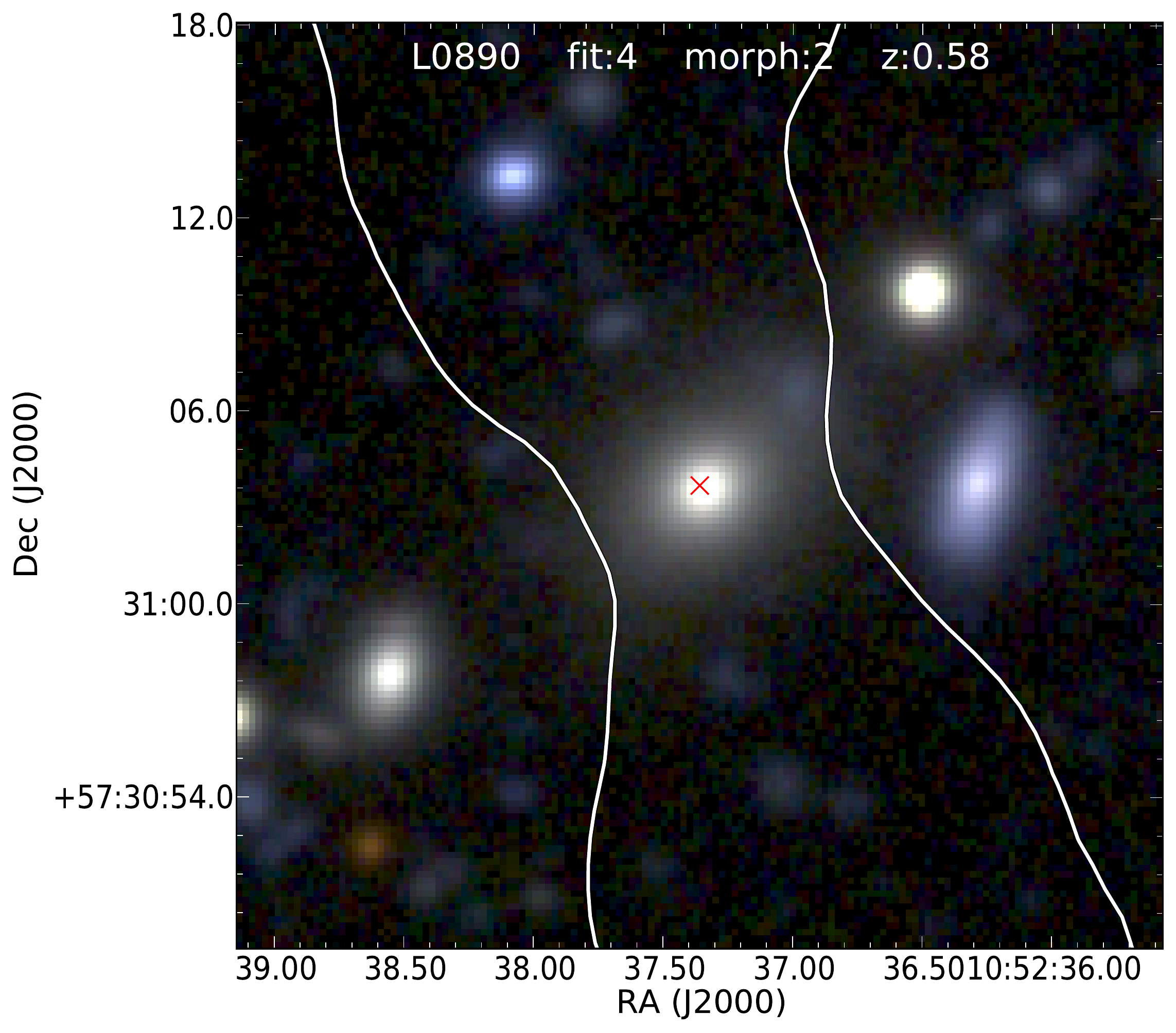}\\
\includegraphics[height=4.5cm]{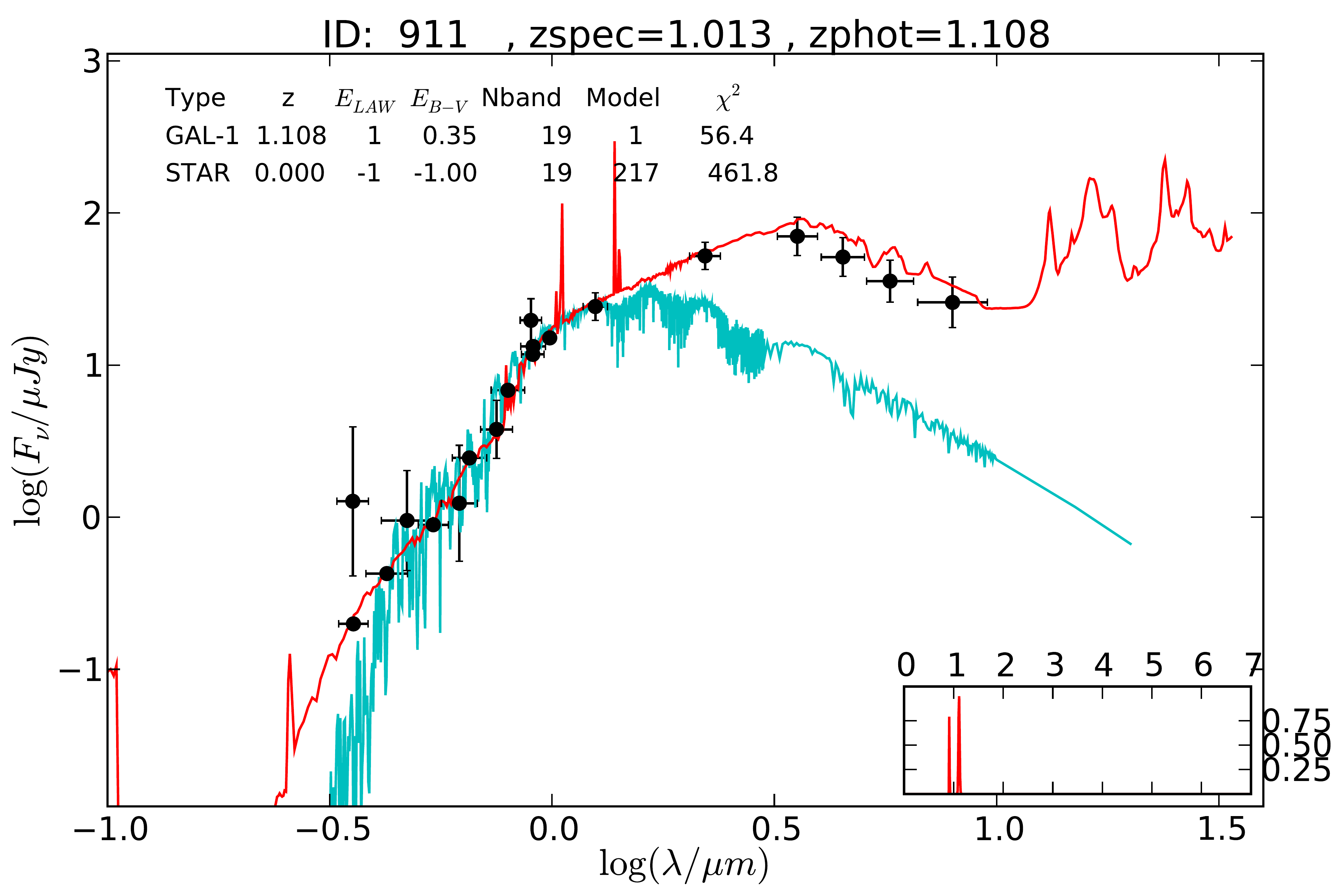}
\includegraphics[height=4.5cm]{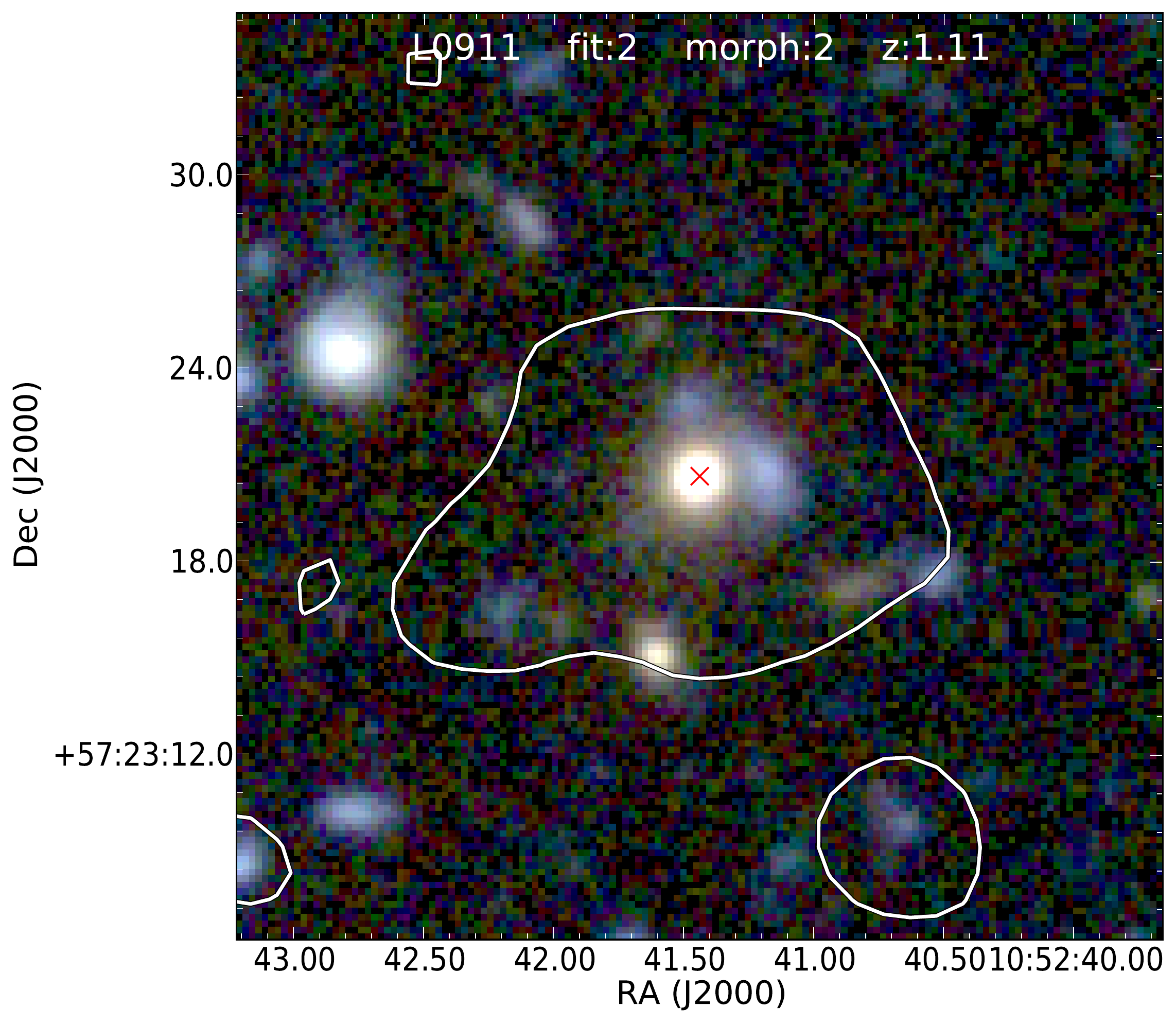}\\
\caption{(Continued)}
\end{figure*}

\begin{figure*}
\ContinuedFloat
\center
\includegraphics[height=4.5cm]{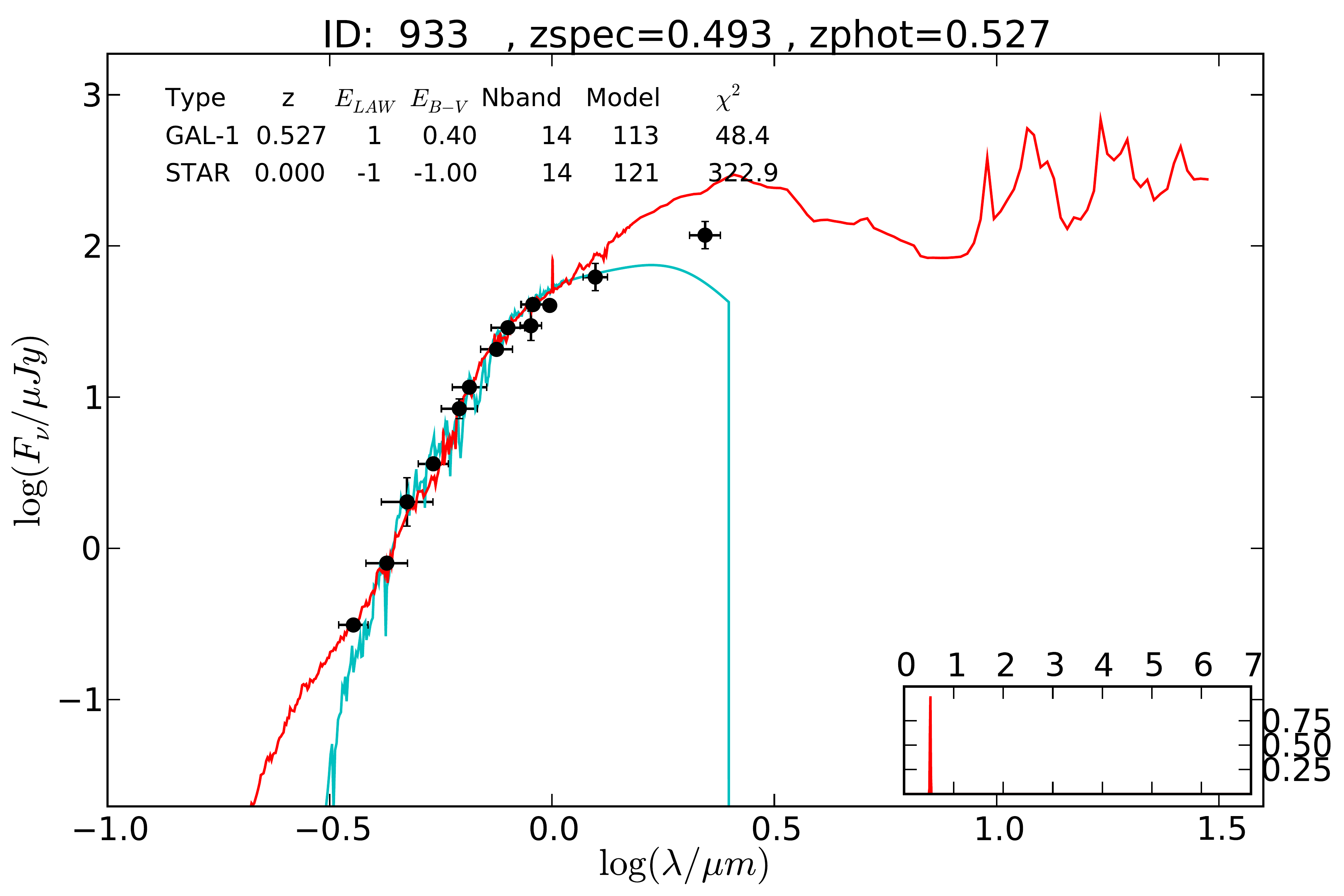}
\includegraphics[height=4.5cm]{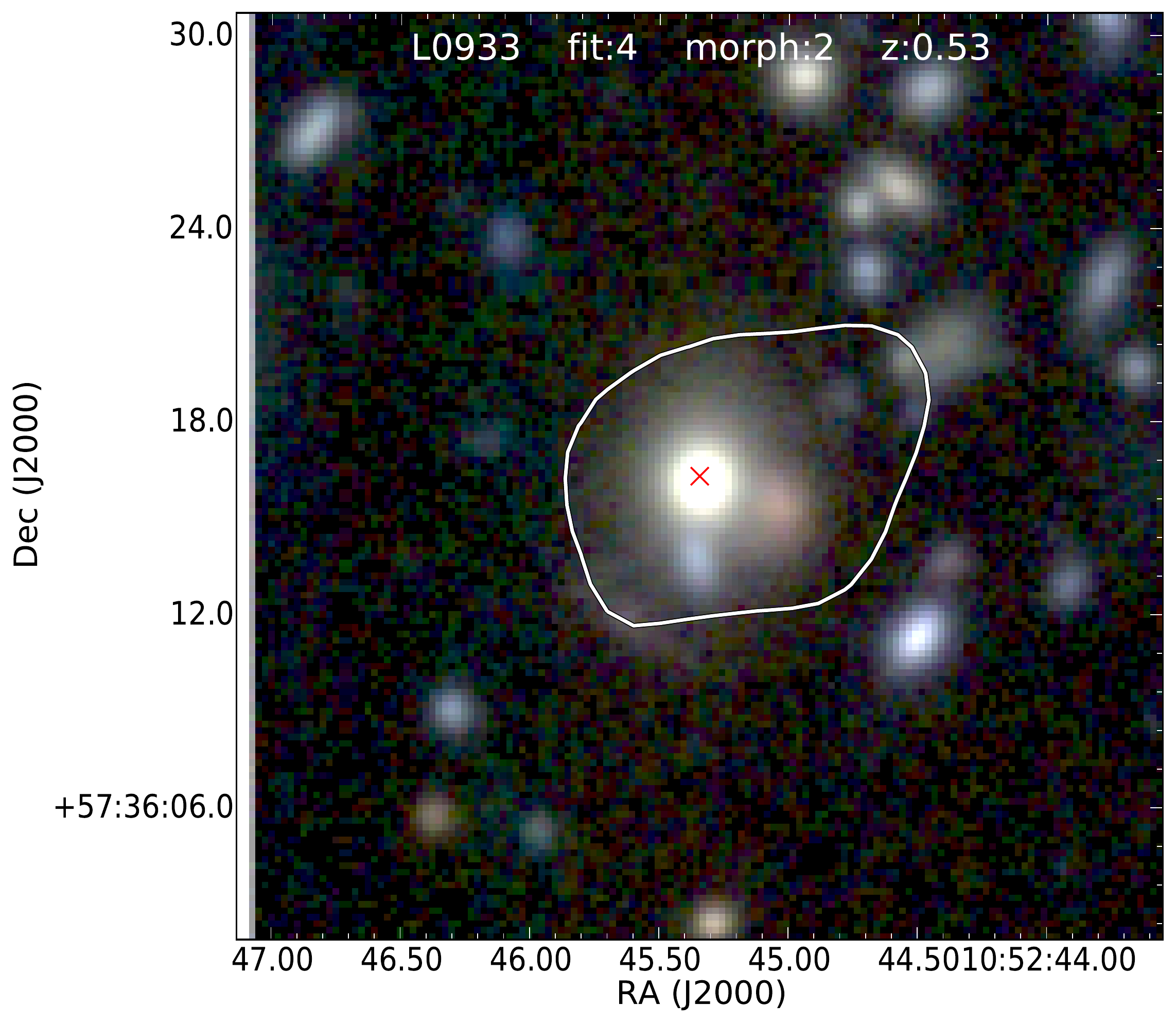}\\
\includegraphics[height=4.5cm]{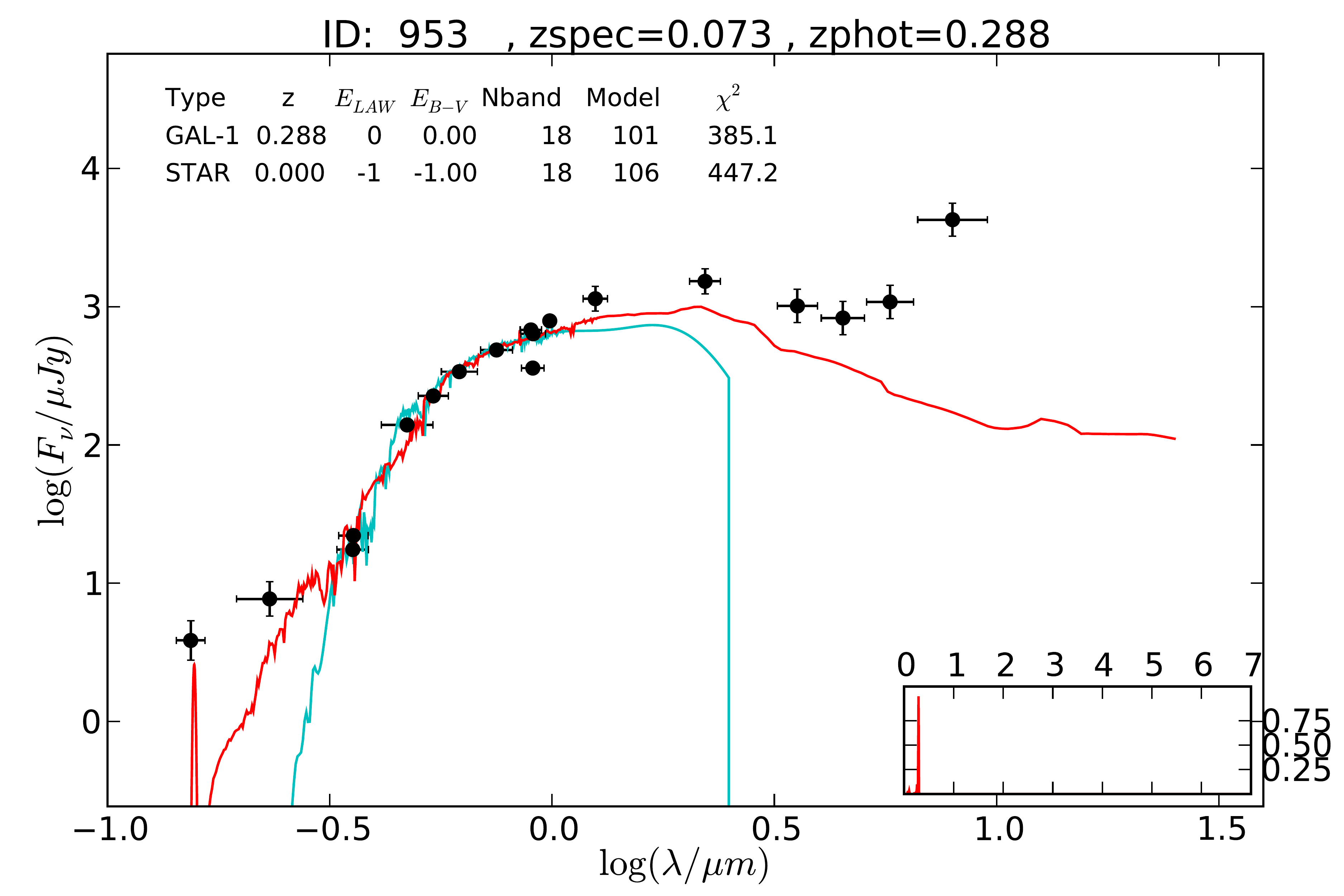}
\includegraphics[height=4.5cm]{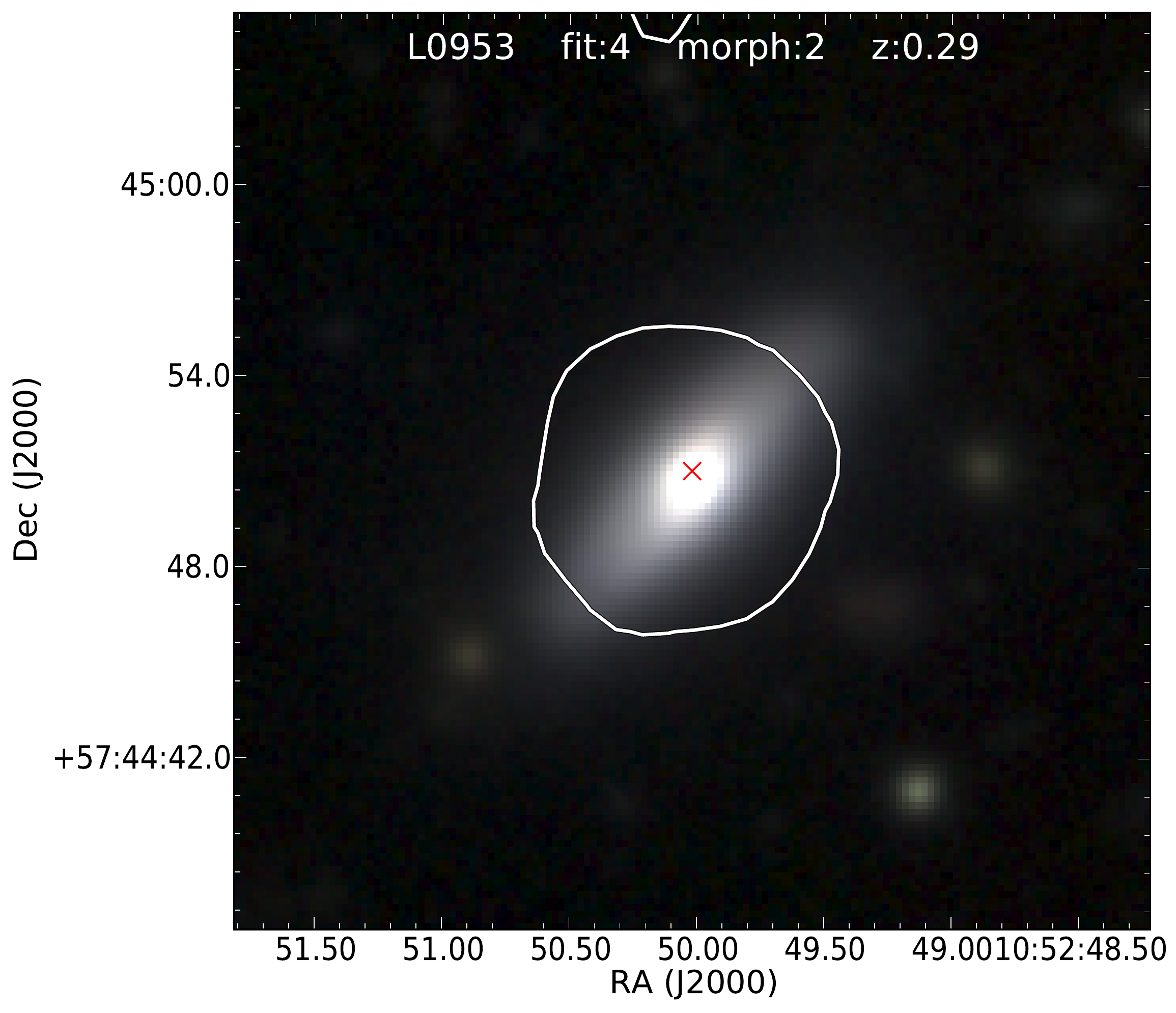}\\
\includegraphics[height=4.5cm]{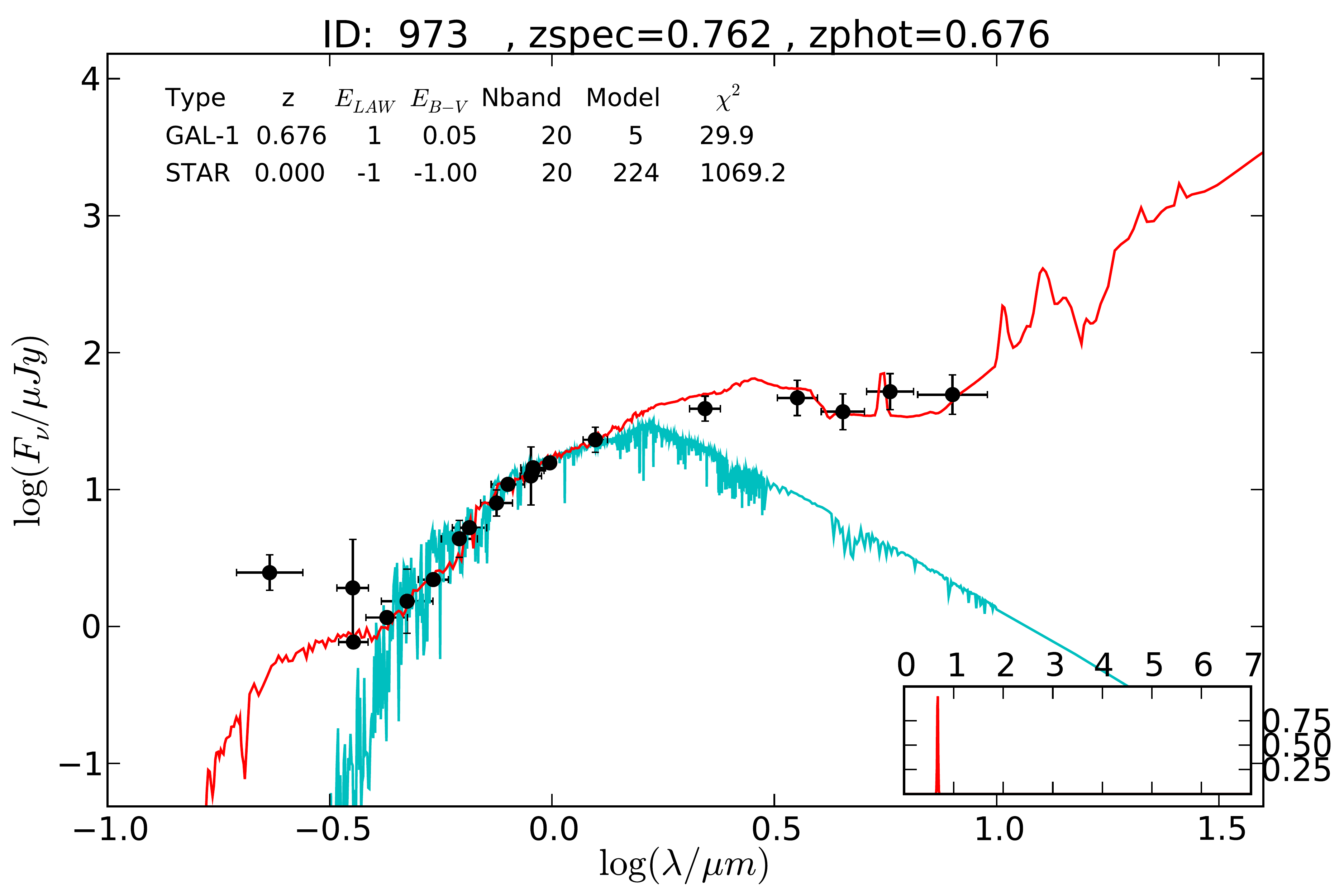}
\includegraphics[height=4.5cm]{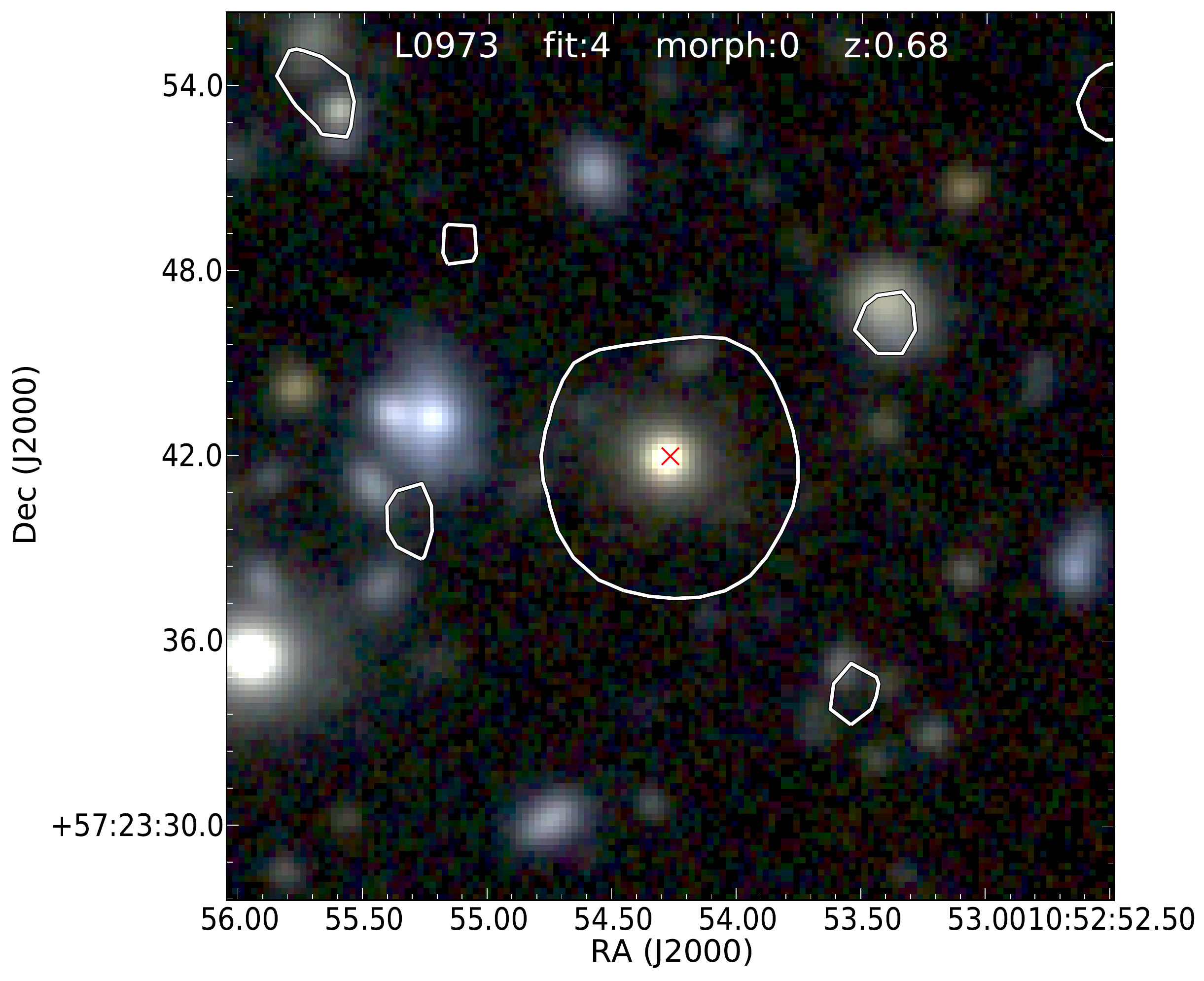}\\
\includegraphics[height=4.5cm]{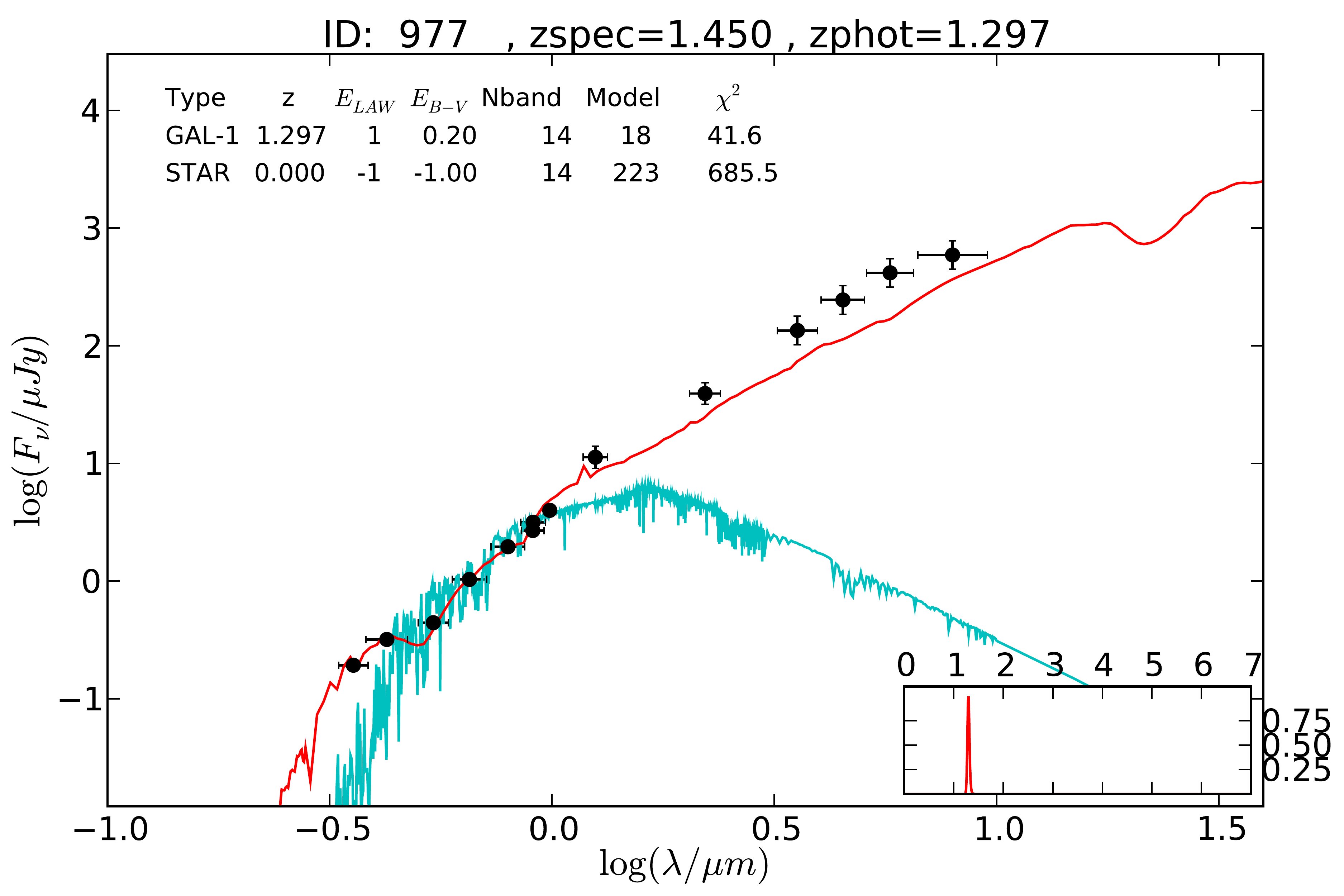}
\includegraphics[height=4.5cm]{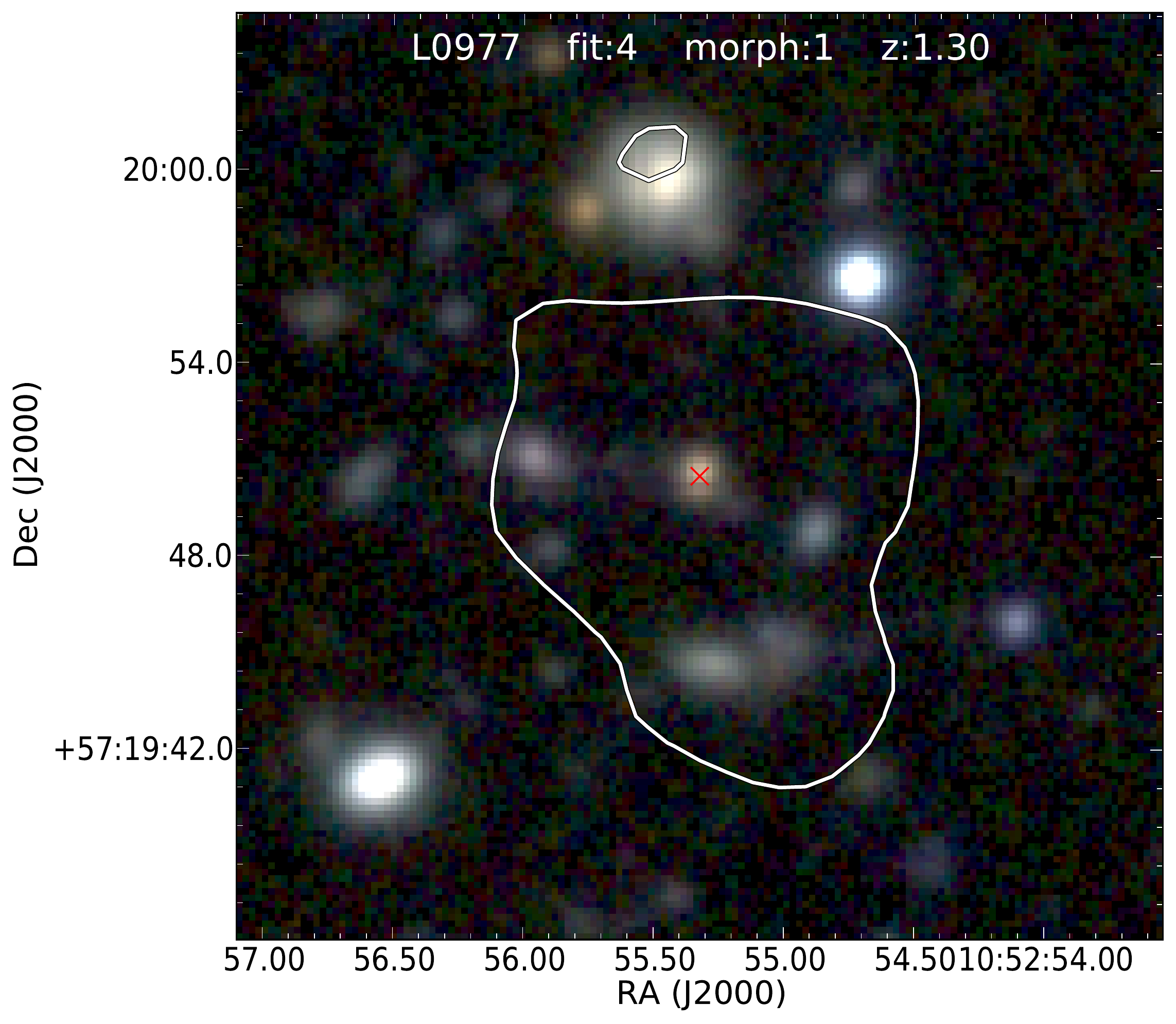}\\
\includegraphics[height=4.5cm]{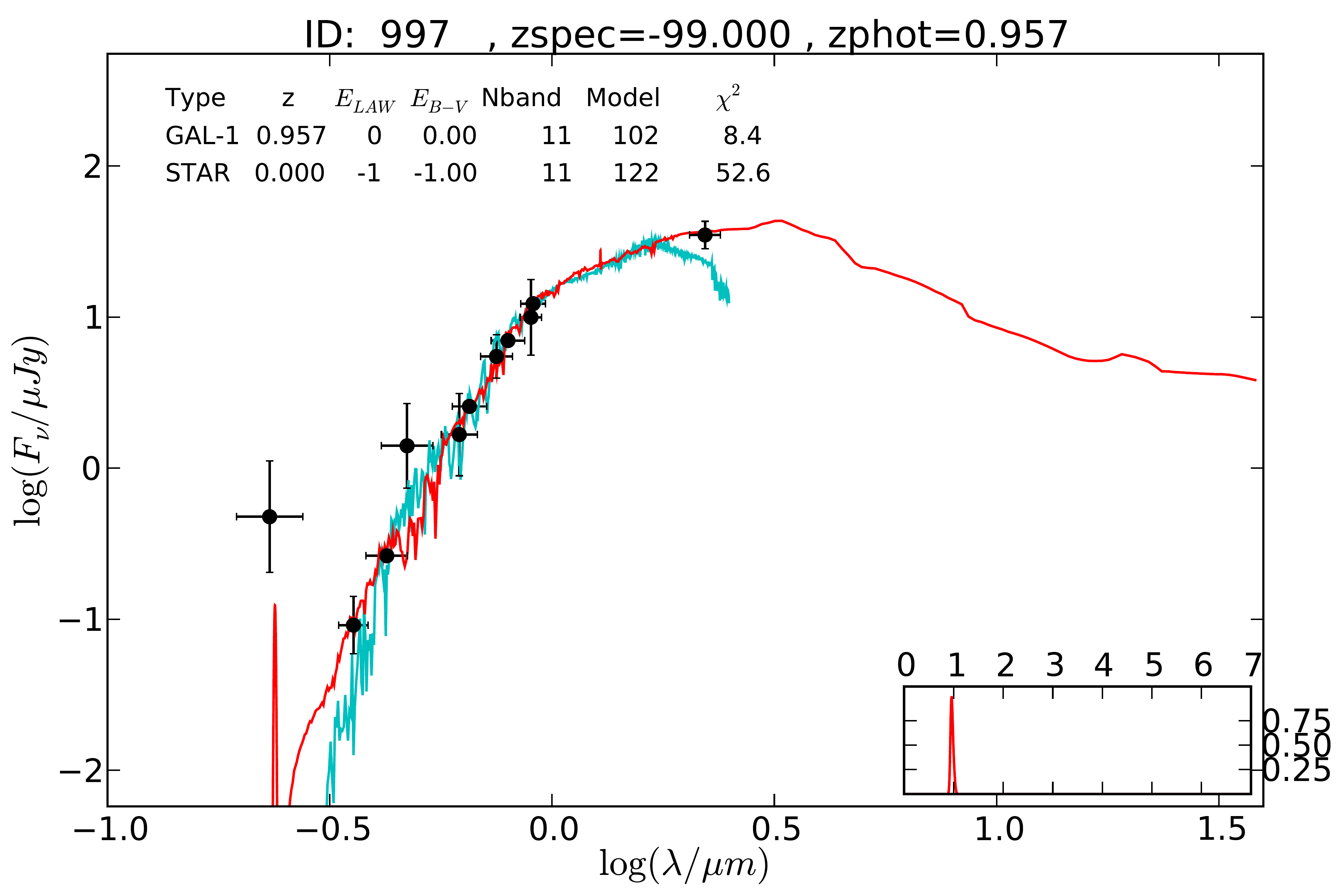}
\includegraphics[height=4.5cm]{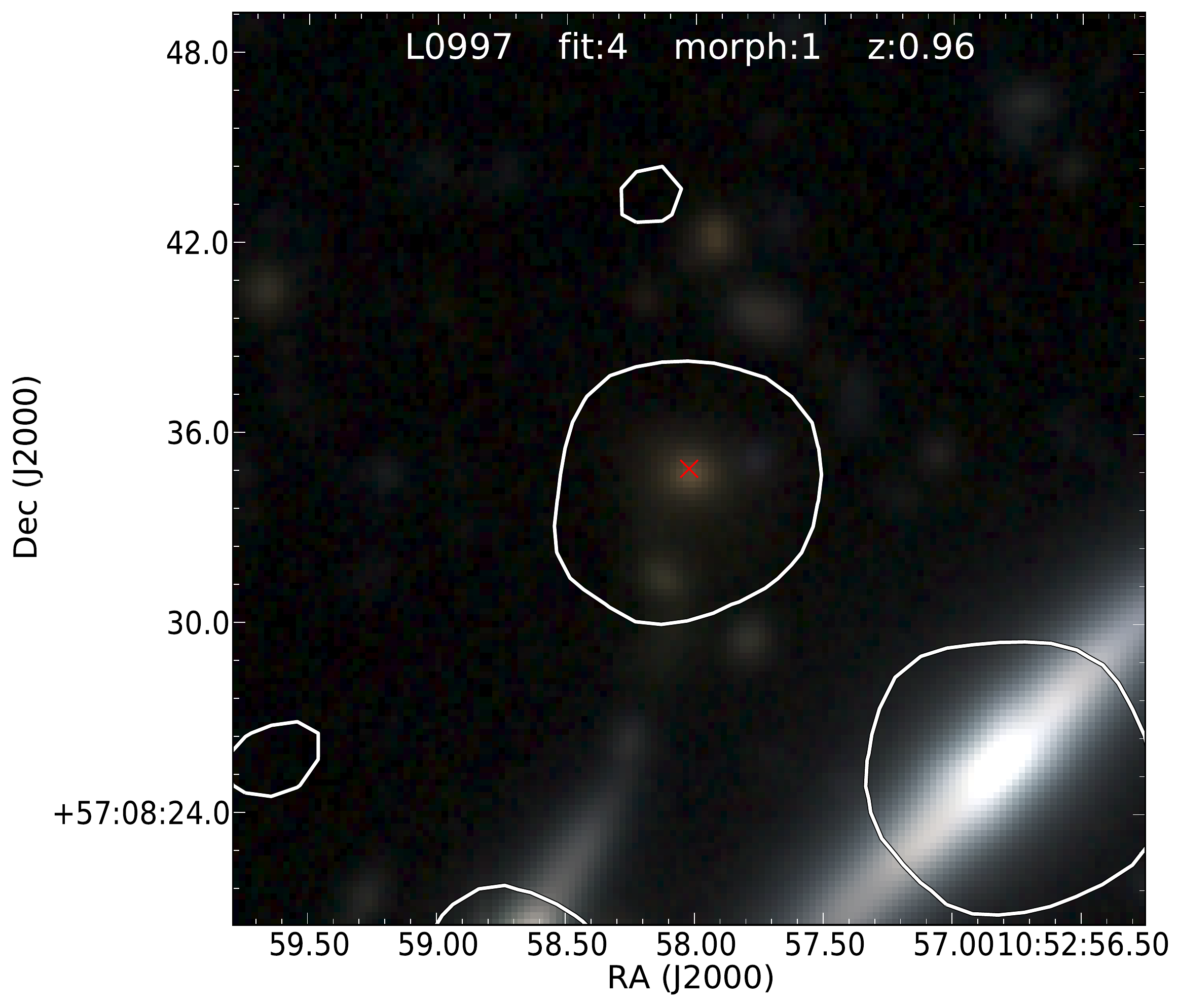}\\
\caption{(Continued)}
\end{figure*}

\begin{figure*}
\ContinuedFloat
\center
\includegraphics[height=4.5cm]{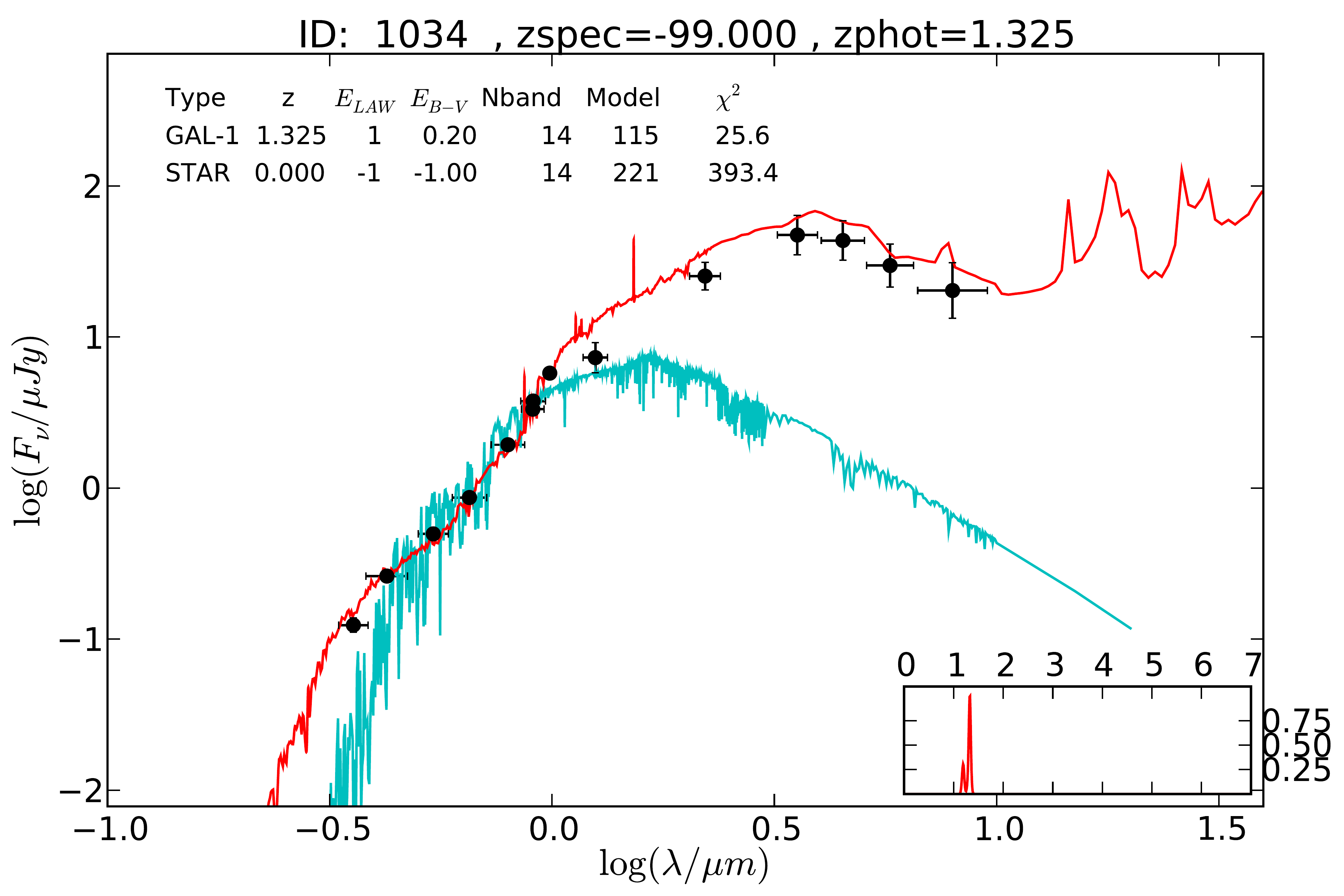}
\includegraphics[height=4.5cm]{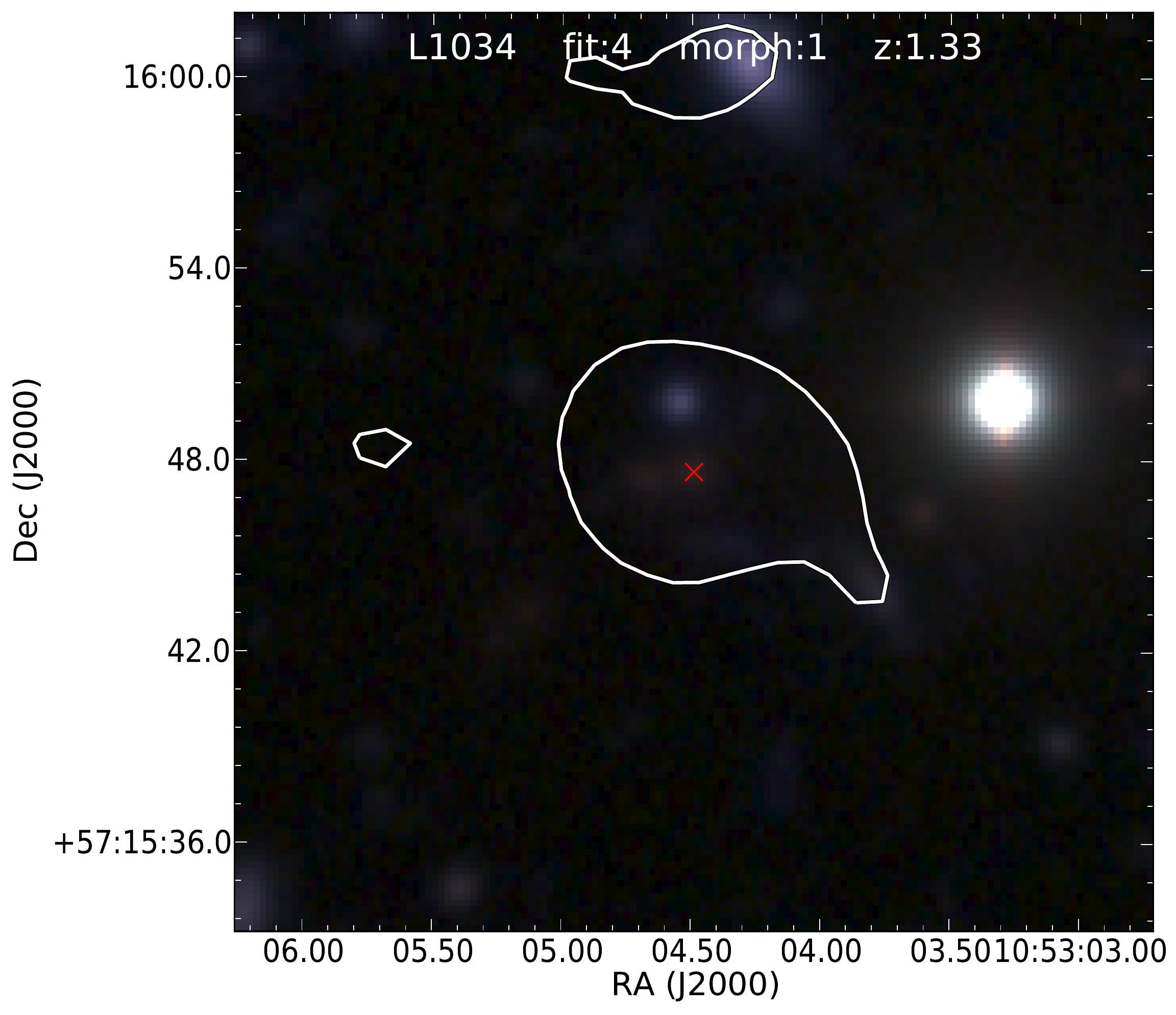}\\
\includegraphics[height=4.5cm]{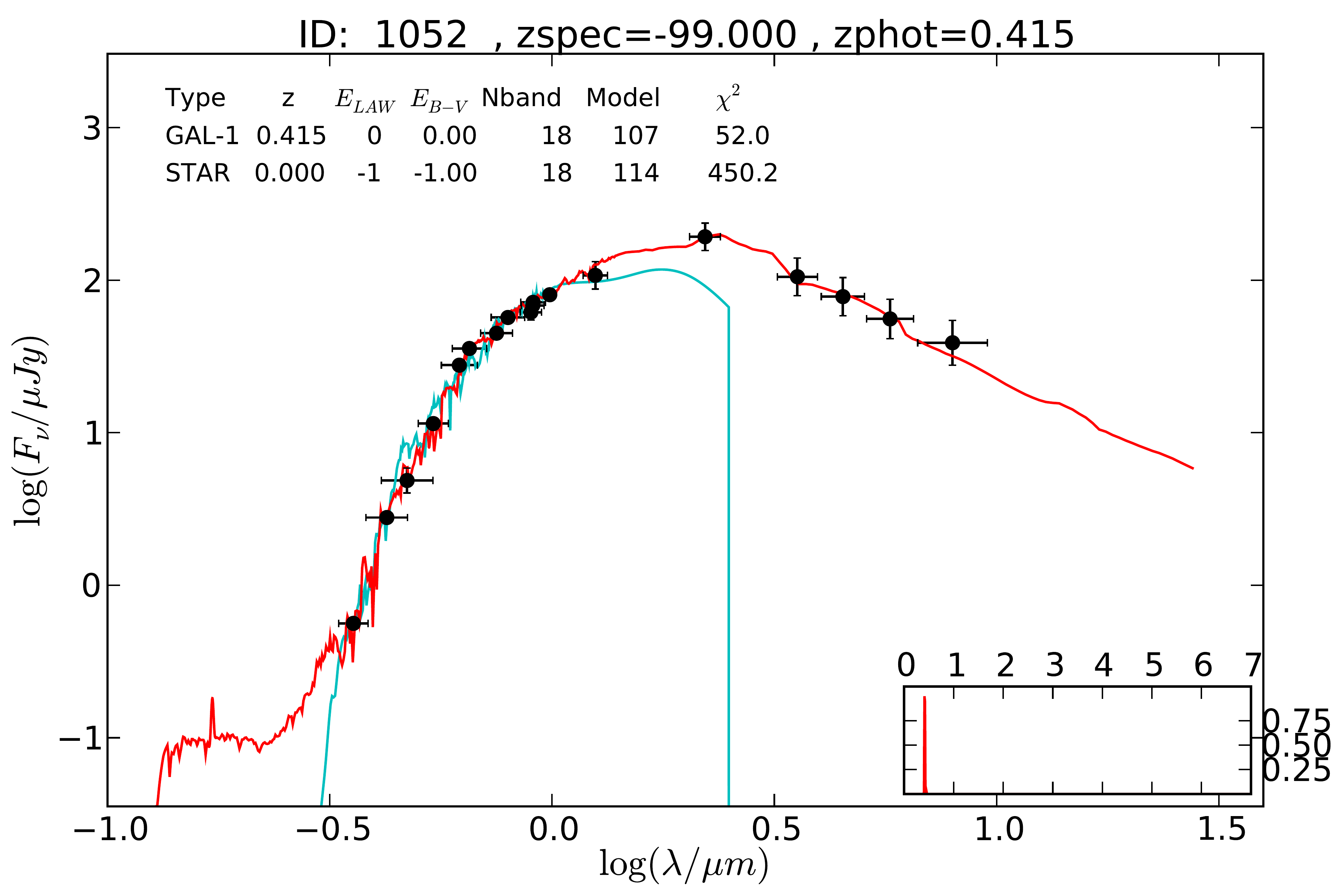}
\includegraphics[height=4.5cm]{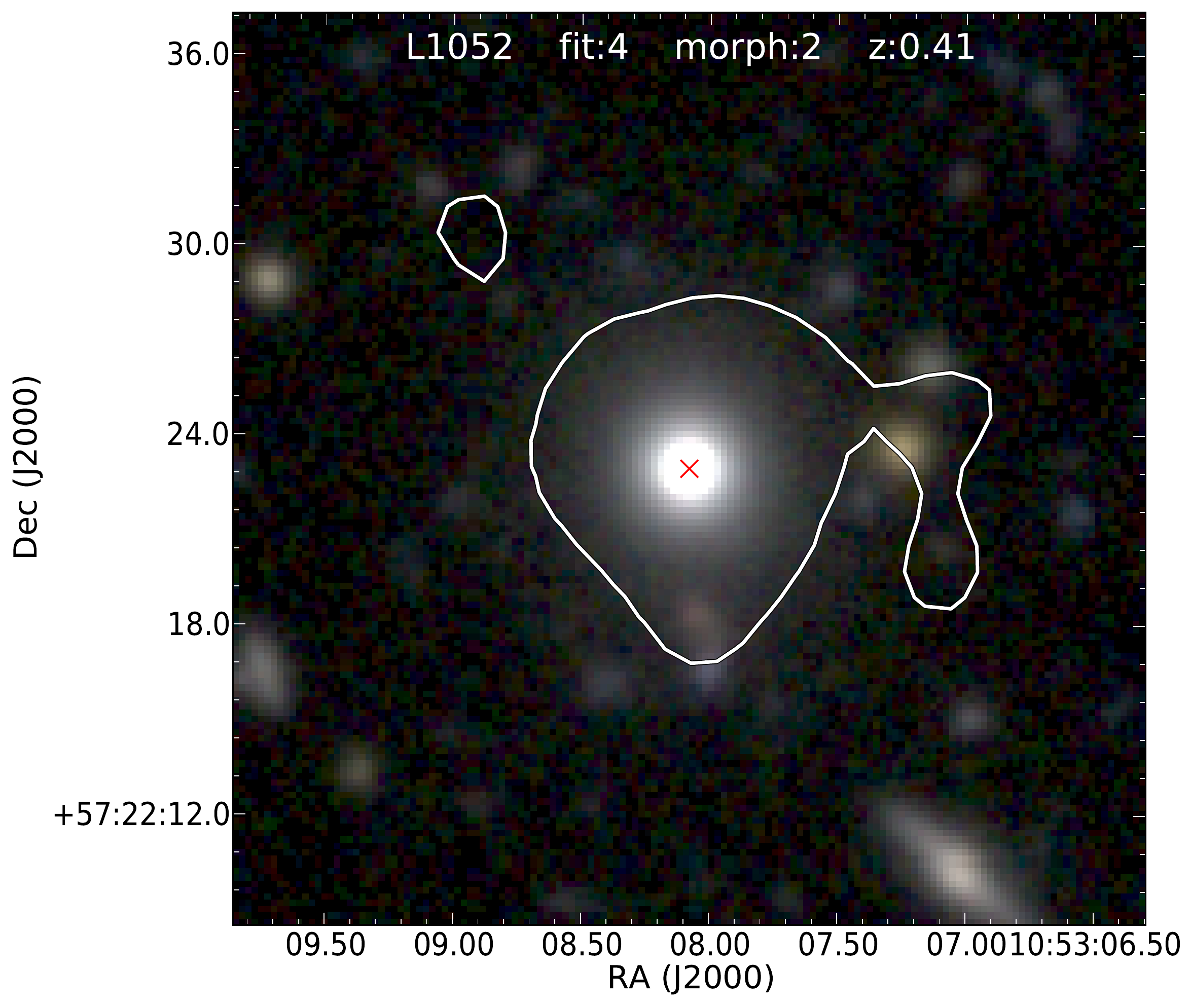}\\
\includegraphics[height=4.5cm]{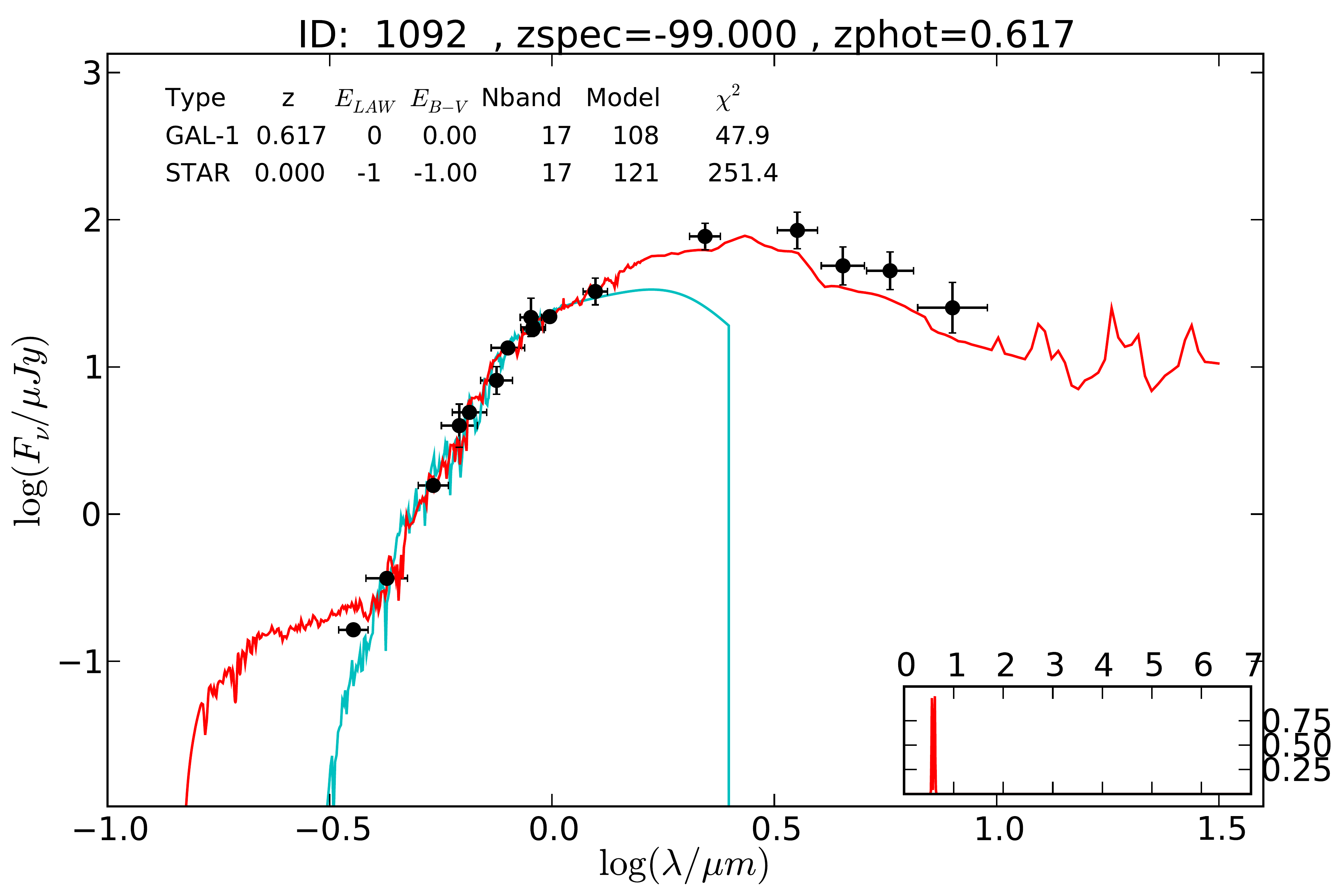}
\includegraphics[height=4.5cm]{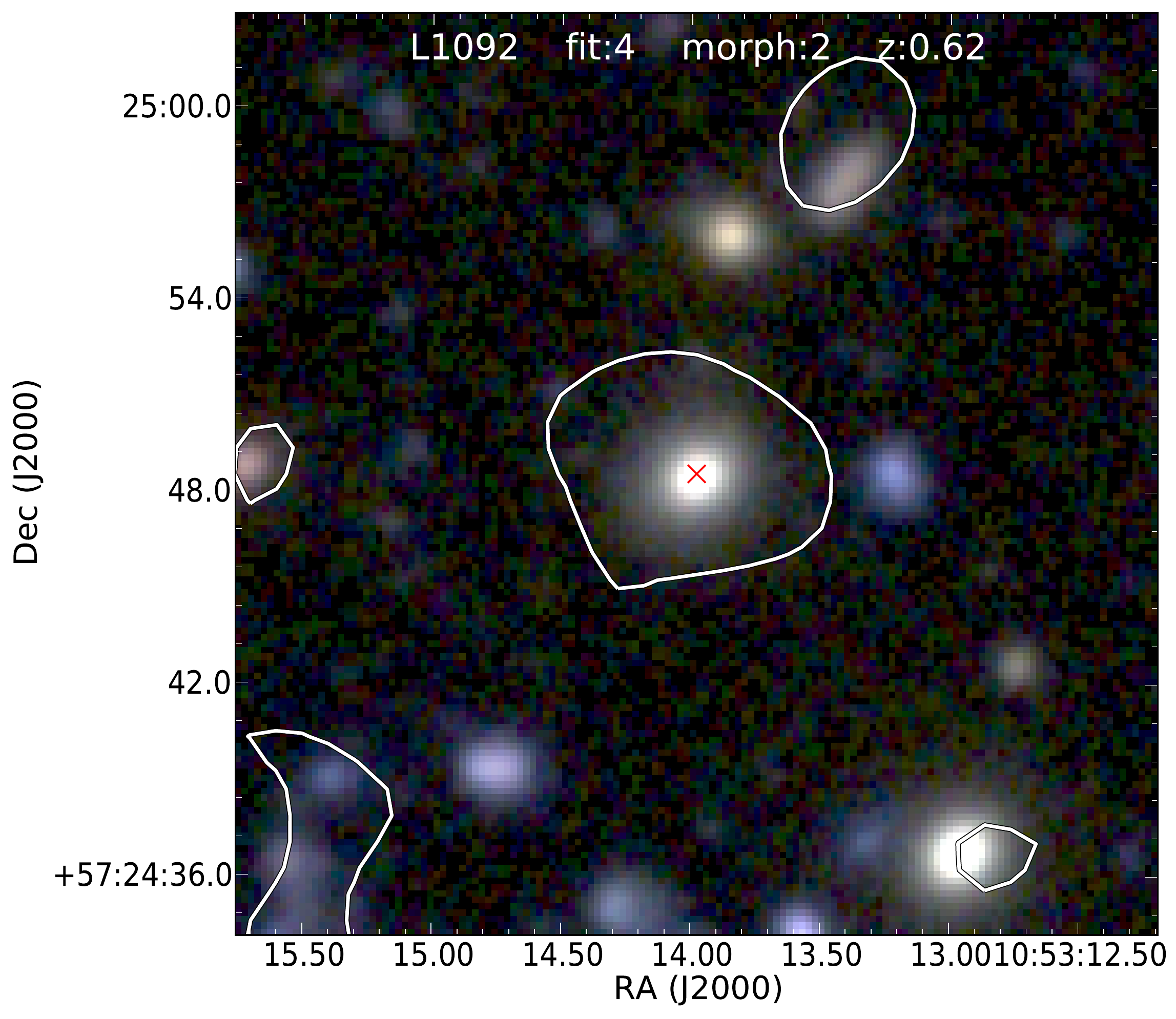}\\
\includegraphics[height=4.5cm]{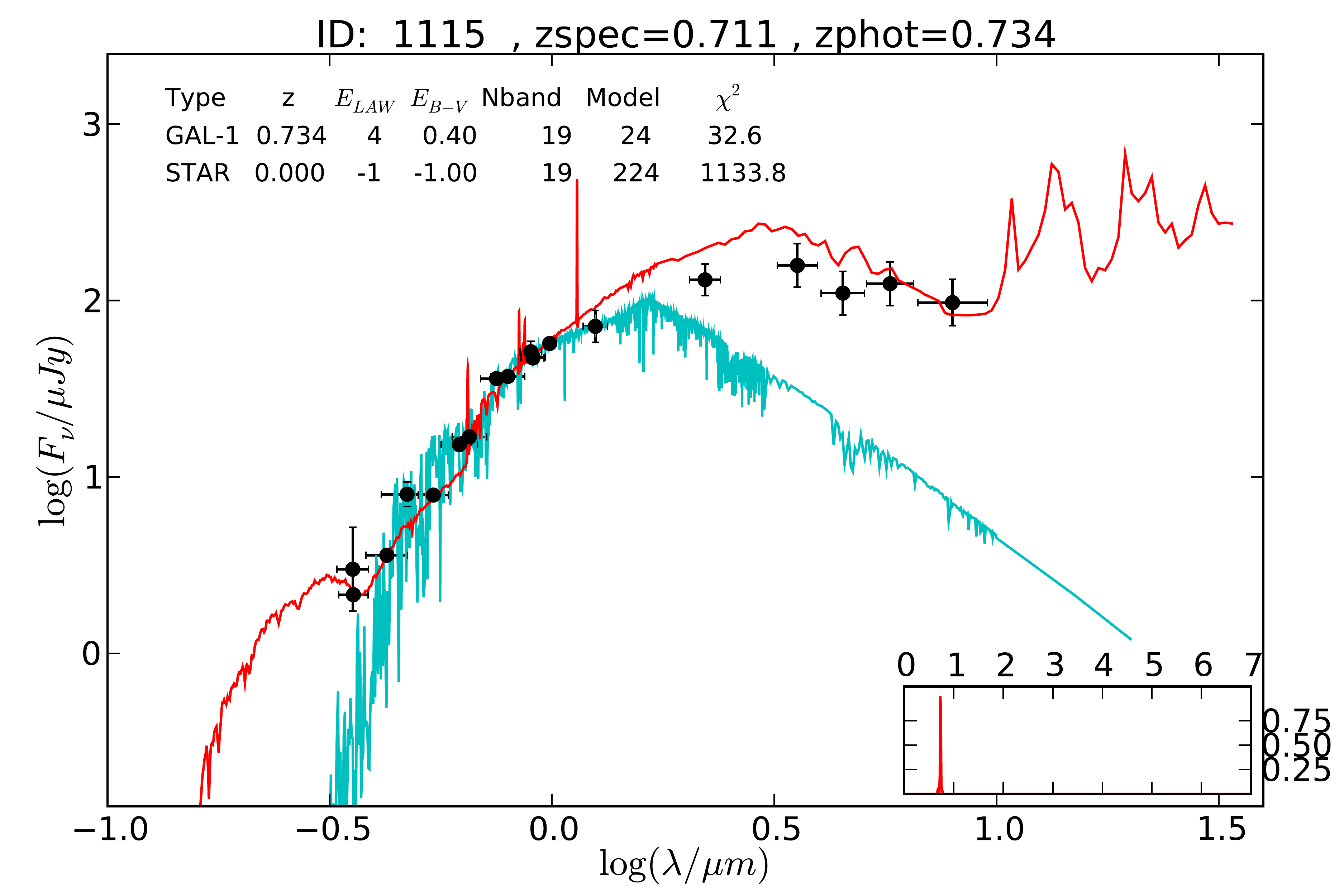}
\includegraphics[height=4.5cm]{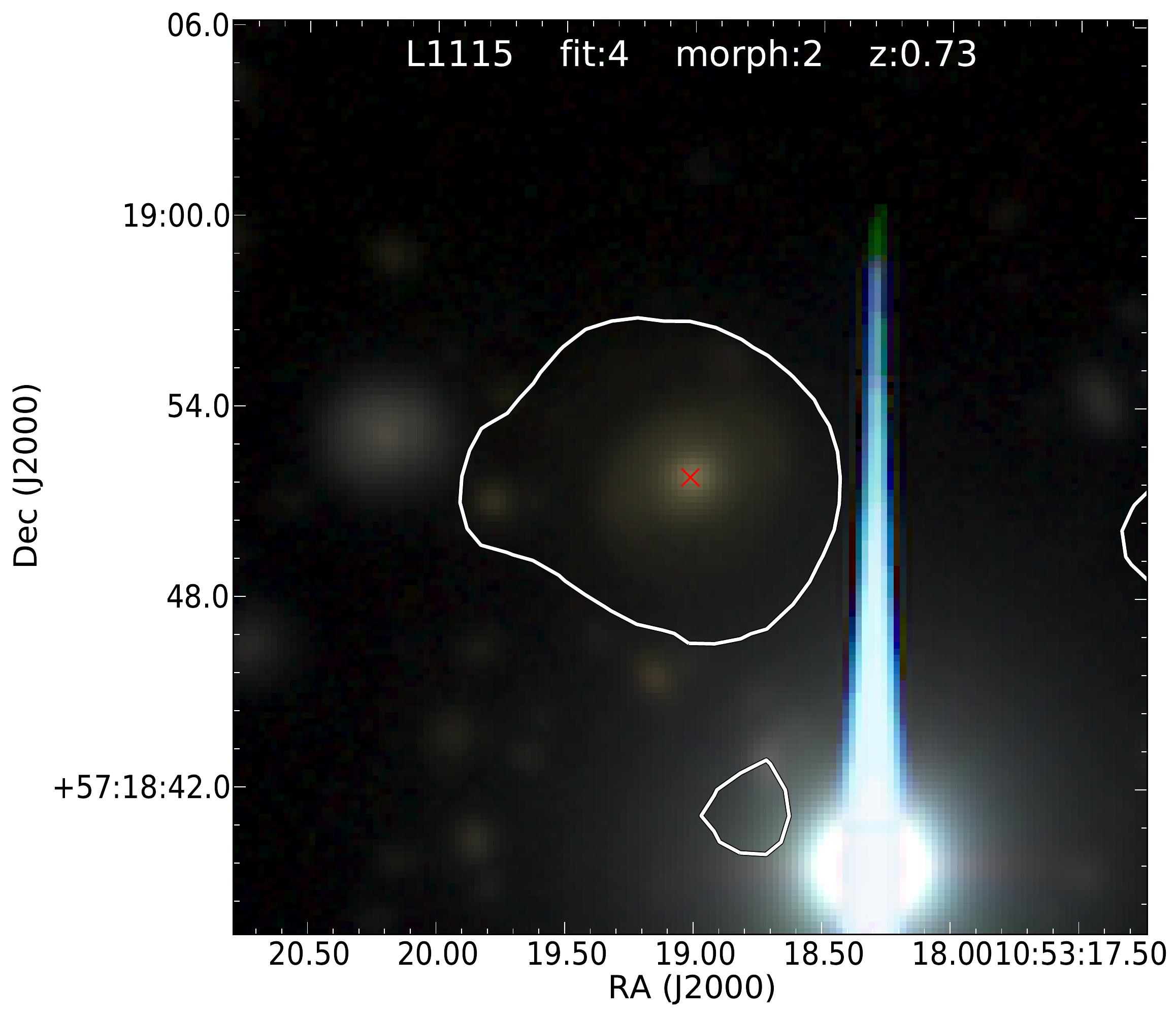}\\
\includegraphics[height=4.5cm]{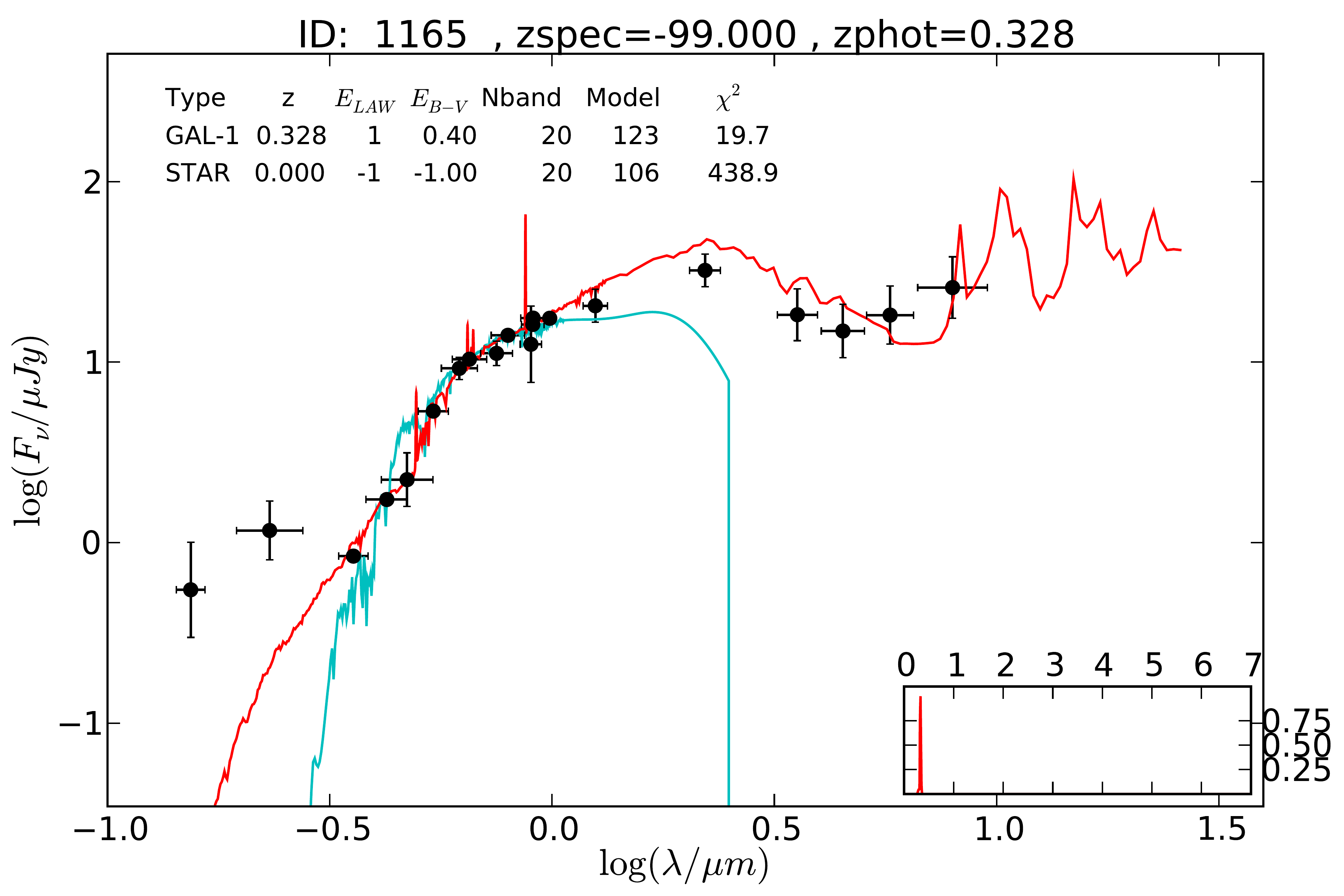}
\includegraphics[height=4.5cm]{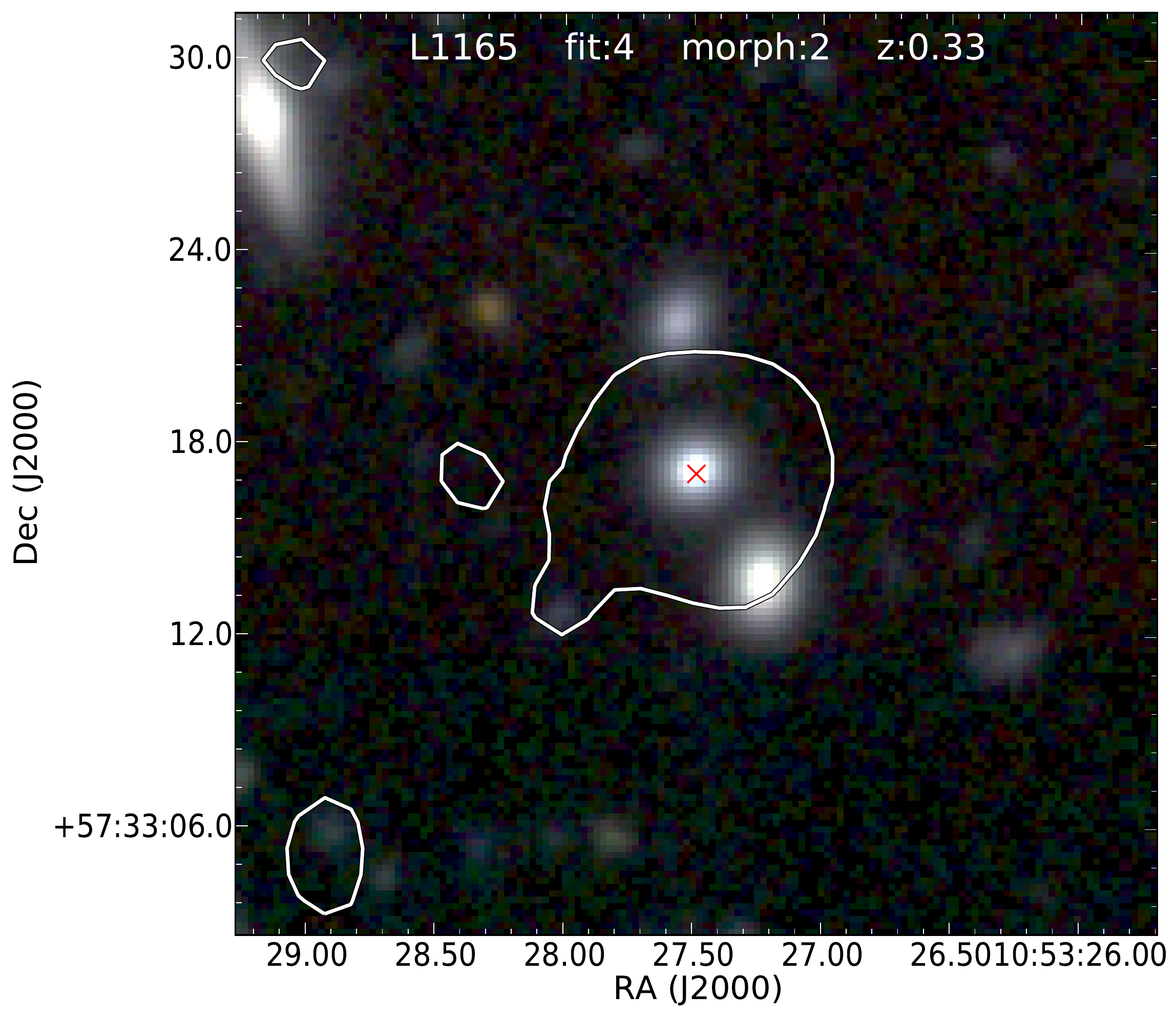}\\
\caption{(Continued)}
\end{figure*}

\begin{figure*}
\ContinuedFloat
\center
\includegraphics[height=4.5cm]{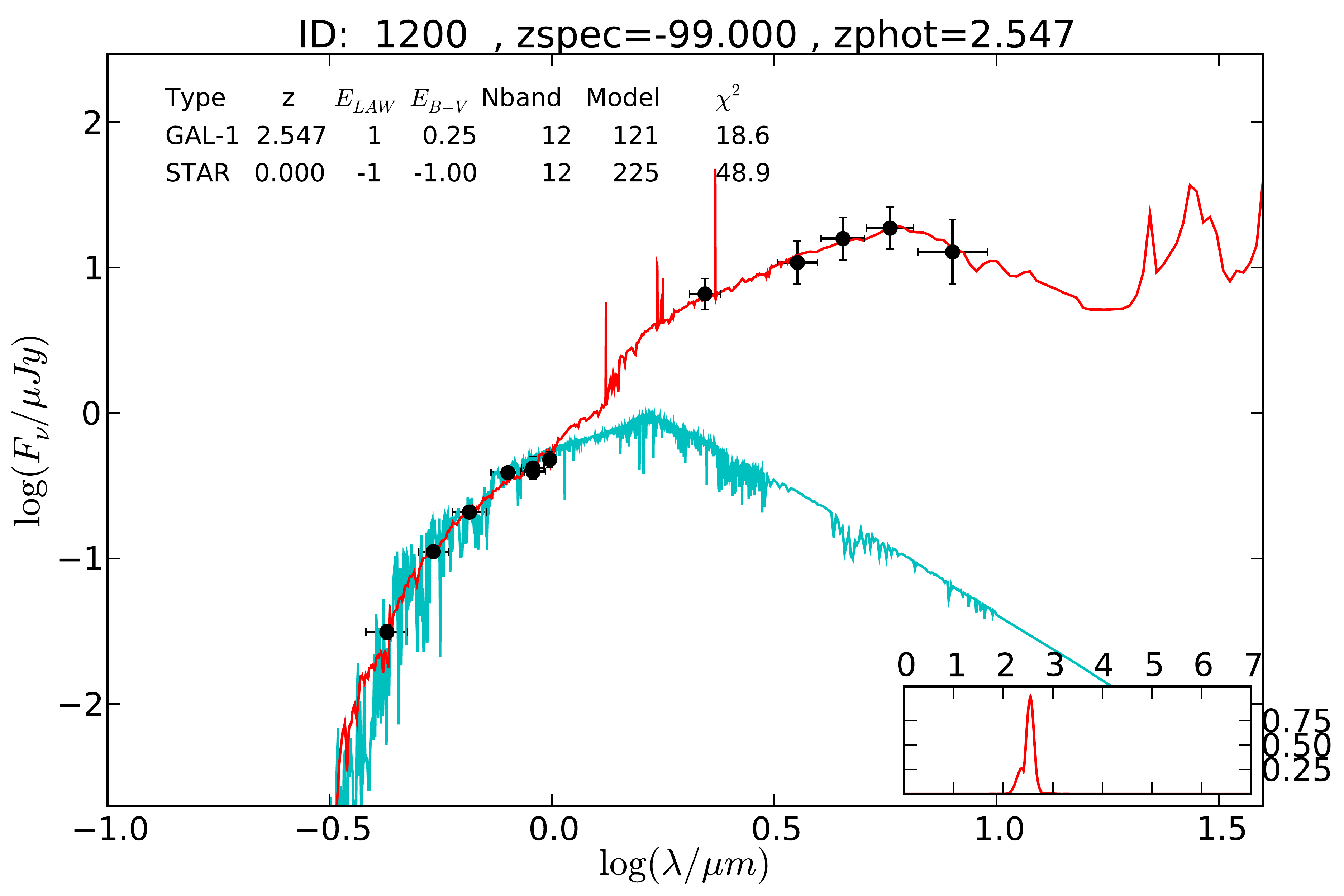}
\includegraphics[height=4.5cm]{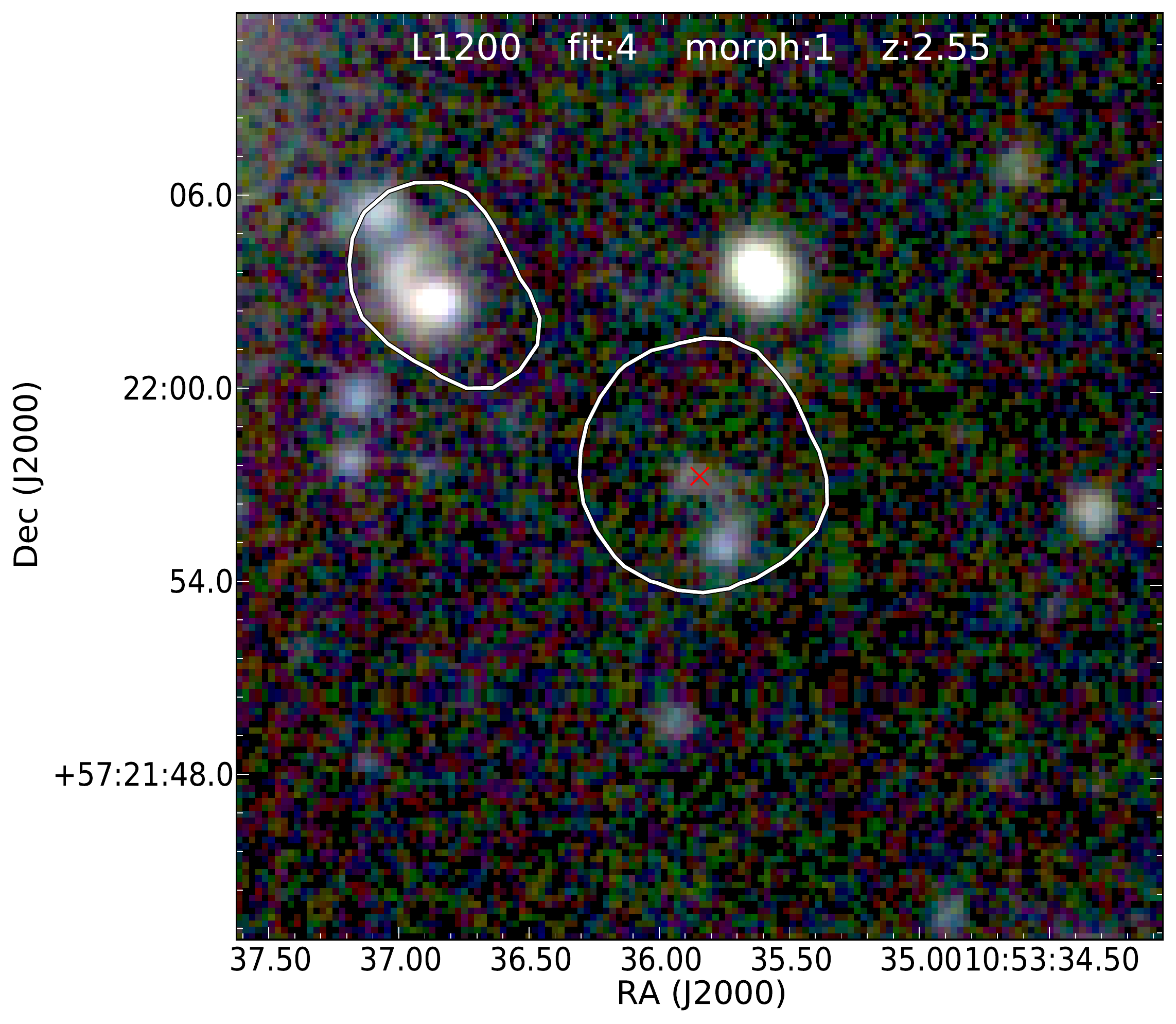}\\
\includegraphics[height=4.5cm]{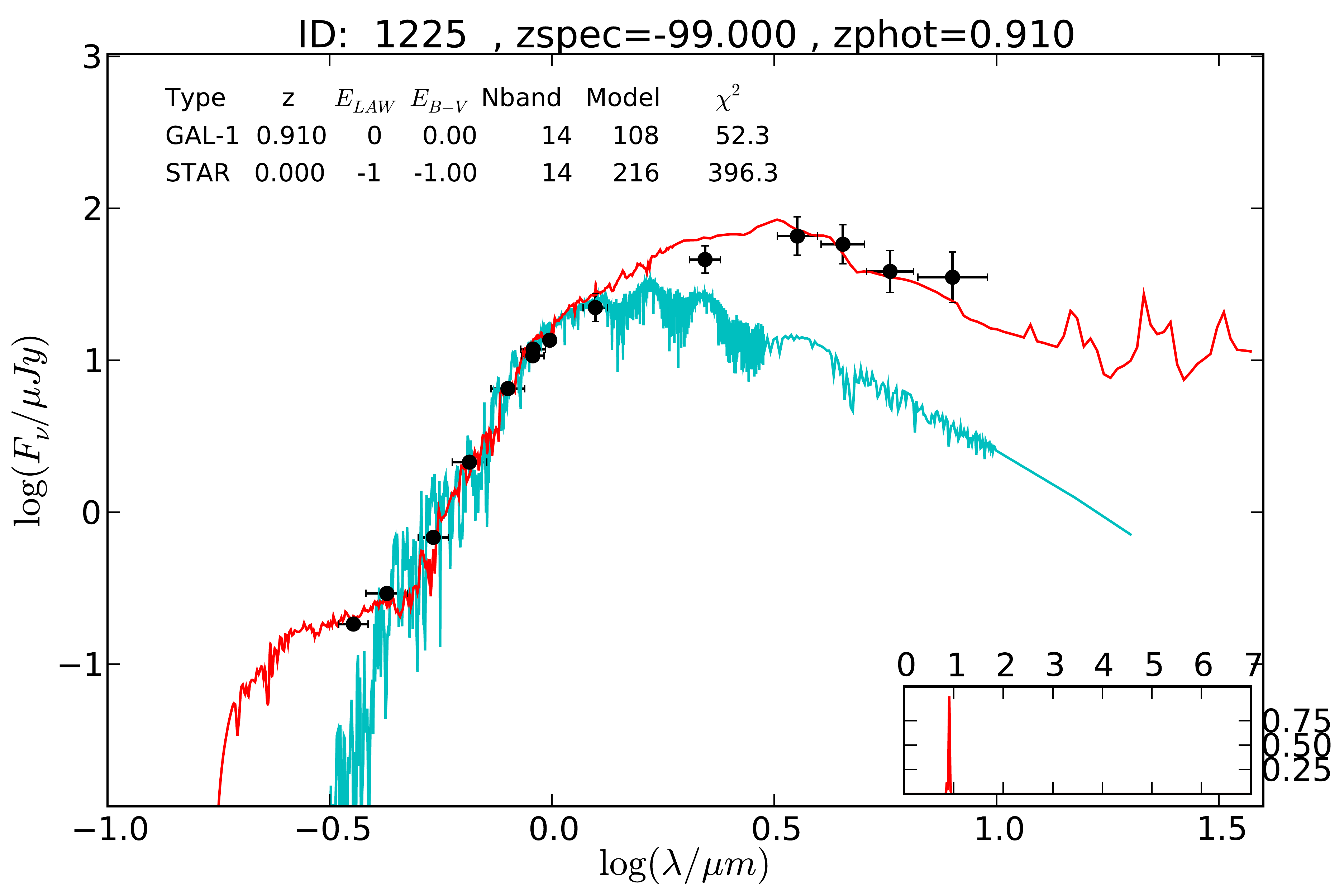}
\includegraphics[height=4.5cm]{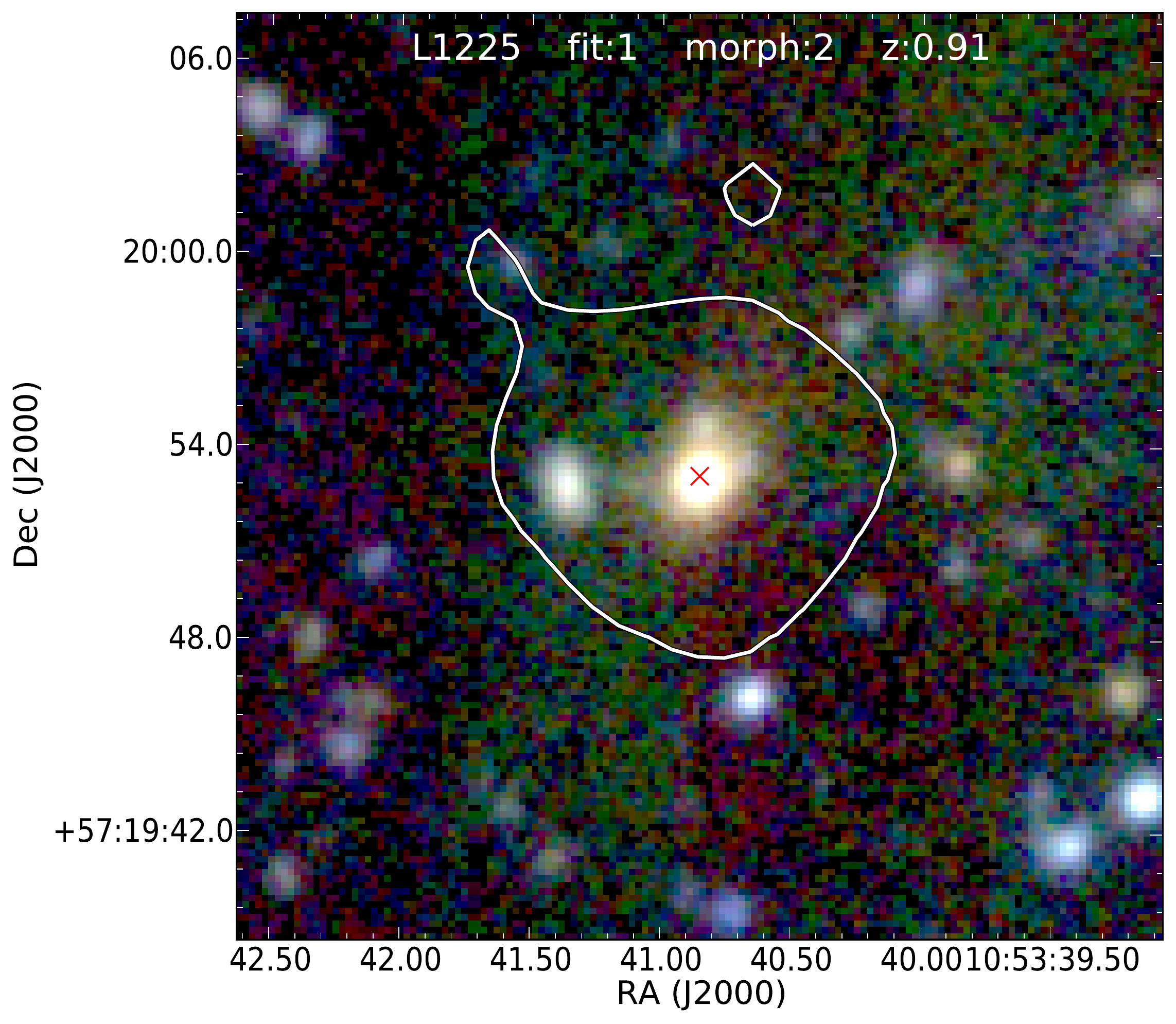}\\
\includegraphics[height=4.5cm]{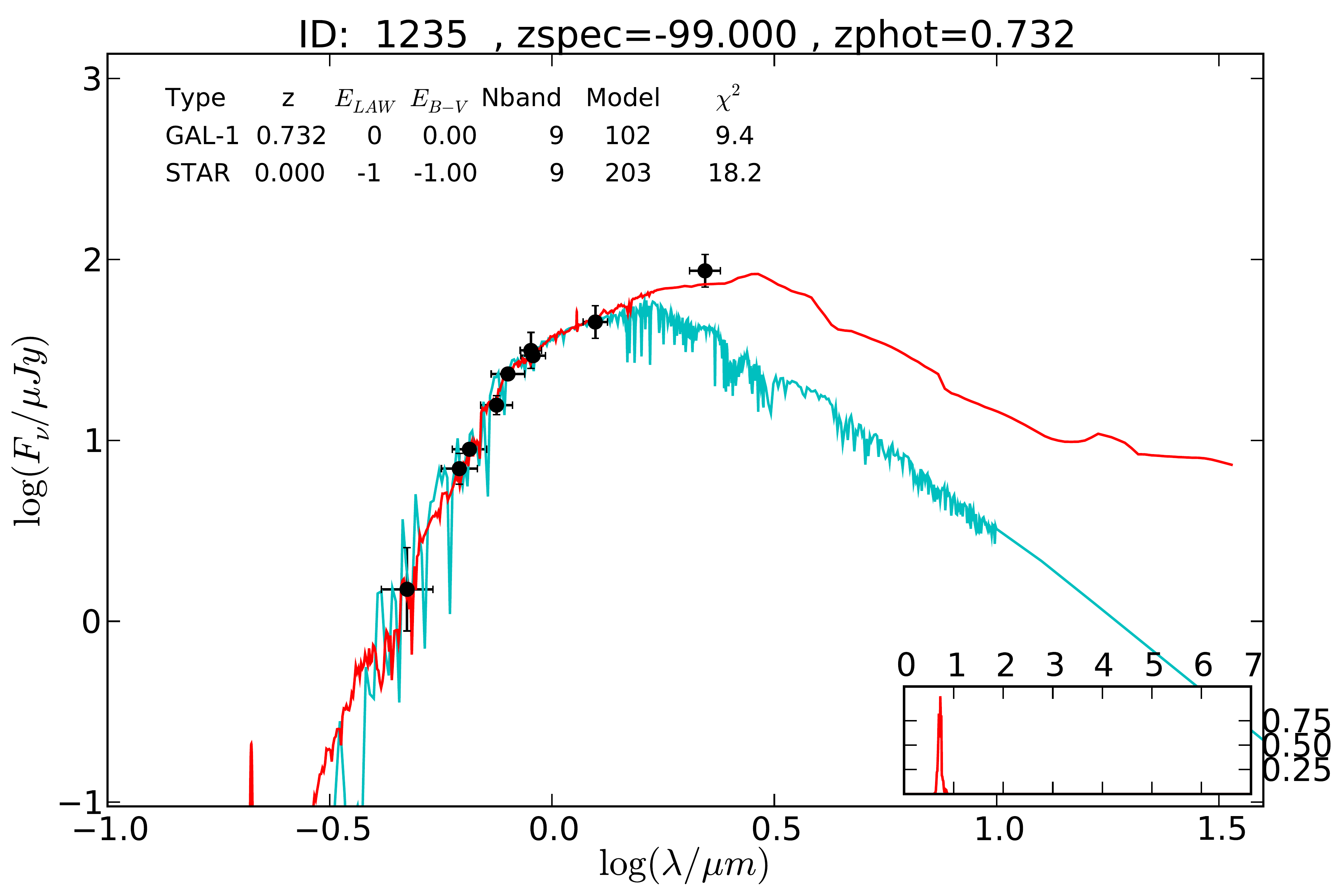}
\includegraphics[height=4.5cm]{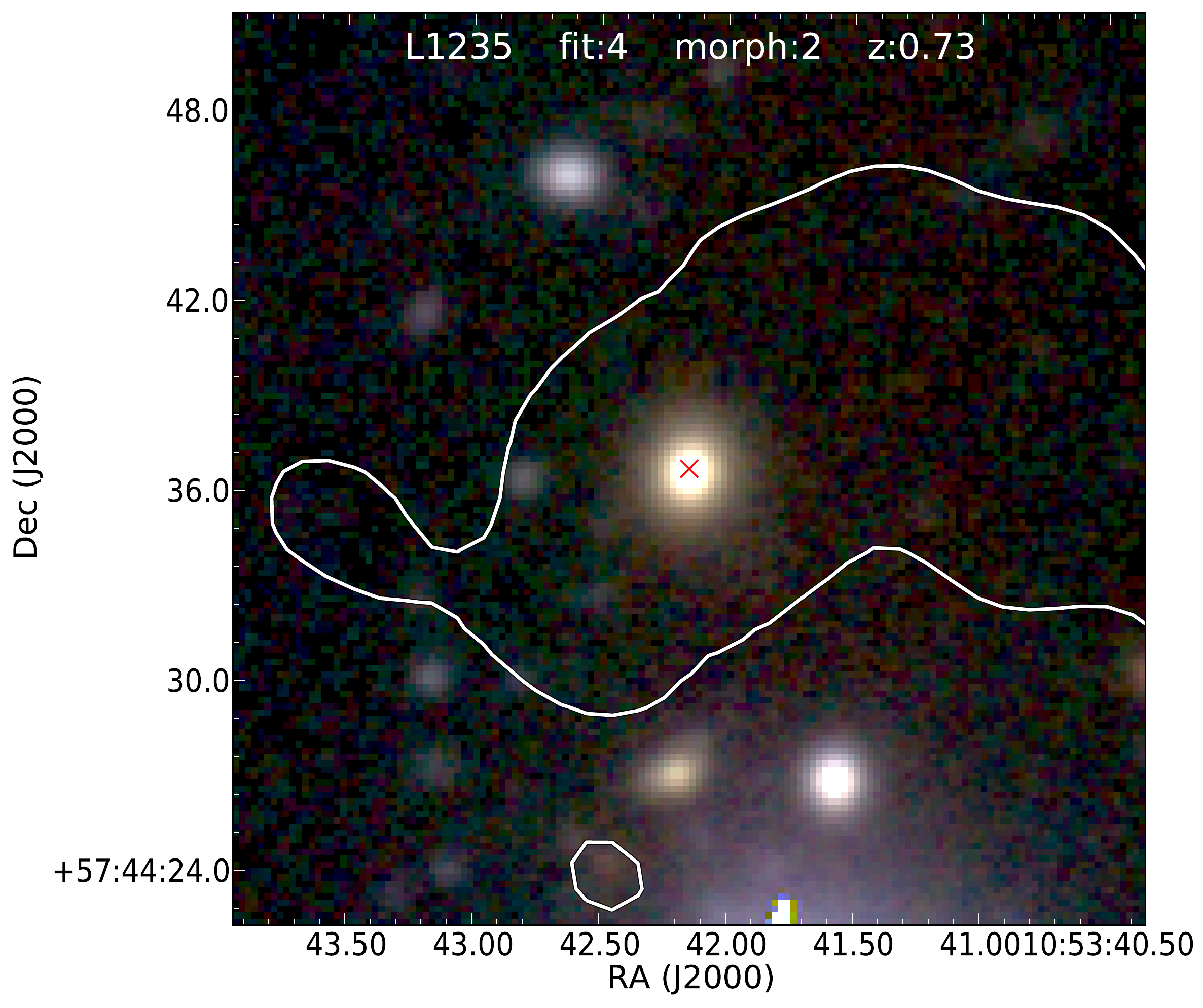}\\
\includegraphics[height=4.5cm]{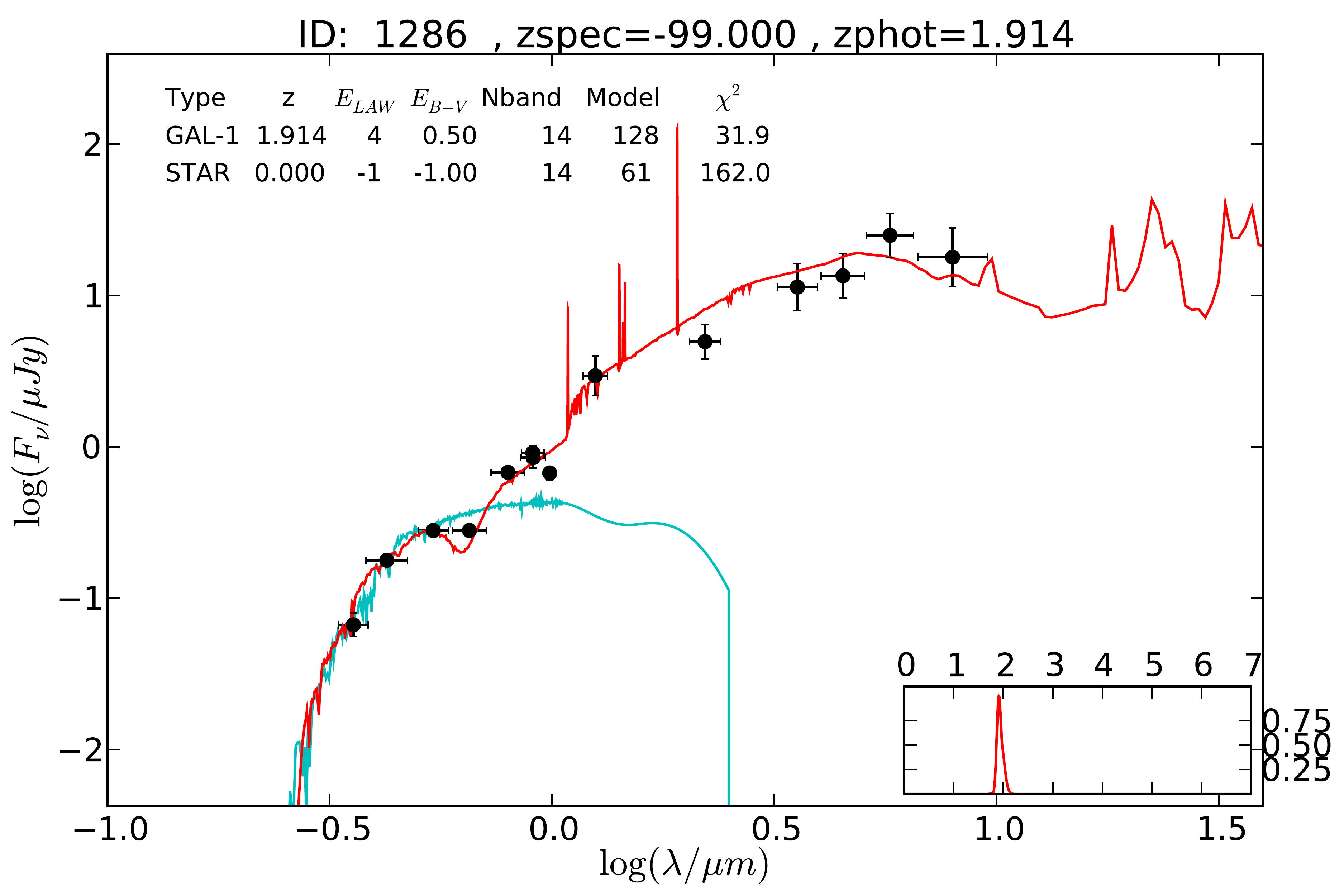}
\includegraphics[height=4.5cm]{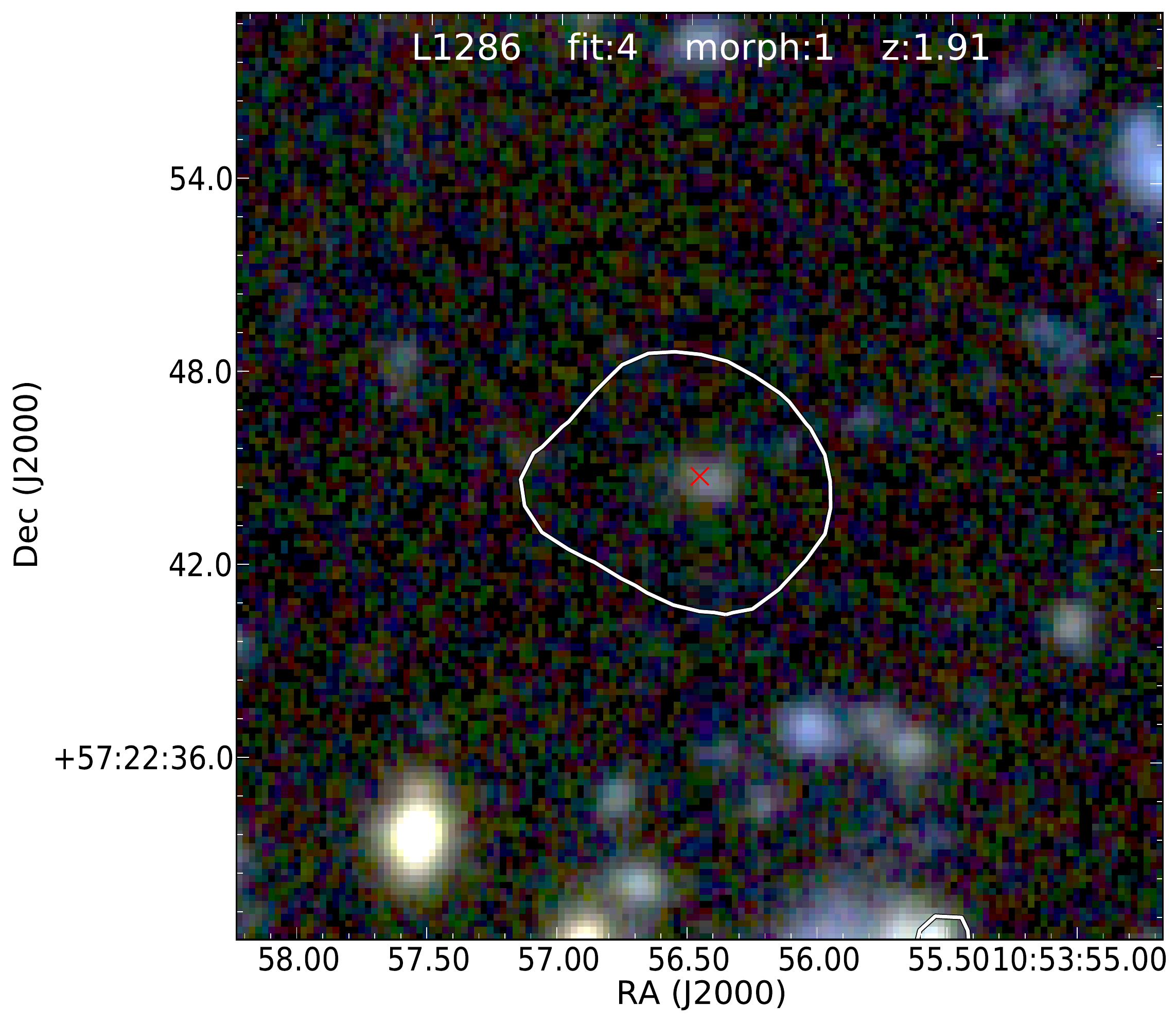}\\
\includegraphics[height=4.5cm]{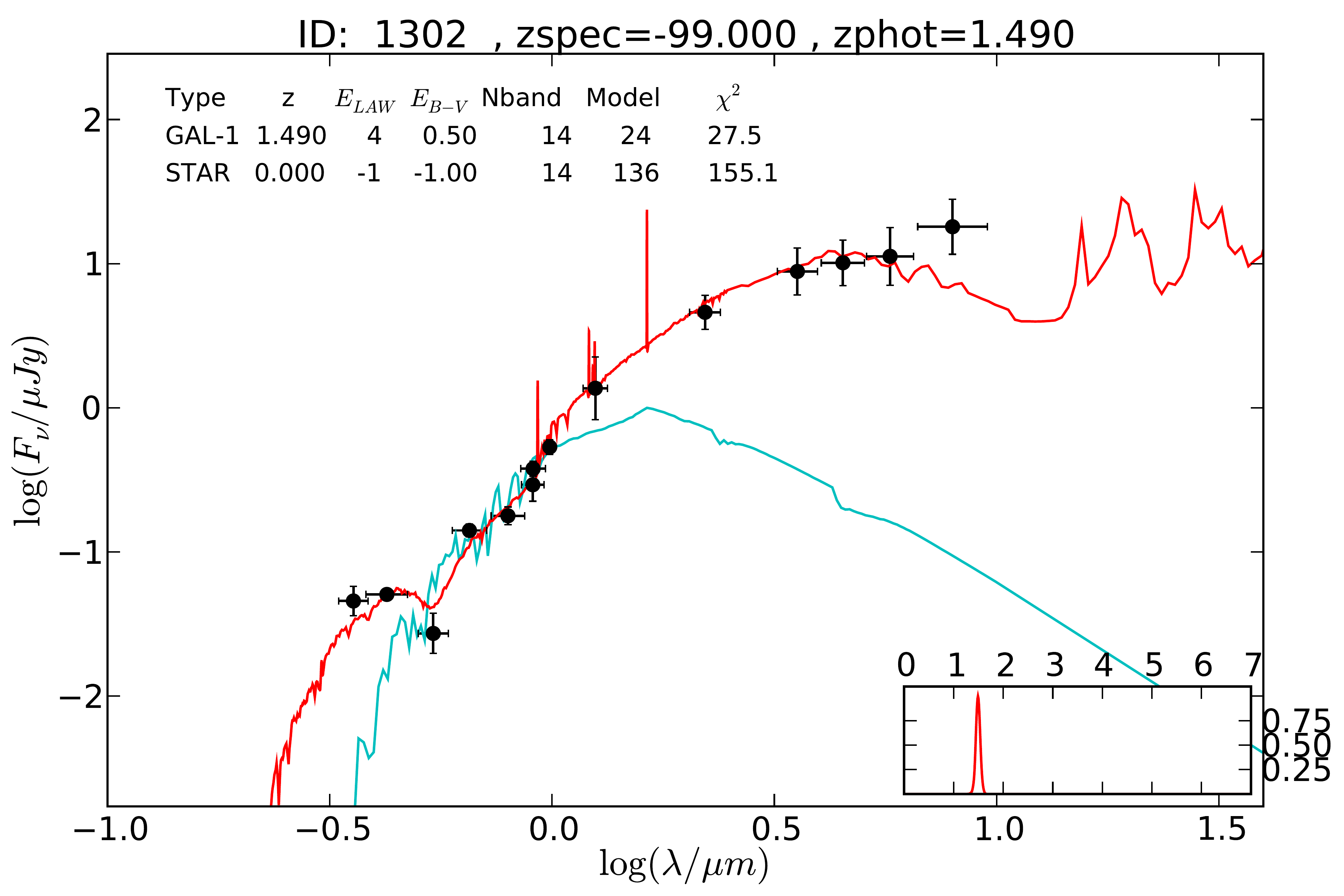}
\includegraphics[height=4.5cm]{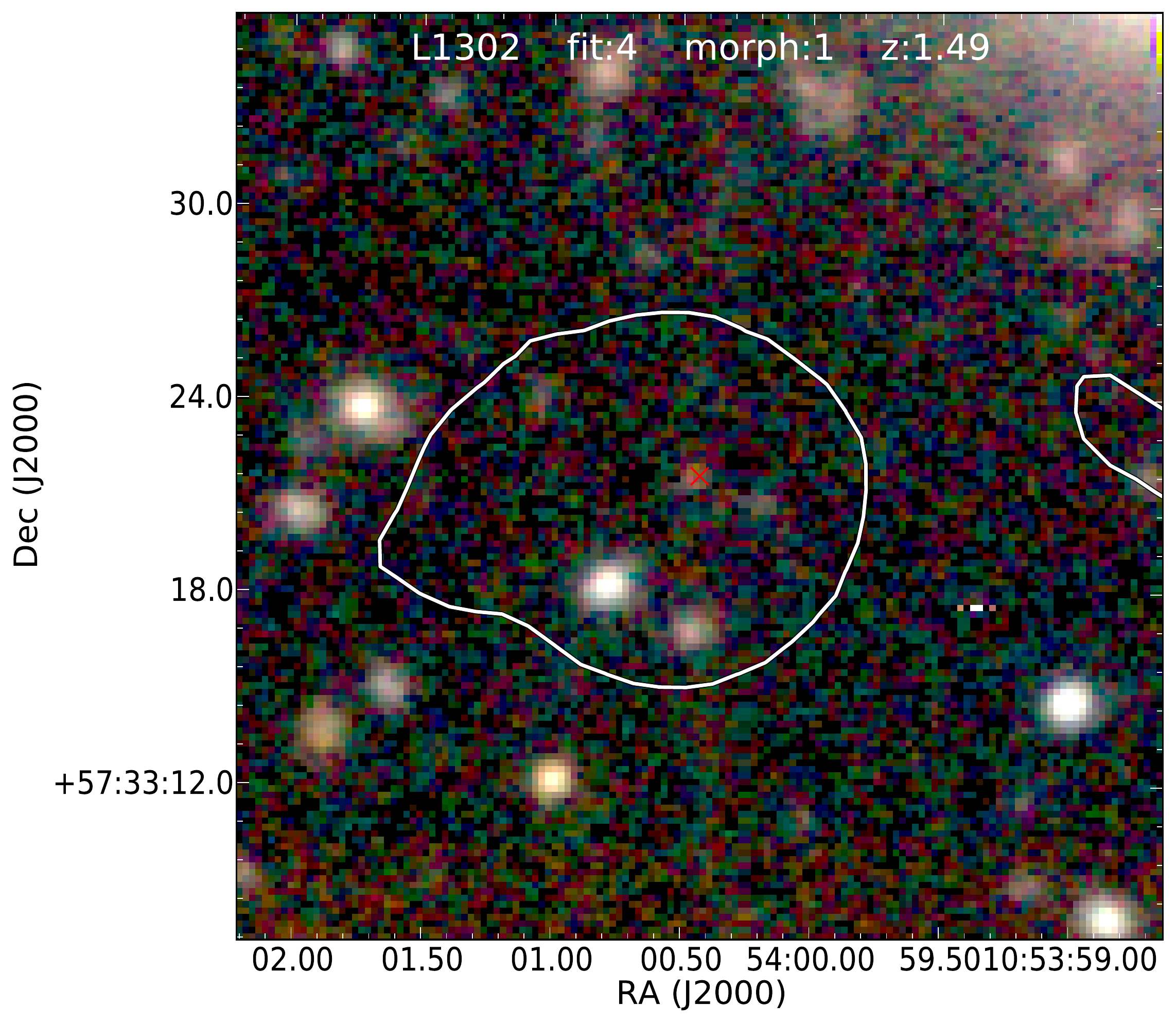}\\
\caption{(Continued)}
\end{figure*}

\begin{figure*}
\ContinuedFloat
\center
\includegraphics[height=4.5cm]{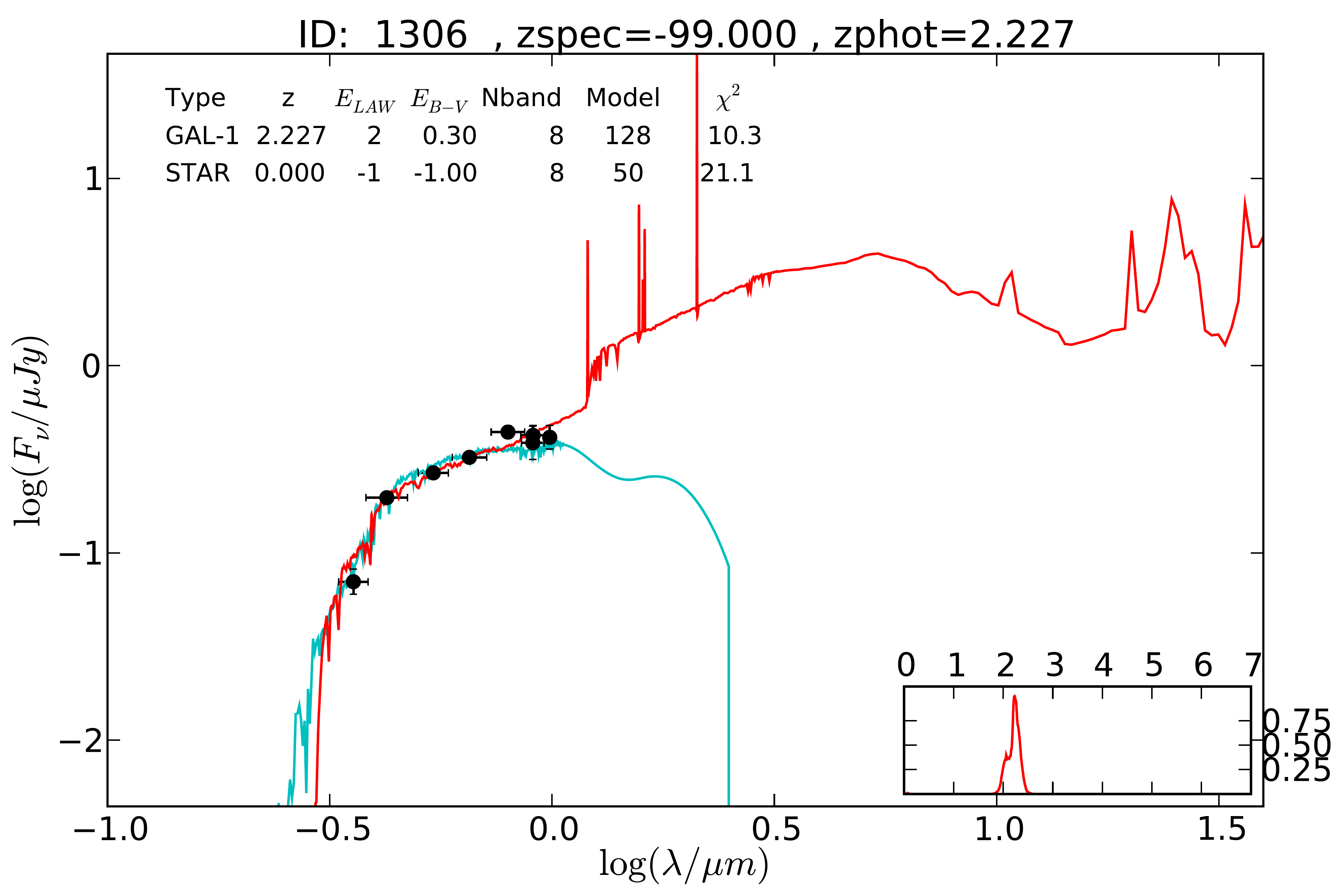}
\includegraphics[height=4.5cm]{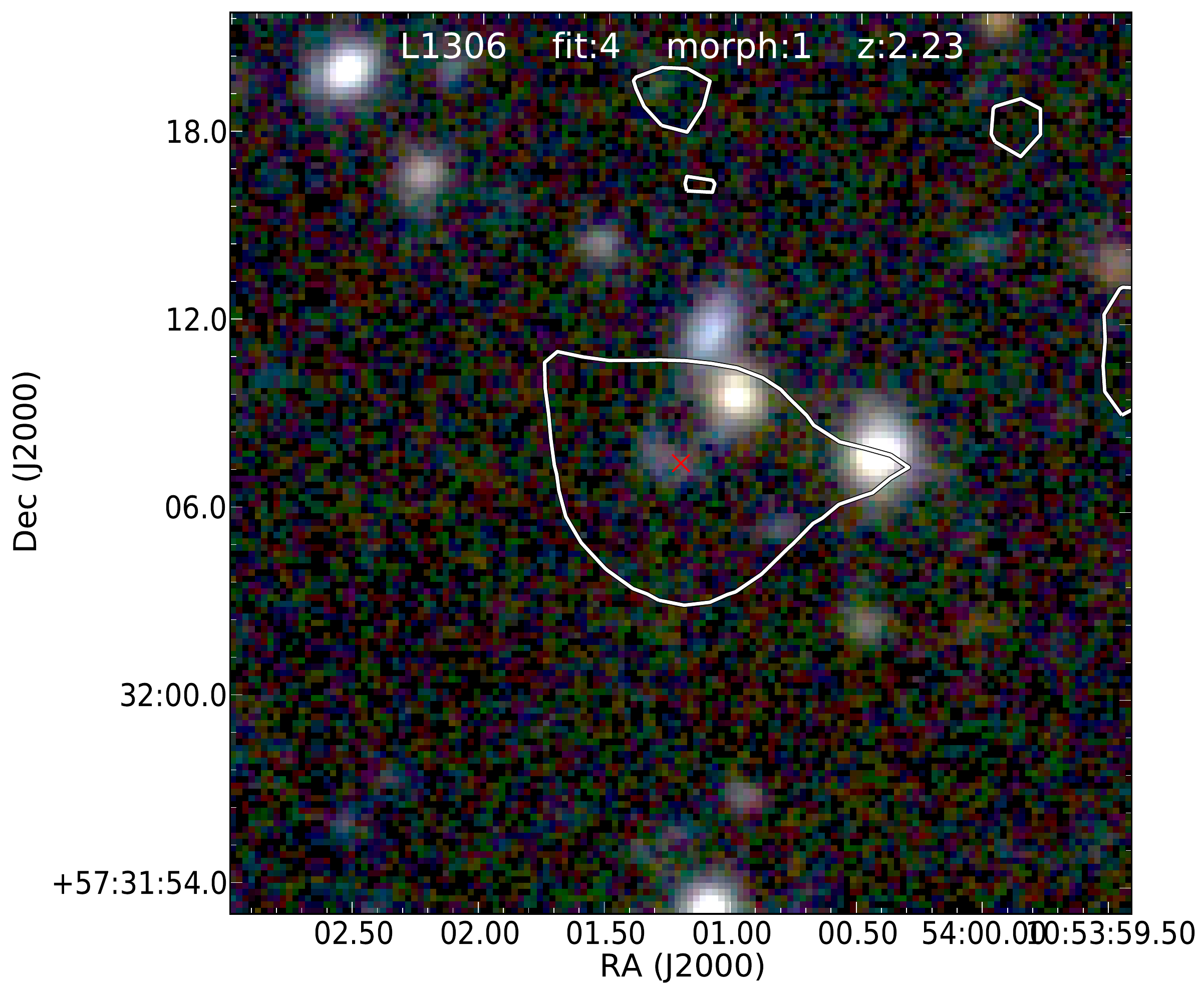}\\
\includegraphics[height=4.5cm]{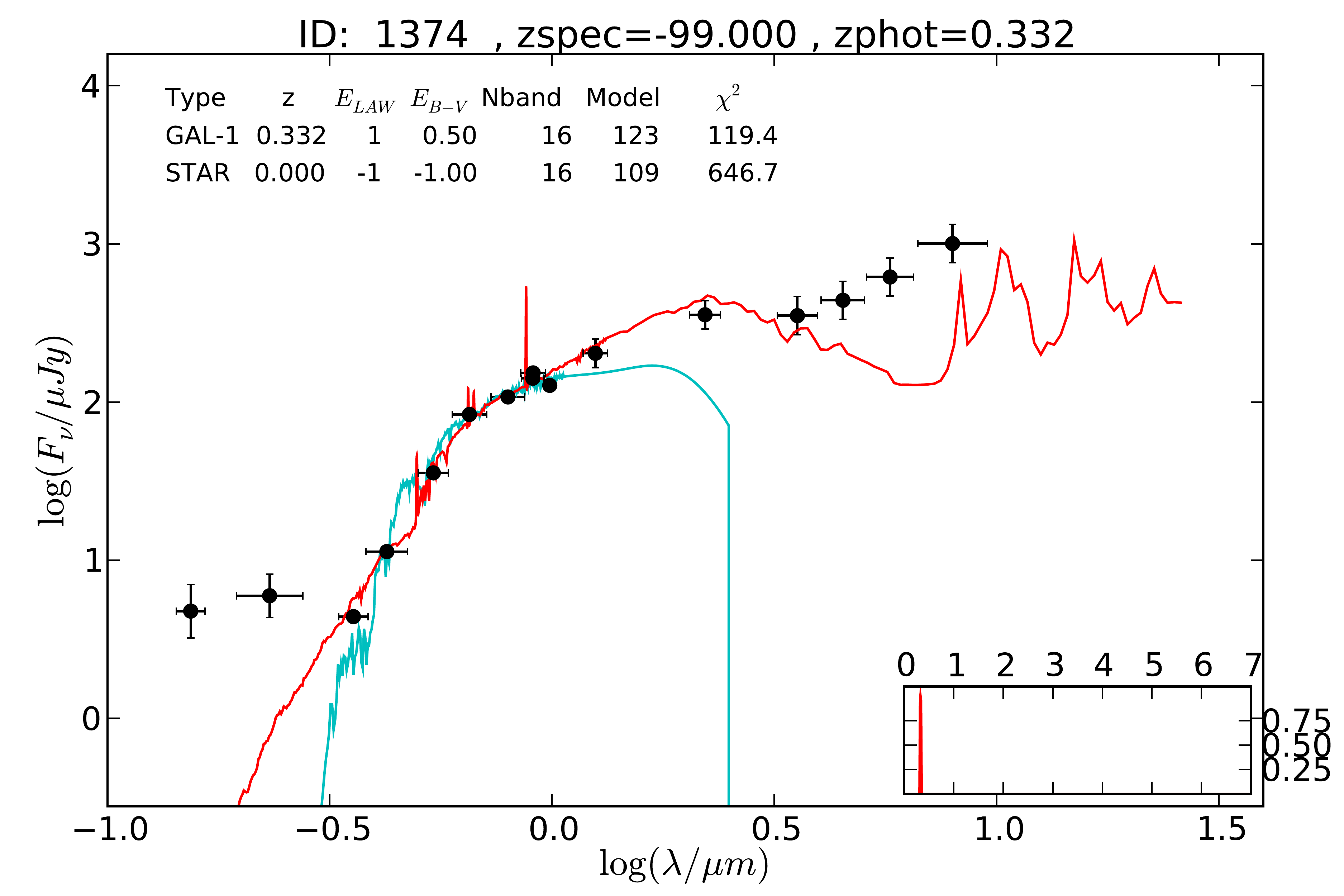}
\includegraphics[height=4.5cm]{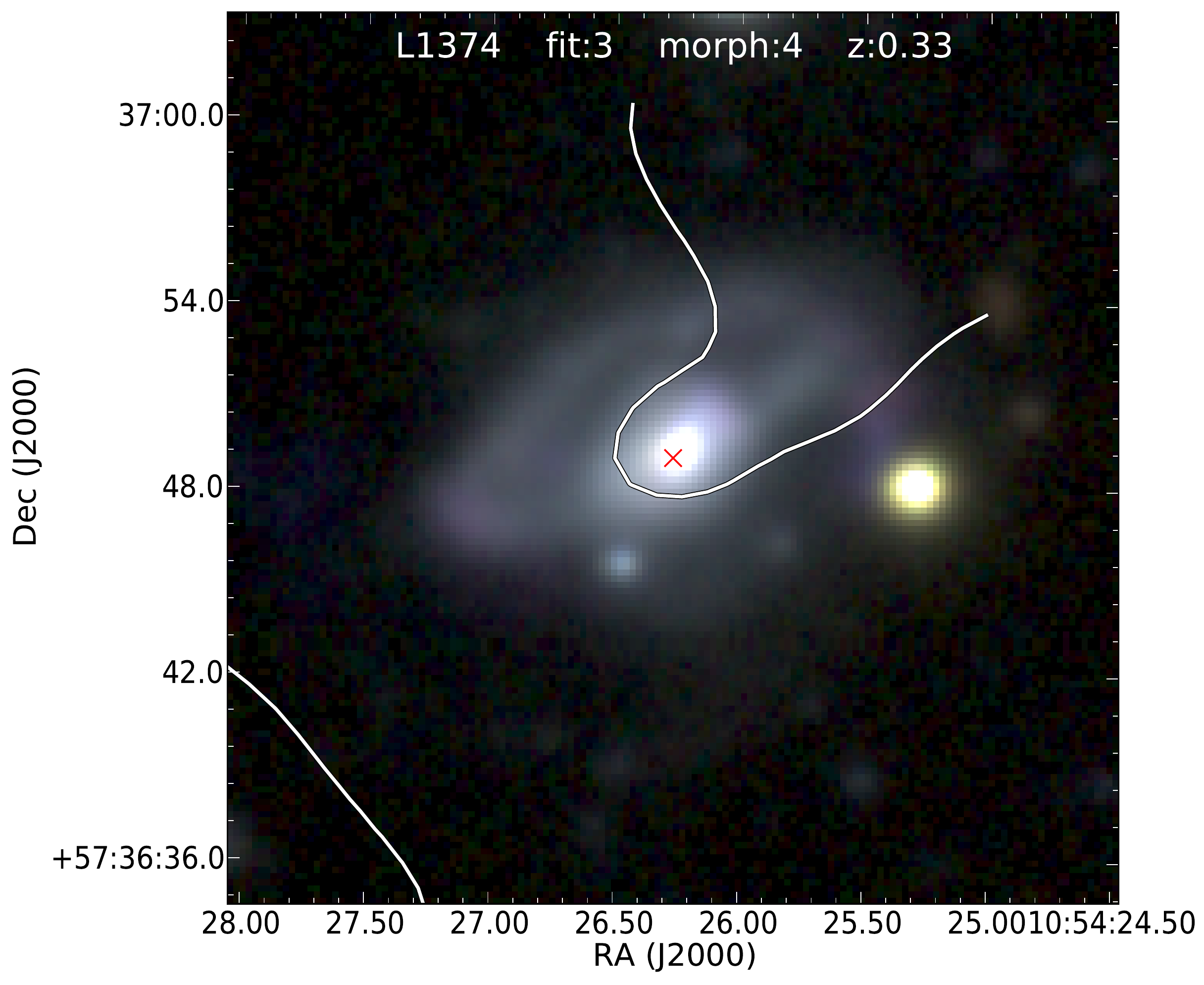}\\
\caption{(Continued)}
\end{figure*}

\begin{figure*}
\center
\includegraphics[height=4.5cm]{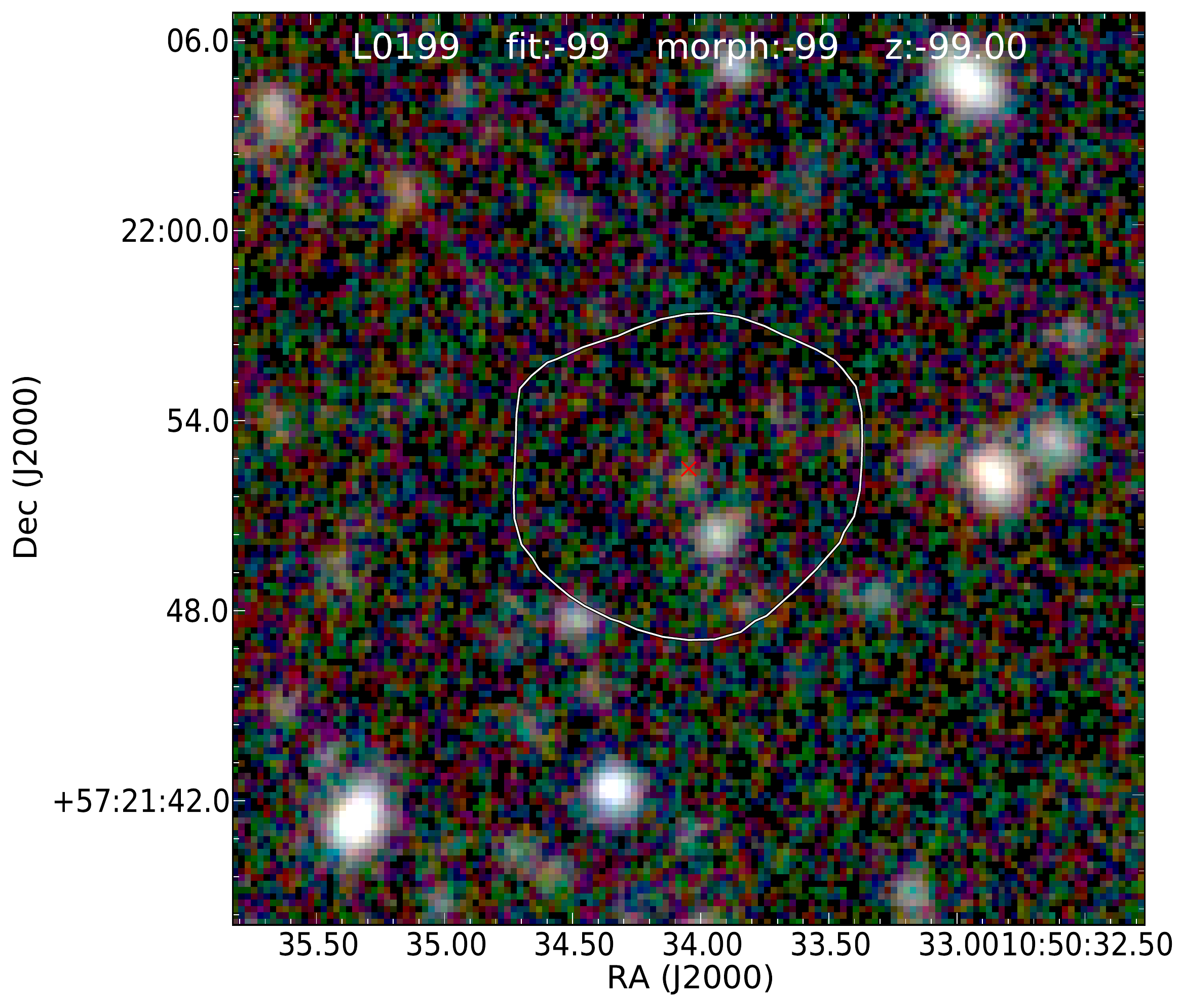}
\includegraphics[height=4.5cm]{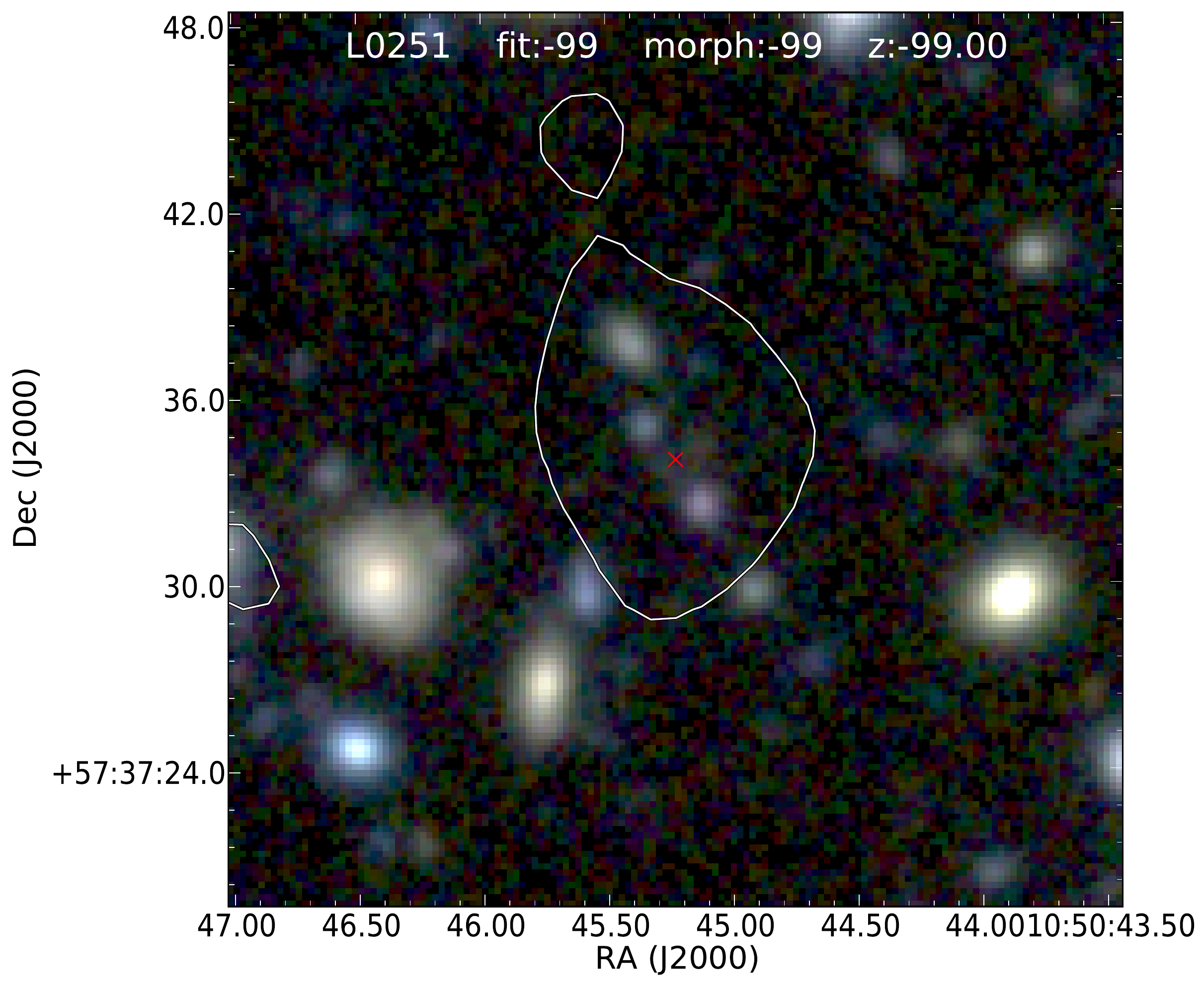}
\includegraphics[height=4.5cm]{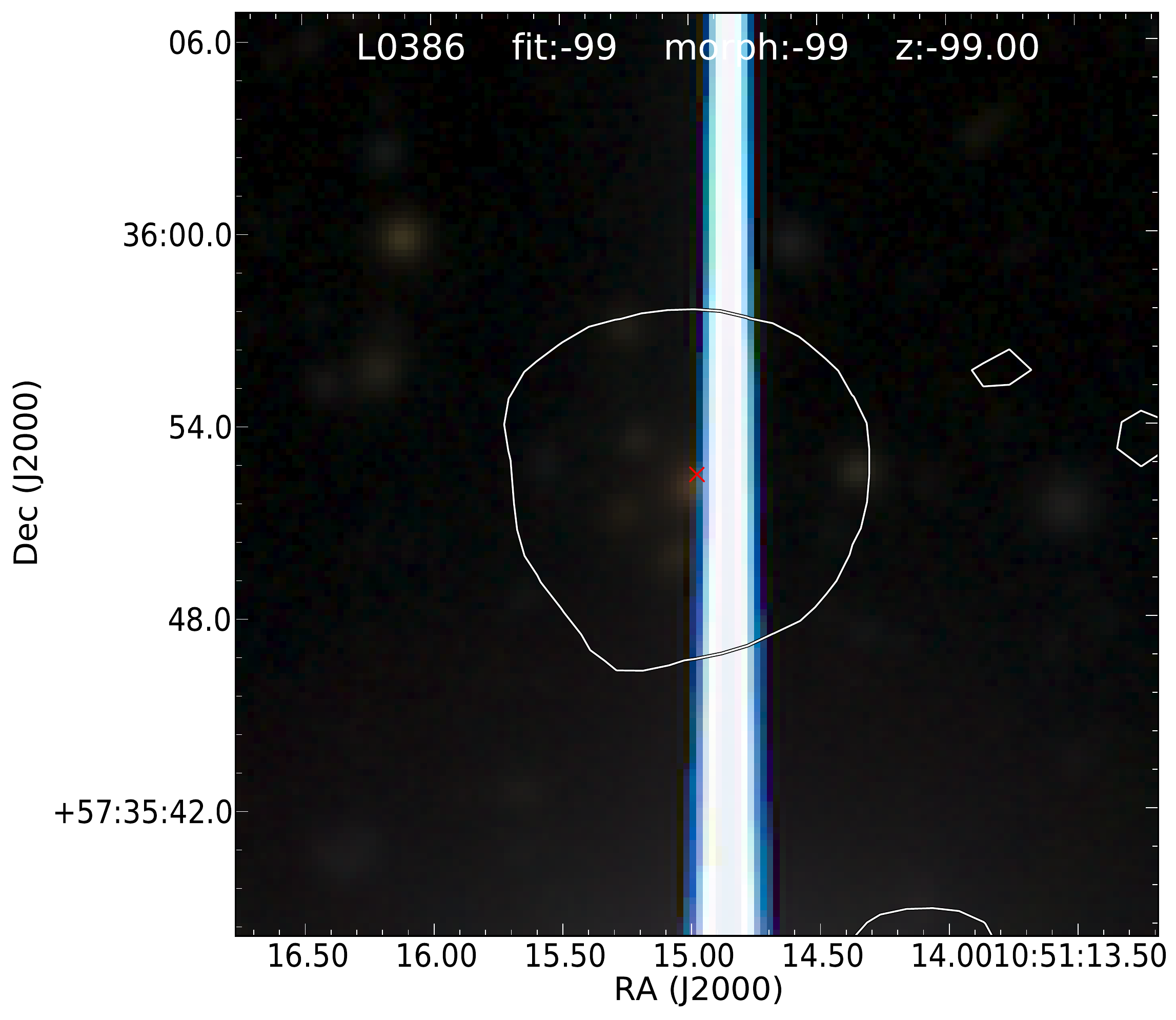}\\
\includegraphics[height=4.5cm]{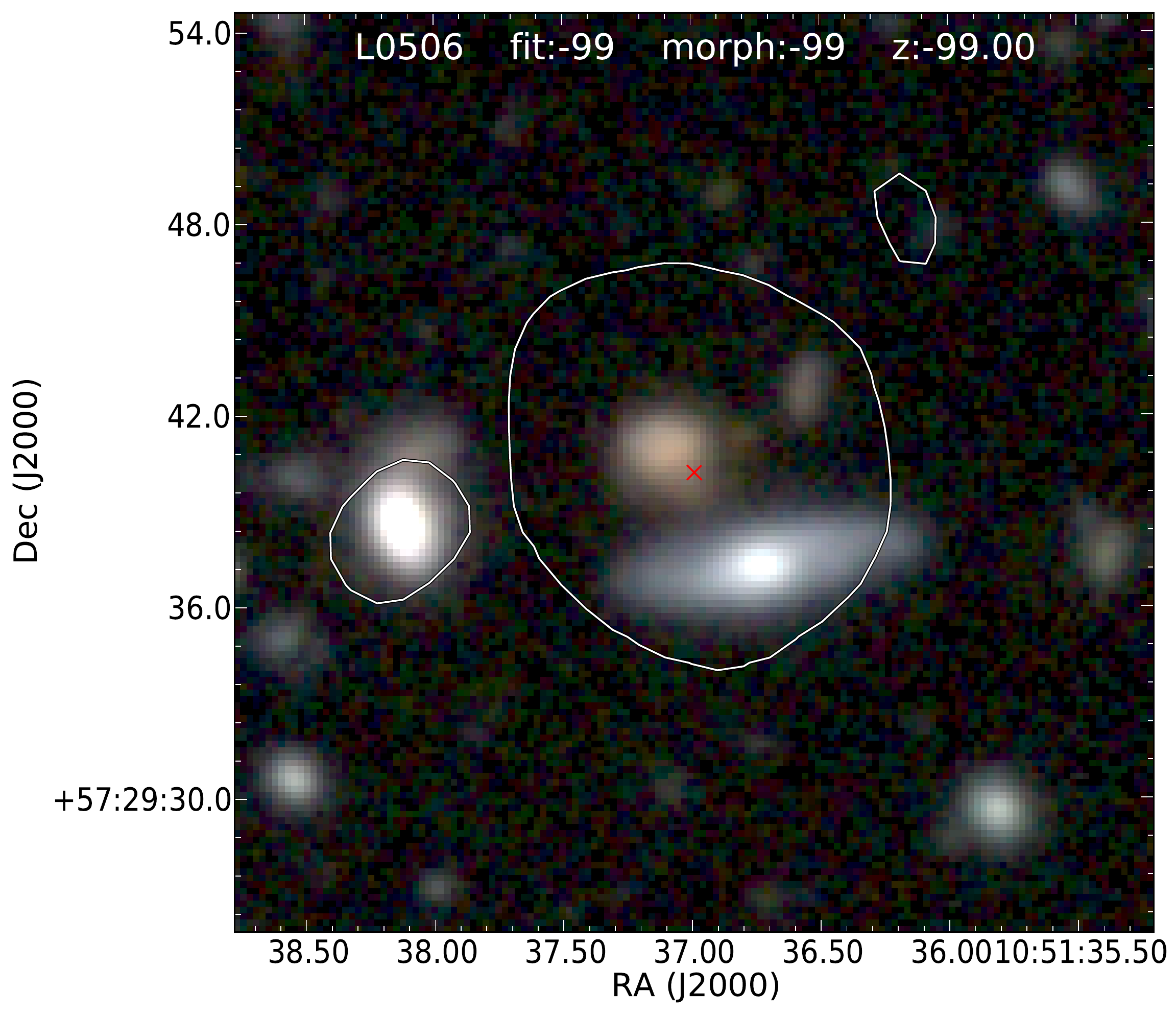}
\includegraphics[height=4.5cm]{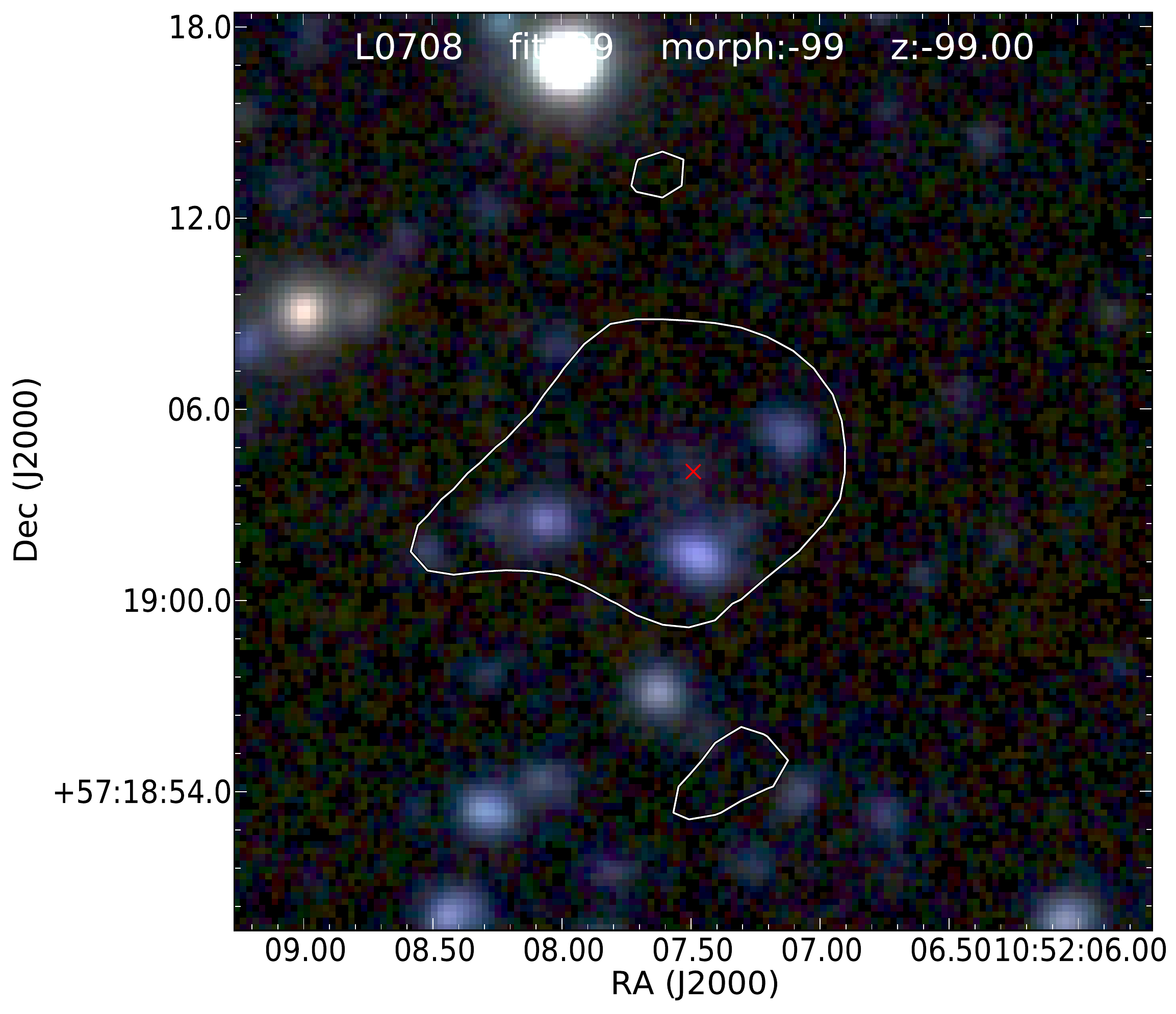}
\includegraphics[height=4.5cm]{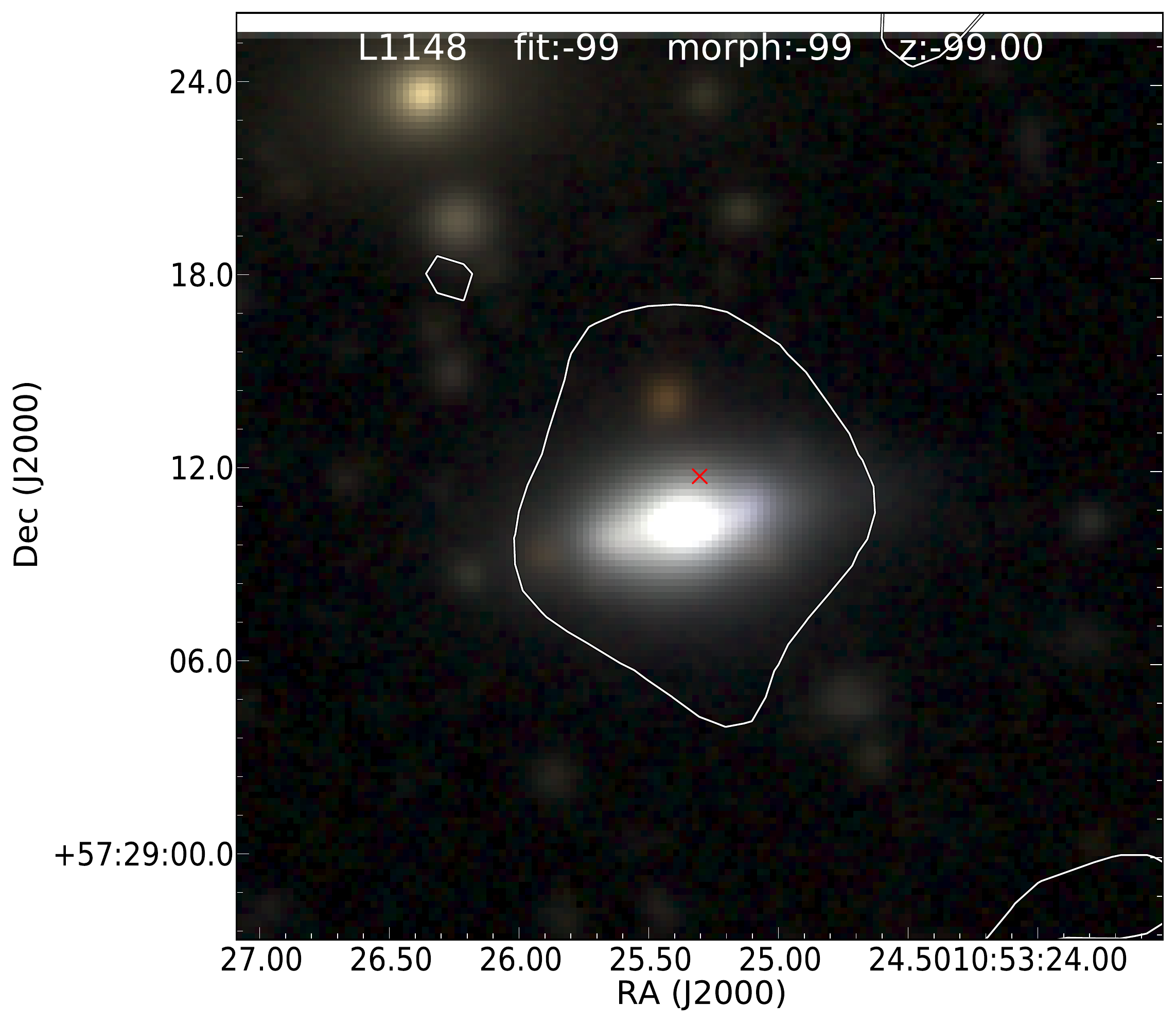}\\
\caption{RGB images of sources with uncatalogued, blended, or
  otherwise unavailable counterparts, made using the same settings as
  used for Fig.~\ref{fig:spectra+rgb}. In L0199, L0251, L0386 and
  L0708 very faint, but uncatalogued counterparts can be seen in the
  images; in L0506 and L1148 the VLBA detection is offset from the
  foreground object by so much that an association was deemed
  unlikely.}
\label{fig:rgb_borked}
\end{figure*}

\begin{figure*}
\center
\includegraphics[height=4.5cm]{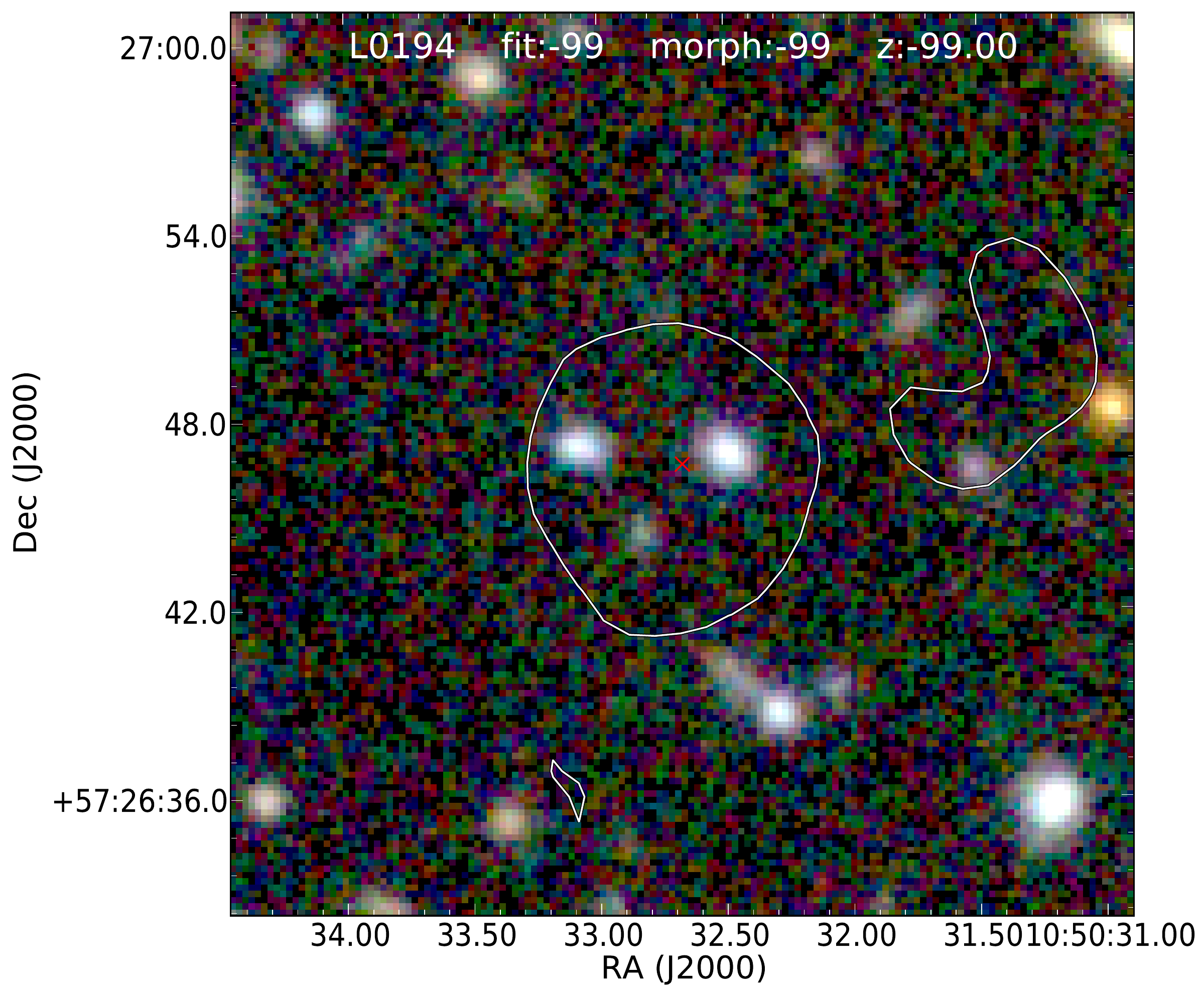}
\includegraphics[height=4.5cm]{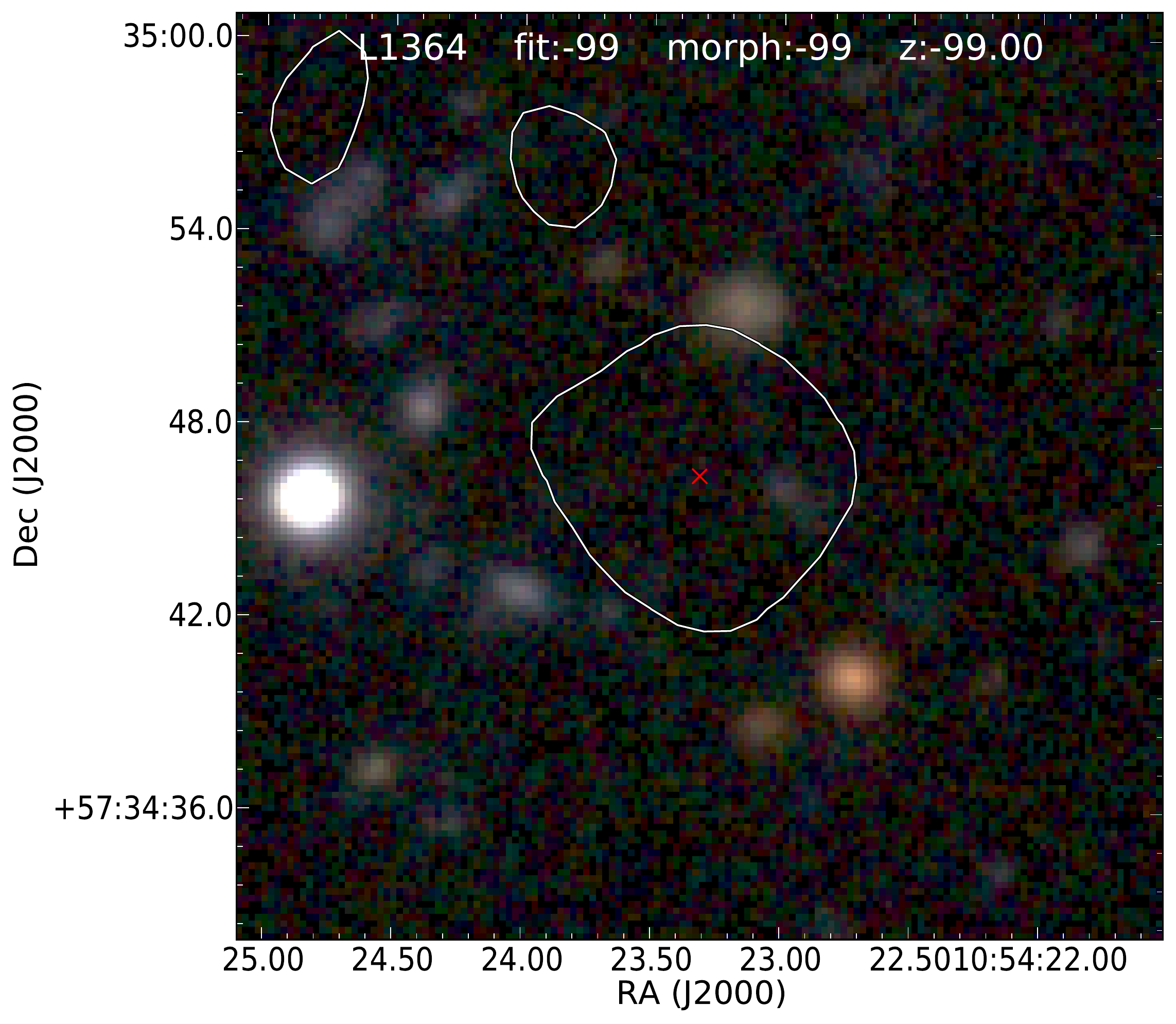}\\
\caption{RGB images of sources with no visible counterparts, made
  using the same settings as used for Fig.~\ref{fig:spectra+rgb}.}
\label{fig:rgb_invisible}
\end{figure*}

\end{document}